\pdfoutput=1
\newcommand*{\ATLASLATEXPATH}{latex/}
\documentclass[cernpreprint,texlive=2011,txfonts,UKenglish]{\ATLASLATEXPATH atlasdoc}
\makeatletter
\DeclareOldFontCommand{\rm}{\normalfont\rmfamily}{\mathrm}
\DeclareOldFontCommand{\sf}{\normalfont\sffamily}{\mathsf}
\DeclareOldFontCommand{\tt}{\normalfont\ttfamily}{\mathtt}
\DeclareOldFontCommand{\bf}{\normalfont\bfseries}{\mathbf}
\DeclareOldFontCommand{\it}{\normalfont\itshape}{\mathit}
\DeclareOldFontCommand{\sl}{\normalfont\slshape}{\@nomath\sl}
\DeclareOldFontCommand{\sc}{\normalfont\scshape}{\@nomath\sc}
\makeatother

\usepackage{\ATLASLATEXPATH atlaspackage}
\usepackage{\ATLASLATEXPATH atlasbiblatex}

\usepackage[process]{\ATLASLATEXPATH atlasphysics}

\graphicspath{{logos/}{figures/}}

\AtlasTitle{Measurement of the $W$-boson mass in pp collisions at $\sqrt{s}=\SI{7}{\tera\electronvolt}$ with the ATLAS detector}

\author{The ATLAS Collaboration}

\PreprintIdNumber{CERN-EP-2016-305}

\AtlasDate{\today}

\AtlasJournalRef{Eur. Phys. J. C 78 (2018) 110}
\AtlasDOI{10.1140/epjc/s10052-017-5475-4}

\AtlasAbstract{%
 A measurement of the mass of the $W$ boson is presented based on
proton--proton collision data recorded in 2011 at a centre-of-mass
energy of 7~TeV with the ATLAS detector at the LHC, and corresponding
to 4.6~fb$^{-1}$ of integrated luminosity.
The selected data sample consists of $7.8\times 10^6$ candidates in
the $W\rightarrow \mu \nu$ channel and $5.9\times 10^6$ candidates in
the $W\rightarrow e \nu$ channel. The $W$-boson mass is obtained from template fits to the reconstructed distributions of
the charged lepton transverse momentum and of the $W$ boson transverse mass in the electron and muon decay channels, yielding
\begin{eqnarray}
\nonumber m_W &=& 80370 \pm 7~(\textrm{stat.}) \pm 11~(\textrm{exp. syst.}) \pm 14~(\textrm{mod. syst.})~\textrm{MeV} \\
\nonumber     &=& 80370 \pm 19~\textrm{MeV}, 
\end{eqnarray}
where the first uncertainty is statistical, the second corresponds to the experimental systematic
uncertainty, and the third to the physics-modelling systematic uncertainty.
A measurement of the mass difference between the $W^+$ and $W^-$ bosons yields $m_{W^+}-m_{W^-} = -29 \pm 28$~MeV.

}

\hypersetup{pdftitle={ATLAS},pdfauthor={The ATLAS Collaboration}} 

\usepackage{booktabs}
\usepackage{dcolumn}
\newcolumntype{d}[1]{D{.}{.}{#1}}
\usepackage{pbox}

\newcommand*{\PowhegPythia}{\textsc{Powheg+Pythia~8}\xspace}
\newcommand*{\PowhegHerwig}{\textsc{Powheg+Herwig~6}\xspace} 
\newcommand*{\PowhegMinlo}{\textsc{Powheg MiNLO+Pythia 8}\xspace}

\newcommand{\mt}{\ensuremath{m_{\mathrm{T}}}}
\newcommand{\avg}[1]{\ensuremath{\left< #1 \right>}} 
\newcommand*{\mptvec}{\ensuremath{\vec{p}_{\text{T}}^{\text{\,miss}}}\xspace}
\newcommand{\ut}{\ensuremath{u_{\mathrm{T}}}}
\newcommand{\utvec}{\ensuremath{\vec{u}_{\mathrm{T}}}}

\makeatletter
\newcommand*{\cropdelims}[6][1]{%
  \begingroup
    \setbox\z@=\hbox{%
      \thinmuskip=\@ne mu %
      \medmuskip=0mu %
      \thickmuskip=0mu %
      \setbox\tw@=\hbox{$#2#3#5#6\m@th$}%
      \kern-\wd\tw@
      $#2#3#5#6{}\m@th$%
    }%
    \ifdim\wd\z@>\z@
      \expandafter\mathinner
    \else
      \expandafter\@firstofone
    \fi
    {%
      \mathopen{}%
      \mathpalette{\cropdelims@aux{#1}{#2#3}{#5#6}}{#4}%
      \mathclose{}%
    }%
  \endgroup
}
\newcommand*{\cropdelims@aux}[5]{%
  \setbox\z@=\hbox{$\m@th#4\vcenter{}$}%
  \dimen@=\ht\z@
  \setbox\z@=\hbox{$#4#5\m@th$}%
  \dimen@=#1\dimexpr\dimen@-.5\dimexpr\ht\z@-\dp\z@\relax\relax
  \raisebox{-\dimen@}{%
    $#4#2%
    \raisebox{\dimen@}{%
      $#4\mathopen{}#5\mathclose{}\m@th$%
    }%
    #3\m@th$%
  }%
}   
\makeatother

\renewcommand{\TeV}{\ifmmode {\mathrm{\ Te\kern -0.1em V}}\else \textrm{Te\kern -0.1em V}\fi}
\renewcommand{\GeV}{\ifmmode {\mathrm{\ Ge\kern -0.1em V}}\else \textrm{Ge\kern -0.1em V}\fi}
\renewcommand{\MeV}{\ifmmode {\mathrm{\ Me\kern -0.1em V}}\else \textrm{Me\kern -0.1em V}\fi}
\renewcommand{\keV}{\ifmmode {\mathrm{\ ke\kern -0.1em V}}\else \textrm{ke\kern -0.1em V}\fi}
\renewcommand{\eV}{\ifmmode  {\mathrm{\ e\kern -0.1em V}}\else \textrm{e\kern -0.1em V}\fi}

\usepackage{color}
\usepackage{soul}

\begin{document}

\maketitle

\clearpage
\newpage

\section{Introduction}

The Standard Model (SM) of particle physics describes the electroweak
interactions as being mediated by the $W$ boson, the $Z$ boson, and the photon, in
a gauge theory based on the ${\mathrm{SU}}(2)_{\mathrm{L}} \times {\mathrm{U}}(1)_{\mathrm{Y}}$
symmetry~\cite{Glashow:1961tr,Salam:1964ry,Weinberg:1967tq}. The
theory incorporates the observed masses of the $W$ and $Z$
bosons through a symmetry-breaking mechanism. In the SM, this mechanism
relies on the interaction of the gauge bosons with a scalar doublet
field and implies the existence of an additional physical state known
as the Higgs boson~\cite{Englert:1964et,Higgs:1964pj,Higgs:1964ia,Guralnik:1964eu}.
The existence of the $W$ and $Z$ bosons was first established at the CERN
SPS in 1983~\cite{Arnison:1983rp,Arnison:1983mk,Banner:1983jy,Bagnaia:1983zx},
and the LHC collaborations ATLAS and CMS reported the discovery of the Higgs
boson in 2012~\cite{HIGG-2012-27,CMS-HIG-12-028}.

At lowest order in the electroweak theory, the $W$-boson mass, $m_W$, can be expressed solely as a
function of the $Z$-boson mass, $m_Z$, the fine-structure constant, $\alpha$, and the Fermi constant, $G_{\mu}$. Higher-order
corrections introduce an additional dependence of the $W$-boson mass on
the gauge couplings and the masses of the heavy particles of the
SM. The mass of the $W$ boson can be expressed in terms of the other
SM parameters as follows:

\begin{equation}
\nonumber m_W^2 \left(1 - \frac{m^2_W}{m^2_Z}\right) = \frac{\pi\alpha}{\sqrt{2}G_{\mu}} (1+\Delta r),
\end{equation}

\noindent where $\Delta r$ incorporates the effect of higher-order
corrections~\cite{Awramik:2003rn,Sirlin:1980nh}. In the SM, $\Delta r$ is in particular sensitive to the top-quark and Higgs-boson masses; in extended theories, $\Delta r$ receives contributions from
additional particles and interactions. These effects can be probed by comparing the measured and predicted values
of $m_W$. In the context of global fits to the SM parameters, constraints on
physics beyond the SM are currently limited by the $W$-boson mass measurement precision~\cite{Baak:2014ora}. Improving the precision of the measurement of $m_W$ is therefore of high importance for
testing the overall consistency of the SM.

Previous measurements of the mass of the $W$ boson were
performed at the CERN SPS proton--antiproton (\ppbar{}) collider with the UA1 and UA2
experiments~\cite{Arnison:1985ut,Alitti:1991dk} at centre-of-mass
energies of $\sqrt{s}=546\GeV$ and $\sqrt{s}=630\GeV$,
at the Tevatron $p\bar{p}$ collider with the CDF
and D0 detectors at
$\sqrt{s}=1.8\TeV$~\cite{Affolder:2000bpa,Abazov:2002bu,Abazov:2003sv} and $\sqrt{s}=1.96\TeV$~\cite{Aaltonen:2012bp,Abazov:2012bv,Aaltonen:2013iut},
and at the LEP electron--positron collider by the ALEPH, DELPHI, L3, and OPAL
collaborations at
$\sqrt{s}=161$--$209\GeV$~\cite{Schael:2006mz,Abdallah:2008ad,Achard:2005qy,Abbiendi:2005eq}.
The current Particle Data Group world average value of $m_W = 80385 \pm 15$~\MeV~\cite{Agashe:2014kda} is dominated
by the CDF and D0 measurements performed at
$\sqrt{s}=1.96\TeV$. Given
the precisely measured values of $\alpha$, $G_{\mu}$ and $m_Z$, and taking
recent top-quark and Higgs-boson mass measurements, the SM prediction of $m_W$ is
$m_W=80358\pm 8$~MeV in Ref. \cite{Baak:2014ora} and $m_W=80362\pm 8$~\MeV~in Ref. \cite{deBlas:2016ojx}.
The SM prediction uncertainty of 8~\MeV{} represents a target for the precision of future measurements of $m_W$.

At hadron colliders, the $W$-boson mass can be determined in
Drell--Yan production~\cite{Drell:1970wh} from $W\to\ell\nu$ decays, where $\ell$ is an
electron or muon.
The mass of the $W$ boson is extracted from the Jacobian edges of the
final-state kinematic distributions, measured in the plane
perpendicular to the beam direction.
Sensitive observables include the transverse momenta of the charged lepton and neutrino
and the $W$-boson transverse mass.

The ATLAS and CMS experiments benefit from large signal and
calibration samples. The numbers of selected $W$- and
$Z$-boson events, collected in a sample corresponding to approximately
4.6~fb$^{-1}$ of integrated luminosity at a centre-of-mass energy of
$7\TeV$, are of the order of $10^7$ for the $W\rightarrow\ell\nu$, and of the order of $10^6$ for the $Z\rightarrow\ell\ell$
processes. The available data sample is therefore larger by an order of magnitude compared to the corresponding samples used for the CDF and D0 measurements.
Given the precisely measured value of the $Z$-boson
mass~\cite{ALEPH:2005ab} and the clean leptonic final state, the $Z\rightarrow\ell\ell$ processes provide the primary constraints for detector
calibration, physics modelling, and validation of the analysis strategy. The sizes of these samples correspond to a statistical 
uncertainty smaller than 10~\MeV{} in the measurement of the $W$-boson mass.

Measurements of $m_W$ at the LHC are affected by significant
complications related to the strong interaction. In particular, in
proton--proton ($pp$) collisions at $\sqrt{s}=7\TeV$, approximately 25\% of
the inclusive $W$-boson production rate is induced by at least one second-generation quark, $s$ or $c$, in the initial state. The amount of
heavy-quark-initiated production has implications for the $W$-boson
rapidity and transverse-momentum distributions~\cite{Krasny:2010vd}. As a consequence, the
measurement of the $W$-boson mass is sensitive to the strange-quark
and charm-quark parton distribution functions (PDFs) of the proton. In
contrast, second-generation quarks contribute only to approximately
5\% of the overall $W$-boson production rate at the Tevatron. 
Other important aspects of the measurement of the $W$-boson mass are
the theoretical description of electroweak corrections, in
particular the modelling of photon radiation from the $W$- and
$Z$-boson decay leptons, and the modelling of the relative
fractions of helicity cross sections in the Drell--Yan
processes~\cite{Mirkes:1992hu}.

This paper is structured as follows.
Section~\ref{sec:strategy} presents an overview of the
measurement strategy.
Section~\ref{sec:atlasdet} describes the ATLAS detector.
Section~\ref{sec:samples} describes the data and simulation
samples used for the measurement.
Section~\ref{sec:objreco} describes
the object reconstruction and the event
selection.
Section~\ref{sec:phymod} summarises the
modelling of vector-boson production and
decay, with emphasis on the QCD effects outlined
above.
Sections~\ref{sec:objcalib} and~\ref{sec:recoilcalib} are
dedicated to the electron, muon, and recoil calibration
procedures. Section~\ref{sec:crosschecks} presents a set of validation
tests of the measurement procedure, performed using the $Z$-boson
event sample. Section~\ref{sec:background} describes the analysis of
the $W$-boson sample. Section~\ref{sec:wfits} presents the extraction
of $m_W$. The results are summarised in Section~\ref{sec:conclusions}.

\section{Measurement overview \label{sec:strategy}}
This section provides the definition of the observables used in the
analysis, an overview of the measurement strategy for the
determination of the mass of the $W$ boson, and a description of the
methodology used to estimate the systematic uncertainties.

\subsection{Observable definitions \label{sec:obsdef}}

ATLAS uses a right-handed coordinate system with its origin at the
nominal interaction point (IP) in the centre of the detector and the $z$-axis along the beam
pipe. The $x$-axis points from the IP to the centre of the LHC ring, and the $y$-axis points
upward. Cylindrical coordinates $(r,\phi)$ are used in the transverse plane, $\phi$ being the
azimuth around the $z$-axis. The pseudorapidity is defined in terms of the polar angle
$\theta$ as $\eta=-\ln\tan(\theta/2)$.

The kinematic properties of charged leptons from $W$- and $Z$-boson decays are
characterised by the measured transverse momentum, $\pt^{\ell}$,
pseudorapidity, $\eta_{\ell}$, and azimuth, $\phi_{\ell}$. The mass of
the lepton, $m_{\ell}$, completes the four-vector.
For $Z$-boson events, the invariant mass, $m_{\ell\ell}$, the rapidity, $y_{\ell\ell}$, and the transverse momentum, $\pt^{\ell\ell}$, are
obtained by combining the four-momenta of the decay-lepton pair.

The recoil in the transverse plane, $\utvec$, is reconstructed from the vector sum of the transverse energy of all clusters reconstructed in the calorimeters (Section~\ref{sec:atlasdet}), excluding energy deposits associated with the decay leptons. It is defined as:
\begin{eqnarray}
\nonumber \utvec = \sum_i \vec{E}_{\mathrm{T},i},
\end{eqnarray}
where $\vec{E}_{\mathrm{T},i}$ is the vector of the
transverse energy of cluster $i$. The
transverse-energy vector of a cluster has magnitude
$E_{\mathrm{T}} = E / \cosh\eta$, with the
energy deposit of the cluster $E$ and its pseudorapidity $\eta$. The
azimuth $\phi$ of the transverse-energy vector is defined from the
coordinates of the cluster in the transverse plane. In $W$- and $Z$-boson events,
$-\utvec$ provides an estimate of the boson
transverse momentum.
The related quantities $u_x$ and $u_y$ are the projections of the
recoil onto the axes of the transverse plane in the ATLAS coordinate system.
In $Z$-boson events, $u_{\parallel}^Z$
and $u_{\perp}^Z$ represent the projections of the recoil
onto the axes parallel and perpendicular to the $Z$-boson transverse
momentum reconstructed from the decay-lepton pair.
Whereas $u_{\parallel}^Z$ can be compared to
$-p_{\mathrm{T}}^{\ell\ell}$ and probes the detector response to the recoil in terms of linearity and resolution, the $u_{\perp}^Z$ distribution satisfies $\avg{u_{\perp}^Z}=0$ and its width provides an estimate of the recoil resolution. 
In $W$-boson events, $u_{\parallel}^\ell$ and $u_{\perp}^\ell$ are the projections of the recoil
onto the axes parallel and perpendicular to the reconstructed charged-lepton
transverse momentum.

The resolution of the recoil is affected by additional event
properties, namely the per-event number of $pp$ interactions per bunch crossing (pile-up) ${\mu}$,
the average number of $pp$ interactions per bunch
crossing $\avg{\mu}$, the total reconstructed transverse energy,
defined as the scalar sum of the transverse energy of all calorimeter
clusters, $\Sigma E_{\mathrm{T}} \equiv \sum_{i} E_{{\mathrm{T}},i}$,
and the quantity $\Sigma E^{*}_{\mathrm{T}} \equiv \Sigma
E_{\mathrm{T}} - |\vec u_{\mathrm{T}}|$. The latter is less correlated with
the recoil than $\Sigma E_{\mathrm{T}}$, and better represents
the event activity related to the pile-up and to the underlying event.

The magnitude and direction of the
transverse-momentum vector of the decay neutrino, $\vec{p}_\textrm{T}^{\,\nu}$, are
inferred from the vector of the missing transverse momentum, $\mptvec$, which
corresponds to the momentum imbalance in the transverse plane and is
defined as:
\begin{eqnarray}
\nonumber \mptvec = -\left(\vec{p}_{\mathrm{T}}^{\,\ell} + \utvec\right).
\end{eqnarray}
The $W$-boson transverse mass, \mt, is derived from $\mpt$ and from
the transverse momentum of the charged lepton as follows:
\begin{eqnarray}
\nonumber \mt = \sqrt{2 \pt^\ell \mpt (1-\cos{\Delta\phi})},
\end{eqnarray}
where $\Delta\phi$ is the azimuthal opening angle between the charged lepton and
the missing transverse momentum.

All vector-boson masses and widths are defined in the running-width
scheme. Resonances are expressed by the relativistic Breit--Wigner
mass distribution:
\begin{eqnarray}
\label{eq:bw} \frac{\textrm{d}\sigma}{\textrm{d}m} \propto \frac{m^2}{(m^2-m_V^2)^2+m^4\Gamma_V^2/m_V^2},
\end{eqnarray}
where $m$ is the invariant mass of the vector-boson decay products, and $m_V$ and
$\Gamma_V$, with $V = W,Z$, are the vector-boson masses and
widths, respectively. This scheme was introduced in Ref.~\cite{Bardin:1988xt}, and is consistent
with earlier measurements of the $W$- and $Z$-boson resonance
parameters~\cite{Aaltonen:2013iut,ALEPH:2005ab}.

\subsection{Analysis strategy\label{subsec:strategy}}

The mass of the $W$ boson is determined from fits to the transverse
momentum of the charged lepton, $\pt^\ell$, and to the transverse mass
of the $W$ boson, \mt.
For $W$ bosons at rest, the transverse-momentum distributions of the $W$ decay leptons
have a Jacobian edge at a value of $m/2$, whereas the distribution of the
transverse mass has an endpoint at the value of
$m$~\cite{Smith:1983aa}, where $m$ is the invariant mass of the
charged-lepton and neutrino system, which is related to $m_W$ through
the Breit--Wigner distribution of Eq.~(\ref{eq:bw}).

The expected final-state distributions, referred to as templates, are
simulated for several values of $m_W$ and include signal and background contributions. The templates are compared to the observed distribution by means of a
$\chi^2$ compatibility test. The $\chi^2$ as a function of $m_W$ is
interpolated, and the measured value is determined by
analytical minimisation of the $\chi^2$ function.
Predictions for different values of $m_W$ are
obtained from a single simulated reference sample, by reweighting the $W$-boson
invariant mass distribution according to the Breit--Wigner
parameterisation of Eq.~(\ref{eq:bw}). The $W$-boson width is scaled accordingly, following
the SM relation $\Gamma_W \propto m_W^3$. 

Experimentally, the $\pt^\ell$ and  $\mpt$
distributions are affected by the lepton energy calibration. The latter is also affected by the calibration of
the recoil. The $\pt^\ell$ and \mpt{} distributions are broadened by the
$W$-boson transverse-momentum distribution, and are
sensitive to the $W$-boson helicity states, which are influenced by the proton
PDFs~\cite{ATL-PHYS-PUB-2014-015}. Compared to $\pt^\ell$, the \mt{} distribution has larger
uncertainties due to the recoil, but smaller sensitivity to such physics-modelling effects. Imperfect modelling of these effects can distort the template distributions, and constitutes a
significant source of uncertainties for the determination of $m_W$.

The calibration procedures described in this paper rely mainly on methods and results published earlier by ATLAS~\cite{PERF-2013-03,PERF-2013-05,PERF-2014-05}, and based on $W$ and $Z$ samples at
$\sqrt{s}=7\TeV$ and $\sqrt{s}=8\TeV$.
The $Z\rightarrow\ell\ell$ event samples are used to calibrate the
detector response. Lepton momentum corrections are derived exploiting
the precisely measured value of the $Z$-boson mass,
$m_Z$~\cite{ALEPH:2005ab}, and the recoil response
is calibrated using the expected momentum balance with $p_{\mathrm{T}}^{\ell\ell}$.
Identification and reconstruction efficiency corrections are
determined from $W$- and $Z$-boson events using the tag-and-probe
method~\cite{PERF-2013-03,PERF-2014-05}. The dependence of these corrections on $\pt^\ell$
is important for the measurement of $m_W$, as it affects the shape of the template
distributions. 

The detector response corrections and the physics modelling are verified in $Z$-boson events by
performing measurements of the $Z$-boson mass with the same method used to
determine the $W$-boson mass, and comparing the results to the LEP combined value of $m_Z$, which is used as input for the lepton calibration.
The determination of $m_Z$ from the lepton-pair invariant mass
provides a first closure test of the lepton energy calibration.
In addition, the extraction of $m_Z$ from the $\pt^\ell$ distribution
tests the $\pt^\ell$-dependence of the efficiency corrections, and the
modelling of the $Z$-boson transverse-momentum distribution and of the
relative fractions of $Z$-boson helicity states.
The \mpt{} and $\mt$ variables are defined in $Z$-boson events by treating one of the reconstructed decay leptons as a neutrino. 
The extraction of $m_Z$ from the $\mt$ distribution provides a test of the recoil calibration.
The combination of the extraction of $m_Z$ from the $m_{\ell\ell}$, $\pt^\ell$ and $\mt$ distributions provides a closure
test of the measurement procedure. The precision of this validation procedure is limited by the finite size of the $Z$-boson sample, which is approximately
ten times smaller than the $W$-boson sample.

The analysis of the $Z$-boson sample does not probe differences in the
modelling of $W$- and $Z$-boson production processes.
Whereas $W$-boson production at the Tevatron is charge
symmetric and dominated by interactions with at least one valence quark, the
sea-quark PDFs play a larger role at the LHC, and 
contributions from processes with heavy quarks in the
initial state have to be modelled properly. The $W^+$-boson production rate exceeds that of $W^-$ bosons
by about 40\%, with a broader rapidity distribution and a softer
transverse-momentum distribution.
Uncertainties in the modelling of these distributions and in the
relative fractions of the $W$-boson helicity states
are constrained using measurements of $W$- and $Z$-boson production
performed with the ATLAS experiment at $\sqrt{s}=7\TeV$ and
$\sqrt{s}=8\TeV$~\cite{WZ2011,Aad:2016izn,STDM-2012-06,STDM-2012-23,STDM-2011-15}.

The final measured value of the $W$-boson mass is obtained from the
combination of various measurements performed in the electron and
muon decay channels, and in charge- and $|\eta_\ell|$-dependent categories,
as defined in Table~\ref{tab:Categories}. The boundaries of the $|\eta_\ell|$ categories are driven mainly by experimental and statistical constraints. The measurements of $m_W$ used in the combination are based on the observed distributions of
$\pt^\ell$ and $\mt$, which are only partially correlated.
Measurements of $m_W$ based on the \mpt{} distributions are performed
as consistency tests, but they are not used in the combination due to
their significantly lower precision.
The consistency of the results in the electron and muon channels provide a further test of the experimental calibrations, whereas the consistency of the results for the different charge and
$|\eta_\ell|$ categories tests the $W$-boson production model.

Further consistency tests are performed by repeating
the measurement in three intervals of $\avg{\mu}$, in two intervals of
\ut{} and $u_{\parallel}^\ell$, and by removing the \mpt\ selection
requirement, which is applied in the nominal signal selection.
The consistency of the values of $m_W$ in these additional categories
probes the modelling of the recoil response, and the
modelling of the transverse-momentum spectrum of the $W$ boson.
Finally, the stability of the result with respect to the
charged-lepton azimuth, and upon variations of the fitting ranges is
verified.

\begin{table*}[tp]
\begin{center}
\begin{tabular}{lcc}
\toprule
Decay channel		&	$W\rightarrow e\nu$				&	$W\rightarrow \mu\nu$		\\
\midrule
Kinematic distributions &	$\pt^\ell$, \mt					&	$\pt^\ell$, \mt				\\
Charge categories	&	$W^+$, $W^-$					&	$W^+$, $W^-$				\\
$|\eta_\ell|$ categories	&	$[0,0.6]$, $[0.6,1.2]$, $[1.8,2.4]$	& $[0,0.8]$, $[0.8,1.4]$, $[1.4,2.0]$, $[2.0,2.4]$	\\
\bottomrule
\end{tabular}
\end{center}
\caption{\label{tab:Categories} Summary of categories and kinematic distributions used in the $m_W$ measurement analysis for the electron and muon decay channels.}
\end{table*}

Systematic uncertainties in the determination of $m_W$ are evaluated
using pseudodata samples produced from the nominal simulated event
samples by varying the parameters corresponding to each source of uncertainty in turn.
The differences between the values of $m_W$ extracted from the
pseudodata and nominal samples are used to estimate the
uncertainty. When relevant, these variations are applied
simultaneously in the $W$-boson signal samples and in the background
contributions. The systematic uncertainties are estimated separately for each source and for fit ranges of
$32<\pt^\ell<45\GeV$ and $66<\mt<99\GeV$. These fit ranges minimise the total expected measurement uncertainty, and are used for the final result as discussed in Section~\ref{sec:wfits}.

In Sections~\ref{sec:phymod}, \ref{sec:objcalib}, \ref{sec:recoilcalib}, and~\ref{sec:background}, which discuss the systematic uncertainties of the $m_W$ measurement, the uncertainties are also given for combinations of measurement categories. This provides information showing the reduction of the systematic uncertainty obtained from the measurement categorisation. For these cases, the combined uncertainties are evaluated including only the expected statistical uncertainty in addition to the systematic uncertainty being considered. 
However, the total measurement uncertainty is estimated by adding all uncertainty contributions in quadrature for each measurement category, and combining the results accounting for correlations across categories. 

During the analysis, an unknown offset was added to the value of $m_W$ used to produce the templates. The offset was randomly selected from a uniform distribution in the range $[-100,100]$~\MeV{}, and
the same value was used for the $W^{+}$ and $W^{-}$ templates. The offset was removed after the $m_W$ measurements performed in all categories were found to be compatible and the analysis procedure was finalised.

\section{The ATLAS detector \label{sec:atlasdet}}

The ATLAS experiment~\cite{PERF-2007-01} is a multipurpose particle detector with a forward-backward symmetric 
cylindrical geometry.
It consists of an inner tracking detector surrounded by a thin superconducting solenoid, electromagnetic and hadronic calorimeters,
and a muon spectrometer incorporating three large superconducting toroid magnets.

The inner-detector system (ID) is immersed in a \SI{2}{\tesla} axial magnetic field 
and provides charged-particle tracking in the range $|\eta| < 2.5$. At
small radii, a high-granularity silicon pixel detector covers the
vertex region and typically provides three measurements per track. It is followed by the silicon 
microstrip tracker, which usually provides
eight measurement points per track. These silicon
detectors are complemented by a gas-filled straw-tube transition
radiation tracker, which enables radially extended track
reconstruction up to $|\eta| = 2.0$. The transition radiation tracker also provides electron
identification information based on the fraction of hits (typically 35
in total) above a higher energy-deposit threshold corresponding to
transition radiation. 

The calorimeter system covers the pseudorapidity range $|\eta| < 4.9$.
Within the region $|\eta|< 3.2$, electromagnetic (EM) calorimetry is
provided by high-granularity lead/liquid-argon (LAr) calorimeters,
with an additional thin LAr presampler covering $|\eta|<1.8$  
to correct for upstream energy-loss fluctuations. The EM calorimeter
is divided into a barrel section covering $|\eta|<1.475$ and two
endcap sections covering $1.375<|\eta|<3.2$. For $|\eta|<2.5$ it is
divided into three layers in depth, which are finely segmented in
$\eta$ and $\phi$. Hadronic calorimetry is provided by a
steel/scintillator-tile calorimeter, segmented into three barrel
structures within $|\eta| < 1.7$ and two copper/LAr hadronic endcap
calorimeters covering $1.5<|\eta|<3.2$. The solid-angle coverage is completed with forward
copper/LAr and tungsten/LAr calorimeter modules in $3.1<|\eta|<4.9$, optimised for
electromagnetic and hadronic measurements, respectively. 

The muon spectrometer (MS) comprises separate trigger and
high-precision tracking chambers measuring the deflection of muons in
a magnetic field generated by superconducting air-core toroids. The
precision chamber system covers the region $|\eta| < 2.7$ with three
layers of monitored drift tubes, complemented by cathode strip
chambers in the forward region. The
muon trigger system covers the range $|\eta| < 2.4$ with resistive
plate chambers in the barrel, and thin gap chambers in the endcap
regions. 

A three-level trigger system is used to select events for offline
analysis~\cite{PERF-2011-02}. The level-1 trigger is implemented in
hardware and uses a subset of detector information to reduce the event
rate to a design value of at most \SI{75}{\kHz}. This is followed by
two software-based trigger levels which together reduce the event rate
to about \SI{300}{\Hz}.

\section{Data samples and event simulation \label{sec:samples}}

The data sample used in this analysis consists of $W$- and $Z$-boson
candidate events, collected in 2011 with the ATLAS detector in
proton--proton collisions at the LHC, at a centre-of-mass energy of
$\sqrt{s}=7$~\TeV. The sample for the electron channel, with all
relevant detector systems operational, corresponds to approximately
$4.6$~fb$^{-1}$ of integrated luminosity.
A smaller integrated luminosity of approximately $4.1$~fb$^{-1}$ is used in the muon channel, as part of the data was discarded due to a timing problem in the resistive plate chambers, which affected the muon trigger efficiency. The relative uncertainty of the integrated luminosity is 1.8\%~\cite{DAPR-2011-01}.
This data set provides approximately 1.4$\times 10^7$ reconstructed $W$-boson events and 1.8$\times 10^6$ $Z$-boson events, after
all selection criteria have been applied.

The \POWHEG MC generator~\cite{Nason:2004rx,Frixione:2007vw,Alioli:2010xd} (v1/r1556) is used for
the simulation of the hard-scattering processes of $W$- and $Z$-boson
production and decay in the electron, muon, and tau channels,
and is interfaced to \PYTHIA 8 (v8.170) for the modelling of the parton
shower, hadronisation, and underlying event~\cite{Sjostrand:2006za,Sjostrand:2007gs},
with parameters set according to the AZNLO tune~\cite{STDM-2012-23}.
The CT10 PDF set~\cite{Lai:2010vv} is used for the hard-scattering
processes, whereas the CTEQ6L1 PDF set~\cite{Pumplin:2002vw} is used
for the parton shower.
In the $Z$-boson samples, the
effect of virtual photon production ($\gamma^*$) and $Z/\gamma^*$ interference is included.
The effect of QED final-state radiation (FSR) is simulated with
\textsc{Photos} (v2.154)~\cite{Golonka:2005pn}.
Tau lepton decays are handled by \PYTHIA 8, taking into account polarisation effects.
An alternative set of samples for $W$- and $Z$-boson production is
generated with \POWHEG interfaced to \HERWIG (v6.520) for the
modelling of the parton shower~\cite{Corcella:2000bw}, and to \JIMMY
(v4.31) for the underlying event~\cite{Butterworth:1996zw}.
The $W$- and $Z$-boson masses are set to $m_W=80.399\GeV$ and
$m_Z=91.1875\GeV$, respectively. During the analysis, the value of
the $W$-boson mass in the $W\to\ell\nu$ and $W\to\tau\nu$ samples was blinded using the
reweighting procedure described in Section~\ref{sec:strategy}.

Top-quark pair production and the single-top-quark processes are modelled
using the \MCatNLO MC generator (v4.01)~\cite{Frixione:2002ik,Frixione:2003ei,Frixione:2005vw}, interfaced to
\HERWIG  and \JIMMY. Gauge-boson pair
production ($WW$, $WZ$, $ZZ$) is simulated with \HERWIG (v6.520).
In all the samples, the CT10 PDF set is used.
Samples of heavy-flavour multijet events ($pp\rightarrow b\bar b +X$ and
$pp\rightarrow c \bar c +X$) are simulated with \PYTHIA 8 to validate the
data-driven methods used to estimate backgrounds with non-prompt
leptons in the final state.

Whereas the extraction of $m_W$ is based on the shape of
distributions, and is not sensitive to the overall normalisation of
the predicted distributions, it is affected by theoretical uncertainties in the relative fractions of
background and signal. The $W$- and $Z$-boson event yields are normalised according to their measured
cross sections, and uncertainties of 1.8\% and 2.3\% are assigned to the
$W^{+}/Z$ and $W^{-}/Z$ production cross-section ratios, respectively~\cite{WZ2011}.
The $\ttbar$ sample is normalised according to its measured cross
section~\cite{TOPQ-2013-04} with an uncertainty of 3.9\%, 
whereas the cross-section predictions for the single-top production
processes of
Refs.~\cite{Kidonakis:2011wy,Kidonakis:2010tc,Kidonakis:2010ux}
are used for the normalisation of the corresponding sample, with an
uncertainty of 7\%. The samples of events with massive gauge-boson pair production
are normalised to the NLO predictions calculated with MCFM~\cite{Campbell:1999ah},
with an uncertainty of 10\% to cover the differences to the NNLO predictions~\cite{Gehrmann:2014fva}.

The response of the ATLAS detector is simulated using
a program~\cite{SOFT-2010-01} based on
\GEANT4~\cite{Agostinelli:2002hh}. The ID and the MS were simulated assuming an ideal detector geometry; alignment corrections are applied to the data during event reconstruction. The description of the detector material incorporates the results of extensive studies of the electron and photon calibration~\cite{PERF-2013-05}. The simulated hard-scattering process is overlaid with additional proton--proton interactions, simulated with
\PYTHIA 8 (v8.165) using the A2 tune~\cite{ATL-PHYS-PUB-2012-003}. The distribution of
the average number of interactions per bunch crossing $\avg{\mu}$ spans the range
$2.5$--$16.0$, with a mean value of approximately $9.0$.

Simulation inaccuracies affecting the distributions of the
signal, the response of the detector, and the underlying-event modelling,
are corrected
as described in the following sections. Physics-modelling corrections,
such as those affecting the $W$-boson transverse-momentum distribution
and the angular decay coefficients, are discussed in
Section~\ref{sec:phymod}. Calibration and detector response
corrections are presented in Sections~\ref{sec:objcalib} and~\ref{sec:recoilcalib}.

\section{Particle reconstruction and event selection \label{sec:objreco}}

This section describes the reconstruction and identification of
electrons and muons, the reconstruction of the recoil, and the
requirements used to select $W$- and $Z$-boson candidate events. The
recoil provides an event-by-event estimate of the $W$-boson transverse momentum. The reconstructed kinematic properties of the leptons and of
the recoil are used to infer the transverse momentum of the neutrino
and the transverse-mass kinematic variables.

\subsection{Reconstruction of electrons, muons and the recoil\label{sec:reco}}

Electron candidates are reconstructed from clusters of energy
deposited in the electromagnetic calorimeter and associated with at least
one track in the ID~\cite{PERF-2013-03,PERF-2013-05}. Quality requirements are
applied to the associated tracks in order to reject poorly
reconstructed charged-particle trajectories.
The energy of the electron is reconstructed from the energy collected
in calorimeter cells within an area of size $\Delta \eta \times \Delta
\phi = 0.075\times0.175$ in the barrel, and $0.125\times0.125$ in the
endcaps. A multivariate regression algorithm, developed and optimised
on simulated events, is used to calibrate the energy reconstruction.
The reconstructed electron energy is corrected to account for the
energy deposited in front of the calorimeter and outside the cluster,
as well as for variations of the energy response as a function of the
impact point of the electron in the calorimeter. The
energy calibration algorithm takes as inputs the energy collected by each calorimeter layer,
including the presampler, the pseudorapidity of the cluster, and the
local position of the shower within the cell of the second layer,
which corresponds to the cluster centroid. The kinematic properties of
the reconstructed electron are inferred from the energy measured
in the EM calorimeter, and from the pseudorapidity and azimuth
of the associated track. Electron candidates are required to have
$\pt > 15\GeV$ and $|\eta|<2.4$ and to fulfil a set of tight identification
requirements~\cite{PERF-2013-03}. The pseudorapidity range
$1.2<|\eta|<1.82$ is excluded from the measurement, as the amount of passive material
in front of the calorimeter and its uncertainty are
largest in this region~\cite{PERF-2013-05}, preventing a sufficiently accurate
description of non-Gaussian tails in the electron energy
response. Additional isolation requirements on the nearby
activity in the ID and calorimeter are applied to improve the
background rejection.
These isolation requirements are implemented by requiring the scalar sum of the \pt{} of tracks in a cone of size $\Delta R \equiv \sqrt{(\Delta\eta)^2+(\Delta\phi)^2} < 0.4$
around the electron, $\pt^{e,\textrm{cone}}$,
and the transverse energy deposited in the calorimeter within a cone of size $\Delta R <0.2$
around the electron, $E_\textrm{T}^\textrm{cone}$, to be small.
The contribution from the electron candidate itself is excluded. The
specific criteria are optimised as a function of electron $\eta$ and
\pt{} to have a combined efficiency of about 95\% in the simulation
for isolated electrons from the decay of a $W$ or $Z$ boson.

The muon reconstruction is performed independently in the ID and in
the MS, and a combined muon candidate is formed from the combination
of a MS track with an ID track, based on the statistical combination
of the track parameters~\cite{PERF-2014-05}.
The kinematic properties of the reconstructed muon are defined using
the ID track parameters alone, which allows a simpler calibration procedure.
The loss of resolution is small (10--15\%) in the transverse-momentum range relevant for the measurement
of the $W$-boson mass.
The ID tracks associated with the muons must satisfy quality
requirements on the number of hits recorded by each
subdetector~\cite{PERF-2014-05}. In order to reject muons from cosmic
rays, the longitudinal coordinate of the point of closest approach of
the track to the beamline is required to be within $10$~mm of the
collision vertex. Muon candidates are required to have $\pt>20\GeV$ and $|\eta|<2.4$.
Similarly to the electrons, the rejection of multijet background
is increased by applying an isolation requirement : the scalar sum of the
\pt{} of tracks in a cone of size $\Delta R < 0.2$ around the muon
candidate, $\pt^{\mu,\textrm{cone}}$, is required to be less than 10\% of the muon \pt.

The recoil, $\utvec$, is reconstructed from the vector sum of the transverse 
energy of all clusters measured in the calorimeters, as defined in Section~\ref{sec:obsdef}. The ATLAS
calorimeters measure energy depositions in the range $|\eta|<4.9$ with
a topological clustering algorithm~\cite{PERF-2014-07}, which starts
from cells with an energy of at least four times the expected noise
from electronics and pile-up. The momentum vector of each cluster is
determined by the magnitude and coordinates of the energy
deposition. Cluster energies are initially measured assuming that the energy deposition occurs only through
electromagnetic interactions, and are then corrected for the different
calorimeter responses to hadrons and electromagnetic particles, for
losses due to dead material, and for energy which is not captured by
the clustering process. The definition of $\utvec$ and the inferred
quantities $\mpt$ and \mt{} do not involve the explicit reconstruction
of particle jets, to avoid possible threshold effects.

Clusters located a distance $\Delta R < 0.2$ from the
reconstructed electron or muon candidates are not used for the
reconstruction of $\utvec$. This ensures that energy
deposits originating from the lepton itself
or from accompanying photons (from FSR or Bremsstrahlung) do not contribute to the recoil measurement.
The energy of any soft particles removed along with the lepton is
compensated for using the total transverse energy measured in a cone of the same size $\Delta R =0.2$, placed at the same absolute
pseudorapidity as the lepton with randomly chosen sign, and at
different $\phi$.
The total transverse momentum measured in this cone is
rotated to the position of the lepton and added to $\utvec$.

\subsection{Event selection \label{sec:eventsel}}

The $W$-boson sample is collected during data-taking with triggers
requiring at least one muon
candidate with transverse momentum larger than $18\GeV$ or at least one 
electron candidate with transverse momentum larger than $20\GeV$.
The transverse-momentum requirement for the electron candidate was
raised to $22\GeV$ in later data-taking periods to cope with the increased
instantaneous luminosity delivered by the LHC. Selected events
are required to have a reconstructed primary vertex with at least three
associated tracks. 

$W$-boson candidate events are selected by
requiring exactly one reconstructed electron or muon with $\pt^\ell >
30\GeV$. The leptons are required to match the corresponding trigger
object.
In addition, the reconstructed recoil is required to be
$\ut < 30\GeV$, the missing transverse momentum $\mpt > 30\GeV$
and the transverse mass $\mt > 60\GeV$.
These selection requirements are optimised to reduce the
multijet background contribution, and to minimise model
uncertainties from $W$ bosons produced at high transverse momentum.
A total of 5.89$\times 10^6$ $W$-boson candidate events are
selected in the $W\to e\nu$ channel, and 7.84$\times 10^6$ events in the
$W\to\mu\nu$ channel.

As mentioned in Section~\ref{sec:strategy}, $Z$-boson events
are extensively used to calibrate the response of the detector to
electrons and muons, and to derive recoil corrections. In addition, $Z$-boson events are used
to test several aspects of the modelling of vector-boson
production. $Z$-boson candidate events are collected with the same trigger
selection used for the $W$-boson sample. The analysis selection
requires exactly two reconstructed leptons with $\pt^\ell > 25\GeV$,
having the same flavour and opposite charges. The events are required to have an invariant mass of the
dilepton system in the range $80 <m_{\ell\ell}<100\GeV$. 
In both channels, selected leptons are required to be isolated in the
same way as in the $W$-boson event selection. In total, 0.58$\times 10^6$ and 1.23$\times 10^6$ $Z$-boson candidate events are selected in the electron and
muon decay channels, respectively.

\section{\label{sec:phymod}Vector-boson production and decay}

Samples of inclusive vector-boson production are produced using the \POWHEG MC generator interfaced to \PYTHIA 8, henceforth
referred to as \PowhegPythia.
The $W$- and $Z$-boson samples are reweighted to include
the effects of higher-order QCD and electroweak (EW) corrections, as well
as the results of fits to measured distributions which improve the agreement of the simulated lepton kinematic distributions with the data. The
effect of virtual photon production
and $Z/\gamma^*$ interference is included in both the predictions and the
\PowhegPythia simulated $Z$-boson samples.
The reweighting procedure used to include the corrections in the
simulated event samples is detailed in Section~\ref{sec:reweight}.

The correction procedure is based on the factorisation of the
fully differential leptonic Drell--Yan cross
section~\cite{Drell:1970wh} into four terms:
\begin{eqnarray}
  \label{eq:decomposition}
  \frac{\textrm{d}\sigma}{\textrm{d}p_1 \, \textrm{d}p_2} = \left[\frac{\textrm{d}\sigma(m)}{\textrm{d}m}\right] \left[\frac{\textrm{d}\sigma(y)}{\textrm{d}y}\right] \left[\frac{\textrm{d}\sigma(\pt, y)}{\textrm{d}\pt\,\textrm{d}y} \left(\frac{\textrm{d}\sigma(y)}{\textrm{d}y}\right)^{-1} \right] \left[(1+\cos^2\theta)+\sum_{i=0}^{7} A_i(\pt,y) P_i(\cos\theta, \phi) \right],
\end{eqnarray}
\noindent where $p_1$ and $p_2$ are the lepton and anti-lepton four-momenta; $m$,
\pt, and $y$ are the invariant mass, transverse momentum, and
rapidity of the dilepton system; $\theta$ and $\phi$ are the
polar angle and azimuth of the lepton\footnote{Here, lepton refers to the negatively charged lepton from a $W^-$ or $Z$ boson, and the neutrino
  from a $W^+$ boson.} in any given rest frame of the
dilepton system; $A_i$ are numerical coefficients, and $P_i$ are spherical
harmonics of order zero, one and two. 

The differential cross section as a function of the invariant mass,
$\textrm{d}\sigma(m)/\textrm{d}m$, is modelled with a Breit--Wigner
parameterisation according to Eq.~(\ref{eq:bw}). In the case of the
$Z$-boson samples, the photon propagator is included using the running electromagnetic coupling
constant; further electroweak corrections are discussed in Section~\ref{sec:physmodew}.
The differential cross section as a function of boson rapidity,
$\textrm{d}\sigma(y)/\textrm{d}y$, and the coefficients $A_i$ are modelled with perturbative QCD fixed-order
predictions, as described in Section~\ref{sec:physmodai}. The transverse-momentum spectrum at a given rapidity,
$\textrm{d}\sigma(\pt,y)/(\textrm{d}\pt\,\textrm{d}y) \cdot
(\textrm{d}\sigma(y)/\textrm{d}y)^{-1}$, is modelled with predictions
based on the \PYTHIA~8 MC generator, as discussed in Section~\ref{sec:physmodptw}. An exhaustive review of available predictions for $W$- and $Z$-boson production at the LHC is given in Ref.~\cite{Alioli:2016fum}.

Measurements of \Wboson- and \Zboson-boson
production are used to validate and constrain the modelling of the
fully differential leptonic Drell--Yan cross section.
The PDF central values and uncertainties, as well as the modelling of
the differential cross section as a function of boson rapidity, are
validated by comparing to the $7\TeV$ \Wboson- and \Zboson-boson rapidity
measurements~\cite{WZ2011}, based on the same data sample. The QCD parameters of the parton
shower model were determined by fits to the transverse-momentum
distribution of the $Z$ boson measured at
$7\TeV$~\cite{STDM-2012-23}. The modelling of the $A_i$
coefficients is validated by comparing the theoretical predictions to
the $8\TeV$ measurement of the angular coefficients in $Z$-boson
decays~\cite{Aad:2016izn}.

\subsection{Electroweak corrections and uncertainties \label{sec:physmodew}}

The dominant source of electroweak corrections to \Wboson- and \Zboson-boson production originates from QED final-state radiation, and is simulated with
\textsc{Photos}. The effect of QED initial-state radiation (ISR) is also included through the \PYTHIA 8 parton shower.
The uncertainty in the modelling of QED FSR is evaluated
by comparing distributions obtained using the default leading-order photon emission matrix elements with predictions obtained using NLO matrix elements, as well as by comparing \textsc{Photos} with an
alternative implementation based on the Yennie--Frautschi--Suura
formalism~\cite{Jadach:1995nk}, which is available in
\textsc{Winhac}~\cite{Placzek:2013moa}. The differences are small in
both cases, and the associated uncertainty is considered negligible.

Other sources of electroweak corrections are not included in the
simulated event samples, and their full effects are considered as systematic
uncertainties. They include the interference between ISR and FSR QED
corrections (IFI), pure weak corrections due to virtual-loop and box
diagrams, and final-state emission of lepton pairs. Complete $O(\alpha)$ electroweak corrections to the $pp\to W+X$, $W\to\ell\nu$ process were initially calculated in
Refs.~\cite{Dittmaier:2001ay,Baur:2004ig}. Combined QCD and EW corrections are however necessary to evaluate the effect of the latter in presence of a realistic $\pt^W$ distribution. Approximate
$O(\alpha_{\mathrm s}\alpha)$ corrections including parton shower effects are available from \textsc{Winhac}, \textsc{Sanc}~\cite{Arbuzov:2005dd} and in the \textsc{Powheg}
framework~\cite{Barze:2012tt,Bernaciak:2012hj,Muck:2016pko}. A complete, fixed-order calculation of $O(\alpha_{\mathrm s}\alpha)$ corrections in the resonance region appeared in Ref.~\cite{Dittmaier:2015rxo}. 

In the present work the effect of the NLO EW corrections are estimated using \textsc{Winhac}, which employs the \textsc{Pythia} 6
MC generator for the simulation of QCD and QED ISR. The corresponding uncertainties are evaluated comparing the final state distributions obtained including QED FSR only with predictions using the
complete NLO EW corrections in the $\alpha(0)$ and $G_\mu$ renormalisation schemes~\cite{Altarelli:116932}. The latter predicts the larger correction and is used to assign the systematic uncertainty.

Final-state lepton pair production, through $\gamma^*\rightarrow\ell\ell$ radiation, is formally a
higher-order correction but constitutes an significant additional
source of energy loss for the $W$-boson decay products. This process is not
included in the event simulation, and the impact on the determination of
$m_W$ is evaluated using \textsc{Photos} and \textsc{Sanc}.

Table~\ref{tab:ewsystwtot} summarises the effect of the uncertainties associated
with the electroweak corrections on the $m_W$ measurements. All comparisons described above were performed at particle level. The impact is larger for the  $\pt^\ell$ distribution than for the $\mt$
distribution, and similar between the electron and muon decay
channels. A detailed evaluation of these uncertainties was performed in Ref.~\cite{CarloniCalame:2016ouw} using \textsc{Powheg}~\cite{Barze:2012tt}, and the results are in fair agreement with
Table~\ref{tab:ewsystwtot}. The study of Ref.~\cite{CarloniCalame:2016ouw} also compares, at fixed order, the effect of the approximate $O(\alpha_{\mathrm s}\alpha)$ corrections with the full
calculation of Ref.~\cite{Dittmaier:2015rxo}, and good agreement is found. The same sources of uncertainty
affect the lepton momentum calibration through their impact on the
$m_{\ell\ell}$ distribution in $Z$-boson events, as discussed in
Section~\ref{sec:objcalib}.

\begin{table}[tp]
  \begin{center}
    \begin{tabular}{lrrrr}
      \toprule
      Decay channel          & \multicolumn{2}{r}{$W \rightarrow e \nu$} & \multicolumn{2}{r}{$W \rightarrow \mu \nu$} \\
      Kinematic distribution & $\pt^\ell$ & $\mt$                         & $\pt^\ell$ & $\mt$ \\
      \midrule
      $\delta m_W$~[\MeV]\\
      \,\,\,\, FSR (real)                     & $<0.1$  & $<0.1$  & $<0.1$  & $<0.1$ \\
      \,\,\,\, Pure weak and IFI corrections  & 3.3     & 2.5     & 3.5     & 2.5 \\
      \,\,\,\, FSR (pair production)          & 3.6     & 0.8     & 4.4     & 0.8 \\
      \midrule
      \,\,\,\, Total                          & 4.9     & 2.6     & 5.6     & 2.6 \\
      \bottomrule
    \end{tabular}
    \caption{Impact on the $m_W$ measurement of systematic uncertainties from higher-order electroweak corrections, for the $\pt^\ell$ and $\mt$ distributions in the electron and muon decay channels.\label{tab:ewsystwtot}}
  \end{center}
\end{table}

\subsection{Rapidity distribution and angular coefficients \label{sec:physmodai}}

At leading order, $W$ and $Z$ bosons are produced with zero transverse momentum,
and the angular distribution of the decay leptons depends solely on the
polar angle of the lepton in the boson rest
frame. Higher-order corrections give rise to sizeable boson transverse
momentum, and to azimuthal asymmetries in the angular distribution of the decay leptons.
The angular distribution of the $W$- and $Z$-boson decay leptons is
determined by the relative fractions of helicity cross sections for the
vector-boson production.
The fully differential leptonic Drell--Yan cross section can be
decomposed as a weighted sum of nine harmonic polynomials, with weights given by the helicity
cross sections. The harmonic polynomials depend on the polar angle, $\theta$, and the
azimuth, $\phi$, of the lepton in a given rest frame of the
boson.
The helicity cross sections depend, in their most general expression,
on the transverse momentum, $\pt$, rapidity, $y$, and invariant mass,
$m$, of the boson.
It is customary to factorise the unpolarised, or angular-integrated,
cross section, $\textrm{d}\sigma/(\textrm{d}p_{\textrm T}^{2} \, \textrm{d}y \, \textrm{d}m)$, and express the
decomposition in terms of dimensionless angular coefficients, $A_{i}$,
which represent the ratios of the helicity cross sections with respect to the
unpolarised cross section~\cite{Mirkes:1992hu}, leading to the
following expression for the fully differential Drell--Yan cross section:

\begin{eqnarray}
\frac{\textrm{d}\sigma}{\textrm{d}\pt^{2}\, \textrm{d}y\, \textrm{d}m\, \textrm{d}\cos\theta\, \textrm{d}\phi} & = & \frac{3}{16\pi}\frac{\textrm{d}\sigma}{\textrm{d}\pt^{2}\, \textrm{d}y\, \textrm{d}m} \times [(1+\cos^{2}  \theta) + A_{0} \, \frac{1}{2}(1-3\cos^{2}\theta) \nonumber \\
& + & A_{1} \, \sin2\theta \cos\phi + A_{2} \, \frac{1}{2}\sin^{2}\theta \cos2\phi + A_{3}\, \sin\theta \cos\phi + A_{4}\, \cos\theta \nonumber \\ 
& + & A_{5}\, \sin^{2}\theta \sin2\phi + A_{6}\, \sin2\theta \sin\phi + A_{7}\,  \sin\theta \sin\phi].
\label{eq:poldecomp}
\end{eqnarray}

\noindent The angular coefficients depend in general on \pt, $y$ and $m$. The $A_{5}$--$A_{7}$ coefficients are non-zero only at order $O(\alpha_{\mathrm s}^2)$ and above. They are small in the \pt{} region relevant for the present analysis, and are not considered further. The angles $\theta$ and $\phi$ are defined in the Collins--Soper (CS) frame~\cite{Collins:1977iv}.

The differential cross section as a function of boson rapidity,
$\textrm{d}\sigma(y)/\textrm{d}y$, and the angular coefficients, $A_i$,
are modelled with
fixed-order perturbative QCD predictions, at $O(\alpha_{\mathrm s}^2)$ in the
perturbative expansion of the strong coupling constant and using the CT10nnlo PDF set~\cite{Gao:2013xoa}. The dependence of the angular coefficients on $m$ is neglected; the effect of this approximation on the measurement of $m_W$ is discussed
in Section~\ref{sec:reweight}. For the calculation of the predictions, an optimised version of
DYNNLO~\cite{Catani:2009sm} is used, which explicitly decomposes the
calculation of the cross section into the different pieces of the
$q_{\mathrm T}$-subtraction formalism, and allows the computation of statistically
correlated PDF variations. In this optimised version of DYNNLO, the
Cuba library~\cite{Hahn:2004fe} is used for the numerical
integration.

The values of the angular coefficients predicted by the
\PowhegPythia samples differ significantly from the
corresponding NNLO predictions.
In particular, large differences are observed in the predictions
of $A_0$ at low values of $\pt^{W,Z}$. Other coefficients, such as $A_1$ and $A_2$, are affected
by significant NNLO corrections at high $\pt^{W,Z}$. In $Z$-boson production, $A_3$ and
$A_4$ are sensitive to the vector couplings between the $Z$ boson and
the fermions, and are predicted assuming the measured value of the
effective weak mixing angle $\sin^2\theta^\ell_{\textrm{eff}}$~\cite{ALEPH:2005ab}.

\subsection{Transverse-momentum distribution \label{sec:physmodptw}}

Predictions of the vector-boson transverse-momentum spectrum cannot rely solely on fixed-order
perturbative QCD. Most \Wboson-boson events used for the analysis have a low transverse-momentum value, 
in the kinematic region $\pt^W < 30\GeV$, where large logarithmic terms
of the type $\log(m_W/\pt^W)$ need to be resummed, and non-perturbative effects must be
included, either with parton showers or with predictions based on analytic
resummation~\cite{Collins:1984kg,Ladinsky:1993zn,Balazs:1997xd,Catani:2015vma,Becher:2011xn}. The
modelling of the transverse-momentum spectrum of vector bosons at
a given rapidity, expressed by the term $\textrm{d}\sigma(\pt,y)/(\textrm{d}\pt\,\textrm{d}y) \cdot (\textrm{d}\sigma(y)/\textrm{d}y)^{-1}$
in Eq.~(\ref{eq:decomposition}), is based on the \PYTHIA
8 parton shower MC generator.
The predictions of vector-boson production in the \PYTHIA 8 MC
generator employ leading-order matrix elements for the $q\bar q'\rightarrow W, Z$ processes and include a reweighting of the first
parton shower emission to the leading-order $V$+jet
cross section~\cite{Miu:1998ju}. The resulting prediction of the
boson \pt\ spectrum is comparable in accuracy to those
of an NLO plus parton shower generator setup such as \PowhegPythia,
and of resummed predictions at next-to-leading logarithmic
order~\cite{Catani:1990rr}. 

The values of the QCD parameters used in \PYTHIA 8 were determined from fits to the $Z$-boson transverse
momentum distribution measured with the ATLAS detector at a
centre-of-mass energy of $\sqrt{s} = 7\TeV$~\cite{STDM-2012-23}.
Three QCD parameters were considered in the fit: the intrinsic transverse momentum of the incoming partons, the value of 
$\alpha_{\mathrm s}(m_Z)$ used for the QCD ISR,
and the value of the ISR infrared cut-off. The resulting
values of the \PYTHIA~8 parameters constitute the AZ tune. The \PYTHIA 8 AZ prediction was found to provide a satisfactory description of the $\pt^Z$ distribution as a function of rapidity, contrarily
to \PowhegPythia~AZNLO; hence the former is chosen to predict the $\pt^W$ distribution. The good consistency of the $m_W$ measurement results in $|\eta_\ell|$ categories, presented in
Section~\ref{sec:wfits}, is also a consequence of this choice.

\begin{figure}
  \begin{center}
    \subfloat[]{\label{fig:CompPTZ}\includegraphics[width=0.49\textwidth]{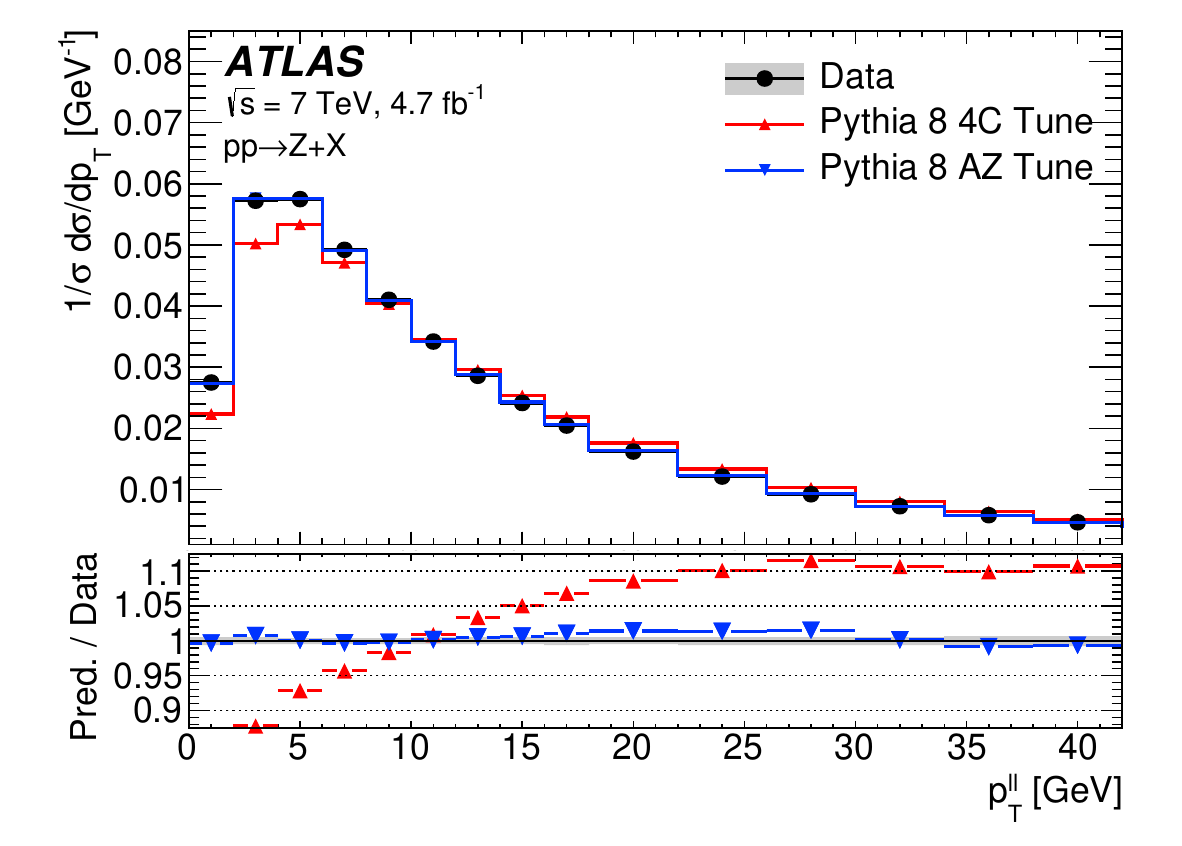}}
    \subfloat[]{\label{fig:CompPTW}\includegraphics[width=0.49\textwidth]{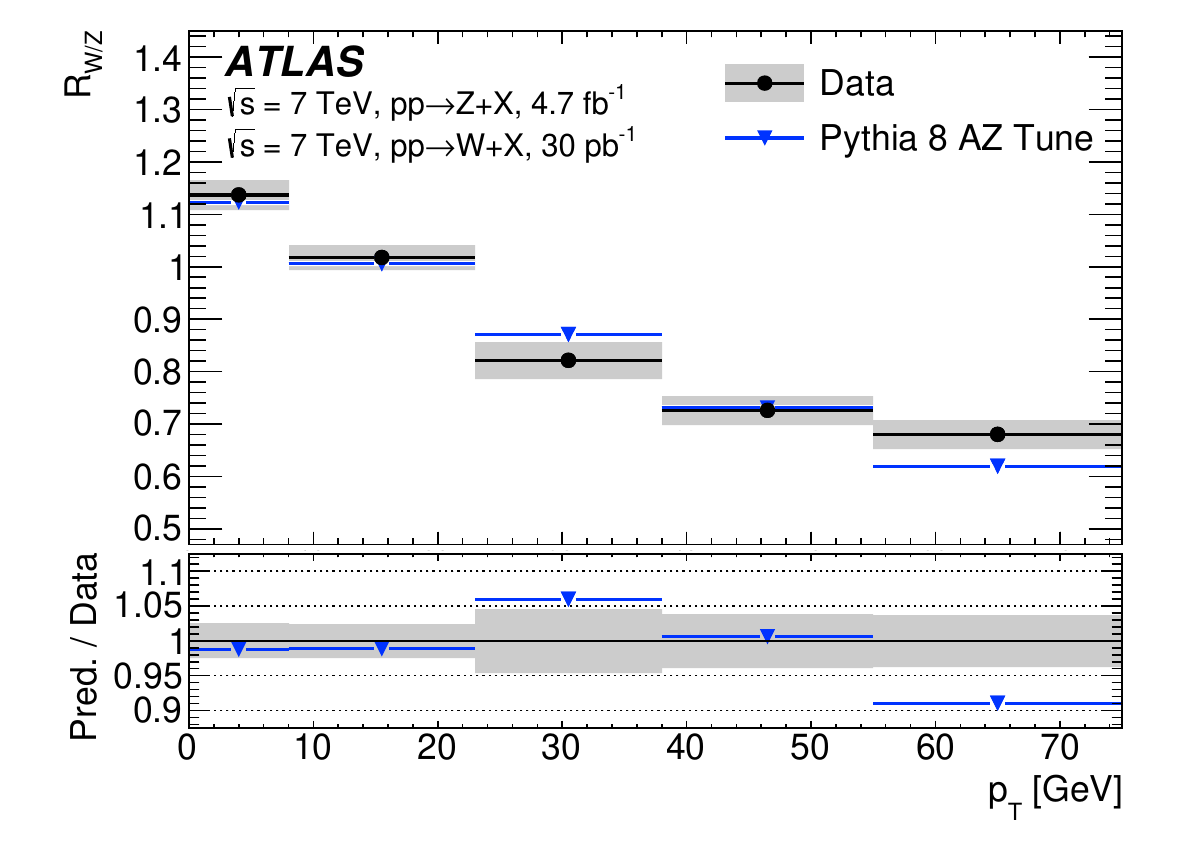}}
  \end{center}
  \caption{(a) Normalised differential cross section as a function of
    $\pt^{\ell\ell}$ in $Z$-boson events~\cite{STDM-2012-23} and (b)
    differential cross-section ratio $R_{W/Z}(\pt)$ as a function of the
    boson $\pt$~\cite{STDM-2012-23,STDM-2011-15}. The measured cross sections are compared to the predictions of the 
    \textsc{Pythia 8} AZ tune and, in (a), of the \textsc{Pythia 8} 4C tune. The shaded bands show the
    total experimental uncertainties.}
  \label{fig:CompPTWZ}
\end{figure}

To illustrate the results of the parameters optimisation, the \PYTHIA 8
AZ and 4C~\cite{Corke:2010yf} predictions of the
$\pt^Z$ distribution are compared in
Figure~\ref{fig:CompPTZ} to the measurement used
to determine the AZ tune. Kinematic requirements
on the decay leptons are applied according to the experimental
acceptance. For further validation, the predicted differential cross-section ratio,
\begin{eqnarray}
\nonumber R_{W/Z}(\pt) = \left(\frac{1}{\sigma_W} \cdot \frac{\textrm{d}\sigma_W(\pt)}{\textrm{d}\pt}\right)  \left(\frac{1}{\sigma_Z} \cdot \frac{\textrm{d}\sigma_Z(\pt)}{\textrm{d}\pt}\right)^{-1},
\end{eqnarray}
is compared to the corresponding ratio of ATLAS measurements of vector-boson
transverse momentum~\cite{STDM-2012-23,STDM-2011-15}. The comparison
is shown in Figure~\ref{fig:CompPTW}, where kinematic requirements
on the decay leptons are applied according to the experimental
acceptance. The measured \Zboson-boson
$\pt$ distribution is rebinned to match the coarser bins of the
\Wboson-boson $\pt$ distribution, which was measured using only 30~pb$^{-1}$ of data.
The theoretical prediction is in agreement with the experimental measurements for the region with
$\pt<30\GeV$, which is relevant for the measurement of the $W$-boson mass.

The predictions of RESBOS~\cite{Ladinsky:1993zn,Balazs:1997xd}, DYRes~\cite{Catani:2015vma} and \PowhegMinlo \cite{Hamilton:2012np,Hamilton:2012rf} are also considered. 
All predict a harder $\pt^W$ distribution for a given $\pt^Z$ distribution, compared to \PYTHIA~8~AZ. Assuming the latter can be adjusted to match the measurement of Ref.~\cite{STDM-2012-23}, the corresponding $\pt^W$ distribution induces a discrepancy with the detector-level $u_{\textrm{T}}$ and $u_{\parallel}^\ell$ distributions observed in the $W$-boson data, as discussed in Section~\ref{sec:ptwxcheck}.
This behaviour is observed using default values for the non-perturbative parameters of these programs, but is not expected to change significantly under variations of these parameters. These
predictions are therefore not used in the determination of $m_W$ or its uncertainty.

\begin{figure}
  \begin{center}
    \subfloat[]{\label{fig:CompPTLMTa}\includegraphics[width=0.49\textwidth]{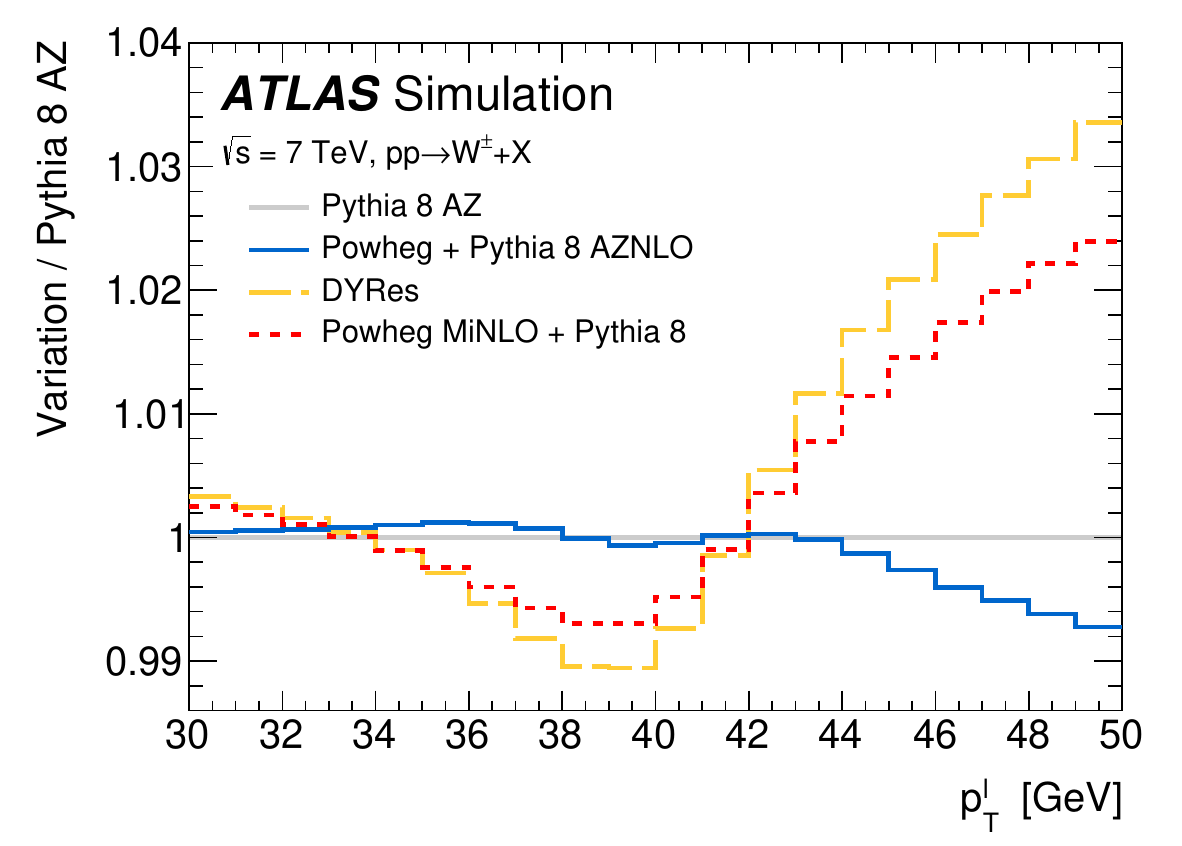}}
    \subfloat[]{\label{fig:CompPTLMTb}\includegraphics[width=0.49\textwidth]{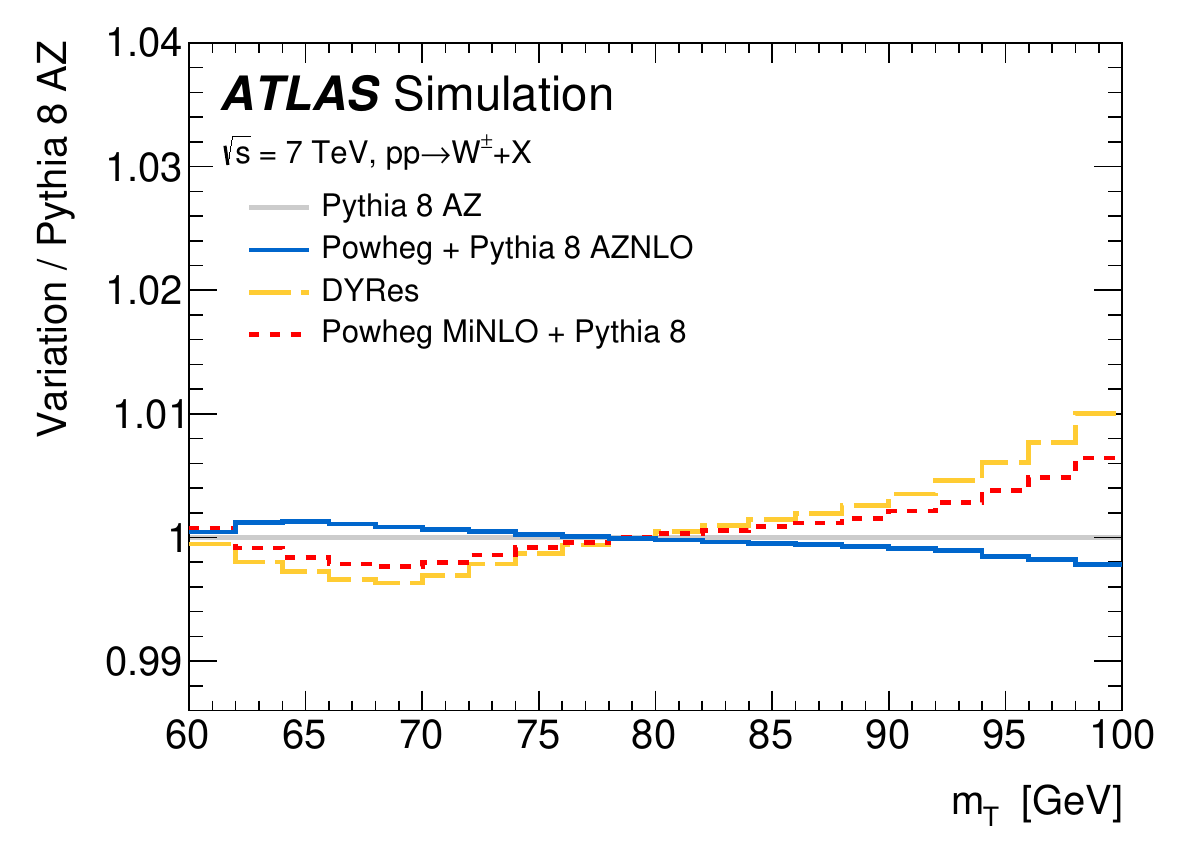}}
  \end{center}
  \caption{Ratios of the reconstruction-level (a) $\pt^\ell$ and (b) \mt\ normalised distributions obtained using \PowhegPythia AZNLO, DYRes and \PowhegMinlo to the baseline normalised distributions obtained using \PYTHIA 8~AZ.}
  \label{fig:CompPTLMT}
\end{figure}

Figure~\ref{fig:CompPTLMT} compares the reconstruction-level $\pt^\ell$ and \mt\ distributions obtained with \PowhegPythia AZNLO, DYRes and \PowhegMinlo to those of \PYTHIA~8~AZ\footnote{Reconstruction-level distributions are obtained from the \PowhegPythia signal sample by
  reweighting the particle-level $\pt^W$ distribution according to the product of the $\pt^Z$ distribution in \PYTHIA~8~AZ, and of $R_{W/Z}(\pt)$ as predicted by \PowhegPythia~AZNLO, DYRes and \PowhegMinlo.}. The effect of varying the $\pt^W$ distribution is largest at high $\pt^\ell$, which explains why the uncertainty due to the $\pt^W$ modelling is reduced when limiting the $\pt^\ell$ fitting range as described in Section~\ref{sec:Unc}.

\subsection{Reweighting procedure\label{sec:reweight}}

The $W$ and $Z$ production and decay model described above is applied to the \PowhegPythia samples through an event-by-event reweighting.
Equation~(\ref{eq:poldecomp}) expresses the factorisation of the cross section
into the three-dimensional boson production phase space, defined by
the variables $m$, $\pt$, and $y$, and the two-dimensional boson decay
phase space, defined by the variables $\theta$ and $\phi$.
Accordingly, a prediction of the kinematic distributions of vector
bosons and their decay products can be transformed into another prediction by
applying separate reweighting of the three-dimensional boson
production phase-space distributions, followed by a reweighting of the
angular decay distributions.

The reweighting is performed in several steps. First, the inclusive rapidity
distribution is reweighted according to the NNLO QCD predictions
evaluated with DYNNLO. Then, at a given rapidity, the vector-boson
transverse-momentum shape is reweighted to the \PYTHIA 8 prediction
with the AZ tune.
This procedure provides the transverse-momentum
distribution of vector bosons predicted by \PYTHIA 8, preserving the rapidity
distribution at NNLO.
Finally, at given rapidity and transverse momentum, the angular variables are reweighted according to:
\begin{eqnarray}
\nonumber w(\cos \theta,\phi, \pt,y) = \frac{1+\cos^{2}\theta+\sum_i \, A'_i(\pt,y) \,
  P_i(\cos \theta,\phi)}{1+\cos^{2}\theta+\sum_i \, A_i(\pt,y) \, P_i(\cos \theta,\phi)},
\end{eqnarray}
where $A'_i$ are the angular coefficients evaluated at
$O(\alpha_{\mathrm s}^2)$, and $A_i$ are the angular coefficients of the \PowhegPythia samples.
This reweighting procedure neglects the small dependence of the two-dimensional (\pt,$y$) distribution and of the angular coefficients on the final state invariant mass.
The procedure is used to include the corrections described in Sections~\ref{sec:physmodai} and~\ref{sec:physmodptw}, as well as to estimate the impact of the QCD modelling uncertainties described in
Section~\ref{sec:modunc}.

The validity of the reweighting procedure is tested at particle level
by generating independent $W$-boson samples using the CT10nnlo and
NNPDF3.0~\cite{Ball:2014uwa} NNLO PDF sets, and the same value of $m_W$. The relevant
kinematic distributions are calculated for both samples and used
to reweight the CT10nnlo sample to the NNPDF3.0 one. The procedure
described in Section~\ref{subsec:strategy} is then used to determine
the value of $m_W$ by fitting the NNPDF3.0 sample using
templates from the reweighted CT10nnlo sample. The fitted value agrees with
the input value within $1.5 \pm 2.0\MeV$. The statistical precision of
this test is used to assign the associated systematic uncertainty.

\begin{figure}
  \begin{center}
    \subfloat[]{\includegraphics[width=0.495\textwidth]{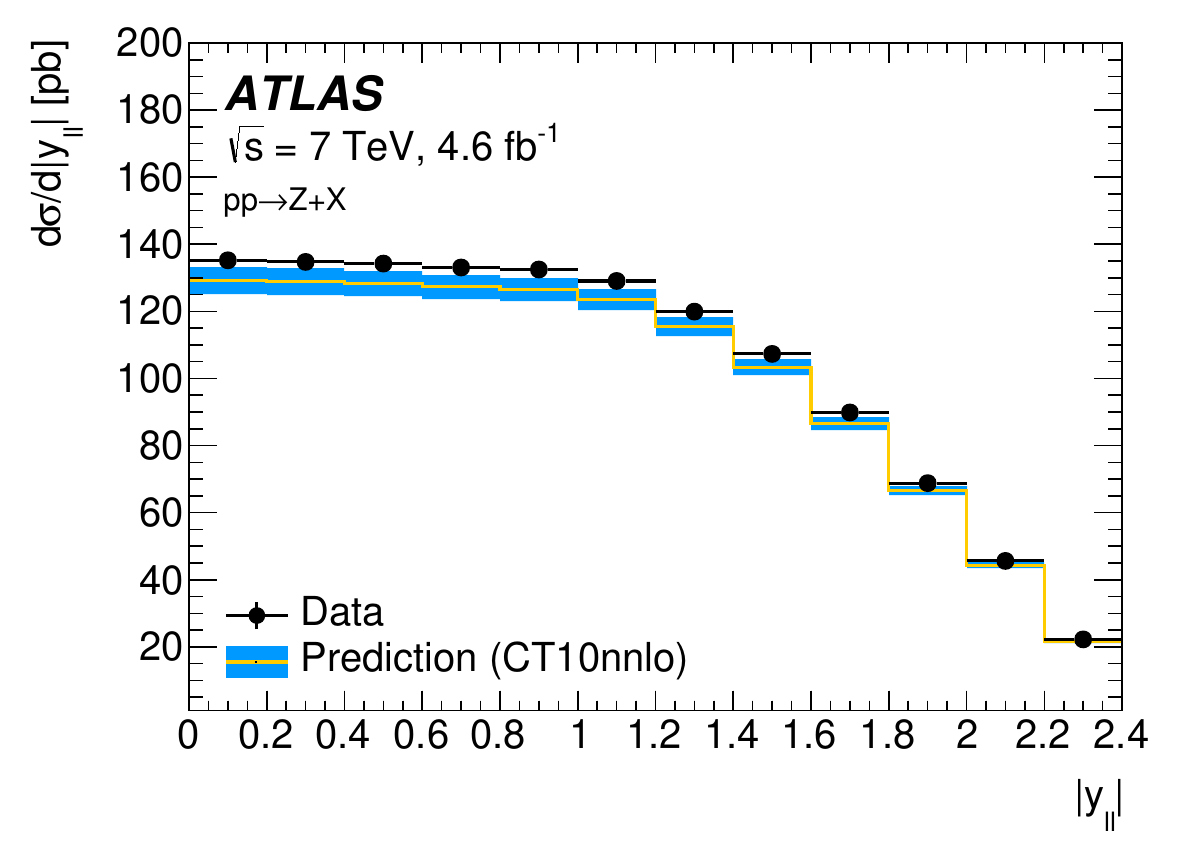}}
    \subfloat[]{\includegraphics[width=0.495\textwidth]{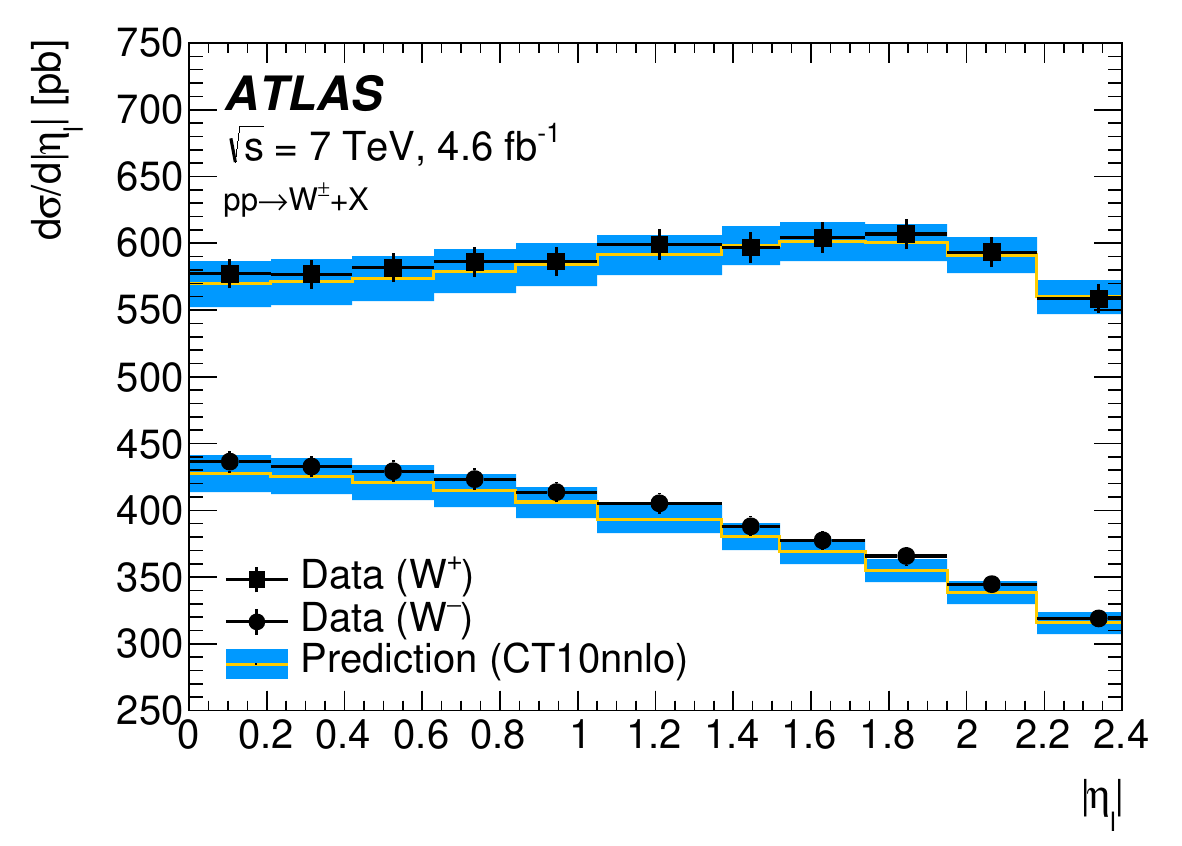}}
  \end{center}
  \caption{(a) Differential $Z$-boson cross section as a
    function of boson rapidity, and (b) differential $W^+$ and $W^-$
    cross sections as a function of charged decay-lepton
    pseudorapidity at $\sqrt{s}=7\TeV$~\cite{WZ2011}. The
    measured cross sections are compared to the \PowhegPythia
    predictions, corrected to NNLO using DYNNLO with the CT10nnlo PDF set.
    The error bars show the
    total experimental uncertainties, including luminosity
    uncertainty, and the bands show the PDF uncertainties of the
    predictions.}
  \label{fig:controlmeasurements}
\end{figure}

The resulting model is tested by comparing the predicted $Z$-boson differential cross section as a
function of rapidity, the $W$-boson differential cross section as a
function of lepton pseudorapidity, and the angular coefficients in $Z$-boson events, to the corresponding ATLAS
measurements~\cite{WZ2011,Aad:2016izn}. The comparison with the measured $W$ and $Z$ cross sections is shown in Figure~\ref{fig:controlmeasurements}. Satisfactory agreement
between the measurements and the theoretical predictions is
observed. A $\chi^2$ compatibility test is performed for the three
distributions simultaneously, including the correlations between the uncertainties. The compatibility test yields a $\chi^2/$dof value of
$45/34$. Other NNLO PDF sets such as NNPDF3.0, CT14~\cite{Dulat:2015mca},
MMHT2014~\cite{Harland-Lang:2014zoa}, and ABM12~\cite{Alekhin:2013nda} are in worse agreement with these
distributions. Based on the quantitative comparisons performed in Ref.~\cite{WZ2011}, only CT10nnlo, CT14 and MMHT2014 are considered further. The better agreement obtained with CT10nnlo can be
ascribed to the weaker suppression of the strange quark density compared to the $u$- and $d$-quark sea densities in this PDF set.

\begin{figure}
  \begin{center}
    \subfloat[]{\includegraphics[width=0.49\textwidth]{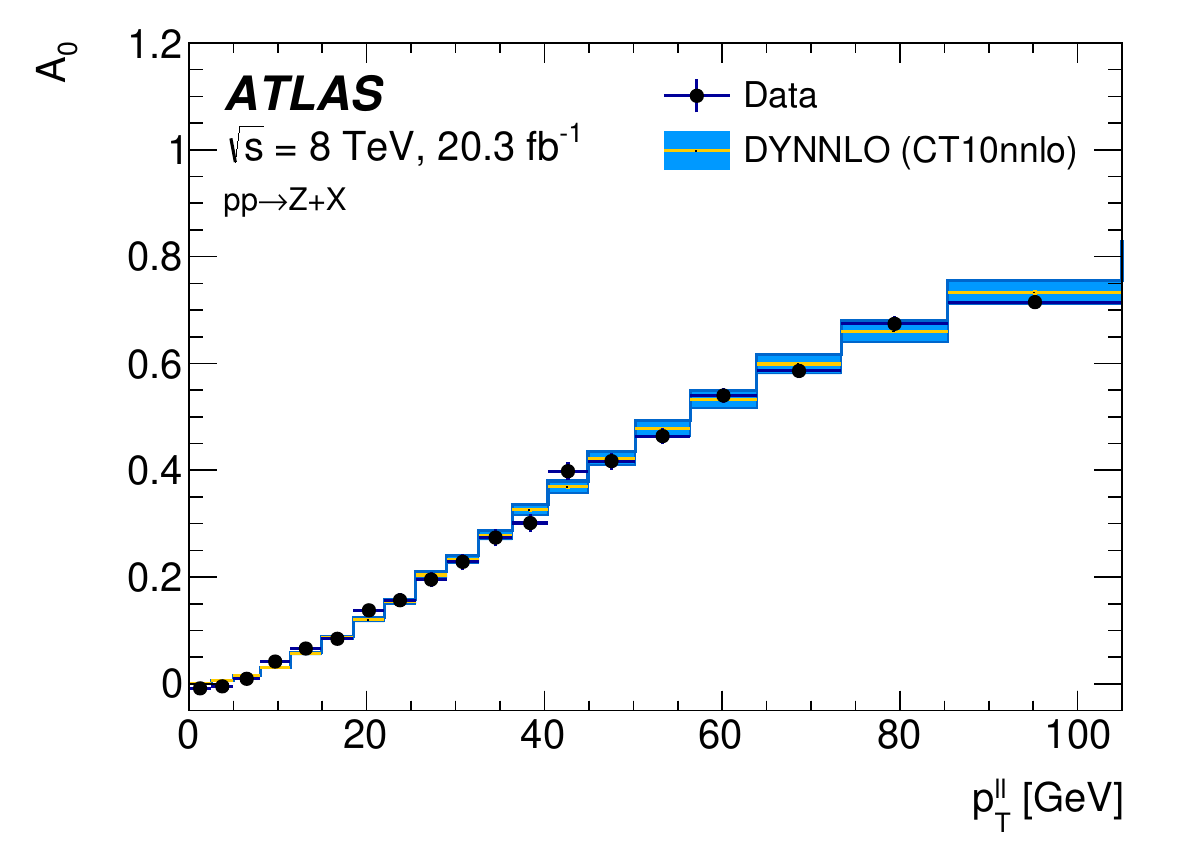}}
    \subfloat[]{\includegraphics[width=0.49\textwidth]{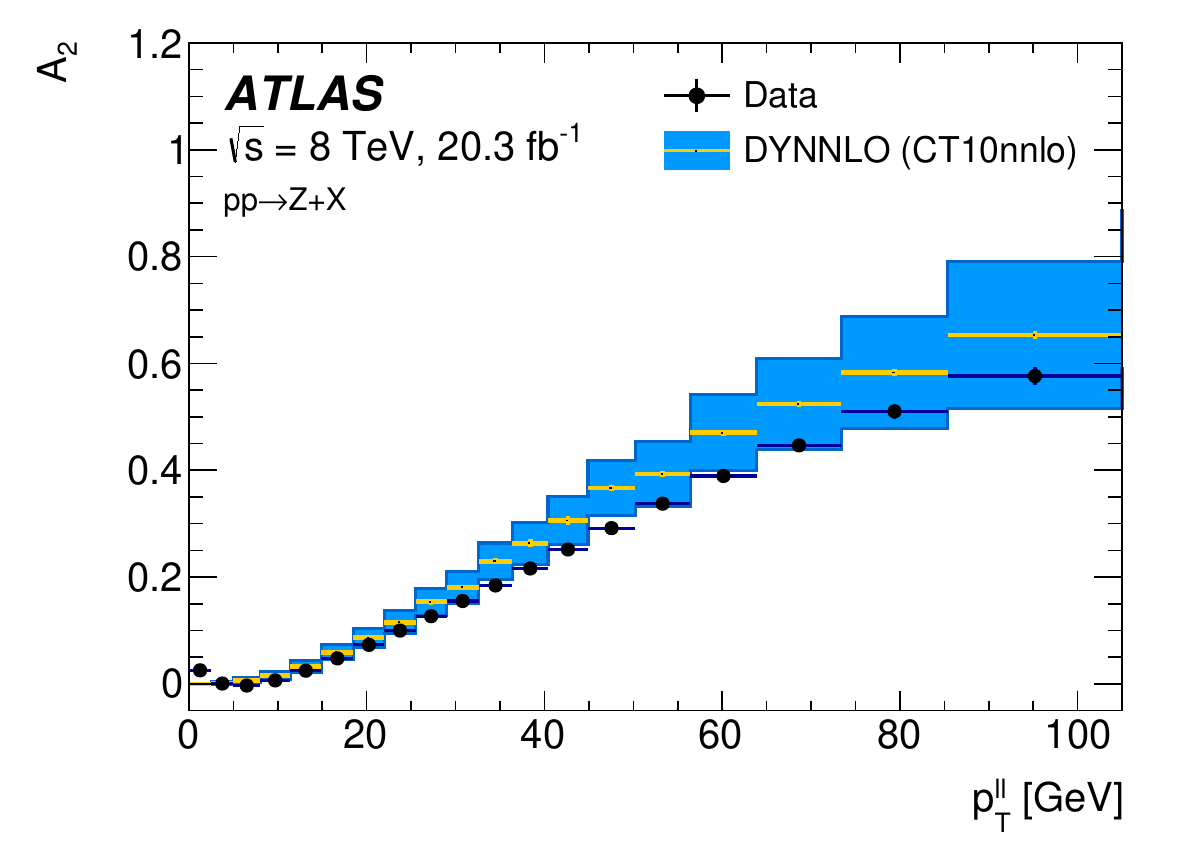}}
  \end{center}
  \caption{The (a) $A_0$ and (b) $A_2$ angular
    coefficients in $Z$-boson events as a function of $\pt^{\ell\ell}$~\cite{Aad:2016izn}.
    The measured coefficients are compared to the DYNNLO predictions using the CT10nnlo PDF set. The error bars
    show the total experimental uncertainties, and the bands show
    the uncertainties assigned to the DYNNLO predictions.
  \label{fig:CompAiCoeff}}
\end{figure}

The predictions of the angular coefficients in $Z$-boson events are compared to the ATLAS measurement at $\sqrt{s}=8\TeV$~\cite{Aad:2016izn}.
Good agreement between the measurements and DYNNLO is observed for the relevant coefficients, except for $A_2$, where the measurement is significantly below the prediction.
As an example, Figure~\ref{fig:CompAiCoeff} shows the comparison for $A_0$ and $A_2$ as a function of $\pt^Z$. For $A_2$, an additional source of uncertainty
in the theoretical prediction is considered to account for the observed disagreement with data, as discussed in Section~\ref{subsubsec:Aiunc}.

\subsection{Uncertainties in the QCD modelling \label{sec:modunc}}

Several sources of uncertainty related to the perturbative and
non-perturbative modelling of the strong interaction affect the
dynamics of the vector-boson production and
decay~\cite{Krasny:2010vd,Krasny:2007cy,Fayette:2008wt,Bozzi:2015hha}.
Their impact on the measurement of $m_W$ is assessed through variations of the model parameters of the
predictions for the differential cross sections as functions of the boson rapidity,
transverse-momentum spectrum at a given rapidity, and angular
coefficients, which correspond to the second, third, and fourth terms
of the decomposition of Eq.~(\ref{eq:decomposition}), respectively.
The parameter variations used to estimate the uncertainties are propagated
to the simulated event samples by means of the reweighting procedure described in
Section~\ref{sec:reweight}. Table~\ref{tab:QCDunc} shows an overview
of the uncertainties due to the QCD modelling which are discussed below.

\begin{table}[tp]
  \begin{center}
    \begin{tabular}{lrrrrrr}
    \toprule
      $W$-boson charge        & \multicolumn{2}{c}{$W^+$} & \multicolumn{2}{c}{$W^-$} & \multicolumn{2}{c}{Combined}\\
      Kinematic distribution & $\pt^\ell$ & $\mt$   & $\pt^\ell$ & $\mt$ & $\pt^\ell$ & $\mt$ \\
      \midrule
      $\delta m_W$~[\MeV]\\
      \,\,\,\, Fixed-order PDF uncertainty      & $13.1$    & $14.9$       & $12.0$  &   $14.2$  & $8.0$ & $8.7$     \\
      \,\,\,\, AZ tune                          &  $3.0$    & $3.4$    & $3.0$  & $3.4$  & $3.0$ & $3.4$ \\ 
      \,\,\,\, Charm-quark mass                 &  $1.2$     & $1.5$       & $1.2$   & $1.5$     & $1.2$   & $1.5$     \\
      \,\,\,\, \pbox{20cm}{Parton shower $\mu_\textrm{F}$ with heavy-flavour decorrelation}            &  $5.0$    & $6.9$     & $5.0$     & $6.9$    & $5.0$     & $6.9$  \\
      \,\,\,\, Parton shower PDF uncertainty    &  $3.6$    & $4.0$     & $2.6$    & $2.4$    & $1.0$    & $1.6$     \\
      \,\,\,\, Angular coefficients            &  $5.8$     & $5.3$        & $5.8$     & $5.3$   & $5.8$     & $5.3$  \\
      \midrule
      \,\,\,\, Total                           &  15.9  &   18.1       &  14.8  &  17.2  &  11.6    &   12.9  \\
      \bottomrule
    \end{tabular}
    \caption{Systematic uncertainties in the $m_W$
      measurement due to QCD modelling, for the different kinematic distributions and $W$-boson charges. Except for the case of PDFs,
      the same uncertainties apply to $W^+$ and $W^-$. The fixed-order PDF uncertainty given for the separate $W^+$ and $W^-$ final states corresponds to the quadrature sum of the CT10nnlo uncertainty variations; the
      charge-combined uncertainty also contains a $3.8\MeV$ contribution from comparing CT10nnlo to CT14 and MMHT2014.\label{tab:QCDunc}} 
  \end{center}
\end{table}

\subsubsection{Uncertainties in the fixed-order predictions\label{subsubsec:FOunc}}

The imperfect knowledge of the PDFs affects the differential cross section as a function
of boson rapidity, the angular coefficients, and the $\pt^W$
distribution. The PDF contribution to the prediction uncertainty is estimated with the CT10nnlo PDF set by using the Hessian
method~\cite{Pumplin:2001ct}. There are 25 error eigenvectors, and a pair of PDF variations associated with
each eigenvector. Each pair corresponds to positive and negative
90\% CL excursions along the corresponding eigenvector.
Symmetric PDF uncertainties are defined as the mean value of the absolute 
positive and negative excursions corresponding to each pair of PDF
variations.
The overall uncertainty of the CT10nnlo PDF set is scaled to 68\% CL
by applying a multiplicative factor of 1/1.645.

The effect of PDF variations on the rapidity distributions and angular
coefficients are evaluated with DYNNLO, while their impact on the
$W$-boson $\pt$ distribution is evaluated using \PYTHIA 8 and by
reweighting event-by-event the PDFs of the hard-scattering process,
which are convolved with the LO matrix elements.
Similarly to other uncertainties which affect the $\pt^W$ distribution (Section~\ref{subsubsec:PSunc}),
only relative variations of the $\pt^W$ and $\pt^Z$ distributions
induced by the PDFs are considered. The PDF variations are applied simultaneously to the
boson rapidity, angular coefficients, and transverse-momentum
distributions, and the overall PDF uncertainty is evaluated with the Hessian method as described above.

Uncertainties in the PDFs are the dominant source of physics-modelling uncertainty,
contributing about $14$ and $13\MeV$ when averaging $\pt^\ell$ and \mt\ fits for $W^+$ and $W^-$, respectively. The
PDF uncertainties are very similar when using $\pt^\ell$ or \mt\ for the
measurement. They are strongly anti-correlated between positively and
negatively charged $W$ bosons, and the uncertainty is reduced to
$7.4\MeV$ on average for  $\pt^\ell$ and \mt\ fits, when combining opposite-charge categories.
The anti-correlation of the PDF uncertainties is due to the fact that the total light-quark sea PDF is
well constrained by deep inelastic scattering data, whereas the $u$-, $d$-, and $s$-quark
decomposition of the sea is less precisely known~\cite{Abramowicz:2015mha}. An increase
in the $\bar{u}$ PDF is at the expense of the $\bar{d}$ PDF, which
produces opposite effects in the longitudinal polarisation of
positively and negatively charged $W$ bosons~\cite{ATL-PHYS-PUB-2014-015}.

Other PDF sets are considered as alternative choices. The envelope of
values of $m_W$ extracted with the MMHT2014 and CT14
NNLO PDF sets is considered as an additional PDF uncertainty of $3.8\MeV$, which is added in quadrature after combining the $W^+$ and $W^-$ categories, leading to overall PDF
uncertainties of $8.0\MeV$ and $8.7\MeV$ for $\pt^\ell$ and \mt\ fits, respectively.

The effect of missing higher-order corrections on the NNLO predictions
of the rapidity distributions of $Z$ bosons, and the pseudorapidity
distributions of the decay leptons of $W$ bosons, is estimated by
varying the renormalisation and factorisation scales by factors of $0.5$ and
$2.0$ with respect to their nominal value $\mu_\textrm{R} = \mu_\textrm{F} = m_V$ in the
DYNNLO predictions. The corresponding relative uncertainty in the normalised
distributions is of the order of 0.1--0.3\%, and significantly
smaller than the PDF uncertainties. These uncertainties are expected
to have a negligible impact on the measurement of $m_W$, and are not considered further.

The effect of the LHC beam-energy uncertainty of 0.65\%~\cite{Wenninger:1546734} on the fixed-order predictions is
studied. Relative variations of 0.65\% around the nominal value of
$3.5\TeV$ are considered, yielding variations of the inclusive $W^+$
and $W^-$ cross sections of 0.6\% and 0.5\%, respectively.
No significant dependence as a function of lepton pseudorapidity is
observed in the kinematic region used for the measurement, and the
dependence as a function of $\pt^\ell$ and \mt{} is expected to be even
smaller. This uncertainty is not considered further.

\subsubsection{Uncertainties in the parton shower predictions\label{subsubsec:PSunc}}

Several sources of uncertainty affect the \PYTHIA 8 parton shower
model used to predict the transverse momentum of the $W$ boson.
The values of the AZ tune parameters, determined by
fits to the measurement of the $Z$-boson transverse momentum, are affected by
the experimental uncertainty of the measurement.
The corresponding uncertainties are propagated to the $\pt^W$ predictions through
variations of the orthogonal eigenvector components of the parameters error
matrix~\cite{STDM-2012-23}. 
The resulting uncertainty in $m_W$ is $3.0\MeV$ for the $\pt^\ell$
distribution, and $3.4\MeV$ for the $\mt$ distribution. In the present analysis, the impact of $\pt^W$ distribution uncertainties is in general smaller when using $\pt^\ell$ than when using \mt, as a
result of the comparatively narrow range used for the $\pt^\ell$ distribution fits.

Other uncertainties affecting predictions of the transverse-momentum
spectrum of the $W$ boson at a given rapidity, are propagated by
considering relative variations of the $\pt^W$ and $\pt^Z$
distributions.
The procedure is based on the assumption that model variations, when
applied to $\pt^Z$, can be largely reabsorbed into new values of the
AZ tune parameters fitted to the $\pt^Z$ data.
Variations that cannot be reabsorbed by the fit are excluded, since
they would lead to a significant disagreement of the prediction with
the measurement of $\pt^Z$.
The uncertainties due to model variations which are largely
correlated between $\pt^W$ and $\pt^Z$ cancel in this procedure.
In contrast, the procedure allows a correct estimation of the uncertainties 
due to model variations which are uncorrelated between $\pt^W$ and
$\pt^Z$, and which represent the only relevant sources of
theoretical uncertainties in the propagation of the QCD modelling from
$\pt^Z$ to $\pt^W$. 

Uncertainties due to variations of parton shower parameters that are
not fitted to the $\pt^Z$ measurement
include variations of the masses of the charm and bottom quarks, and
variations of the factorisation scale used for the QCD ISR.
The mass of the charm quark is varied in \textsc{Pythia 8}, conservatively, by $\pm 0.5\GeV$
around its nominal value of $1.5\GeV$. The resulting uncertainty
contributes $1.2\MeV$ for the $\pt^\ell$ fits, and $1.5\MeV$ for the $\mt$ fits.
The mass of the bottom quark is varied in \textsc{Pythia 8}, conservatively, by $\pm 0.8\GeV$
around its nominal value of $4.8\GeV$. The resulting variations have
a negligible impact on the transverse-momentum distributions of $Z$
and $W$ bosons, and are not considered further.

The uncertainty due to higher-order QCD corrections to the parton shower
is estimated through variations of the factorisation scale, $\mu_\textrm{F}$,
in the QCD ISR by factors of $0.5$ and $2.0$ with respect to the central
choice $\mu_\textrm{F}^2 = p_{\textrm{T},0}^2 + \pt^2$, where $p_{\textrm{T},0}$ is an infrared
cut-off, and \pt{} is the evolution variable of the parton shower~\cite{Sjostrand:2004ef}.
Variations of the renormalisation scale in the QCD ISR are equivalent
to a redefinition of $\alpha_{\mathrm s}(m_Z)$ used for the QCD ISR, which is fixed
from the fits to the $\pt^Z$ data. As a consequence, variations of the ISR
renormalisation scale do not apply when estimating the uncertainty in the
predicted $\pt^W$ distribution.

Higher-order QCD corrections are expected to be largely correlated between
$W$-boson and $Z$-boson production induced by the light quarks, $u$,
$d$, and $s$, in the initial state.
However, a certain degree of decorrelation between $W$- and $Z$-boson
transverse-momentum distributions is expected, due to the different
amounts of heavy-quark-initiated production, where heavy refers to
charm and bottom flavours.
The physical origin of this decorrelation can be ascribed to the
presence of independent QCD scales corresponding to the three-to-four
flavours and four-to-five flavours matching scales $\mu_c$ and $\mu_b$
in the variable-flavour-number scheme PDF evolution~\cite{Bonvini:2016fgf}, which are
of the order of the charm- and bottom-quark masses, respectively.
To assess this effect, the variations of $\mu_\textrm{F}$ in
the QCD ISR are performed simultaneously for all light-quark $q\bar{q} \to W,Z$ processes,
with $q = u,d,s$, but independently for each of the $c\bar{c} \to Z$,
$b\bar{b} \to Z$, and $c\bar{q} \to W$ processes, where $q = d,s$. The effect of the $c\bar{q} \to W$ variations on the determination of $m_W$ is reduced by a factor of two, to account for the
presence of only one heavy-flavour quark in the initial state. 
The resulting uncertainty in $m_W$ is $5.0\MeV$ for the $\pt^\ell$
distribution, and $6.9\MeV$ for the $\mt$ distribution. Since the
$\mu_\textrm{F}$ variations affect all the branchings of the shower
evolution and not only vertices involving heavy quarks, this procedure is expected to
yield a sufficient estimate of the $\mu_{c,b}$-induced decorrelation between the $W$- and
$Z$-boson $\pt$ distributions.
Treating the $\mu_\textrm{F}$ variations as correlated between all quark
flavours, but uncorrelated between $W$- and $Z$-boson production,
would yield a systematic uncertainty in $m_W$ of approximately
30~\MeV.

The predictions of the \textsc{Pythia 8} MC generator include a
reweighting of the first parton shower emission to the leading-order
$W$+jet cross section, and do not include matching corrections to the
higher-order $W$+jet cross section. As discussed in
Section~\ref{sec:ptwxcheck}, predictions matched to the NLO $W$+jet
cross section, such as \PowhegMinlo and DYRes, are in disagreement
with the observed $u^\ell_\parallel$ distribution and cannot be used
to provide a reliable estimate of the associated uncertainty.
The $u^\ell_\parallel$ distribution, on the other hand, validates the \PYTHIA8 AZ prediction and its uncertainty, which gives confidence that missing
higher-order corrections to the $W$-boson \pt{} distribution are small
in comparison to the uncertainties that are already included, and can
be neglected at the present level of precision.

The sum in quadrature of the experimental uncertainties of the AZ tune
parameters, the variations of the mass of the charm quark, and the
factorisation scale variations, leads to uncertainties on $m_W$ of $6.0\MeV$ and
$7.8\MeV$ when using the $\pt^\ell$ distribution and the $\mt$ distribution, respectively.
These sources of uncertainty are taken as fully correlated between the
electron and muon channels, the positively and negatively charged
$W$-boson production, and the $|\eta_\ell|$ bins.

The \PYTHIA 8 parton shower simulation employs the CTEQ6L1 leading-order PDF
set. An additional independent source of PDF-induced uncertainty in the
$\pt^W$ distribution is estimated by comparing several choices of the
leading-order PDF used in the parton shower, corresponding to the CT14lo, MMHT2014lo and NNPDF2.3lo~\cite{Ball:2012cx} PDF sets.
The PDFs which give the largest deviation from the nominal ratio of the
$\pt^W$ and $\pt^Z$ distributions are used to estimate the uncertainty.
This procedure yields an uncertainty of about $4\MeV$ for $W^+$, and of about $2.5\MeV$ for $W^-$.
Similarly to the case of fixed-order PDF uncertainties, there is a
strong anti-correlation between positively and negatively charged $W$
bosons, and the uncertainty is reduced to about $1.5\MeV$ when combining positive- and
negative-charge categories.

The prediction of the $\pt^W$ distribution relies on the \pt-ordered
parton shower model of the \PYTHIA 8 MC generator. In order to assess
the impact of the choice of parton shower model on the determination of $m_W$, the
\PYTHIA 8 prediction of the ratio of the $\pt^W$ and $\pt^Z$
distributions is compared to the corresponding prediction of the
\textsc{Herwig 7} MC generator~\cite{Bellm:2015jjp,Bahr:2008pv}, which
implements an angular-ordered parton shower model.
Differences between the \PYTHIA 8 and \textsc{Herwig 7} predictions
are smaller than the uncertainties in the \PYTHIA 8
prediction, and no additional uncertainty is considered.

\subsubsection{Uncertainties in the angular coefficients\label{subsubsec:Aiunc}}

The full set of angular coefficients can only be measured precisely
for the production of $Z$ bosons.
The accuracy of the NNLO predictions of the angular coefficients is
validated by comparison to the $Z$-boson measurement, and extrapolated to $W$-boson
production assuming that NNLO predictions have similar accuracy for the $W$- and $Z$-boson processes.
The ATLAS measurement of the angular coefficients in $Z$-boson
production at a centre-of-mass energy of $\sqrt{s} =
8\TeV$~\cite{Aad:2016izn} is used for this validation.
The $O(\alpha_{\mathrm s}^2)$ predictions, evaluated with DYNNLO, are in
agreement with the measurements of the angular coefficients within the
experimental uncertainties, except for the measurement of $A_2$ as a
function of $Z$-boson \pt.

Two sources of uncertainty affecting the modelling of the angular
coefficients are considered, and propagated to the $W$-boson predictions.
One source is defined from the experimental uncertainty of the
$Z$-boson measurement of the angular coefficients which is used to validate the NNLO predictions.
The uncertainty in the corresponding $W$-boson predictions is
estimated by propagating the experimental uncertainty of the $Z$-boson
measurement as follows.
A set of pseudodata distributions are obtained by fluctuating the angular coefficients
within the experimental uncertainties,
preserving the correlations between the different measurement bins for the different coefficients.
For each pseudoexperiment, the differences in the $A_i$ coefficients
between fluctuated and nominal $Z$-boson measurement results are
propagated to the corresponding coefficient in $W$-boson
production. The corresponding uncertainty is defined from the standard
deviation of the $m_W$ values as estimated from the pseudodata
distributions.

The other source of uncertainty is considered to account for the
disagreement between the measurement and the NNLO QCD predictions
observed for the $A_2$ angular coefficient as a function of the
$Z$-boson \pt (Figure~\ref{fig:CompAiCoeff}).
The corresponding uncertainty in $m_W$ is estimated by propagating the difference in $A_2$ between the $Z$-boson measurement and the theoretical prediction to the corresponding coefficient in $W$-boson production.
The corresponding uncertainty in the measurement of $m_W$ is
$1.6\MeV$ for the extraction from the $\pt^\ell$ distribution.
Including this contribution, total uncertainties of $5.8\MeV$ and $5.3\MeV$ due to
the modelling of the angular coefficients are estimated in the
determination of the $W$-boson mass from the $\pt^\ell$ and \mt{}
distributions, respectively. The uncertainty is dominated by the
experimental uncertainty of the $Z$-boson measurement used
to validate the theoretical predictions.

\section{Calibration of electrons and muons \label{sec:objcalib}}

Any imperfect calibration of the detector response to electrons and
muons impacts the measurement of the $W$-boson
mass, as it affects the position and shape of the Jacobian
edges reflecting the value of $m_W$. In addition, the $\pt^\ell$ and $\mt$ distributions are broadened by
the electron-energy and muon-momentum resolutions. Finally, the
lepton-selection efficiencies depend on the lepton pseudorapidity and
transverse momentum, further modifying these
distributions. Corrections to the detector response are derived from
the data, and presented below. In most cases, the corrections are
applied to the simulation, with the exception of the muon sagitta bias corrections and electron energy response corrections, which
are applied to the data. Backgrounds to the selected $Z\rightarrow \ell\ell$ samples are taken into account using the same procedures as discussed 
in Section~\ref{sec:crosschecks}. Since the $Z$ samples are used separately for momentum calibration and efficiency measurements, as well as for the recoil response corrections discussed in
Section~\ref{sec:recoilcalib}, correlations among the corresponding uncertainties can appear. These correlations were investigated and found to be negligible.

\subsection{Muon momentum calibration \label{sec:muonscale}}

As described in Section~\ref{sec:reco}, the kinematic parameters of
selected muons are determined from the associated inner-detector tracks.
The accuracy of the momentum measurement is limited by imperfect knowledge of the detector alignment and resolution, of the
magnetic field, and of the amount of passive material in the detector.

Biases in the reconstructed muon track momenta are classified as radial or sagitta biases.
The former originate from detector movements along the particle trajectory 
and can be corrected by an $\eta$-dependent, charge-independent momentum-scale correction. 
The latter typically originate from curl distortions or linear twists of the
detector around the $z$-axis~\cite{ATLAS-CONF-2012-141}, 
and can be corrected with $\eta$-dependent correction factors proportional
to $q\times \pt^\ell$, where $q$ is the charge of the muon. The momentum scale and resolution corrections are applied to the simulation, while the sagitta bias correction is applied to the data:

\begin{eqnarray}\label{eq:id_cor}
  \nonumber \pt^{\textrm{MC,corr}} &=&  \pt^{\textrm{MC}} \times \left[1 + \alpha(\eta,\phi)\right] \times \left[1 + \beta_{\textrm{curv}}(\eta) \cdot G(0,1) \cdot p_{\textrm{T}}^{\textrm{MC}}\right],\\
  \nonumber \pt^{\textrm{data,corr}} &=&  \frac  {\pt^{\textrm{data}}}{1 + q \cdot \delta(\eta,\phi) \cdot \pt^{\textrm{data}}},
\end{eqnarray}

where $\pt^{\textrm{data,MC}}$ is the uncorrected muon
transverse momentum in data and simulation, $G(0,1)$ are
normally distributed random variables with mean zero and unit
width, and $\alpha$, $\beta_{\textrm{curv}}$, and $\delta$ represent
the momentum scale, intrinsic resolution and sagitta bias corrections, respectively.
Multiple-scattering contributions to the resolution are relevant at low \pt, and the
corresponding corrections are neglected.

Momentum scale and resolution corrections are derived using
$Z\to\mu\mu$ decays, following the method described in
Ref.~\cite{PERF-2014-05}. Template histograms of the dimuon invariant
mass are constructed from the simulated event samples, including momentum
scale and resolution corrections in narrow steps within a range
covering the expected uncertainty. The optimal values of $\alpha$ and
$\beta_{\mathrm{curv}}$ are determined by means of a $\chi^2$
minimisation, comparing data and simulation in the range of twice the standard deviation on each side of the mean value of the invariant mass distribution.
In the first step, the corrections are derived
by averaging over $\phi$, and for 24 pseudorapidity bins in the range
$-2.4 < \eta_\ell < 2.4$. In the second iteration, $\phi$-dependent
correction factors are evaluated in coarser bins of $\eta_\ell$.
The typical size of $\alpha$ varies from $-$0.0005 to $-$0.0015 depending on 
$\eta_\ell$, while $\beta_{\textrm{curv}}$ values increase from $0.2 \TeV^{-1}$ in the barrel 
to $0.6 \TeV^{-1}$ in the high $\eta_\ell$ region. Before the correction, the $\phi$-dependence 
has an amplitude at the level of 0.1\%.

The $\alpha$ and $\beta_{\mathrm{curv}}$ corrections are sensitive to the 
following aspects of the calibration procedure, which are considered for the systematic uncertainty: the choice of the fitting
range, methodological biases, background contributions, theoretical
modelling of $Z$-boson production, non-linearity of the corrections,
and material distribution in the ID.
The uncertainty due to the choice of fitting range is
estimated by varying the range by $\pm10\%$, and repeating the procedure. 
The uncertainty due to the fit methodology is estimated by comparing the template fit results with an
alternative approach, based on an iterative $\chi^2$ minimisation.
Background contributions from gauge-boson pair and top-quark pair
production are estimated using the simulation. The uncertainty in
these background contributions is evaluated by varying their normalisation within the theoretical uncertainties on the production cross sections.
The uncertainty in the theoretical modelling of $Z$-boson
production is evaluated by propagating the effect of electroweak
corrections to QED FSR, QED radiation of fermion pairs, and other NLO
electroweak corrections described in Section~\ref{sec:physmodew}.
The experimental uncertainty in the value of the $Z$-boson mass used
as input is also accounted for. These sources of uncertainty are
summed in quadrature, yielding an uncertainty $\delta\alpha$ in the muon momentum
scale correction of approximately $0.5 \times 10^{-4}$;
these sources are considered fully correlated across muon pseudorapidity.

\begin{figure}
  \centering
  \subfloat[]{\includegraphics[width=0.49\textwidth]{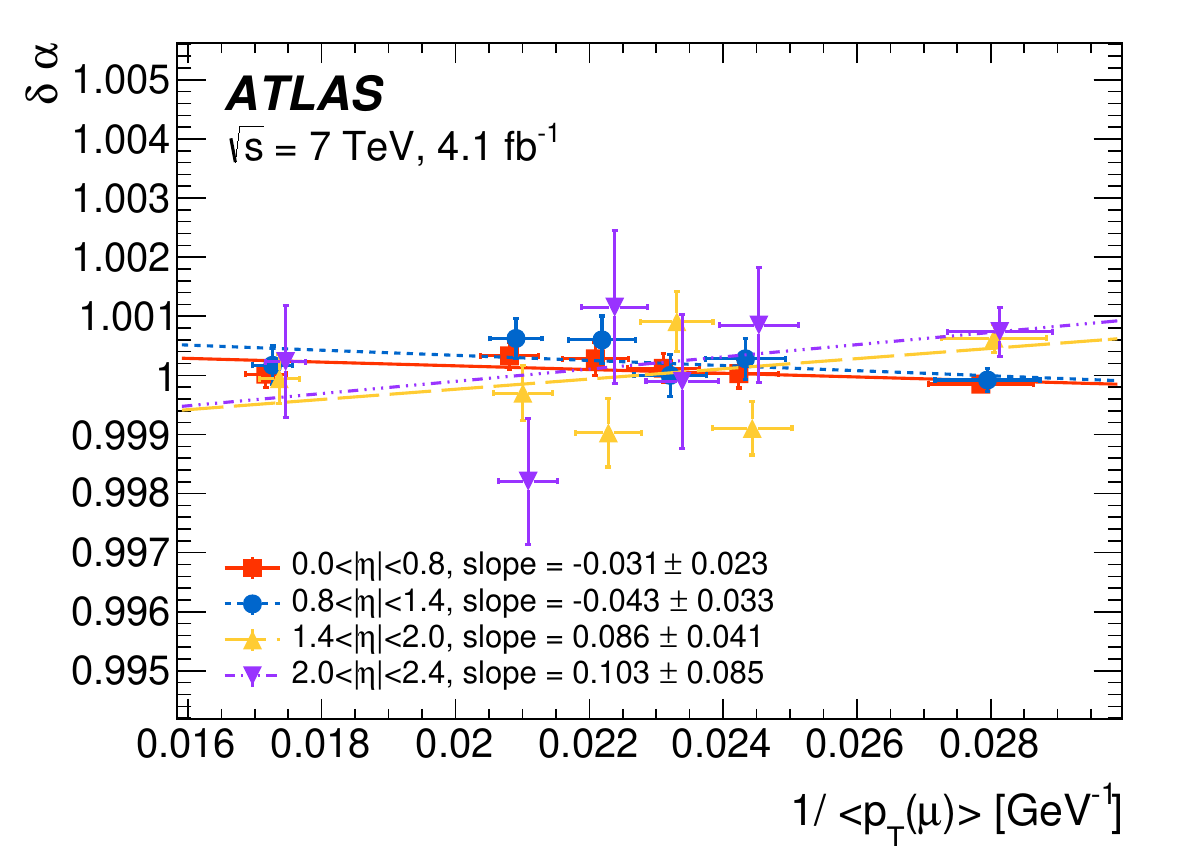}\label{fig:MuonCalib1b}}
  \subfloat[]{\includegraphics[width=0.49\textwidth]{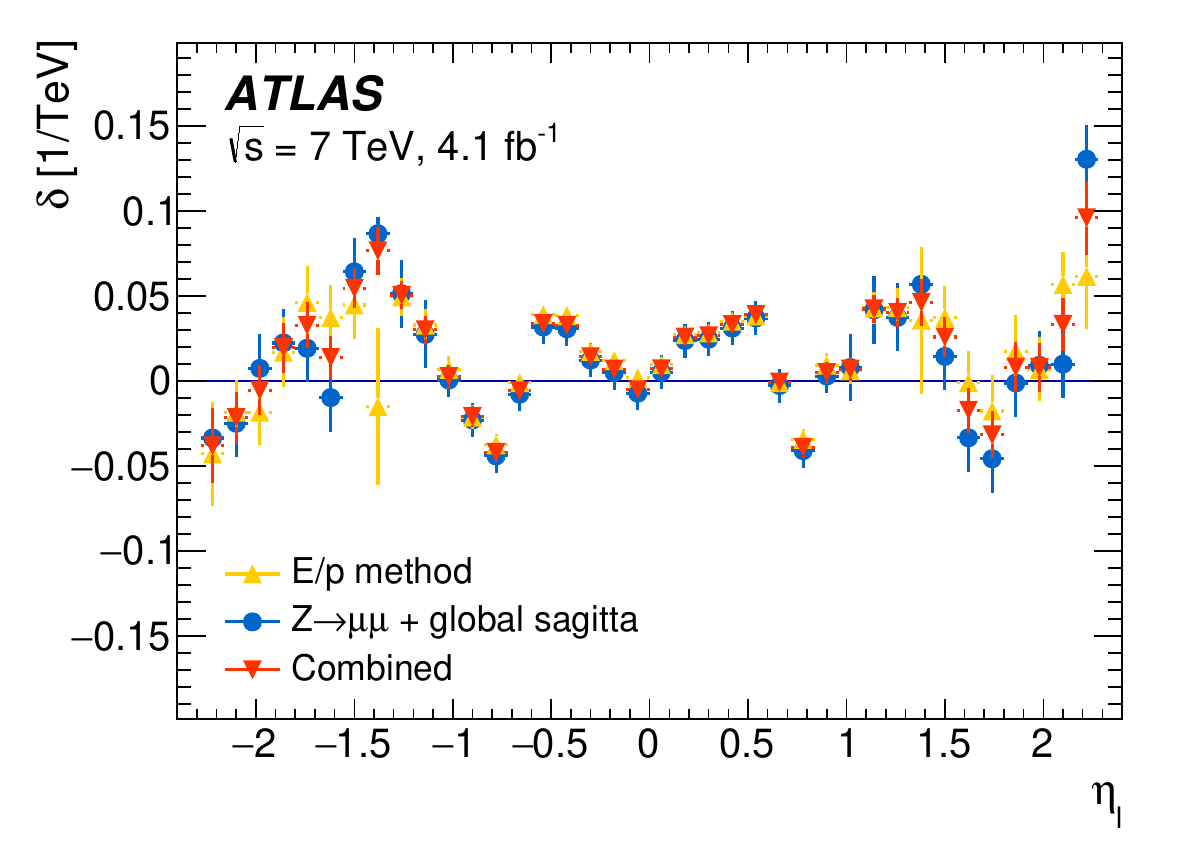}\label{fig:MuonCalib1a}}
  \caption{(a) Residual muon momentum scale corrections as a function of muon $1/p_{\textrm{T}}$ in four pseudorapidity regions, obtained with $Z\rightarrow \mu\mu$
    events. The points are fitted using a linear function which parameterises the extrapolation of the muon momentum scale correction from $Z$ to $W$ events, as explained in the text. The error bars on
    the points show statistical uncertainties only. (b) Sagitta bias, $\delta$, as a function of $\eta_\ell$ averaged over $\phi_\ell$. The results are obtained with the $Z\rightarrow \mu\mu$ and $E/p$ methods and the combination of the
    two. The results obtained with the $Z\rightarrow \mu\mu$ method are corrected for the global sagitta bias. The $E/p$ method uses electrons from $W\rightarrow e\nu$ decays. The two measurements are combined assuming they are uncorrelated. The error bars on the
    points show statistical uncertainties only.} 
  \label{fig:MuonCalib1}
\end{figure}

The systematic uncertainty in the muon momentum scale
due to the extrapolation from the $Z\to\mu\mu$ momentum range to the
$W\to\mu\nu$ momentum range is estimated by evaluating momentum-scale
corrections as a function of $1/\pt$ for muons in various $|\eta|$ ranges. The
extrapolation uncertainty $\delta \alpha$ is parameterised as follows:
\begin{eqnarray}
\nonumber \delta\alpha = p_0 +\frac{p_1}{\cropdelims[1]\left<{\pt^\ell(W)}\right>} ,
\end{eqnarray}
where $\cropdelims[1]\left<{\pt^\ell(W)}\right>$ is the average \pt of muons in $W$-boson events,
and $p_0$ and $p_1$ are free parameters.
If the momentum-scale corrections are independent of $1/\pt$, the fitting parameters
are expected to be $p_0=1$ and $p_1=0$. Deviations of $p_1$ from zero indicate a possible momentum
dependence.
The fitted values of $\delta\alpha$ are shown in Figure~\ref{fig:MuonCalib1b}, and are consistent with one, within
two standard deviations of the statistical error. The corresponding
systematic uncertainty in $m_W$ is defined assuming, in each bin of
$|\eta|$, a momentum non-linearity given by the larger of the fitted
value of $p_1$ and its uncertainty.
This source of uncertainty is considered uncorrelated across
muon pseudorapidity given that $p_1$ is dominated by statistical fluctuations.
The effect of the imperfect knowledge of the material in the ID is studied using simulated event samples including an increase of the ID material by 10\%, according to the uncertainty estimated in Ref.~\cite{PERF-2011-08}. The impact of this variation is found to be negligible in comparison with the uncertainties discussed above.

Two methods are used for the determination of the sagitta bias $\delta$. The first method exploits \Zmm{} events.
Muons are categorised according to their charge and pseudorapidity,
and for each of these categories, the position of the peak in the dimuon invariant
mass distribution is determined for data and simulation. The procedure
allows the determination of the charge dependence of the momentum
scale for $p_{\mathrm T}$ values of approximately $42\GeV$, which corresponds to the average
transverse momentum of muons from $Z$-boson decays.
The second method exploits identified electrons in a sample of
$W\rightarrow e\nu$ decays. It is based on the ratio of the measured
electron energy deposited in the calorimeter, $E$, to the electron
momentum, $p$, measured in the ID. A clean sample of $W\rightarrow
e\nu$ events with tightly identified electrons~\cite{PERF-2013-03} is selected.
Assuming that the response of the electromagnetic calorimeter is
independent of the charge of the incoming particle, charge-dependent
ID track momentum biases are extracted from the average differences in
$E/p$ for electrons and positrons~\cite{ATLAS-CONF-2012-141}. 
This method benefits from a larger event sample compared to the first method, and allows the determination of
charge-dependent corrections for $p_{\mathrm T}$ values of approximately $38\GeV$, which corresponds to the average transverse
momentum of muons in $W$-boson decays. The sagitta bias correction factors are derived using both methods separately in 40 $\eta$ bins and 40 $\phi$ bins. The results are found to agree within uncertainties and are combined, as illustrated in Figure~\ref{fig:MuonCalib1a}. The combined correction uncertainty
is dominated by the finite size of the event samples.

\begin{figure}
  \centering
  \includegraphics[width=0.7\textwidth]{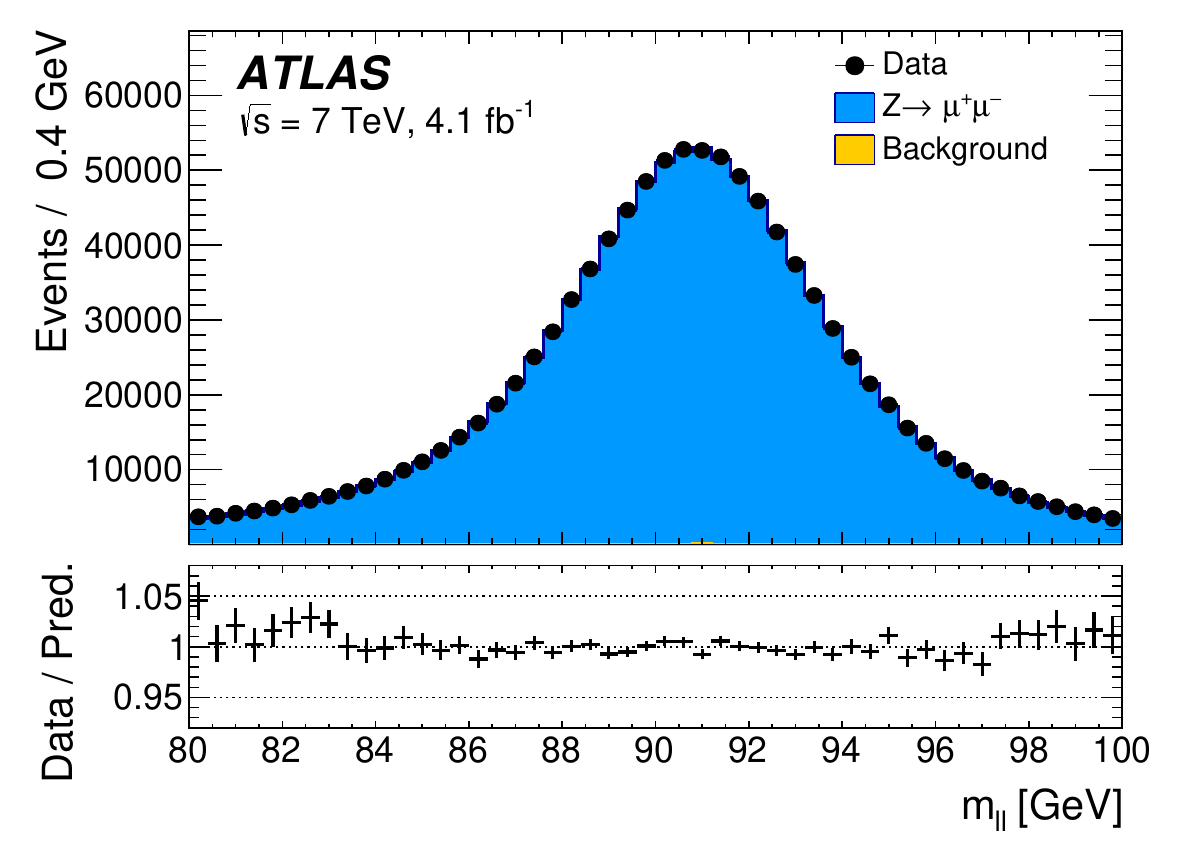}
  \caption{Dimuon invariant mass distribution in $Z\rightarrow \mu\mu$ events. The data are compared to the simulation including signal and background contributions.
    Corrections for momentum scale and resolution, and for reconstruction, isolation, and trigger
    efficiencies are applied to the muons in the simulated events. Background events contribute less than 0.2\% of the observed distribution. The lower panel shows the data-to-prediction ratio,
    with the error bars showing the statistical uncertainty.}
  \label{fig:MuonCalib2}
\end{figure}

Figure~\ref{fig:MuonCalib2} shows the dimuon invariant mass
distribution of \Zmm{} decays in data and simulation, after applying all corrections. Table~\ref{tab:muoncalibsummary} summarises the effect of the muon momentum scale and resolution
uncertainties on the determination of $m_W$. 
The dominant systematic uncertainty in the momentum scale is due to the 
extrapolation of the correction from the $Z$-boson
momentum range to the $W$-boson momentum range. The
extrapolation uncertainty $\delta\alpha$ is $(2$--$5)\times
10^{-5}$ for $|\eta_\ell|<2.0$, and $(4$--$7)\times 10^{-4}$ for $|\eta_\ell|>2.0$.
Systematic uncertainties from other sources are relatively small.
The systematic uncertainty of the resolution corrections is dominated
by the statistical uncertainty of the $Z$-boson event sample, and
includes a contribution from the imperfect closure of the method. The
latter is defined from the residual difference between the standard
deviations of the dimuon invariant mass in data and simulation,
after applying resolution corrections.

\subsection{Muon selection efficiency}

The selection of muon candidates in $W\rightarrow\mu\nu$ and
$Z\rightarrow\mu\mu$ events requires an isolated track
reconstructed in the inner detector and in the muon spectrometer.
In addition, the events are required to pass the muon trigger selection.
Differences in the efficiency of the reconstruction and selection
requirements between data and simulation can introduce a
systematic shift in the measurement of the $W$-boson mass, and have to
be corrected. In particular, the extraction of $m_W$ is sensitive to the dependence
of the trigger, reconstruction and isolation efficiencies on the
muon \pt{} and on the projection of the recoil on the
lepton transverse momentum, $u^\ell_\parallel$.

For muons with \pt{} larger than approximately $15\GeV$ the detector simulation
predicts constant efficiency as a function of $\pt^\ell$, both for the muon trigger selection and the track reconstruction.
In contrast, the efficiency of the isolation
requirement is expected to vary as a function of $\pt^\ell$ and
$u^\ell_\parallel$. The efficiency corrections also affect the muon selection inefficiency, and hence the
estimation of the $Z\rightarrow\mu\mu$ background, which contributes to the
$W\rightarrow\mu\nu$ selection when one of the decay muons fails the
muon reconstruction or kinematic selection requirements.

Corrections to the muon reconstruction, trigger and isolation efficiencies are estimated by applying the
tag-and-probe method~\cite{PERF-2014-05} to $Z\rightarrow\mu\mu$ events in data and simulation. 
Efficiency corrections are defined as the ratio of efficiencies
evaluated in data to efficiencies evaluated in simulated events.
The corrections are evaluated as functions of two variables,
$\pt^\ell$ and $u^\ell_\parallel$, and in various regions of the detector. The
detector is segmented into regions corresponding to the $\eta$ and
$\phi$ coverage of the muon spectrometer. The subdivision accounts for
the geometrical characteristics of the detector, such as the presence of
uninstrumented or transition regions.
The dependence of the efficiencies on $u^\ell_\parallel$ agree in data and simulation. Therefore, the muon efficiency corrections are 
evaluated only as a function of $\pt^\ell$ and $\eta_\ell$, separately
for positive and negative muon charges. The final efficiency correction factors are linearly
interpolated as a function of muon \pt. No significant \pt-dependence
of the corrections is observed in any of the detector regions.

\begin{figure}
  \centering
  \subfloat[]{\includegraphics[width=0.49\textwidth]{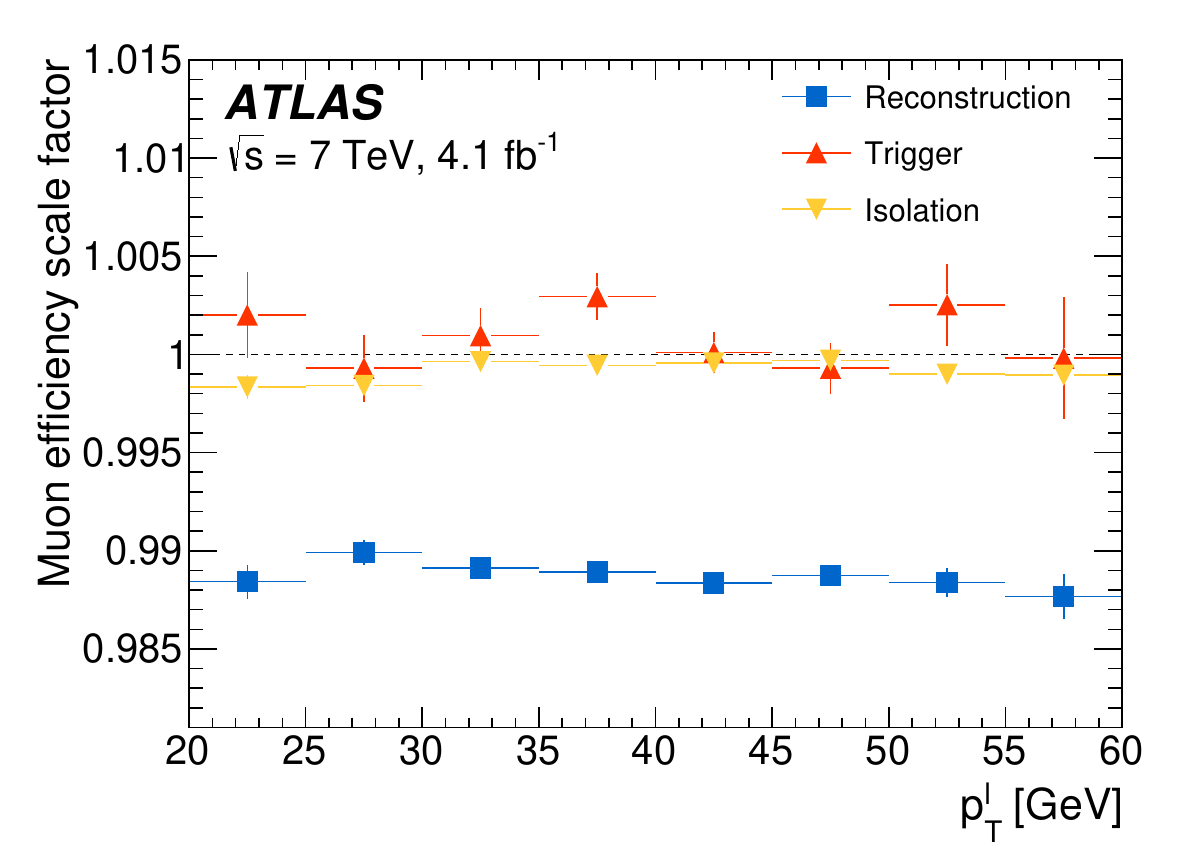}\label{fig:MuonCalib3a}}
  \subfloat[]{\includegraphics[width=0.49\textwidth]{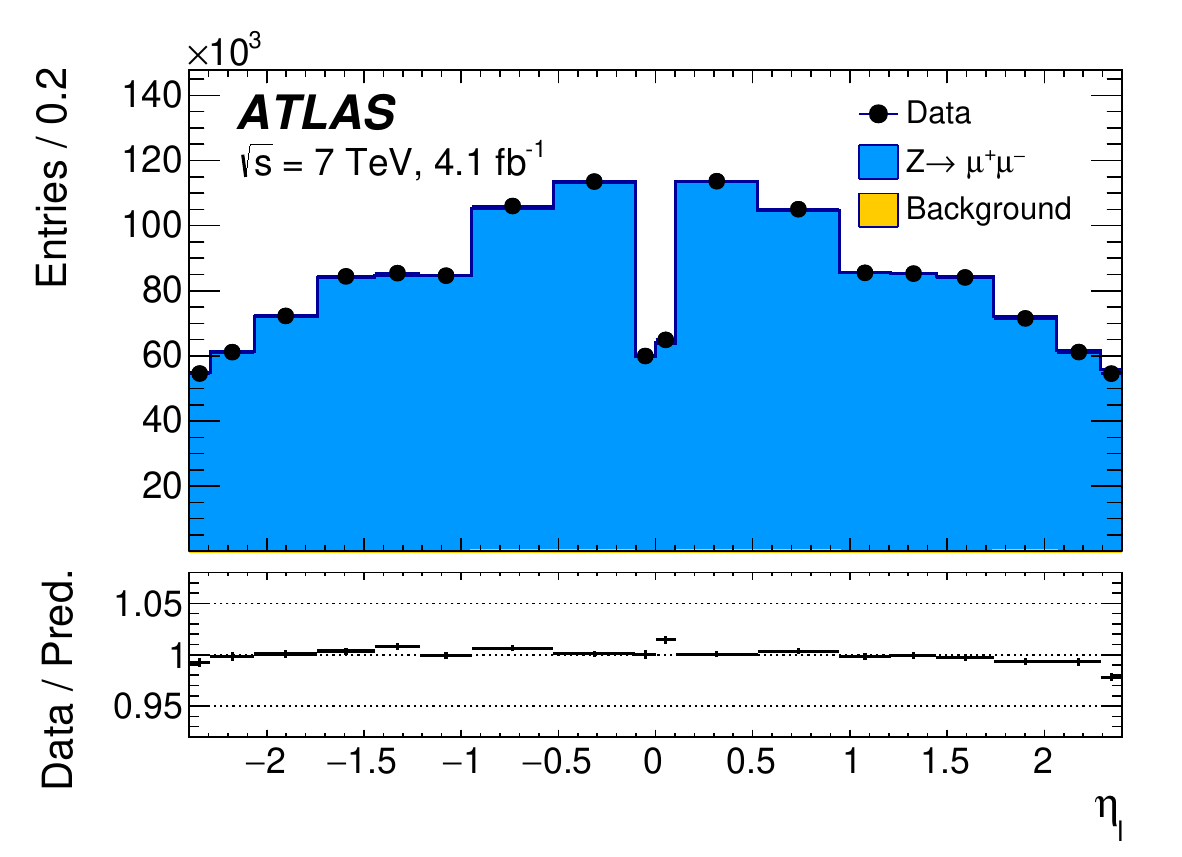}\label{fig:MuonCalib3b}}
  \caption{(a) Scale factors for the muon reconstruction, trigger and isolation efficiency obtained with the tag and probe method as a function of the muon $p_{\mathrm T}$. Scale factors for the trigger
    efficiency are averaged over two data-taking periods as explained in the text. The error bars on the points show statistical uncertainties only. (b) Distribution of the reconstructed muons $\eta$ in
    \Zmm{} events. The data are compared to the simulation including signal and background contributions. Corrections for momentum scale and resolution, and for reconstruction, isolation, and trigger
    efficiencies are applied to the muons in the simulated events. Background events contribute less than 0.2\% of the observed distribution. The lower panel shows the data-to-prediction ratio, with the error bars showing the statistical uncertainty.}
  \label{fig:MuonCalib3}
\end{figure}

The selection of tag-and-probe pairs from $Z\rightarrow\mu\mu$ events is based on the kinematic requirements described in Section~\ref{sec:eventsel}. The tag muon is required to be a combined and energy-isolated muon candidate (see Section~\ref{sec:reco}) which fulfils the muon
trigger requirements. The selection requirements applied to the
probe muon candidate differ for each efficiency
determination: the selection requirement for which the efficiency is
determined is removed from the set of requirements applied to the
probe muon.
All the efficiency corrections are derived inclusively for the full
data set, with the exception of the trigger, for which they
are derived separately for two different data-taking periods. The resulting scale factors are shown as a function of $\pt^\ell$ and averaged over $\eta_\ell$ in Figure~\ref{fig:MuonCalib3a}. The
trigger and isolation efficiency corrections are typically below 0.3\%, while the reconstruction efficiency correction is on average about 1.1\%. The corresponding impact on muon selection
inefficiency reaches up to about 20\%.

The quality of the efficiency corrections is evaluated by applying the corrections to the $Z\rightarrow\mu\mu$ simulated sample, and comparing the simulated kinematic
distributions to the corresponding distributions in data. Figure~\ref{fig:MuonCalib3b} illustrates this procedure for the $\eta_\ell$ distribution. Further distributions are shown in Section~\ref{sec:crosschecks}.

The dominant source of uncertainty in the determination of the muon
efficiency corrections is the statistical uncertainty of the
$Z$-boson data sample. The largest sources of systematic uncertainty
are the multijet background contribution and the momentum-scale
uncertainty. The corresponding uncertainty in the measurement of
$m_W$ is approximately 5~\MeV. The ID tracking efficiencies for muon candidates are above
99.5\% without any significant $p_{\mathrm T}$ dependence, and the associated
uncertainties are not considered further.
An overview of the uncertainties associated with the muon efficiency
corrections is shown in Table~\ref{tab:muoncalibsummary}.

\begin{table*}[tp]
  \centering
  \begin{tabular}[h!]{lrrrrrrrrrr}
    \toprule
    $|\eta_\ell|$ range  & \multicolumn{2}{r}{$[0.0,0.8]$} & \multicolumn{2}{r}{$[0.8,1.4]$} & \multicolumn{2}{r}{$[1.4,2.0]$} & \multicolumn{2}{r}{$[2.0,2.4]$} & \multicolumn{2}{r}{Combined}\\
    Kinematic distribution      & $\pt^\ell$ & $\mt$ & $\pt^\ell$ & $\mt$ & $\pt^\ell$ & $\mt$ & $\pt^\ell$ & $\mt$ & $\pt^\ell$ & $\mt$ \\
    \midrule
    $\delta m_W$~[\MeV]\\
    \,\,\,\, Momentum scale                        &  8.9 &  9.3 & 14.2 & 15.6 & 27.4 & 29.2 & 111.0 & 115.4 &  8.4 &  8.8 \\
    \,\,\,\, Momentum resolution                   &  1.8 &  2.0 &  1.9 &  1.7 &  1.5 &  2.2 &   3.4 &   3.8 &  1.0 &  1.2 \\
    \,\,\,\, Sagitta bias                          &  0.7 &  0.8 &  1.7 &  1.7 &  3.1 &  3.1 &   4.5 &   4.3 &  0.6 &  0.6 \\
    \,\,\,\, Reconstruction and \\
    \,\,\,\, isolation efficiencies                &  4.0 &  3.6 &  5.1 &  3.7 &  4.7 &  3.5 &   6.4 &   5.5 &  2.7 &  2.2 \\
    \,\,\,\, Trigger efficiency                    &  5.6 &  5.0 &  7.1 &  5.0 & 11.8 &  9.1 &  12.1 &   9.9 &  4.1 &  3.2 \\
    \midrule
    \,\,\,\, Total                                 & 11.4 & 11.4 & 16.9 & 17.0 & 30.4 & 31.0 & 112.0 & 116.1 &  9.8 &  9.7 \\
    \bottomrule
  \end{tabular}
  \caption{Systematic uncertainties in the $m_W$ measurement from muon calibration and efficiency corrections, for the different kinematic
    distributions and $|\eta_\ell|$ categories, averaged over lepton charge. The momentum-scale uncertainties include the effects of both the momentum scale and linearity corrections. Combined uncertainties are evaluated as described in Section~\ref{subsec:strategy}. \label{tab:muoncalibsummary}}
\end{table*}

\subsection{Electron energy response\label{sec:ElectronResponse}}

The electron-energy corrections and uncertainties are largely based on
the ATLAS Run~1 electron and photon calibration
results~\cite{PERF-2013-05}.
The correction procedure starts with the intercalibration of the first and second layers of
the EM calorimeter for minimum-ionising particles, using the energy deposits of muons in
$Z\rightarrow\mu\mu$ decays. After the intercalibration of the calorimeter layers,
the longitudinal shower-energy profiles of electrons and photons are used to determine the presampler energy scale and 
probe the passive material in front of the EM calorimeter, leading to
an improved description of the detector material distribution and providing estimates of the
residual passive material uncertainty. Finally, a dependence of the cell-level energy measurement on the read-out gain is observed in the second layer and corrected 
for. After these preliminary corrections, an overall energy-scale correction
is determined as a function of $\eta_\ell$ from $Z\rightarrow ee$ decays,
by comparing the reconstructed mass distributions in data and
simulation. Simultaneously, an effective constant term for the
calorimeter energy resolution is extracted by adjusting the width of
the reconstructed dielectron invariant mass distribution in
simulation to match the distribution in data. 

Uncertainties in the energy-response corrections arise from the
limited size of the $Z\rightarrow ee$ sample, from the physics modelling of the
resonance and from the calibration algorithm itself. Physics-modelling
uncertainties include uncertainties from missing higher-order
electroweak corrections (dominated by the absence of lepton-pair emissions in the
simulation) and from the experimental uncertainty in $m_Z$; these effects are taken fully correlated with the muon channel. Background contributions are
small and the associated uncertainty is considered
to be negligible. Uncertainties related to the calibration procedure are estimated by varying the
invariant mass range used for the calibration, and with a closure test. For the closure test, a
pseudodata sample of $Z\rightarrow ee$ events is obtained from the nominal sample by rescaling the electron energies by known $\eta$-dependent factors; the calibration algorithm is then applied, and the
measured energy corrections are compared with the input rescaling factors.

These sources of uncertainty constitute a subset of those listed
in Ref.~\cite{PERF-2013-05}, where additional variations were considered in
order to generalise the applicability of the $Z$-boson calibration
results to electrons and photons spanning a wide energy range. The effect of these uncertainties is averaged within the different $\eta_\ell$ categories. The overall relative energy-scale uncertainty, averaged over $\eta_\ell$, is
$9.4\times 10^{-5}$ for electrons from $Z$-boson decays.

In addition to the uncertainties in the energy-scale corrections
arising from the $Z$-boson calibration procedure, possible
differences in the energy response between electrons from $Z$-boson
and $W$-boson decays constitute a significant source of
uncertainty. The linearity of the response is affected by
uncertainties in the intercalibration of the layers and in the passive material and calorimeter
read-out corrections mentioned above. Additional uncertainties are assigned to
cover imperfect electronics pedestal subtraction affecting the energy
measurement in the cells of the calorimeter, and to the modelling of the
interactions between the electrons and the detector material in
\textsc{Geant4}. The contribution from these sources to the relative energy-scale uncertainty is $(3$--$12)\times
10^{-5}$ in each $\eta$ bin, and $5.4\times 10^{-5}$ when averaged
over the full $\eta$ range after taking into account the correlation between the $\eta$ bins.

\begin{figure}
  \centering
  \subfloat[]{\includegraphics[width=0.49\textwidth]{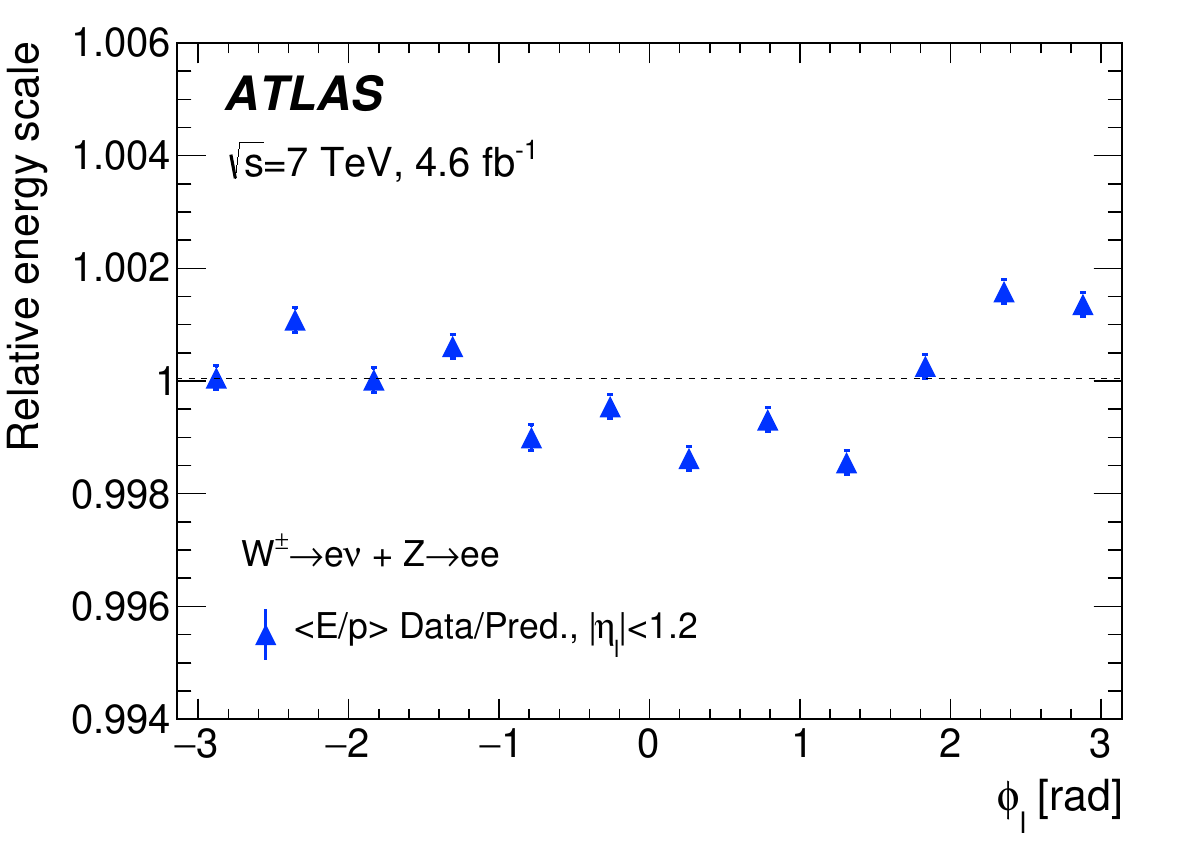}\label{fig:eleccalib1a}}
  \subfloat[]{\includegraphics[width=0.49\textwidth]{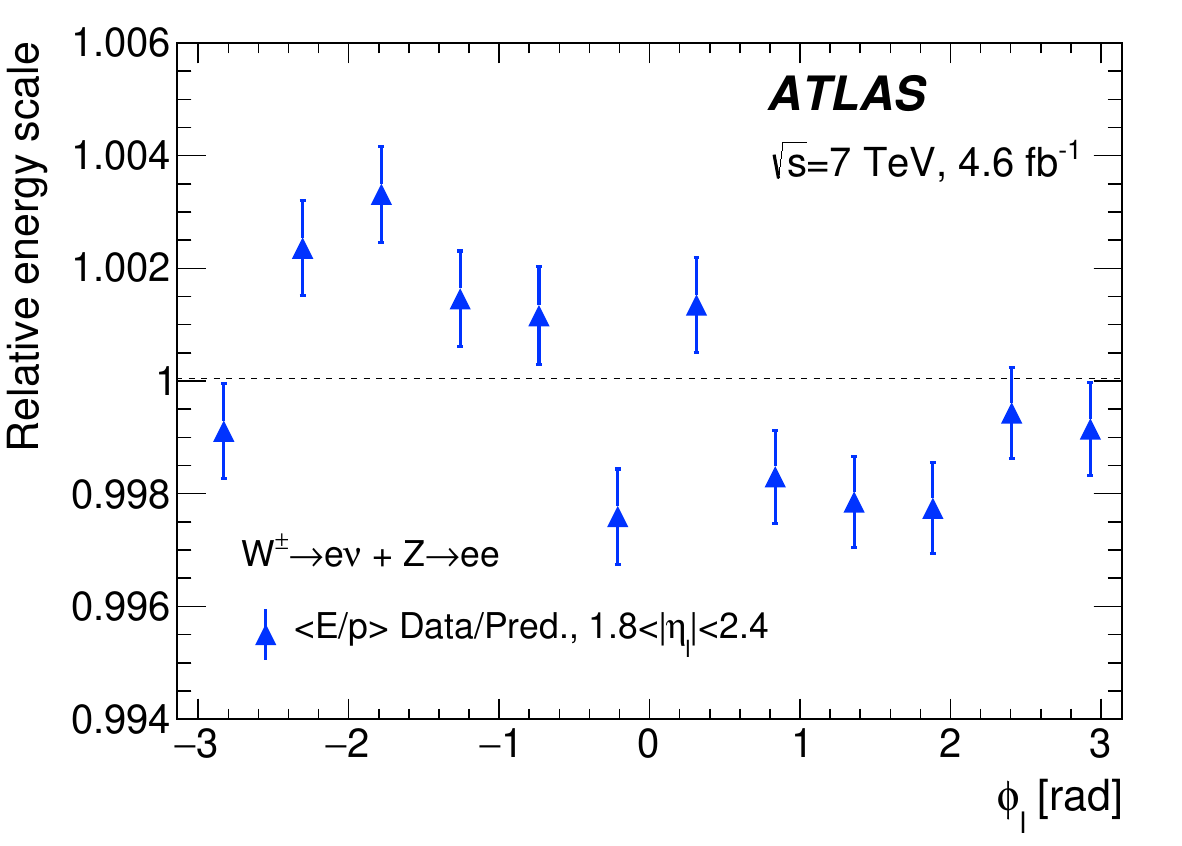}\label{fig:eleccalib1b}}
  \caption{Azimuthal variation of the data-to-prediction ratio of $\avg{E/p}$ in $W$ and $Z$ events, for electrons in (a) $|\eta_\ell| < 1.2$ and (b) $ 1.8 < |\eta_\ell| < 2.4$. The electron energy
    calibration based on $Z\rightarrow ee$ events is applied, and the track $p$ is corrected for the momentum scale, resolution and sagitta bias. The mean for the $E/p$ distribution integrated in
    $\phi$ is normalised to unity.  The error bars are statistical only.}
  \label{fig:eleccalib1}
\end{figure}

Azimuthal variations of the electron-energy response are expected from gravity-induced mechanical deformations of the EM calorimeter, and are observed especially in the endcaps, as illustrated in
Figure~\ref{fig:eleccalib1}. As the $Z$-boson calibration averages over $\phi_\ell$ and the azimuthal distributions of the selected electrons differ in the two processes, a small residual effect from this
modulation is expected when applying the calibration results to the $W\rightarrow e\nu$ sample. Related effects are discussed in
Section~\ref{sec:recoilcalib}. A dedicated correction is derived
using the azimuthal dependence of the mean of the electron
energy/momentum ratio, $\avg{E/p}$, after correcting $p$ for the momentum
scale and curvature bias discussed in Section~\ref{sec:muonscale}. The
effect of this correction is a relative change of the average energy
response of $3.8\times 10^{-5}$ in $W$-boson events, with negligible
uncertainty.

The $E/p$ distribution is also used to test the modelling of
non-Gaussian tails in the energy response. An excess of events is observed
in data at low values of $E/p$, and interpreted as the result of the
mismodelling of the lateral development of EM showers in the
calorimeter. Its impact is evaluated by
removing the electrons with $E/p$ values in the region where
the discrepancy is observed. The effect of this removal is
compatible for electrons from $W$- and $Z$-boson decays within
$4.9\times 10^{-5}$, which corresponds to the statistical uncertainty
of the test and is considered as an additional systematic
uncertainty.

\begin{figure}
  \centering
  \includegraphics[width=0.7\textwidth]{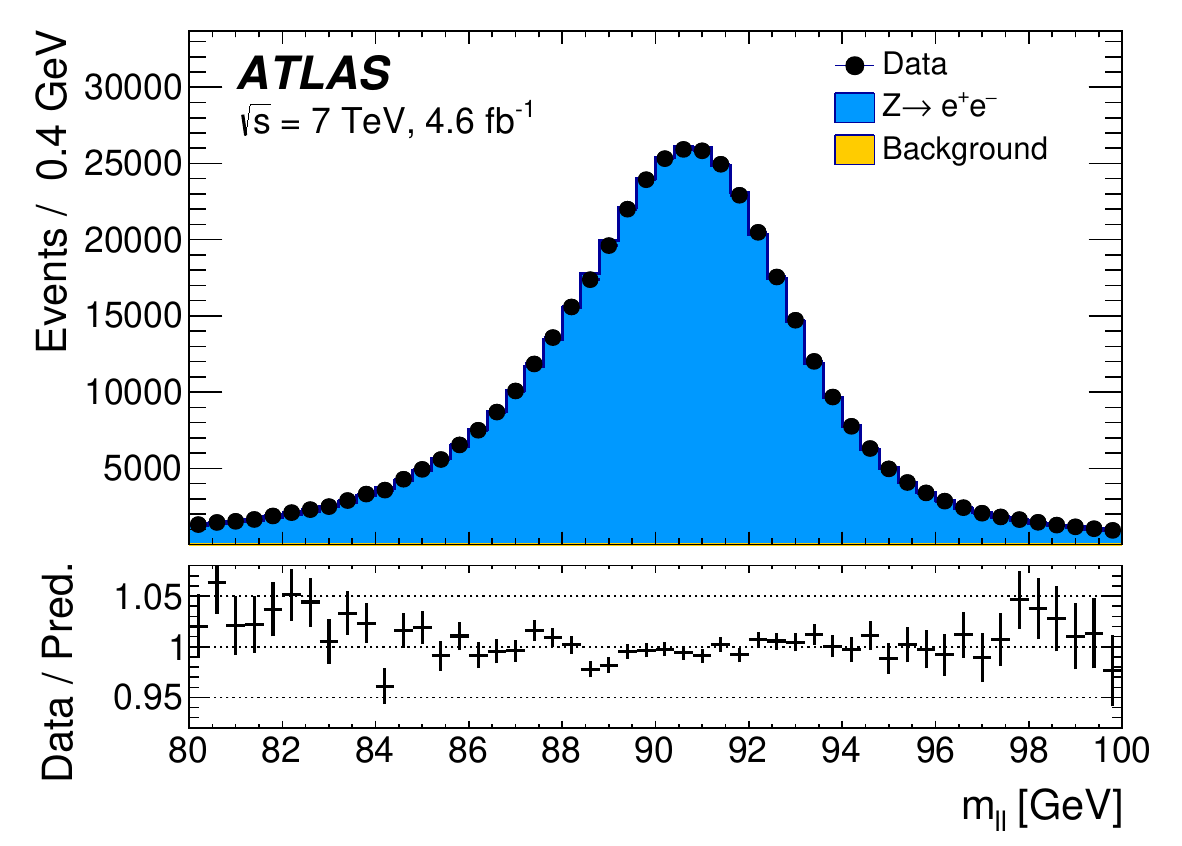}
  \caption{Dielectron invariant mass distribution in \Zee{} events. The data are compared to the simulation including signal and backgrounds.
    Corrections for energy resolution, and for reconstruction, identification, isolation and trigger efficiencies are applied to the simulation; energy-scale corrections are applied to the
    data. Background events contribute less than 0.2\% of the observed distribution. The
    lower panel shows the data-to-prediction ratio, with the error bars showing the statistical uncertainty.}
  \label{fig:eleccalib2}
\end{figure}

The result of the complete calibration procedure is illustrated in
Figure~\ref{fig:eleccalib2}, which shows the comparison of the dielectron invariant mass distribution for
$Z\rightarrow ee$ events in data and simulation.
The impact of the electron-energy calibration uncertainties on the $m_W$ measurement is summarised in
Table~\ref{tab:eleccalibsummary}.

\subsection{Electron selection efficiency}

Electron efficiency corrections are determined using samples
of $W\rightarrow e\nu$, $Z\rightarrow ee$, and
$J/\psi\rightarrow ee$ events, and measured separately for
electron reconstruction, identification and trigger
efficiencies~\cite{PERF-2013-03}, as a function of electron $\eta$ and
\pt. In the \pt{} range relevant for the measurement of the $W$-boson
mass, the reconstruction and identification efficiency corrections
have a typical uncertainty of 0.1--0.2\% in the barrel, and 0.3\% in
the endcap. The trigger efficiency corrections have an uncertainty
smaller than 0.1\%, and are weakly dependent on $\pt^\ell$.

For a data-taking period corresponding to approximately 20\% of the
integrated luminosity, the LAr calorimeter suffered from six front-end
board failures. During this period, electrons could not be
reconstructed in the region of $0<\eta<1.475$ and $-0.9<\phi<-0.5$.
The data-taking conditions are reflected in the simulation for
the corresponding fraction of events. 
However, the trigger acceptance loss is not perfectly
simulated, and dedicated efficiency corrections are derived as a function
of $\eta$ and $\phi$ to correct the mismodelling, and applied in
addition to the initial corrections. 

As described in Section~\ref{sec:objreco}, isolation
requirements are applied to the identified electrons. Their efficiency
is approximately 95\% in the simulated event samples, and energy-isolation efficiency
corrections are derived as for the reconstruction, identification, and
trigger efficiencies. The energy-isolation efficiency corrections
deviate from unity by less than 0.5\%, with an uncertainty smaller
than 0.2\% on average. 

Finally, as positively and negatively charged $W$-boson events have
different final-state distributions, the $W^+$ contamination in the
$W^-$ sample, and vice versa, constitutes an additional source
of uncertainty. The rate of electron charge mismeasurement in
simulated events rises from about 0.2\% in the barrel
to 4\% in the endcap. Estimates of charge mismeasurement in data
confirm these predictions within better than 0.1\%, apart from the
high $|\eta|$ region where differences up to 1\% are observed.
The electron charge mismeasurement induces a systematic uncertainty in
$m_W$ of approximately 0.5~\MeV{} in the regions of $|\eta_\ell|<0.6$
and $0.6<|\eta_\ell|<1.2$, and of 5~\MeV{} in the region of
$1.8<|\eta_\ell|<2.4$, separately for $W^+$ and $W^-$.
Since the $W^+$ and $W^-$ samples contaminate each other, the effect is
anti-correlated for the $m_W$ measurements in the two different charge
categories, and cancels in their combination, up to the asymmetry in the $W^+/W^-$
production rate. After combination, the residual uncertainty in $m_W$
is 0.2~\MeV{} for $|\eta_\ell|<1.2$, and 1.5~\MeV{} for
$1.8<|\eta_\ell|<2.4$, for both the $\pt^\ell$ and \mt{}
distributions. The uncertainties are considered as uncorrelated across
pseudorapidity bins.

\begin{figure}
  \centering
  \includegraphics[width=0.7\textwidth]{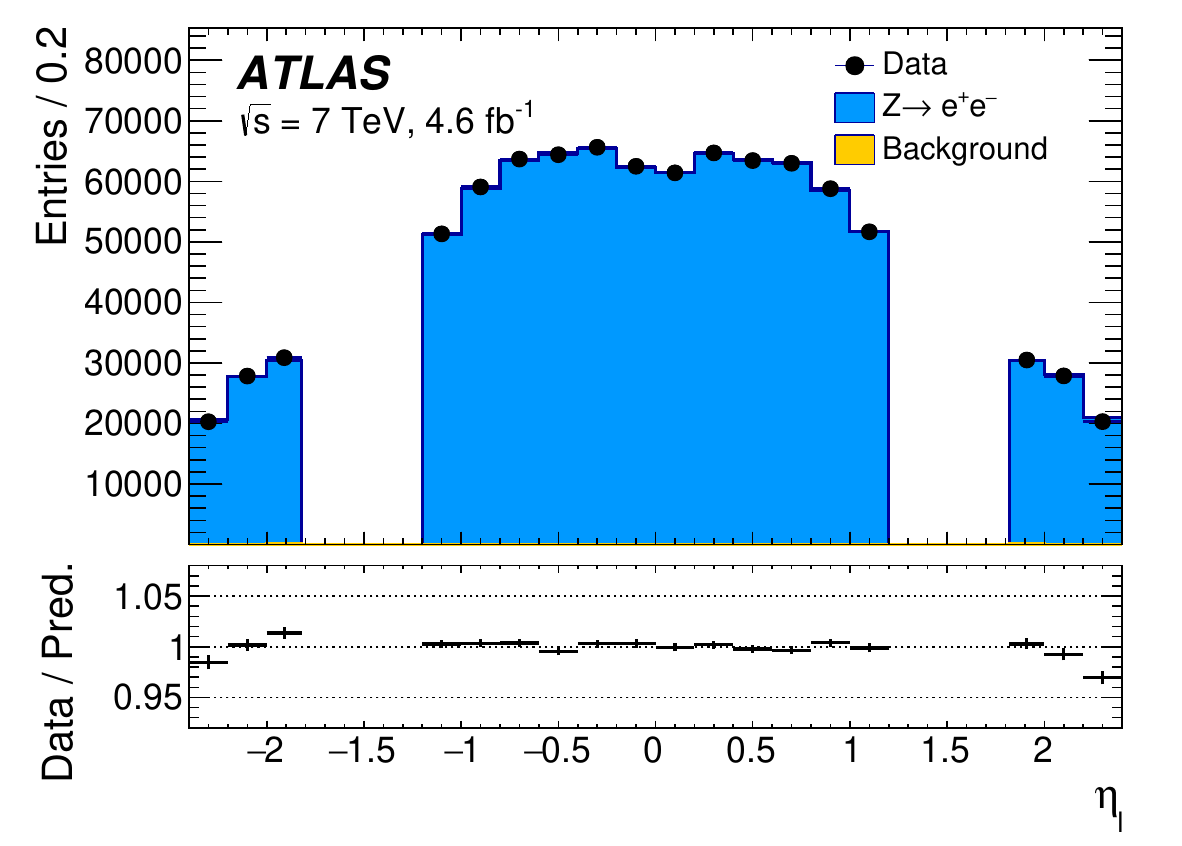}
  \caption{Distribution of reconstructed electrons $\eta$ in \Zee{} events. The data are compared to the simulation including signal and background contributions.
    Corrections for energy resolution, and for reconstruction, identification, isolation and trigger efficiencies are applied to the simulation; energy-scale corrections are applied to the
    data. Background events contribute less than 0.2\% of the observed distribution. The
    lower panel shows the data-to-prediction ratio, with the error bars showing the statistical uncertainty.}
  \label{fig:eleccalib3}
\end{figure}

Figure~\ref{fig:eleccalib3} compares the $\eta_\ell$ distribution in data and simulation for $Z\rightarrow ee$ events, after applying the efficiency corrections discussed above. The corresponding uncertainties
in $m_W$ due to the electron efficiency corrections are shown in Table~\ref{tab:eleccalibsummary}. 

\begin{table*}[tp]
  \centering
  \begin{tabular}{lrrrrrrrr}
    \toprule
    $|\eta_\ell|$ range  & \multicolumn{2}{r}{$[0.0,0.6]$} & \multicolumn{2}{r}{$[0.6,1.2]$} & \multicolumn{2}{r}{$[1.82,2.4]$} & \multicolumn{2}{r}{Combined}\\
    Kinematic distribution      & $\pt^\ell$ & $\mt$ & $\pt^\ell$ & $\mt$ & $\pt^\ell$ & $\mt$ & $\pt^\ell$ & $\mt$ \\
    \midrule
    $\delta m_W$~[\MeV]\\
    \,\,\,\, Energy scale                             & 10.4 & 10.3 & 10.8 & 10.1 & 16.1 & 17.1 & 8.1  & 8.0 \\
    \,\,\,\, Energy resolution                        &  5.0 & 6.0  &  7.3 &  6.7 & 10.4 & 15.5 & 3.5  & 5.5 \\
    \,\,\,\, Energy linearity                         &  2.2 & 4.2  &  5.8 &  8.9 &  8.6 & 10.6 & 3.4  & 5.5 \\
    \,\,\,\, Energy tails                             &  2.3 & 3.3  &  2.3 &  3.3 &  2.3 &  3.3 & 2.3  & 3.3 \\
    \,\,\,\, Reconstruction efficiency                & 10.5 & 8.8  &  9.9 &  7.8 & 14.5 & 11.0 & 7.2  & 6.0  \\
    \,\,\,\, Identification efficiency                & 10.4 & 7.7  & 11.7 &  8.8 & 16.7 & 12.1 & 7.3  & 5.6  \\
    \,\,\,\, Trigger and isolation efficiencies       &  0.2 & 0.5  &  0.3 &  0.5 &  2.0 &  2.2 &  0.8 &  0.9 \\
    \,\,\,\, Charge mismeasurement                   &  0.2 & 0.2  & 0.2  & 0.2  & 1.5  & 1.5 & 0.1 & 0.1 \\
    \midrule
    \,\,\,\, Total                                    & 19.0 & 17.5 & 21.1 & 19.4 & 30.7 & 30.5 & 14.2 & 14.3 \\
    \bottomrule
  \end{tabular}
  \caption{Systematic uncertainties in the $m_W$ measurement due to electron energy calibration, efficiency corrections and 
     charge mismeasurement, for the different kinematic
    distributions and $|\eta_\ell|$
    regions, averaged over lepton charge. Combined uncertainties are evaluated as described in Section~\ref{subsec:strategy}.\label{tab:eleccalibsummary}}
\end{table*}

\section{Calibration of the recoil \label{sec:recoilcalib}}

The calibration of the recoil, \ut, affects the
measurement of the $W$-boson mass through its impact on the \mt{} distribution, which is used to extract $m_W$.
In addition, the recoil calibration affects the $\pt^\ell$ and \mt{} distributions
through the $\mpt$, \mt, and $u_{\mathrm{T}}$  event-selection requirements.
The calibration procedure proceeds in two steps. First, the
dominant part of the \ut{} resolution mismodelling is
addressed by correcting the modelling of the overall event activity in simulation. These
corrections are derived separately in the $W$- and $Z$-boson
samples. Second, corrections for residual differences in the recoil
response and resolution are derived using $Z$-boson events in data, and
transferred to the $W$-boson sample.

\subsection{Event activity corrections}

The pile-up of multiple proton--proton interactions has a significant impact
on the resolution of the recoil.
As described in Section~\ref{sec:samples}, the pile-up is modelled by overlaying the
simulated hard-scattering process with additional $pp$
interactions simulated using \PYTHIA 8 with the A2 tune.
The average number of interactions per bunch crossing is defined, for
each event, as $\avg{\mu}=\mathcal{L}
\sigma_\textrm{in}/f_{\textrm{BC}}$, where $\mathcal{L}$ is the
instantaneous luminosity, $\sigma_\textrm{in}$ is the total $pp$
inelastic cross section and $f_{\textrm{BC}}$ is the average
bunch-crossing rate. The distribution of $\avg{\mu}$ in the simulated
event samples is reweighted to match the corresponding distribution in data.
The distribution of $\avg{\mu}$ is affected in particular by the uncertainty in the cross section and properties of inelastic collisions.
In the simulation, $\avg{\mu}$ is scaled by a factor $\alpha$ to optimise the modelling of
observed data distributions which are relevant to the modelling of \ut.
A value of $\alpha=1.10\pm 0.04$ is determined by minimising
the $\chi^2$ function of the compatibility test between data and
simulation for the $\Sigma E^{*}_{\textrm{T}}$ and $u_{\perp}^Z$
distributions, where the uncertainty accounts for differences in the
values determined using the two distributions.

After the correction applied to the average number of pile-up
interactions, residual data-to-prediction differences in the $\Sigma E^{*}_{\textrm{T}}$ distribution are
responsible for most of the remaining \ut{} resolution mismodelling.
The $\Sigma E^{*}_{\textrm{T}}$ distribution is corrected by means of
a Smirnov transform, which is a mapping $x \to x'(x)$
such that a function $f(x)$ is transformed into another target function
$g(x)$ through the relation $f(x) \to f(x') \equiv g(x)$~\cite{SmirnovTrans}.
Accordingly, a mapping $\Sigma E^{*}_{\textrm{T}}\rightarrow\Sigma {E^{*}_{\textrm{T}}}'$ is defined such that the distribution of $\Sigma E^{*}_{\textrm{T}}$ in simulation,
$h_{\mathrm{MC}}(\Sigma E^{*}_{\textrm{T}})$, is transformed into
$h_{\mathrm{MC}}(\Sigma {E^{*}_{\textrm{T}}}')$ to match the
$\Sigma E^{*}_{\textrm{T}}$ distribution in data,
$h_{\mathrm{data}}(\Sigma E^{*}_{\textrm{T}})$.
The correction is derived for $Z$-boson events in bins of
$p_{\mathrm{T}}^{\ell\ell}$, as the observed differences in the
$\Sigma E^{*}_{\textrm{T}}$ distribution depend on the $Z$-boson
transverse momentum. The result of this procedure
is illustrated in Figure~\ref{Fig:HRRecoilPhiB_a}.
The modified distribution is used to parameterise the recoil response corrections discussed in the next section.

In $W$-boson events, the transverse momentum of the boson can only be inferred
from \ut, which has worse resolution
compared to $p_{\mathrm{T}}^{\ell\ell}$ in $Z$-boson events. To overcome this limitation, a \pt-dependent correction is defined assuming that the \pt\ dependence of differences between data and simulation in the
$\Sigma E^{*}_{\textrm{T}}$ distribution in $W$-boson events follows
the corresponding differences observed in $Z$-boson events.
The $\Sigma E^{*}_{\textrm{T}}$ distribution to be matched by the simulation is defined as follows for $W$-boson events:
\begin{equation}
\label{eq:sigmaet} \tilde{h}^W_{\textrm{data}}(\Sigma E^{*}_{\textrm{T}}, \pt^W) \equiv \, h_{\textrm{data}}^Z(\Sigma E^{*}_{\textrm{T}}, \pt^{\ell\ell}) \left(\frac{h^W_{\textrm{data}}(\Sigma E^{*}_{\textrm{T}})}{h^W_{\textrm{MC}}(\Sigma E^{*}_{\textrm{T}})}\  \Big/\ \frac{h^Z_{\textrm{data}}(\Sigma E^{*}_{\textrm{T}})}{h^Z_{\textrm{MC}}(\Sigma E^{*}_{\textrm{T}})}\right),
\end{equation}
where $\pt^W$ is the particle-level $W$-boson transverse momentum, and
$\pt^{\ell\ell}$ the transverse momentum measured from the decay-lepton pair, used as an approximation of the particle-level $\pt^Z$. The superscripts $W$ and $Z$ refer to $W$- or $Z$-boson event
samples, and the double ratio in the second term accounts for the differences
between the inclusive distributions in $W$- and $Z$-boson events. This correction is defined separately for
positively and negatively charged $W$ bosons, so as to incorporate the
dependence of the $\pt^W$ distribution on the charge of the $W$
boson. Using $\tilde{h}^W_{\textrm{data}}(\Sigma E^{*}_{\textrm{T}}, \pt^W) $
defined in Eq.~(\ref{eq:sigmaet}) as the target distribution, the
$\pt^W$-dependent Smirnov transform of the
$\Sigma E^{*}_{\textrm{T}}$ distribution in $W$-boson events is
defined as follows:
\begin{equation}
\nonumber h^W_{\mathrm{MC}}(\Sigma E^{*}_{\textrm{T}}; \pt^W) \, \rightarrow \, h_{\mathrm{MC}}^{W}(\Sigma {E^{*}_{\textrm{T}}}'; \pt^W) \, \equiv \, \tilde{h}^W_{\mathrm{data}}(\Sigma E^{*}_{\textrm{T}};
\pt^W).
\label{eq:smirnovW}
\end{equation}
The validity of the approximation introduced in Eq.~(\ref{eq:sigmaet})
is verified by comparing
$h^W_{\textrm{data}}(\Sigma E^{*}_{\textrm{T}})/h^W_{\textrm{MC}}(\Sigma E^{*}_{\textrm{T}})$
and $h^Z_{\textrm{data}}(\Sigma E^{*}_{\textrm{T}})/h^Z_{\textrm{MC}}(\Sigma E^{*}_{\textrm{T}})$
in broad bins of \ut. The associated systematic uncertainties
are discussed in Section~\ref{sec:recoilsyst}.

\subsection{Residual response corrections\label{sec:residcorr}}

\begin{figure}
  \begin{center}
    \subfloat[]{\includegraphics[width=0.49\textwidth]{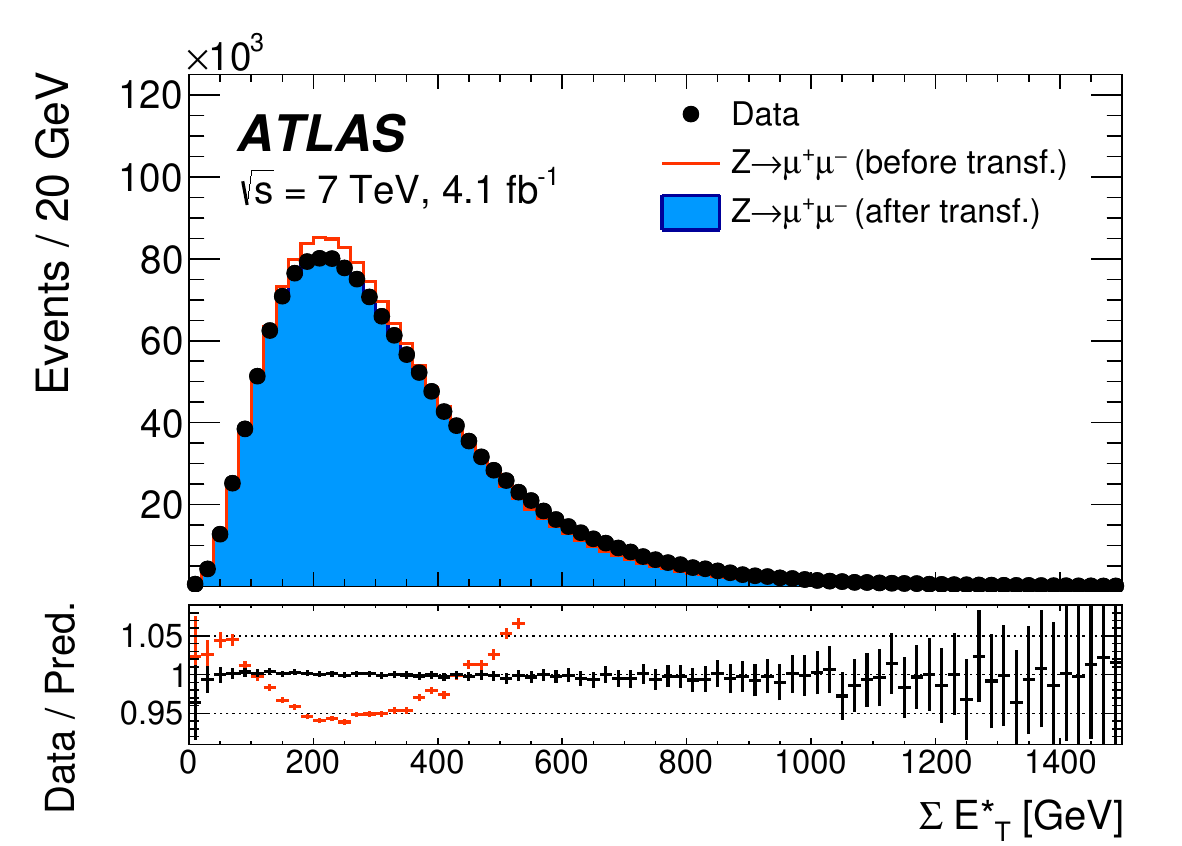}\label{Fig:HRRecoilPhiB_a}}
    \subfloat[]{\includegraphics[width=0.49\textwidth]{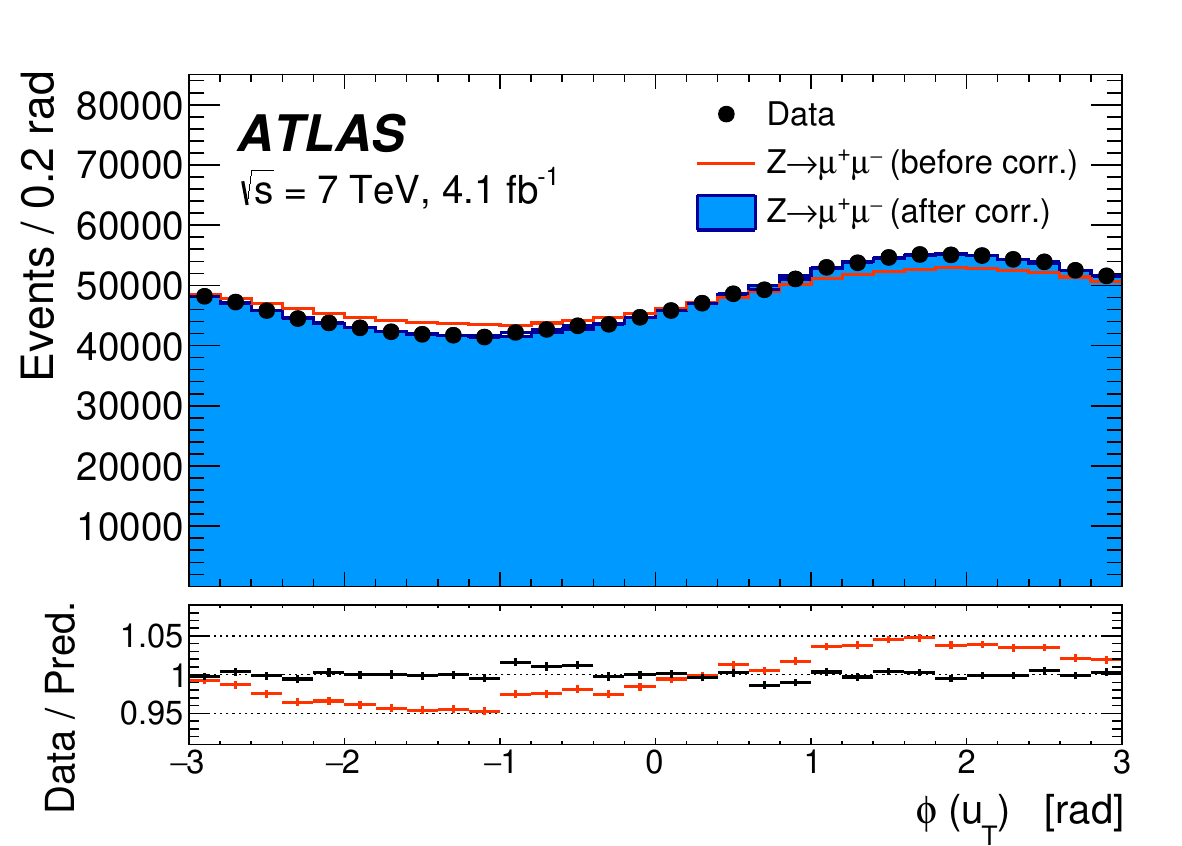}\label{Fig:HRRecoilPhiB_b}}\\
  \end{center}
  \caption{Distributions of (a) $\Sigma E^{*}_{\textrm{T}}$ and (b)
    azimuth $\phi$ of the recoil in data and simulation
    for $Z\to\mu\mu$ events. The $\Sigma E^{*}_{\textrm{T}}$ distribution is shown before and after applying the Smirnov-transform correction, and the $\phi$ 
    distribution is shown before and after the $u_{x,y}$ correction. The lower panels show the data-to-prediction ratios, with the
    vertical bars showing the statistical uncertainty.}
  \label{Fig:HRRecoilPhiB}
\end{figure}

In the ideal case of beams coinciding with the $z$-axis, the physical transverse momentum of $W$ and $Z$ bosons is uniformly distributed
in $\phi$. However, an offset of the interaction point with respect to
the detector centre in the transverse plane, the non-zero crossing angle between the proton beams, and $\phi$-dependent response of the calorimeters generate anisotropies in the
reconstructed recoil distribution.
Corresponding differences between data and simulation are addressed by effective corrections applied to $u_{x}$ and $u_{y}$ in simulation:
\begin{eqnarray}
\nonumber u^{\prime}_{x} &=& u_{x} + \cropdelims[0.5]\left({\cropdelims[0.5]\left<{u_{x}}\right>_{\textrm{data}} - \cropdelims[0.5]\left<{u_{x}}\right>_{\textrm{MC}}}\right),\\
\nonumber u^{\prime}_{y} &=& u_{y} + \cropdelims[0.5]\left({\cropdelims[0.5]\left<{u_{y}}\right>_{\textrm{data}} - \cropdelims[0.5]\left<{u_{y}}\right>_{\textrm{MC}}}\right),
\end{eqnarray}
where $\cropdelims[0.5]\left<{u_{x,y}}\right>_{\textrm{data}}$ and
$\cropdelims[0.5]\left<{u_{x,y}}\right>_{\textrm{MC}}$ are the mean
values of these distributions in data and simulation,
respectively. The corrections are evaluated in $Z$-boson events and parameterised as
a function of $\Sigma E^{*}_{\textrm{T}}$. The effect of these corrections on the recoil $\phi$ distribution is illustrated in Figure~\ref{Fig:HRRecoilPhiB_b}.

The transverse momentum of $Z$ bosons can be reconstructed from the
decay-lepton pair with a resolution of $1$--$2\GeV$,
which is negligible compared to the recoil energy resolution. The recoil response can thus be calibrated from comparisons with the reconstructed $\pt^{\ell\ell}$ in data and simulation.
Recoil energy scale and resolution corrections are derived in bins of $\Sigma E^{*}_{\textrm{T}}$
and $\pt^{\ell\ell}$ at reconstruction level, and are applied in simulation as a
function of the particle-level vector-boson momentum $\pt^V$ in both
the $W$- and $Z$-boson samples.
The energy scale of the recoil is calibrated by comparing the
$u_{\parallel}^Z+\pt^{\ell\ell}$ distribution in data and simulation, whereas
resolution corrections are evaluated from the $u_{\perp}^Z$
distribution.
Energy-scale corrections $b(\pt^{V},\Sigma {E^{*}_{\textrm{T}}}')$
are defined as the difference between the average values of the $u_{\parallel}^Z+\pt^{\ell\ell}$ distributions in
data and simulation, and the energy-resolution correction factors
$r(\pt^{V},\Sigma {E^{*}_{\textrm{T}}}')$ as the ratio of the standard
deviations of the corresponding $u_{\perp}^Z$ distributions.

\begin{figure}
  \begin{center}
    \subfloat[]{\includegraphics[width=0.49\textwidth]{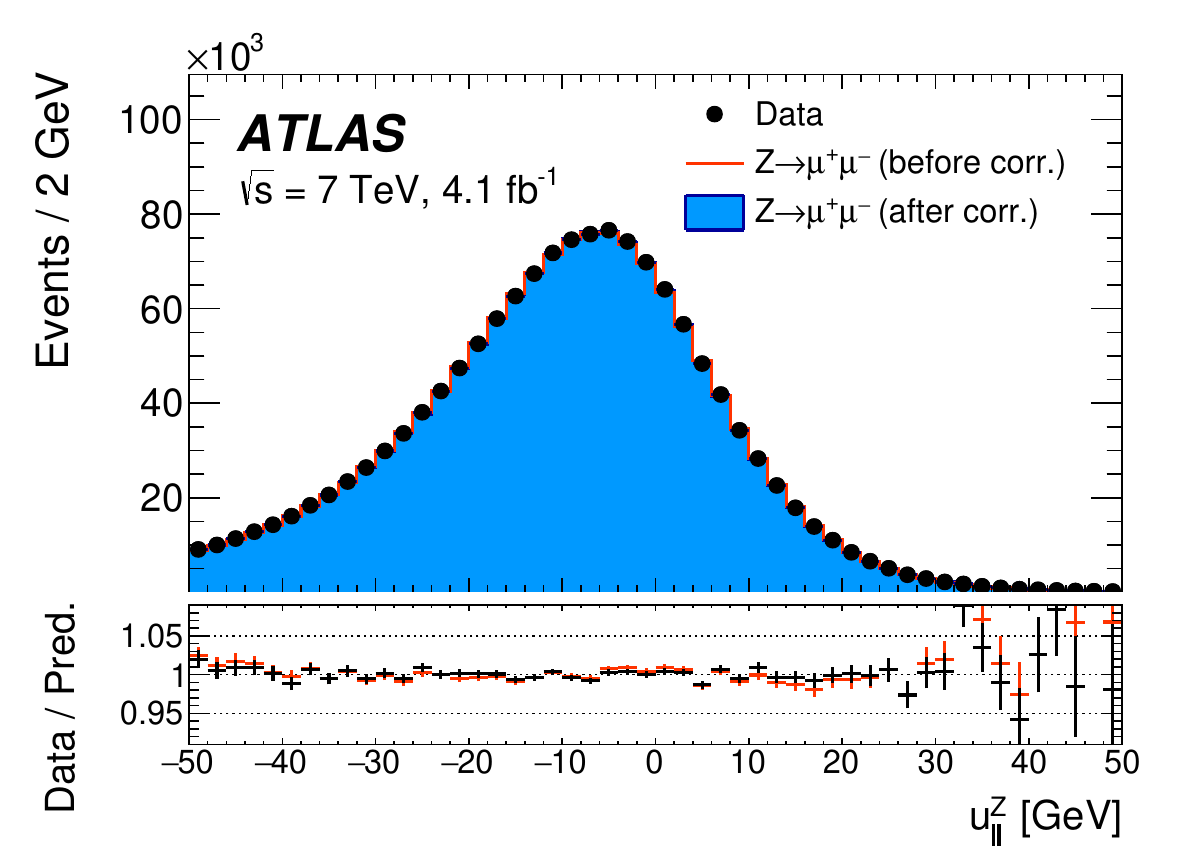}}
    \subfloat[]{\includegraphics[width=0.49\textwidth]{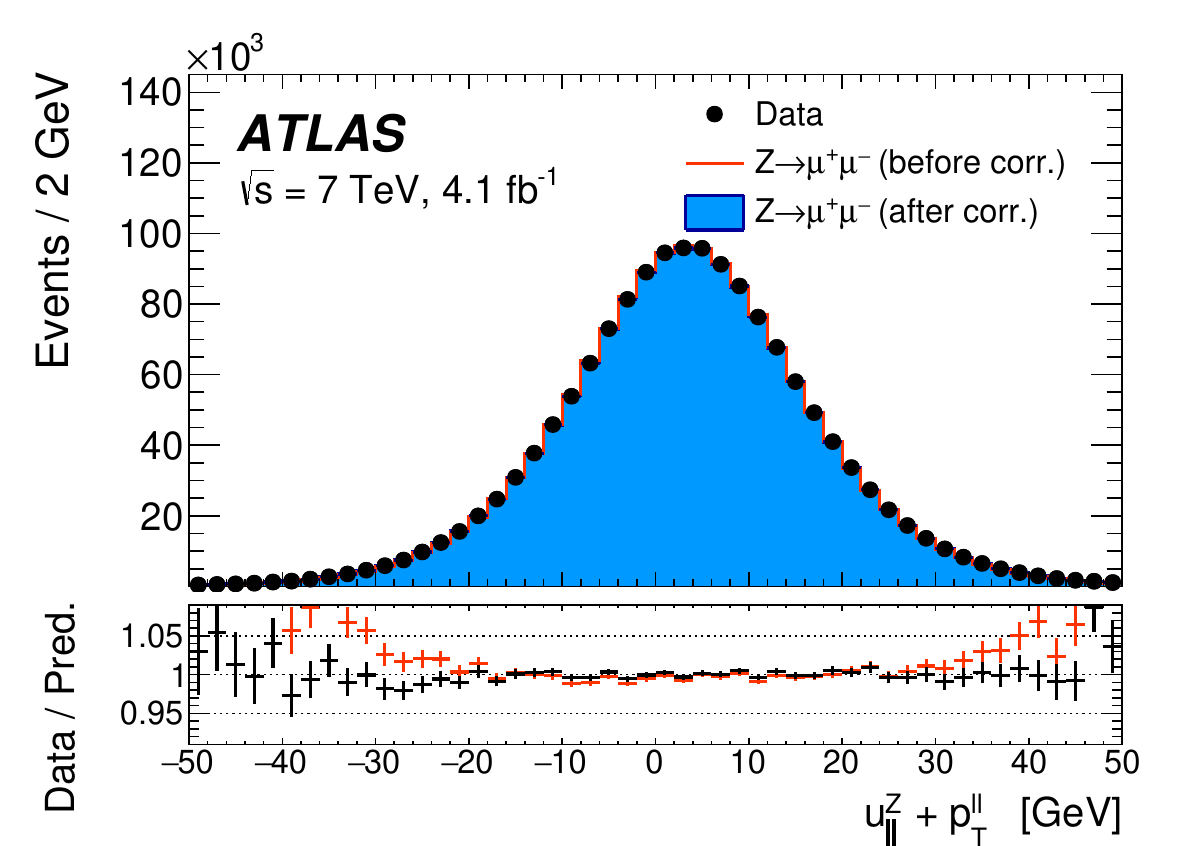}}\\
    \subfloat[]{\includegraphics[width=0.49\textwidth]{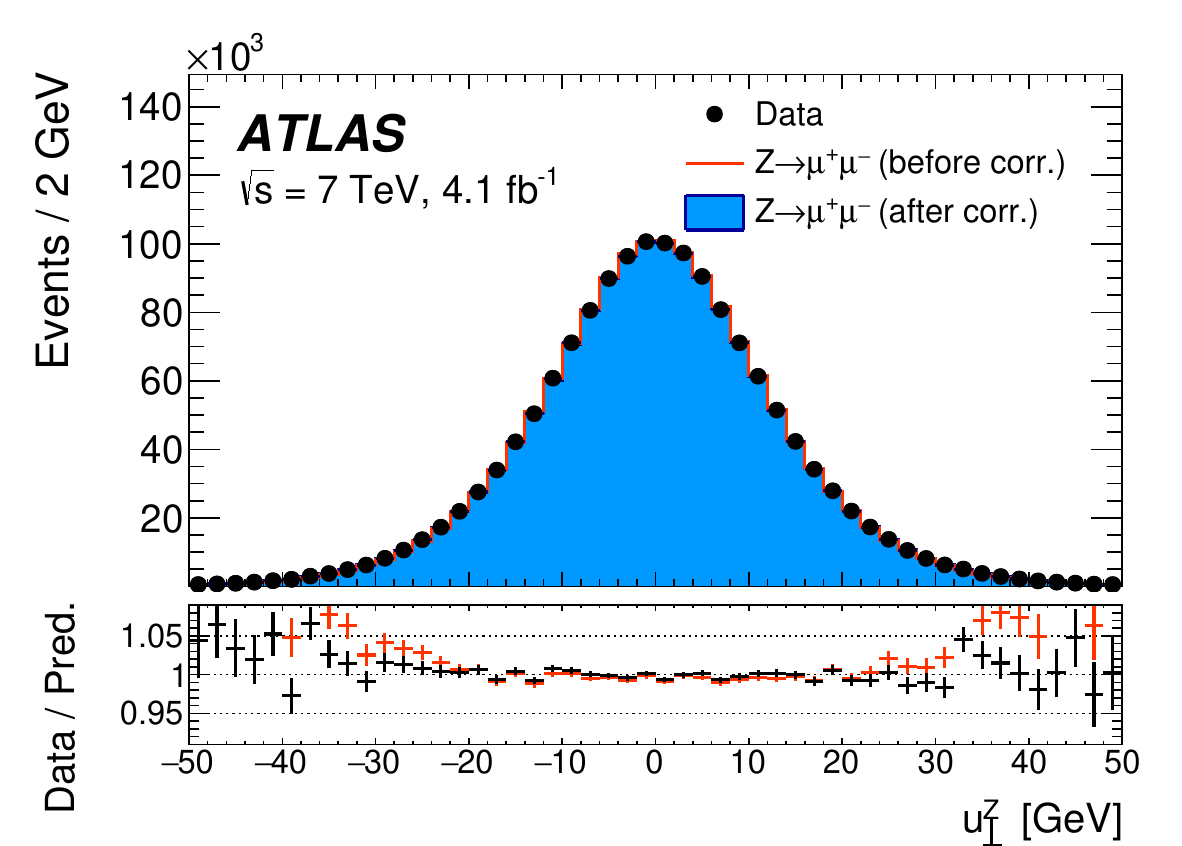}}
    \subfloat[]{\includegraphics[width=0.49\textwidth]{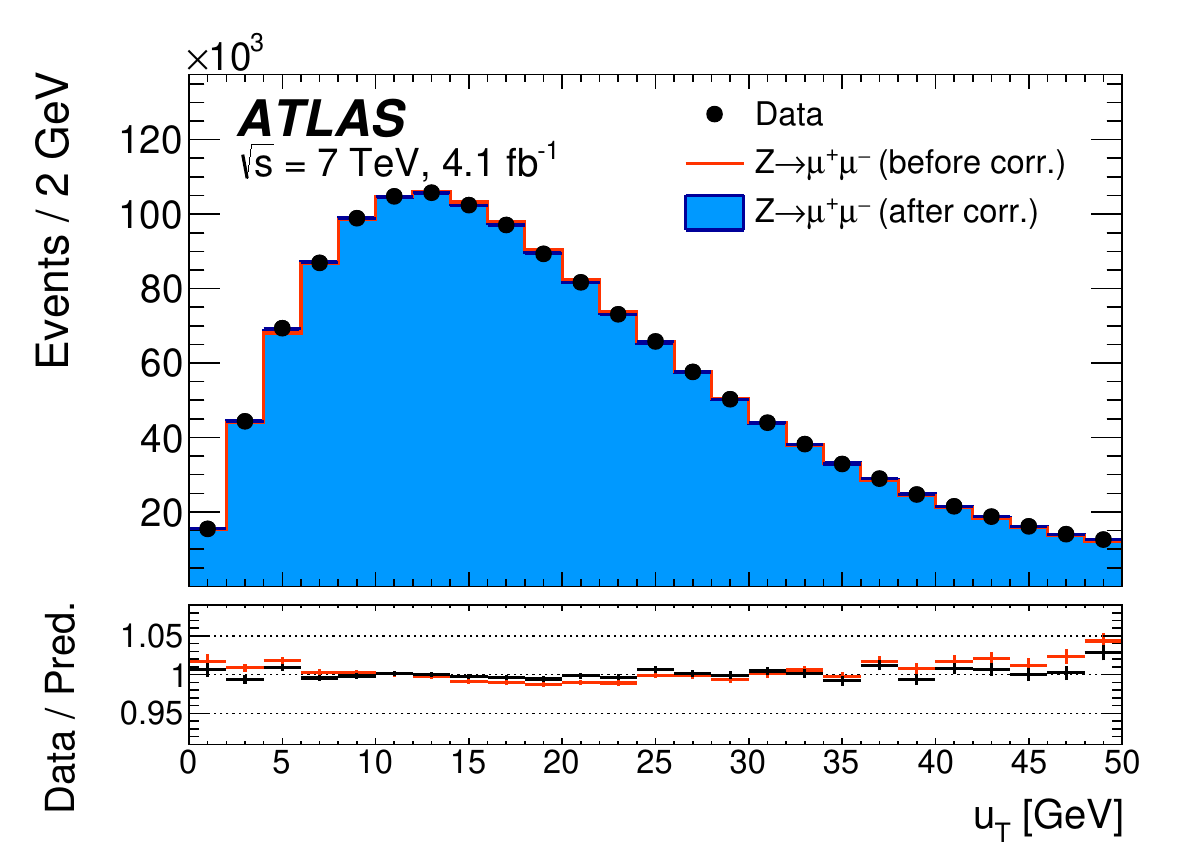}}
  \end{center}
  \caption{Recoil distributions for (a) $u_{\parallel}^Z$, (b)
    $u_{\parallel}^Z+\pt^{\ell\ell}$, (c) $u_{\perp}^Z$, and (d)
    $u_{\textrm{T}}$ in \Zmm{} events. The data are compared
    to the simulation before and after applying the
    recoil corrections described in the text. The lower panels show
    the data-to-prediction ratios, with the vertical bars showing the
    statistical uncertainty.}
  \label{Fig:recoil_RecoilCor}
\end{figure}

The parallel component of \ut{} in
simulated events is corrected for energy scale and resolution,
whereas the perpendicular component is corrected for energy
resolution only.
The corrections are defined as follows:
\begin{eqnarray}
\label{eq:reccorr1} u^{V,\textrm{corr}}_{\parallel} &=& \left[u^{V,\textrm{MC}}_{\parallel} - \avg{u^{Z,\textrm{data}}_{\parallel}}\!(\pt^V,\Sigma {E^{*}_{\textrm{T}}}')\right] \cdot r(\pt^V,\Sigma
      {E^{*}_{\textrm{T}}}') \, + \avg{u^{Z,\textrm{data}}_{\parallel}}\!(\pt^V,\Sigma {E^{*}_{\textrm{T}}}') + \, b(\pt^V,\Sigma {E^{*}_{\textrm{T}}}'),\\
\label{eq:reccorr2} u^{V,\textrm{corr}}_{\perp} &=& u^{V,\textrm{MC}}_{\perp} \cdot r (\pt^V,\Sigma {E^{*}_{\textrm{T}}}');
\end{eqnarray}
where $V=W,Z$, $u^{V,\textrm{MC}}_{\parallel}$ and $u^{V,\textrm{MC}}_{\perp}$
are the parallel and perpendicular components of \ut{} in the
simulation, and $u^{V,\textrm{corr}}_{\parallel}$ and
$u^{V,\textrm{corr}}_{\perp}$ are the corresponding corrected values.
As for $b$ and $r$, the average $\avg{u^{Z,\textrm{data}}_{\parallel}}$ is mapped as a function of the reconstructed $\pt^{\ell\ell}$ in $Z$-boson data, and used as a function of $\pt^V$ in both $W$- and $Z$-boson simulation.
Since the resolution of \ut{} has a sizeable dependence on the amount
of pile-up, the correction procedure is defined in three bins of
$\avg{\mu}$, corresponding to low, medium, and high pile-up
conditions, and defined by the ranges of $\avg{\mu}\, \in [2.5,6.5]$,
$\avg{\mu}\, \in [6.5,9.5]$, and $\avg{\mu}\, \in [9.5,16.0]$,
respectively. Values for $b(\pt^{V},\Sigma {E^{*}_{\textrm{T}}}')$ are typically $O(100\MeV)$, and $r(\pt^{V},\Sigma {E^{*}_{\textrm{T}}}')$ deviates from unity by 2\% at most.
The effect of the calibration is shown in Figure~\ref{Fig:recoil_RecoilCor} for
$Z\rightarrow\mu\mu$ events. The level of agreement obtained after corrections is satisfactory, and similar performance is observed for $Z\rightarrow ee$
events.

\begin{figure}
  \begin{center}
    \subfloat[]{\includegraphics[width=0.49\textwidth]{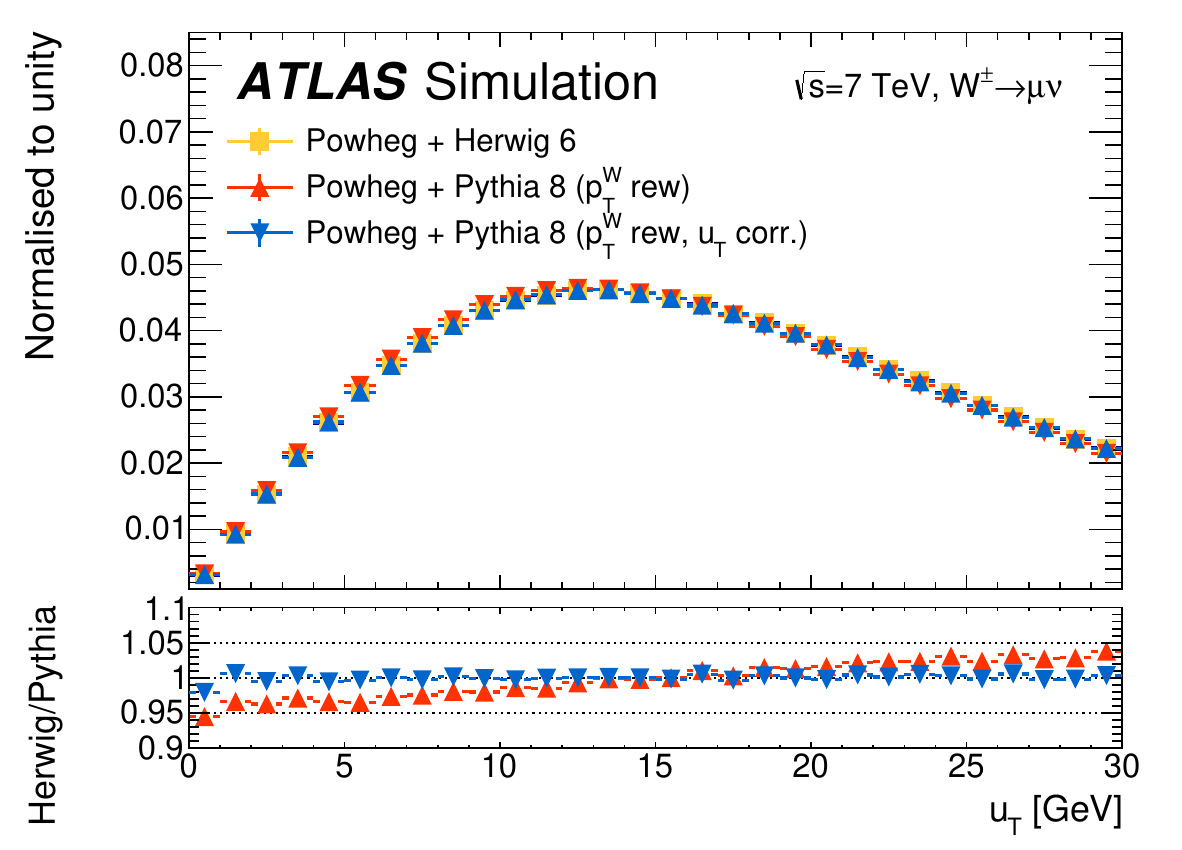}}
    \subfloat[]{\includegraphics[width=0.49\textwidth]{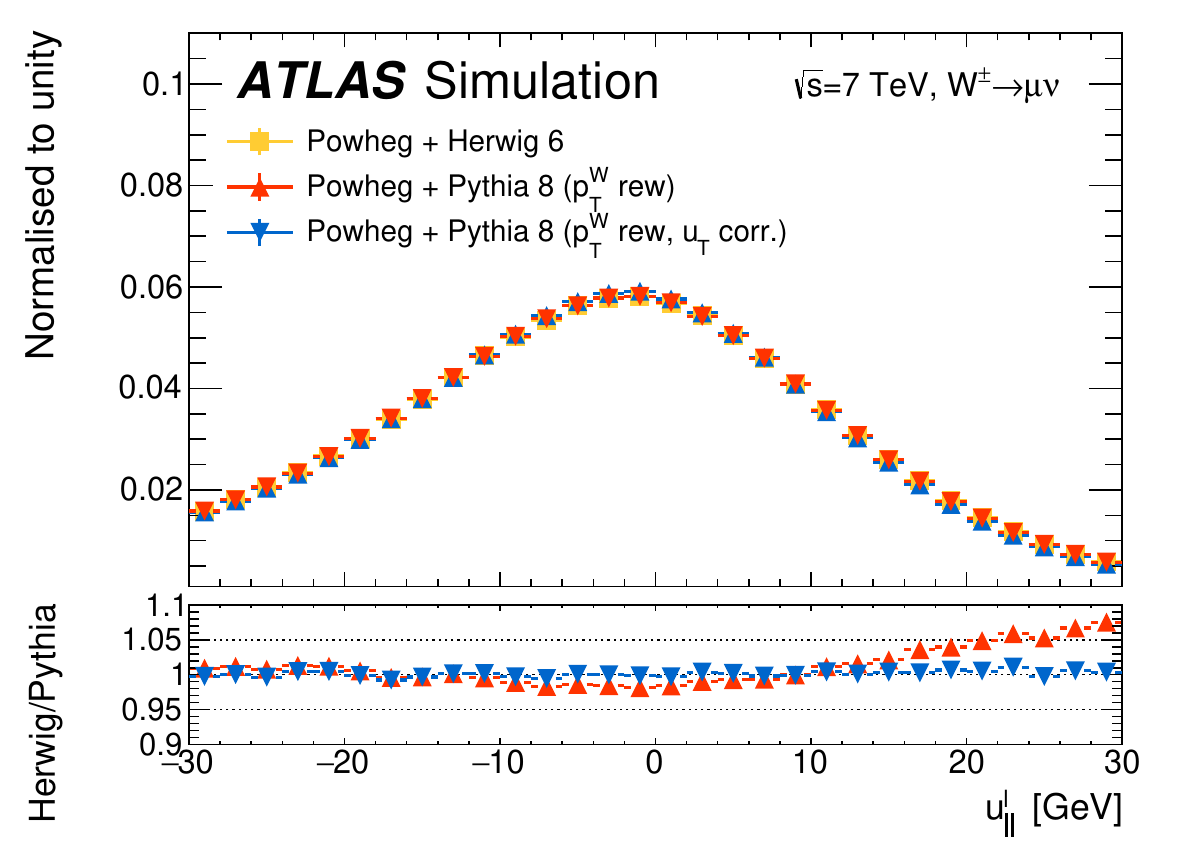}}
  \end{center}
  \caption{Distributions of (a) $u_{\textrm{T}}$ and (b) $u_{\parallel}^\ell$ in $W$ events simulated using \PowhegPythia and \PowhegHerwig. The recoil response in \PowhegPythia is corrected to
    the \PowhegHerwig response using simulated $Z$ events following the method described in the text. The $\pt^W$ distribution in \PowhegPythia is reweighted to the \PowhegHerwig prediction. The lower panels show the ratios of \PowhegHerwig to \PowhegPythia, with and without the response correction in the \PowhegPythia sample.}
  \label{Fig:recoil_RecoilClosure}
\end{figure}

A closure test of the applicability of $Z$-based corrections to $W$ production is performed using $W$ and $Z$ samples simulated with \PowhegHerwig, which provide an alternative model for the
description of hadronisation and the underlying event. The procedure described above is used to correct the recoil response from \PowhegPythia to \PowhegHerwig, where the latter is treated as
pseudodata. As shown in Figure~\ref{Fig:recoil_RecoilClosure}, the corrected $W$ recoil distributions in \PowhegPythia match the corresponding distributions in \PowhegHerwig. For this study, the effect of the
different particle-level $\pt^W$ distributions in both samples is removed by reweighting the \PowhegPythia prediction to \PowhegHerwig. This study is performed applying the standard lepton selection cuts, but
avoiding further kinematic selections in order to maximize the statistics available for the test.

\subsection{Systematic uncertainties\label{sec:recoilsyst}}

The recoil calibration procedure is sensitive to the following sources of systematic uncertainty: the uncertainty of the scale
factor applied to the $\avg{\mu}$ distribution, uncertainties due to
the Smirnov transform of the $\Sigma {E^{*}_{\textrm{T}}}$ distribution, uncertainties in the correction of the
average value of the $u_{x,y}$ distributions, statistical
uncertainties in the residual correction factors and their \pt
dependence, and expected differences in the recoil response between
$Z$- and $W$-boson events.

The uncertainty from the $\avg{\mu}$ scale-factor $\alpha$ is evaluated
by varying it by its uncertainty and repeating all
steps of the recoil calibration procedure. These variations affect the
determination of $m_W$ by less than $1\MeV$.

The systematic uncertainty related to the dependence of the $\Sigma {E^{*}_{\textrm{T}}}$
correction on \pt{} is estimated by comparing with the results of a
\pt-inclusive correction. This source contributes, averaging over $W$-boson charges, an uncertainty of
approximately $1\MeV$ for the extraction of $m_W$ from the
$p_{\textrm{T}}^\ell$ distribution, and $11\MeV$ when using the
$m_{\textrm{T}}$ distribution.

The recoil energy scale and resolution corrections of
Eqs.~(\ref{eq:reccorr1}) and~(\ref{eq:reccorr2}) are derived from the
$Z$-boson sample and applied to $W$-boson events. Differences in the detector
response to the recoil between $W$- and $Z$-boson processes are considered as
a source of systematic uncertainty for these corrections.
Differences between the $u_{\perp}^W$ and $u_{\perp}^Z$ distributions originating from different vector-boson kinematic properties,
different ISR and FSR photon emission, and from different selection
requirements are, however, discarded as they are either accurately
modelled in the simulation or already incorporated in the correction
procedure.

To remove the effect of such differences, the two-dimensional
distribution $h^W_\textrm{MC}(\pt,\Sigma E^{*}_{\textrm{T}})$ in
$W$-boson simulated events is corrected to match the corresponding
distribution in $Z$-boson simulated events, treating the neutrinos in $W$-boson decays as charged leptons to calculate $u_{\textrm{T}}$ as in $Z$-boson events.
Finally, events containing a particle-level photon from final-state
radiation are removed.
After these corrections, the standard deviation of the $u_{\perp}$
distribution agrees within 0.03\% between simulated $W$-
and $Z$-boson events.
This difference is equivalent to 6\% of the size of the residual resolution correction, which increases the standard deviation of the $u_{\perp}$ distribution by 0.5\%.
Accordingly, the corresponding systematic uncertainty due to the
extrapolation of the recoil calibration from $Z$- to $W$-boson events
is estimated by varying the energy resolution parameter $r$ of
Eqs.~(\ref{eq:reccorr1}) and~(\ref{eq:reccorr2}) by 6\%.
The impact of this uncertainty on the extraction of $m_W$ is
approximately $0.2\MeV$ for the $p_{\textrm{T}}^\ell$ distribution, and
$5.1\MeV$ for the $m_{\textrm{T}}$ distribution. The extrapolation uncertainty of the energy-scale correction $b$ was found to be negligible in comparison. 

In addition, the statistical uncertainty of the correction factors contributes $2.0\MeV$
for the $p_{\textrm{T}}^\ell$ distribution, and $2.7\MeV$ for the $m_{\textrm{T}}$ distribution.
Finally, instead of using a binned correction, a smooth interpolation
of the correction values between the bins is performed.
Comparing the binned and interpolated correction parameters $b(\pt^V,\Sigma {E^{*}_{\textrm{T}}}')$ and
$r(\pt^V,\Sigma {E^{*}_{\textrm{T}}}')$ leads to a systematic
uncertainty in $m_W$ of $1.4\MeV$ and $3.1\MeV$ for the
$\pt^\ell$ and $\mt$ distributions, respectively.
Systematic uncertainties in the $u_{x,y}$ corrections are found to be small compared to the other systematic uncertainties, and are neglected.

The impact of the uncertainties of the recoil calibration on the extraction of the $W$-boson mass from the $\pt^\ell$ and \mt{}
distributions are summarised in Table~\ref{Tab:sysTotal}. The determination of $m_W$ from the $\pt^\ell$ distribution is only slightly affected by the 
uncertainties of the recoil calibration, whereas larger uncertainties
are estimated for the \mt{} distribution. The largest uncertainties
are induced by the $\Sigma {E^{*}_{\textrm{T}}}$ corrections and by the
extrapolation of the recoil energy-scale and energy-resolution
corrections from $Z$- to $W$-boson events. The systematic
uncertainties are in general smaller for $W^-$ events than for $W^+$ events, as the
$\Sigma {E^{*}_{\textrm{T}}}$ distribution in $W^-$ events is
closer to the corresponding distribution in $Z$-boson events.

\begin{table*}[tp]
  \centering
  \begin{tabular}{lrrrrrr}
    \toprule
    $W$-boson charge  & \multicolumn{2}{c}{$W^+$} & \multicolumn{2}{c}{$W^-$} & \multicolumn{2}{c}{Combined} \\
    Kinematic distribution    & $\pt^\ell$ & $\mt$ & $\pt^\ell$ & $\mt$ & $\pt^\ell$ & $\mt$ \\
    \midrule
    $\delta m_W$~[\MeV]\\
    \,\,\,\, $\avg{\mu}$ scale factor                         &  0.2    &  1.0    &  0.2    & 1.0   & 0.2   & 1.0 \\
    \,\,\,\, $\Sigma E^{*}_{\textrm{T}}$ correction            &   0.9     &  12.2   &  1.1    & 10.2   & 1.0   & 11.2 \\
    \,\,\,\, Residual corrections (statistics)                &  2.0    &  2.7    &  2.0    & 2.7   & 2.0   &  2.7 \\
    \,\,\,\, Residual corrections (interpolation)             &  1.4    &  3.1    &  1.4    & 3.1   & 1.4   &  3.1 \\
    \,\,\,\, Residual corrections ($Z\to W$ extrapolation)    &  0.2    &  5.8    &  0.2    & 4.3   & 0.2   & 5.1 \\
    \midrule
    \,\,\,\, Total                                             &  2.6    &  14.2    &  2.7   & 11.8    & 2.6   & 13.0  \\
    \bottomrule
  \end{tabular}
  \caption{Systematic uncertainties in the $m_W$ measurement due to recoil
    corrections, for the different kinematic
    distributions and $W$-boson charge categories. Combined uncertainties are evaluated as described in Section~\ref{subsec:strategy}. \label{Tab:sysTotal}}
\end{table*}

\section{Consistency tests with $Z$-boson events \label{sec:crosschecks}}

The $Z\rightarrow \ell \ell$ event sample allows several
validation and consistency tests of the $W$-boson analysis to be performed.
All the identification requirements of Section~\ref{sec:reco}, the
calibration and efficiency corrections of Sections~\ref{sec:objcalib} and~\ref{sec:recoilcalib},
as well as the physics-modelling corrections described in
Section~\ref{sec:phymod}, are applied consistently in the $W$- and
$Z$-boson samples.
The $Z$-boson sample differs from the $W$-boson sample in the
selection requirements, as described in Section~\ref{sec:eventsel}.
In addition to the event-selection requirements described there, the
transverse momentum of the dilepton system, $\pt^{\ell\ell}$, is required to be smaller than $30\GeV$.

The missing transverse momentum in $Z$-boson events is defined by treating
one of the two decay leptons as a neutrino and ignoring its transverse momentum when defining the event kinematics.
This procedure allows the \mpt{} and $\mt$ variables to be defined in the $Z$-boson sample in close analogy to their definition
in the $W$-boson sample. The procedure is repeated, removing the positive and negative lepton in turn.

In the $Z$-boson sample, the background contribution arising from top-quark
and electroweak production is estimated using Monte Carlo samples. Each process is normalised using the
corresponding theoretical cross sections, evaluated at NNLO in the
perturbative expansion of the strong coupling constant.
This background contributes a 0.12\% fraction in each channel. 
In the muon channel, the background contribution from multijet
events is estimated to be smaller than 0.05\% using simulated event samples of $b\bar{b}$ and
$c\bar{c}$ production, and neglected.
In the electron channel, a data-driven estimate of the multijet
background contributes about a 0.1\% fraction, before applying the
isolation selections, which reduce it to a negligible level.

\begin{figure}
  \begin{center}
    \subfloat[]{\includegraphics[width=0.49\textwidth]{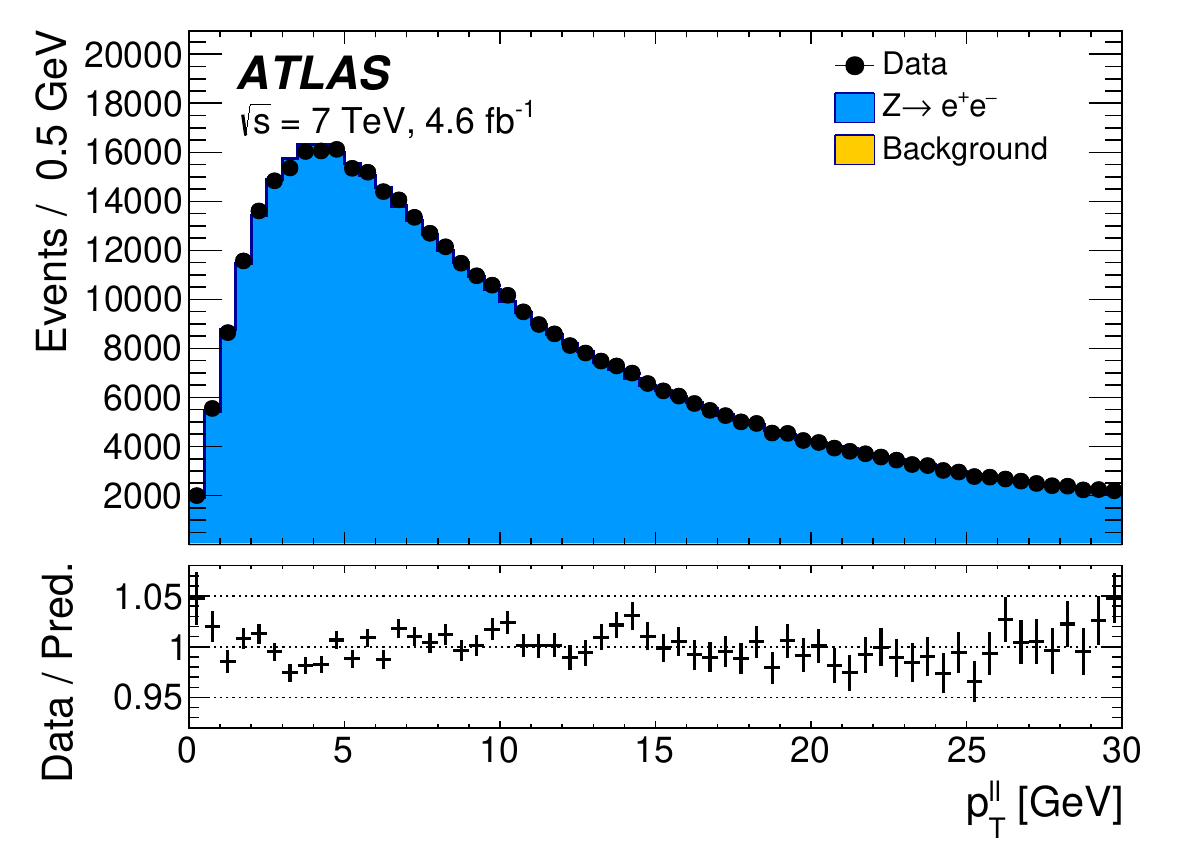}}
    \subfloat[]{\includegraphics[width=0.49\textwidth]{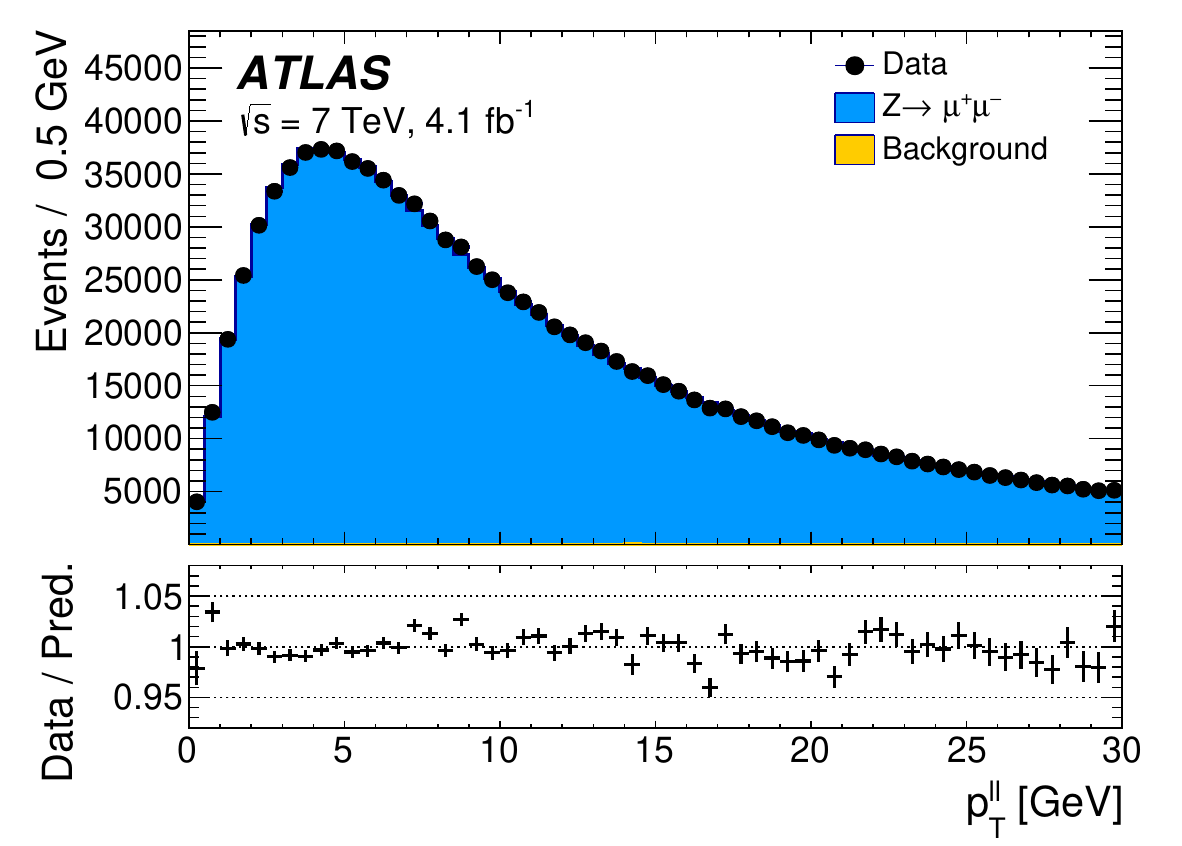}}\\
    \subfloat[]{\includegraphics[width=0.49\textwidth]{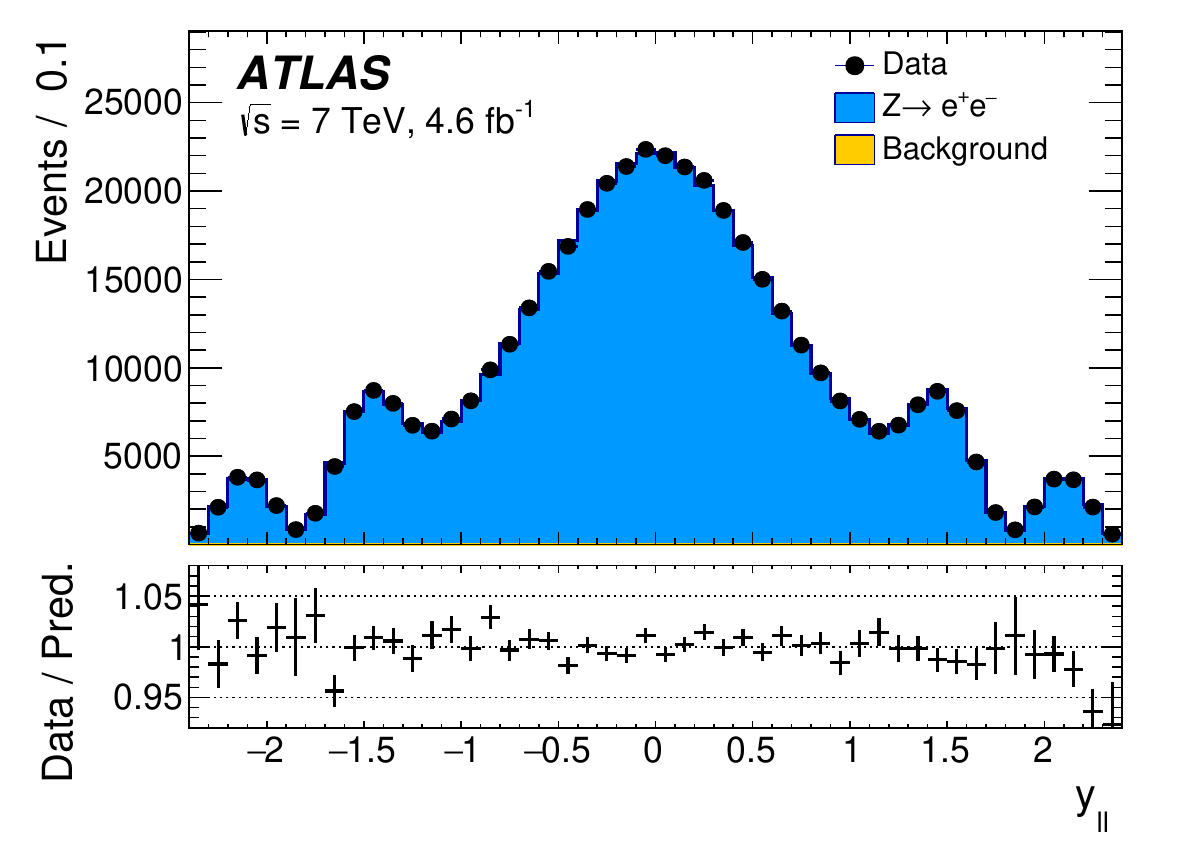}}
    \subfloat[]{\includegraphics[width=0.49\textwidth]{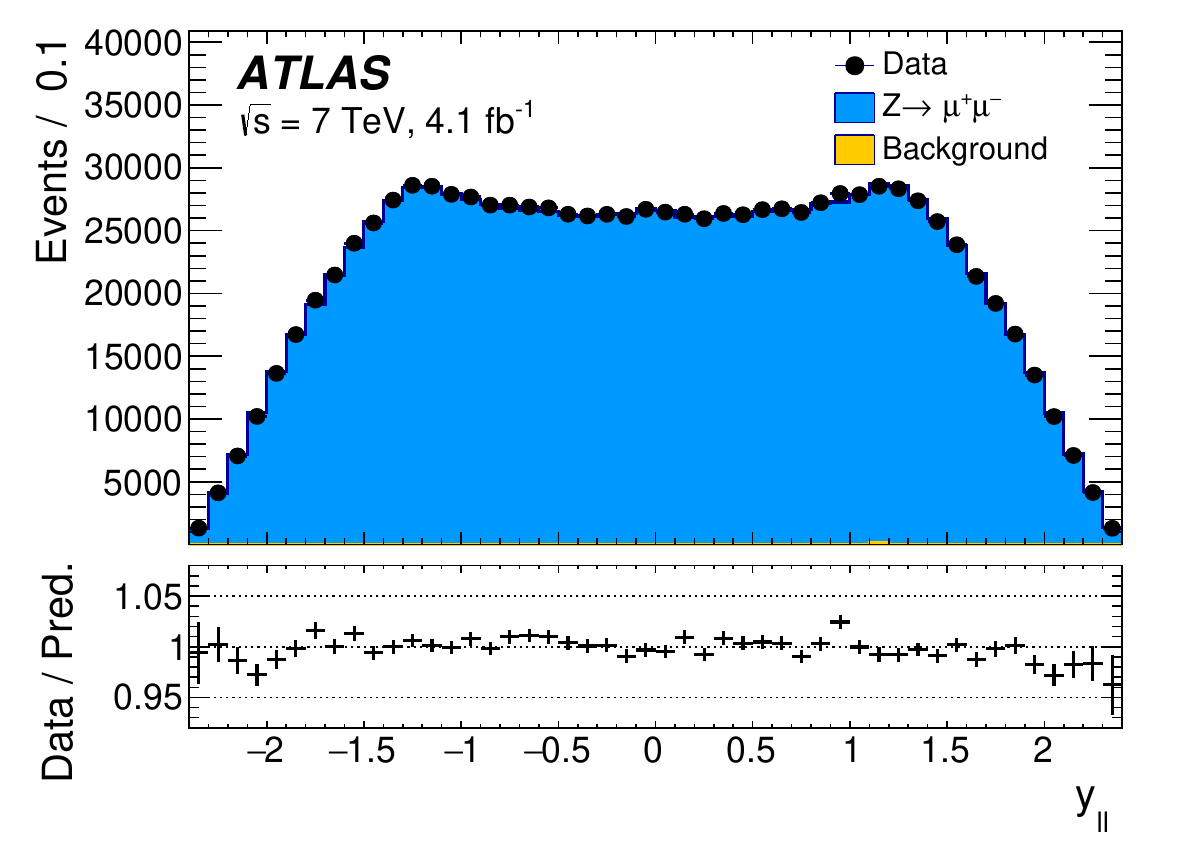}}
  \end{center}
  \caption{The (a,b) $\pt^{\ell\ell}$ and (c,d) $y_{\ell\ell}$ distributions in $Z$-boson events for the (a,c) electron and (b,d) muon decay channels.
    The data are compared to the simulation including signal and backgrounds.
    Detector calibration and physics-modelling corrections are applied
    to the simulated events. Background events contribute less than 0.2\% of the observed distributions. The lower panels show the
    data-to-prediction ratios, with the error bars showing the
    statistical uncertainty.}
  \label{fig:ZBosonControl1}
\end{figure}

\begin{figure}
  \begin{center}
    \subfloat[]{\includegraphics[width=0.49\textwidth]{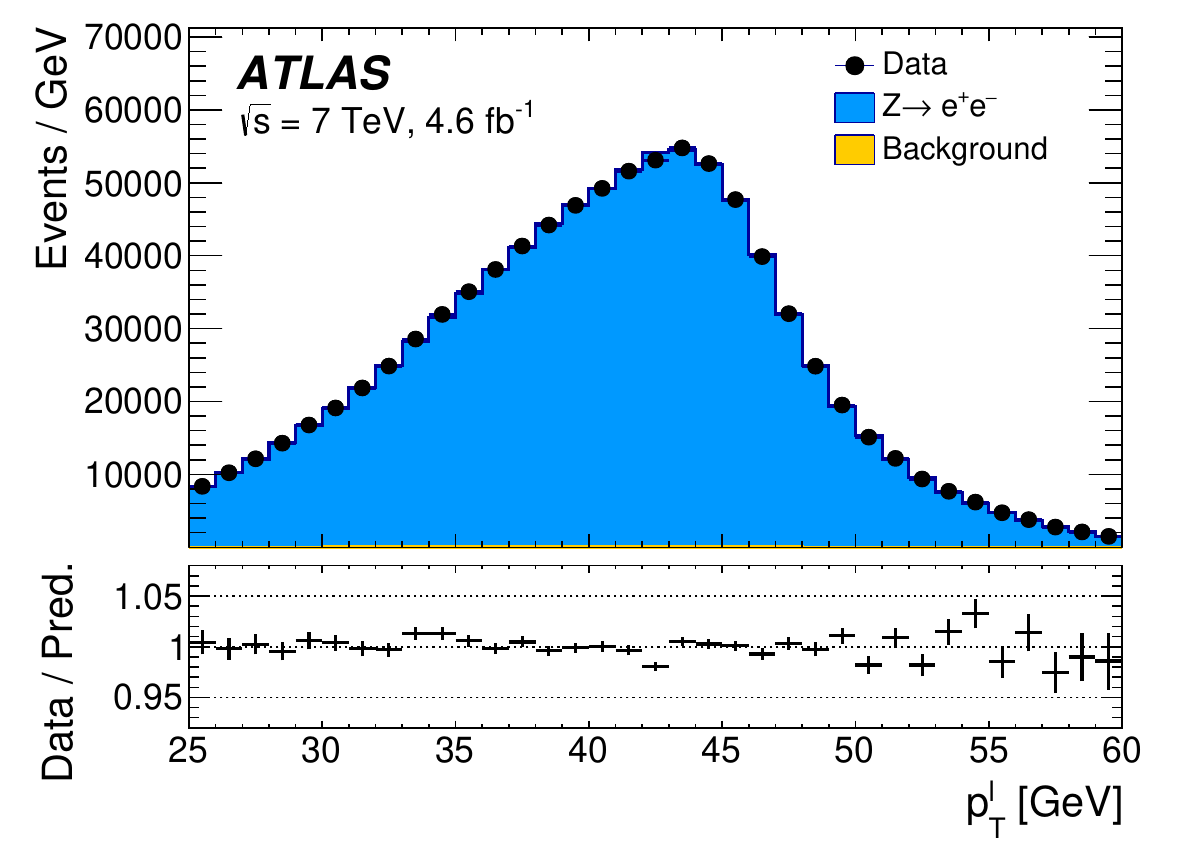}}
    \subfloat[]{\includegraphics[width=0.49\textwidth]{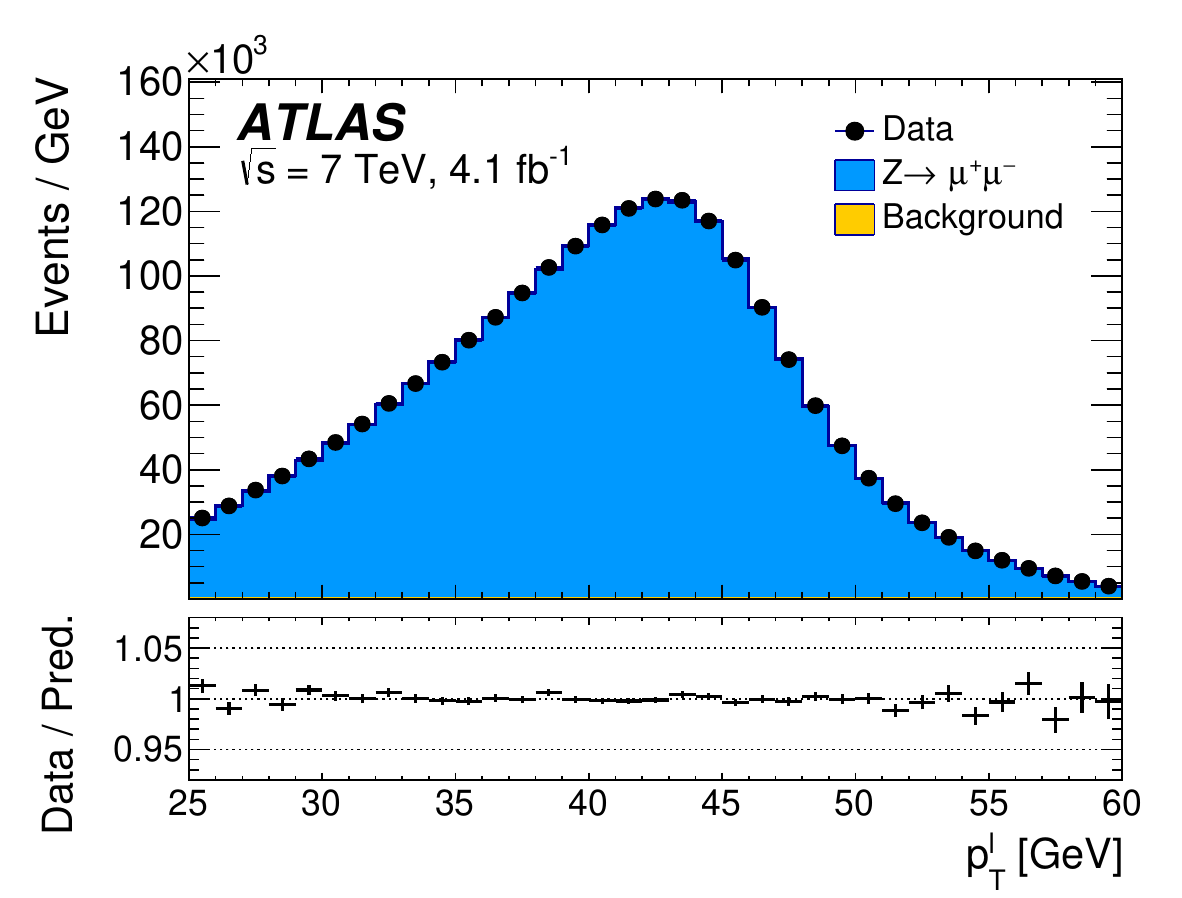}}\\
    \subfloat[]{\includegraphics[width=0.49\textwidth]{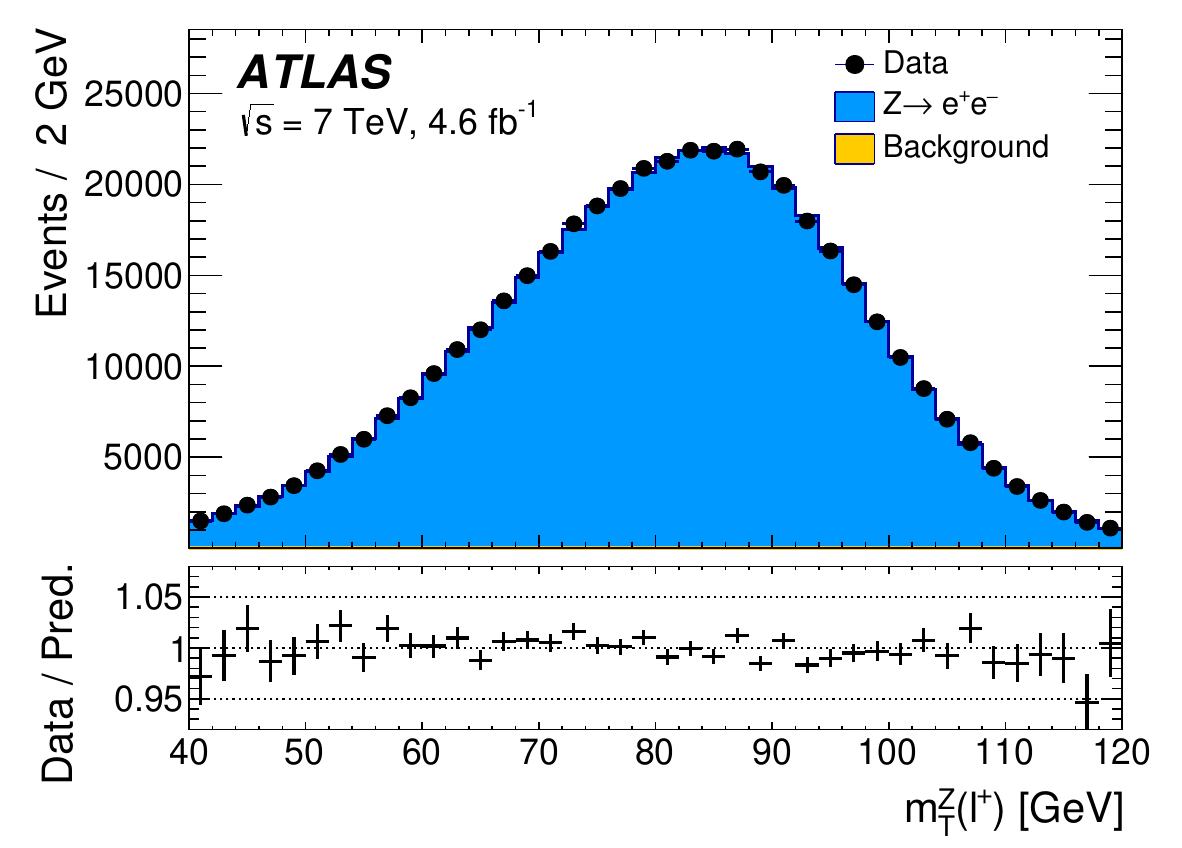}}
    \subfloat[]{\includegraphics[width=0.49\textwidth]{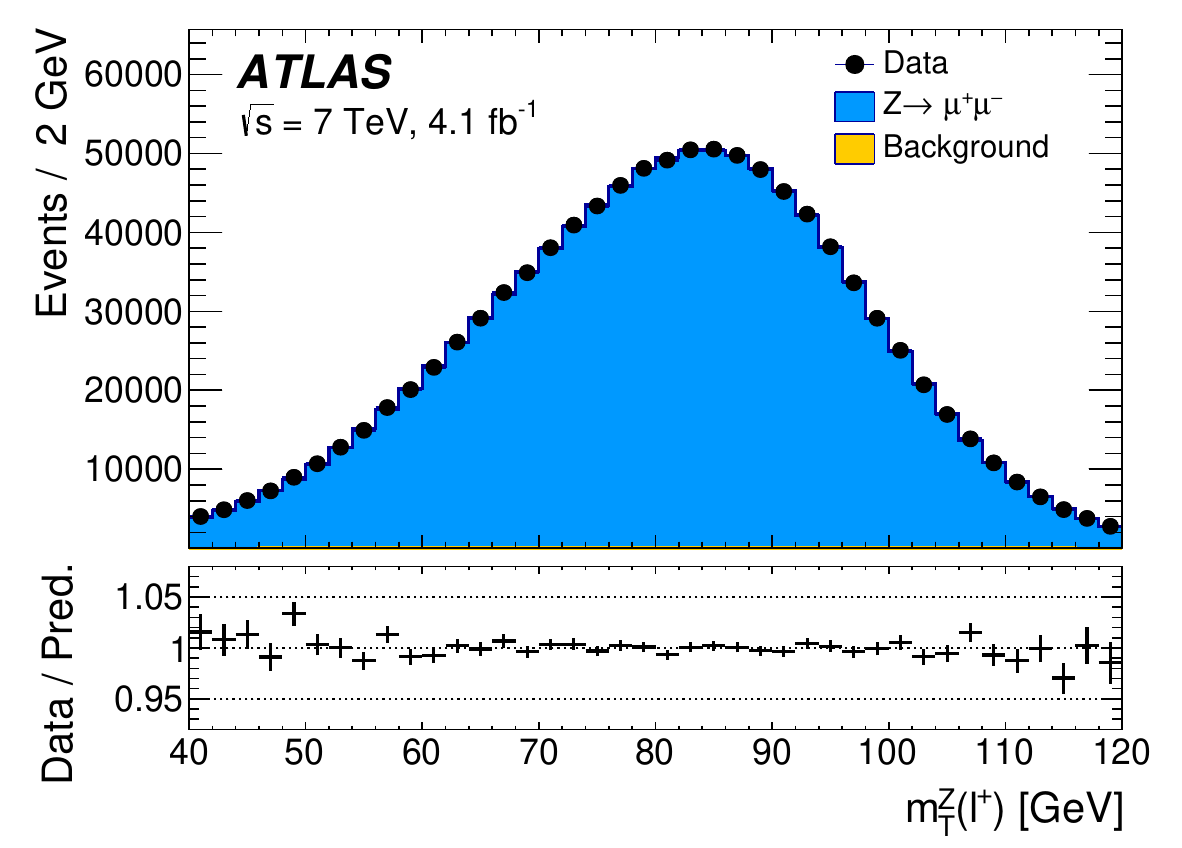}} \\
    \subfloat[]{\includegraphics[width=0.49\textwidth]{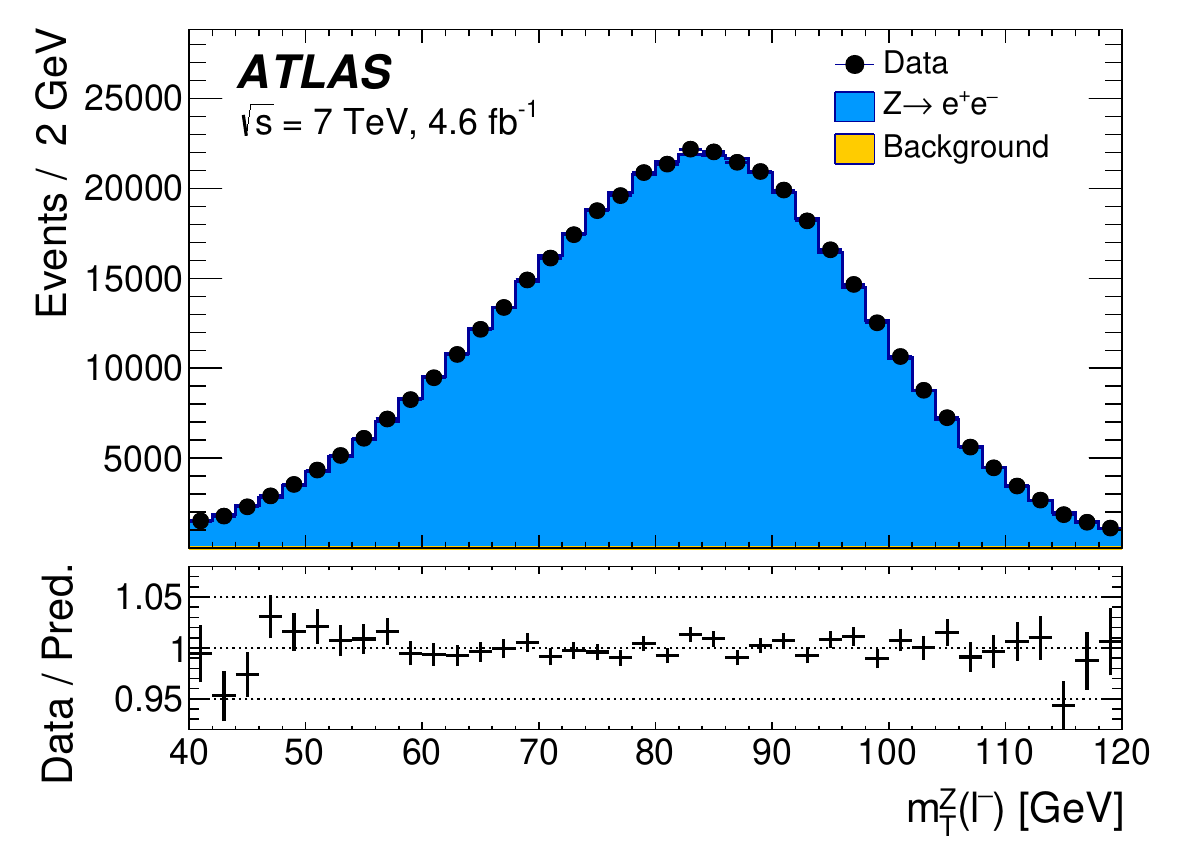}}
    \subfloat[]{\includegraphics[width=0.49\textwidth]{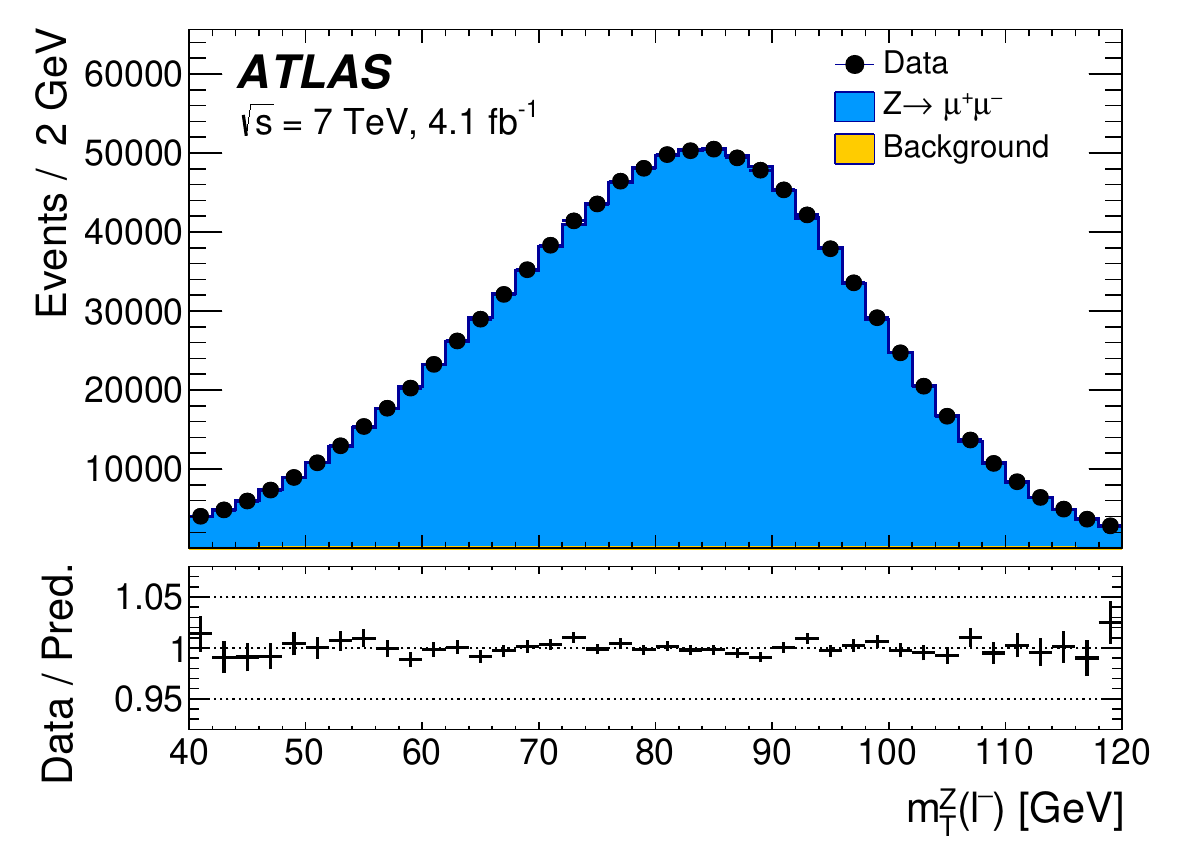}} \\
  \end{center}
  \caption{
    The $\pt^{\ell}$ distribution in the (a) electron and (b) muon
    channels, and \mt{} distributions in the (c,e) electron and (d,f) muon
    decay channels for $Z$ events when the (c,d) negatively charged, or
    (e,f) positively charged lepton is removed.
    The data are compared to the simulation including signal and backgrounds.
    Detector calibration and physics-modelling corrections are applied
    to the simulated events. Background events contribute less than 0.2\% of the observed distributions. The lower panels show the
    data-to-prediction ratios, with the error bars showing the
    statistical uncertainty.}
  \label{fig:ZBosonControl2}
\end{figure}

Figure~\ref{fig:ZBosonControl1} shows the reconstructed distributions
of $\pt^{\ell\ell}$ and $y_{\ell\ell}$ in selected $Z$-boson events;
these distributions are not sensitive to the value of
$m_Z$. Figure~\ref{fig:ZBosonControl2} shows the corresponding
distributions for $\pt^{\ell}$ and $\mt$, variables which are sensitive
to $m_Z$. Data and simulation agree at the level of 1--2\% percent in
all the distributions.

The mass of the $Z$ boson is extracted with template fits to the
$m_{\ell\ell}$, $\pt^{\ell}$, and $\mt$ kinematic distributions.
The extraction of the $Z$-boson mass from the dilepton invariant mass distribution
is expected to yield, by construction, the value of $m_Z$ used as
input for the muon-momentum and electron-energy calibrations,
providing a closure test of the lepton calibration procedures.
The $\pt^\ell$ distribution is very sensitive to the physics-modelling
corrections described in Section~\ref{sec:phymod}. The comparison of
the value of $m_Z$ extracted from the $\pt^\ell$ distribution with the
value used as input for the calibration tests the physics modelling and efficiency corrections. Finally, $m_Z$ measurements from the \mt{}
distribution provides a test of the recoil calibration.

Similarly to the $W$-boson mass, the value of $m_Z$ is determined by
minimising the $\chi^2$ function of the compatibility test between the
templates and the measured distributions. The templates are generated with values
of $m_Z$ in steps of $4$ to $25\MeV$ within a range of $\pm 450\MeV$,
centred around a reference value corresponding to the LEP combined value,
$m_Z= 91187.5\MeV$~\cite{ALEPH:2005ab}.
The $\chi^2$ function is interpolated with a second order polynomial. The
minimum of the $\chi^2$ function yields the extracted value of $m_Z$,
and the difference between the extracted value of $m_Z$ and the reference
value is defined as $\Delta m_{Z}$.
The ranges used for the extraction are $[80, 100]\GeV$
for the $m_{\ell\ell}$ distributions, $[30,55]\GeV$ for the
$\pt^\ell$ distribution, and $[40,120]\GeV$ for the \mt{}
distribution.
The extraction of $m_Z$ from the \mt{} distribution is performed
separately for positively and negatively charged leptons in the event, by
reconstructing \mt{} from the kinematic properties of one of the two
charged leptons and of the recoil reconstructed by treating the other
as a neutrino.

\begin{figure*}
  \centering
  \includegraphics[width=0.85\textwidth]{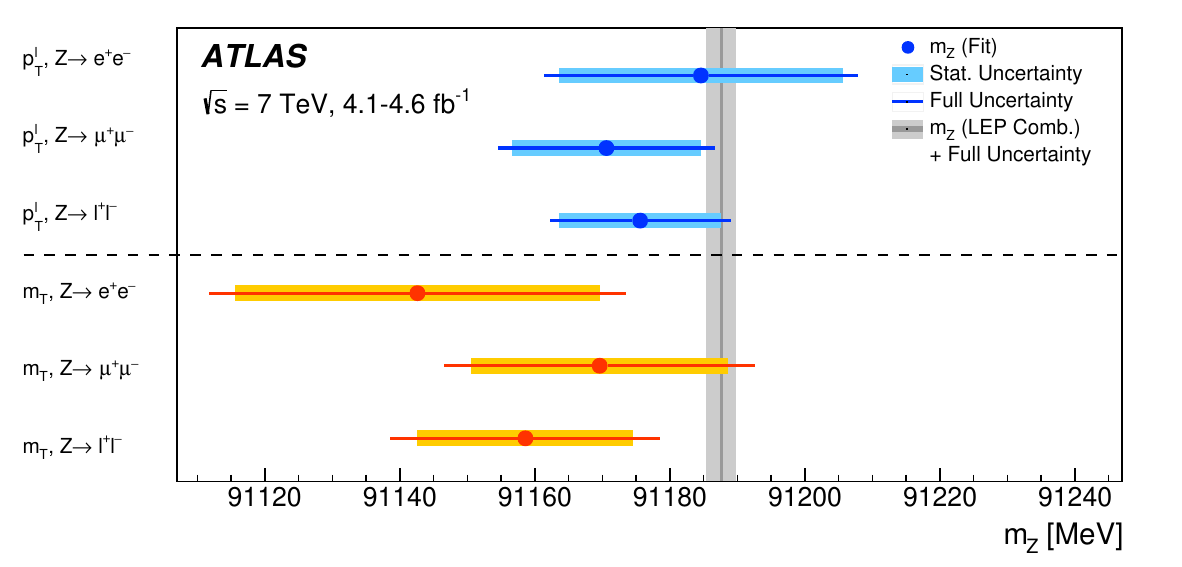} 
  \caption{Summary of the $m_Z$ determinations from the $\pt^\ell$ and
    $\mt$ distributions in the muon and electron decay channels. The
    LEP combined value of $m_Z$, which is used as input for the
    detector calibration, is also indicated. The horizontal and vertical bands
    show the uncertainties of the $m_Z$ determinations and of the
    LEP combined value, respectively.}
  \label{fig:MzfitResults}
\end{figure*}

$Z$-boson mass fits are performed using the
\mt{} and $\pt^\ell$ distributions in the electron and muon decay channels, inclusively in $\eta$ and separately for positively
and negatively charged leptons. The results of the fits are summarised in
Figure~\ref{fig:MzfitResults} and Table~\ref{tab:zfitspt}. The $\pt^\ell$ fit results
include all lepton reconstruction systematic uncertainties except the $Z$-based energy or momentum scale calibration uncertainties; the $\mt$ fit results include recoil calibration systematic
uncertainties in addition. Physics-modelling uncertainties are neglected.

The value of $m_Z$ measured from positively charged leptons is correlated with the corresponding
extraction from the negatively charged leptons.
The $\pt^\ell$ distributions for positively and negatively charged leptons are statistically
independent, but the \mt{} distributions share the same reconstructed recoil event by event, and are statistically correlated. 
In both cases, the decay of the $Z$-boson induces a kinematical
correlation between the distributions of positively and negatively charged leptons. The correlation is estimated by constructing
two-dimensional $\ell^+$ and $\ell^-$ distributions, separately for $\pt^{\ell}$ and $\mt$, fluctuating the bin contents
of these distributions within their uncertainties, and repeating the
fits for each pseudodata sample.
The correlation values are $-7\%$ for the $\pt^{\ell}$
distributions, and $-12\%$ for the \mt{} distributions.

Accounting for the experimental uncertainties as described above, the combined extraction of $m_Z$ from the $\pt^\ell$ distribution yields a 
result compatible with the reference value within $0.9$ standard deviations.
The difference between the $m_Z$ extractions from positively and negatively charged
lepton distributions is compatible with zero within $1.4$ standard deviations.
For the extraction from the \mt{} distribution, the compatibility with
the reference value of $m_Z$ is at the level of $1.5$ standard deviations.
Fits using the lepton pair invariant mass distribution agree with the
reference, yielding $\Delta m_Z = 1\pm 3\MeV$ in the muon channel and $\Delta m_Z =
3\pm 5 \MeV$ in the electron channel, as expected from the calibration
procedure. In summary, the consistency tests based on the $Z$-boson
sample agree with the expectations within the experimental uncertainties.

\begin{table}[tp]
\begin{small}
\centering
\resizebox{\textwidth}{!}{
\begin{tabular}{lrrrrrr}
\toprule
Lepton charge 	& \multicolumn{2}{c}{\hspace{1.5cm}$\ell^+$}		& \multicolumn{2}{c}{\hspace{1.5cm}$\ell^-$}		& \multicolumn{2}{c}{\hspace{1.5cm}Combined} \\
Distribution 	& $\pt^{\ell}$		& $\mt$         & $\pt^{\ell}$		& $\mt$         & $\pt^{\ell}$		& $\mt$ \\
\midrule
$\Delta m_Z$~[\MeV]\\
\,\,\,\, $Z \to ee$   	& $ 13 \pm 31\pm 10$ 	        & $-93 \pm 38\pm15$ 	& $-20 \pm 31\pm 10 $ 	& $ 4 \pm 38\pm15$  	& $ -3 \pm 21\pm10$ 	& $ -45 \pm 27\pm15$	\\
\,\,\,\, $Z \to \mu \mu$	& $1 \pm 22\pm\;\,8$ 		& $-35 \pm 28\pm13$  	& $-36 \pm 22\pm\;\,8$ 	& $-1 \pm 27\pm13$   	& $-17 \pm 14\pm\;\,8$ 	& $-18 \pm 19\pm13$	\\ 
\,\,\,\, Combined	& $ 5 \pm 18\pm\;\,6$ 		& $ -58  \pm 23\pm12$ 	& $-31 \pm 18\pm\;\,6$ 	& $ 1 \pm 22\pm12$ 	& $ -12 \pm 12\pm\;\,6$ 	& $ -29 \pm 16\pm12$	\\
\bottomrule
\end{tabular}}
\caption{ Difference between $Z$-boson mass, extracted from $\pt^{\ell}$ and \mt\ distributions, and the LEP combined value. 
The results are shown separately for the electron and muon decay
  channels, and their combination. The first quoted uncertainty is
  statistical, the second is the experimental systematic uncertainty, which includes lepton efficiency and recoil
  calibration uncertainties where applicable. Physics-modelling uncertainties are neglected.\label{tab:zfitspt}}
\end{small}
\end{table}

\section{Backgrounds in the $W$-boson sample \label{sec:background}}
The $W$-boson event sample, selected as described in Section~\ref{sec:eventsel},
includes events from various background processes. Background contributions
from $Z$-boson, $W\rightarrow\tau\nu$, boson pair, and top-quark production are estimated using simulation. Contributions from multijet production are
estimated with data-driven techniques.

\subsection{Electroweak and top-quark backgrounds}

The dominant sources of background contribution in the
$W\rightarrow\ell\nu$ sample are $Z \rightarrow \ell\ell$ events,
in which one of the two leptons escapes detection, and
$W\rightarrow\tau\nu$ events, where the $\tau$ decays to an electron or
muon. These background contributions are estimated using the
\textsc{Powheg+Pythia 8} samples after applying the modelling
corrections discussed in Section~\ref{sec:phymod}, which include NNLO
QCD corrections to the angular coefficients and rapidity
distributions, and corrections to the vector-boson transverse
momentum.
The $Z\rightarrow ee$ background represents 2.9\% of the $W^+\to e\nu$
sample and 4.0\% of the $W^-\to e\nu$ sample. In the muon channel, the
$Z\rightarrow \mu\mu$ background represents 4.8\% and~6.3\% of the
$W^+\to \mu\nu$ and $W^-\to \mu\nu$ samples, respectively.
The $W\to\tau\nu$ background represents 1.0\% of the selected sample in
both channels, and the $Z\to\tau\tau$ background contributes approximately 0.12\%. The
normalisation of these processes relative to the $W$-boson signal and
the corresponding uncertainties are discussed in
Section~\ref{sec:samples}.
A relative uncertainty of 0.2\% is assigned to the normalisation of the
$W\rightarrow\tau\nu$ samples with respect to the $W$-boson signal
sample, to account for the uncertainty in the $\tau$-lepton branching fractions to electrons and muons. In the determination of the $W$-boson mass, the
variations of $m_W$ are propagated to the $W\rightarrow\tau\nu$
background templates in the same way as for the signal.

Similarly, backgrounds involving top-quark (top-quark pairs and single top-quark)
production, and boson-pair production are estimated using simulation,
and normalisation uncertainties are assigned as discussed in
Section~\ref{sec:samples}. These processes represent 0.11\% and 0.07\%
of the signal event selection, respectively.

Uncertainties in the distributions of the $W\rightarrow\tau\nu$ and $Z\rightarrow \ell\ell$ processes are described by the physics-modelling uncertainties discussed in
Section~\ref{sec:phymod}, and are treated as fully correlated with the signal. Shape uncertainties for boson-pair production and top-quark production are considered negligible compared to the
uncertainties in their cross sections, given the small contributions of these processes to the signal event selection.

\subsection{Multijet background}

Inclusive multijet production in strong-interaction processes constitutes a
significant source of background.
A fraction of multijet events contains semileptonic decays of bottom
and charm hadrons to muons or electrons and neutrinos, and can pass
the $W$-boson signal selection. In addition, inclusive jet production
contributes to the background if one jet is misidentified as electron
or muon, and sizeable missing transverse momentum is reconstructed in the
event.
In-flight decays of pions or kaons within the tracking region can
mimic the $W$-boson signal in the muon channel. In the electron
channel, events with photon conversions and hadrons misidentified as electrons can
be selected as $W$-boson events.
Due to the small selection probability for multijet events, their
large production cross section, and the relatively complex modelling
of the hadronisation processes, the multijet background contribution
cannot be estimated precisely using simulation, and a data-driven
method is used instead.

The estimation of the multijet background contribution follows similar
procedures in the electron and muon decay channels, and relies on
template fits to kinematic distributions in background-dominated regions. 
The analysis uses the distributions of \mpt{}, \mt, and the $\pt^\ell/\mt$ ratio,
where jet-enriched regions are obtained by relaxing a subset of the
signal event-selection requirements. The first kinematic region, denoted FR1, is
defined by removing the \mpt\ and \mt\ requirements from the event
selection. A second kinematic region, FR2, is defined in the same way
as FR1, but by also removing the requirement on \ut. Multijet background
events, which tend to have smaller values of \mpt{} and \mt\ than the
signal, are enhanced by this selection. The $\pt^\ell/\mt$
distribution is sensitive to the angle between the $\pt^{\ell}$ and
\mpt\ vectors in the transverse plane. Whereas $W$-boson events are
expected to peak at values of $\pt^\ell/\mt=0.5$, relatively large
tails are observed for multijet events.

Templates of the multijet background distributions for these
observables are obtained from data by inverting the lepton energy-isolation
requirements. Contamination of these control regions by electroweak
and top production is estimated using simulation and subtracted.
In the muon channel, the anti-isolation requirements 
are defined from the ratio of the scalar sum of the \pt{} of
tracks in a cone of size $\Delta R < 0.2$ around the reconstructed muon to
the muon \pt. The isolation variable $\pt^{\mu,\textrm{cone}}$,
introduced in Section~\ref{sec:reco}, is required to satisfy $c_1 <
\pt^{\mu,\textrm{cone}} / \pt^\ell < c_2$, where the anti-isolation
boundaries $c_1$ and $c_2$ are varied as discussed below. In order to
avoid overlap with the signal region, the lower boundary $c_1$ is
always larger than $0.1$.  
In the electron channel, the scalar sum of the \pt{} of tracks in a cone of size $\Delta R < 0.4$ around the reconstructed electron, defined as $\pt^{e,\textrm{cone}}$ in Section~\ref{sec:reco}, is used to define the templates, while the requirements on the calorimeter isolation are omitted.

\begin{figure}[tpb]
\begin{center}
\subfloat[]{\includegraphics[width=0.49\textwidth]{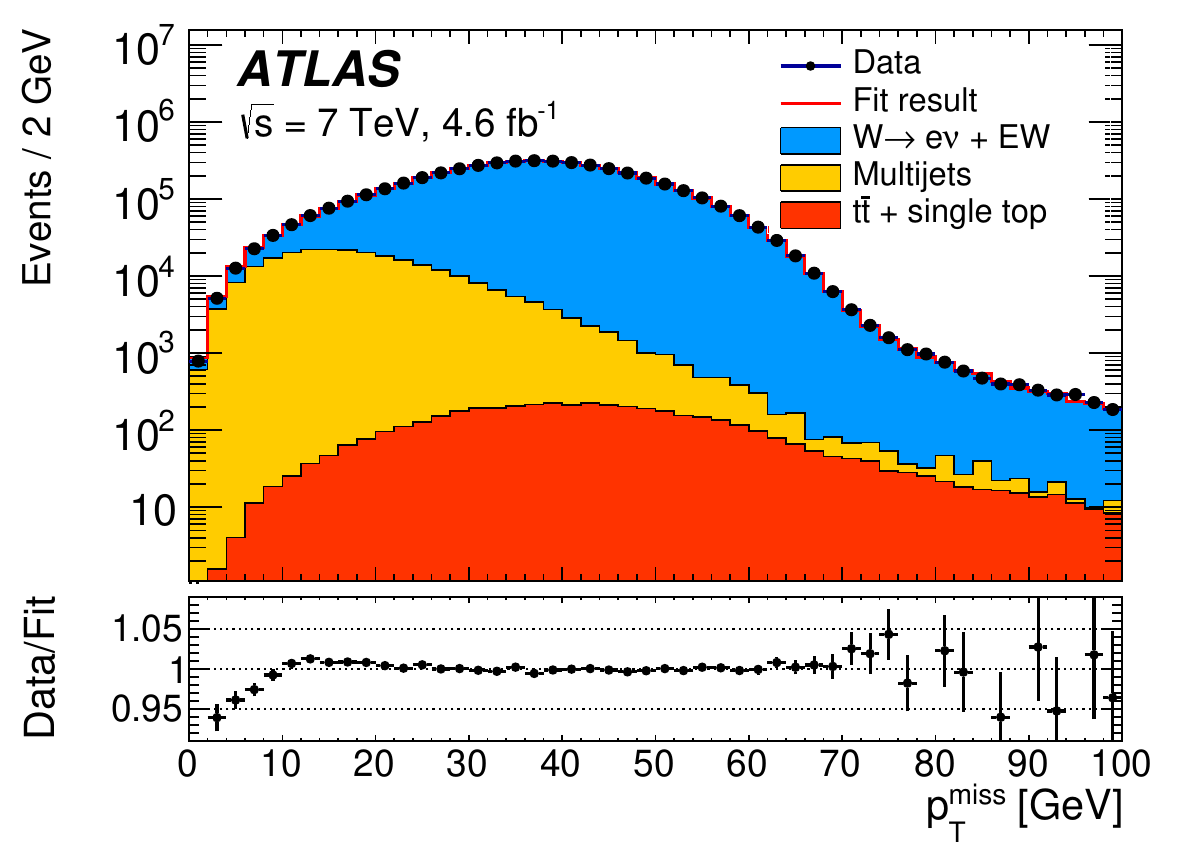}}
\subfloat[]{\includegraphics[width=0.49\textwidth]{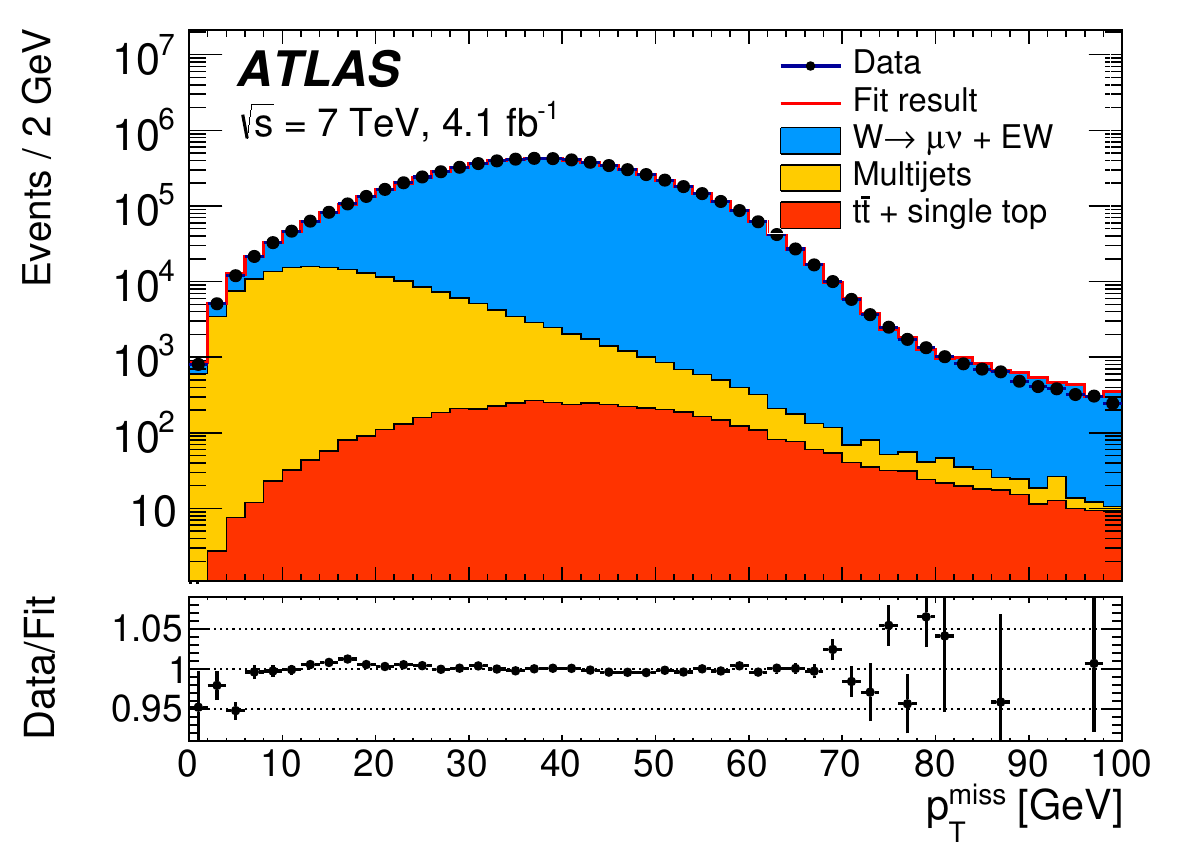}}\\
\subfloat[]{\includegraphics[width=0.49\textwidth]{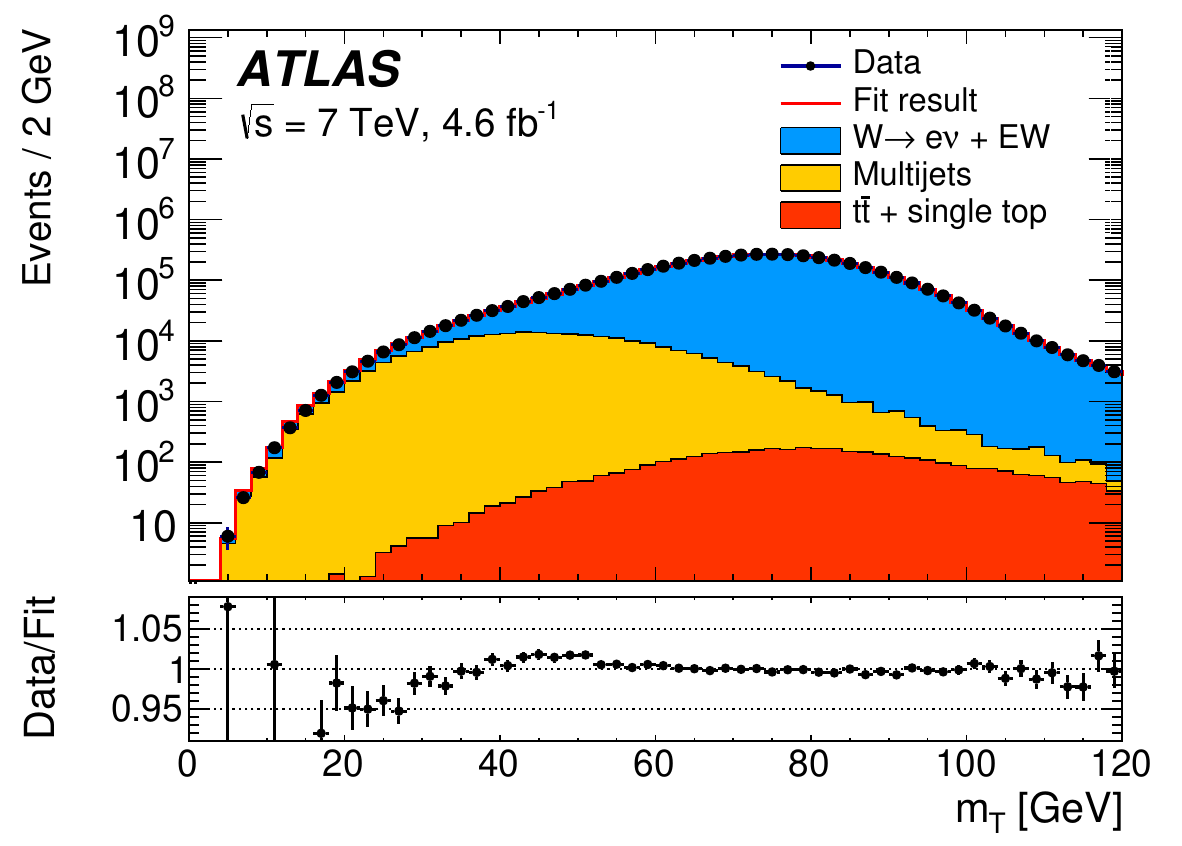}}
\subfloat[]{\includegraphics[width=0.49\textwidth]{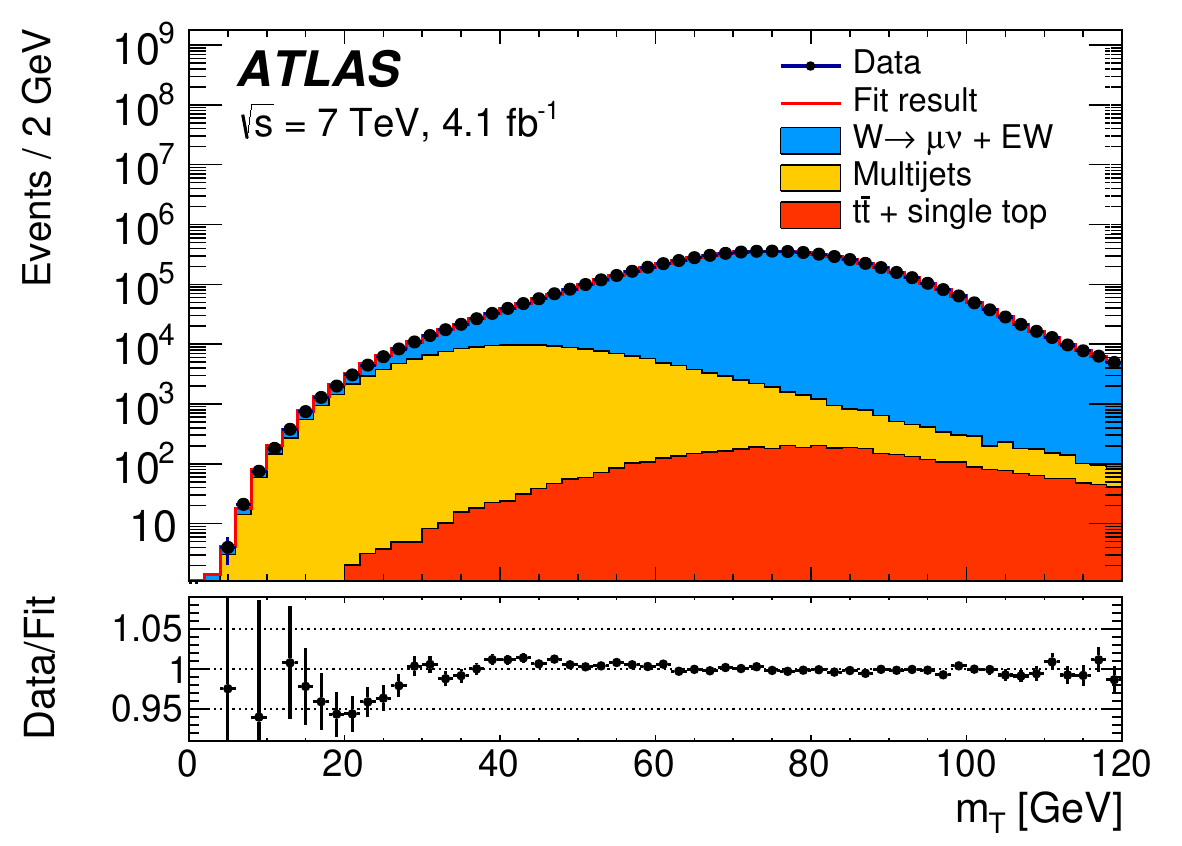}}\\
\subfloat[]{\includegraphics[width=0.49\textwidth]{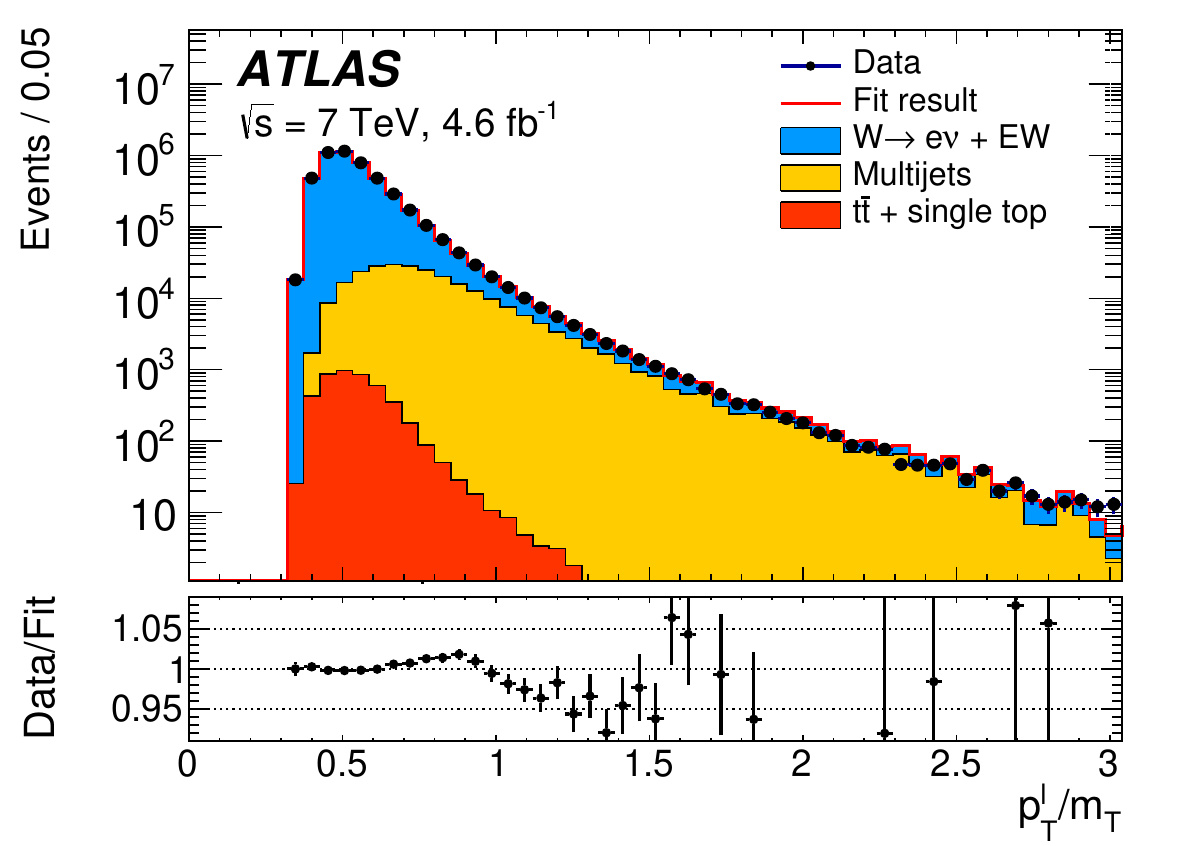}}
\subfloat[]{\includegraphics[width=0.49\textwidth]{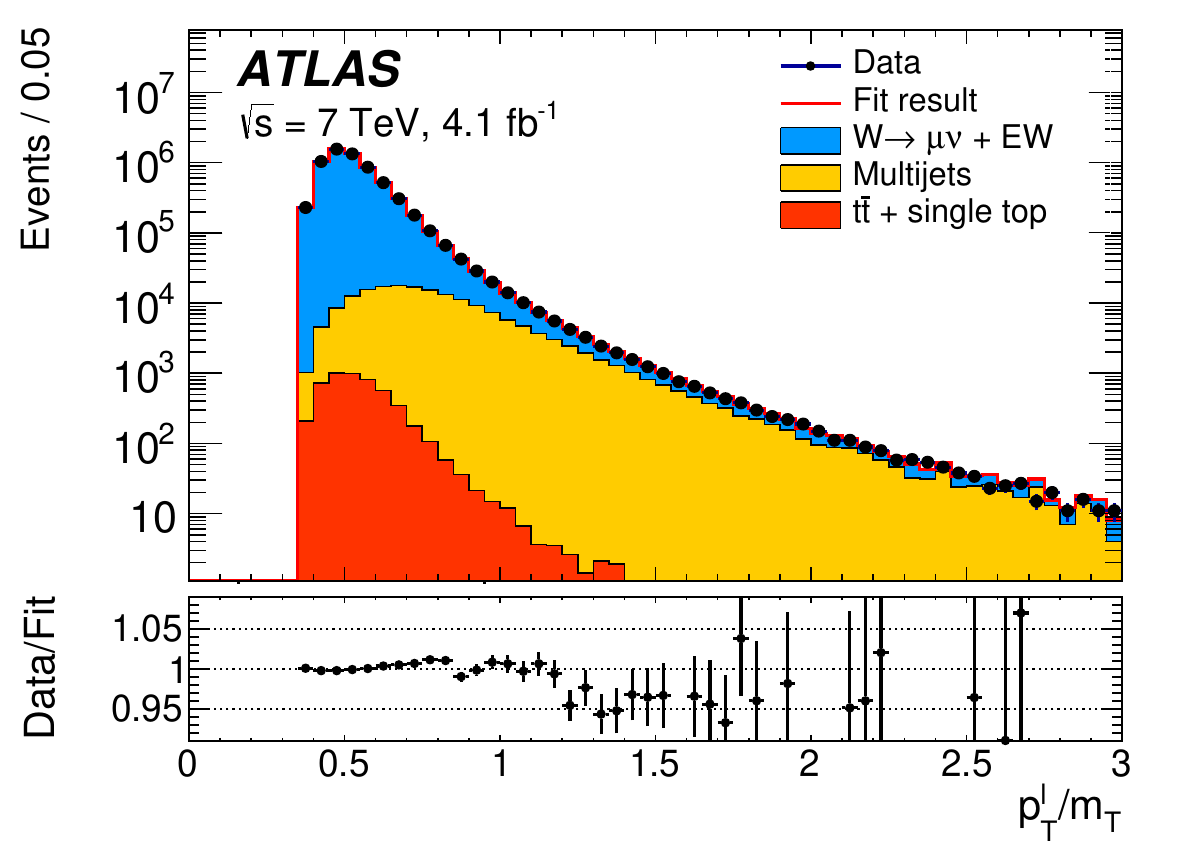}}
\caption{Example template
  fits to the (a,b) \mpt, (c,d) \mt, and (e,f) $\pt^\ell/\mt$
  distributions in the FR1 kinematic region, in the (a,c,e) electron 
  and (b,d,f) muon decay channels. Multijet templates are derived
  from the data requiring $4\GeV<\pt^{e,\textrm{cone}}<8\GeV$ in the electron channel, and $0.2<\pt^{\mu,\textrm{cone}} / \pt^\ell<0.4$ in the muon channel. The data are compared to the simulation including signal and background contributions.}
 \label{fig:QCDfits}
\end{center}
\end{figure}

\begin{figure}
\begin{center}
  \subfloat[]{\includegraphics[width=0.49\textwidth]{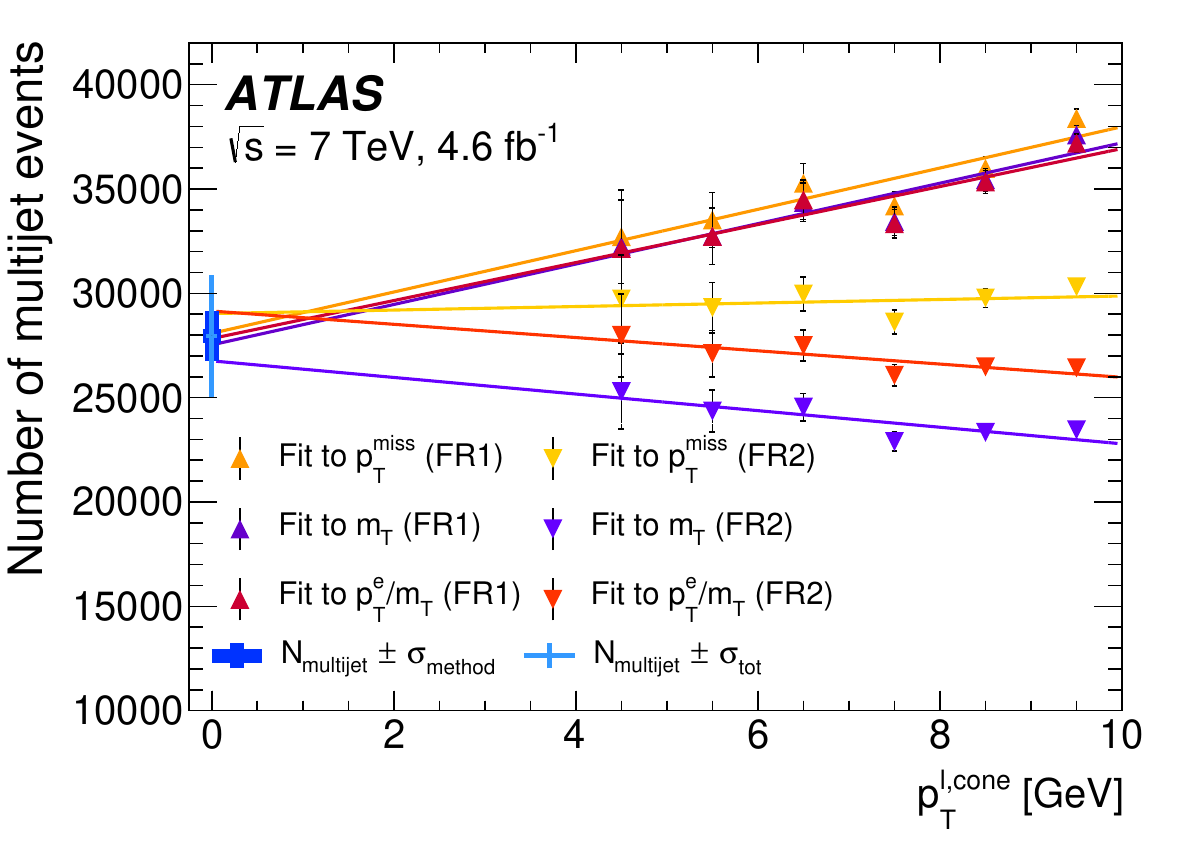}}
  \subfloat[]{\includegraphics[width=0.49\textwidth]{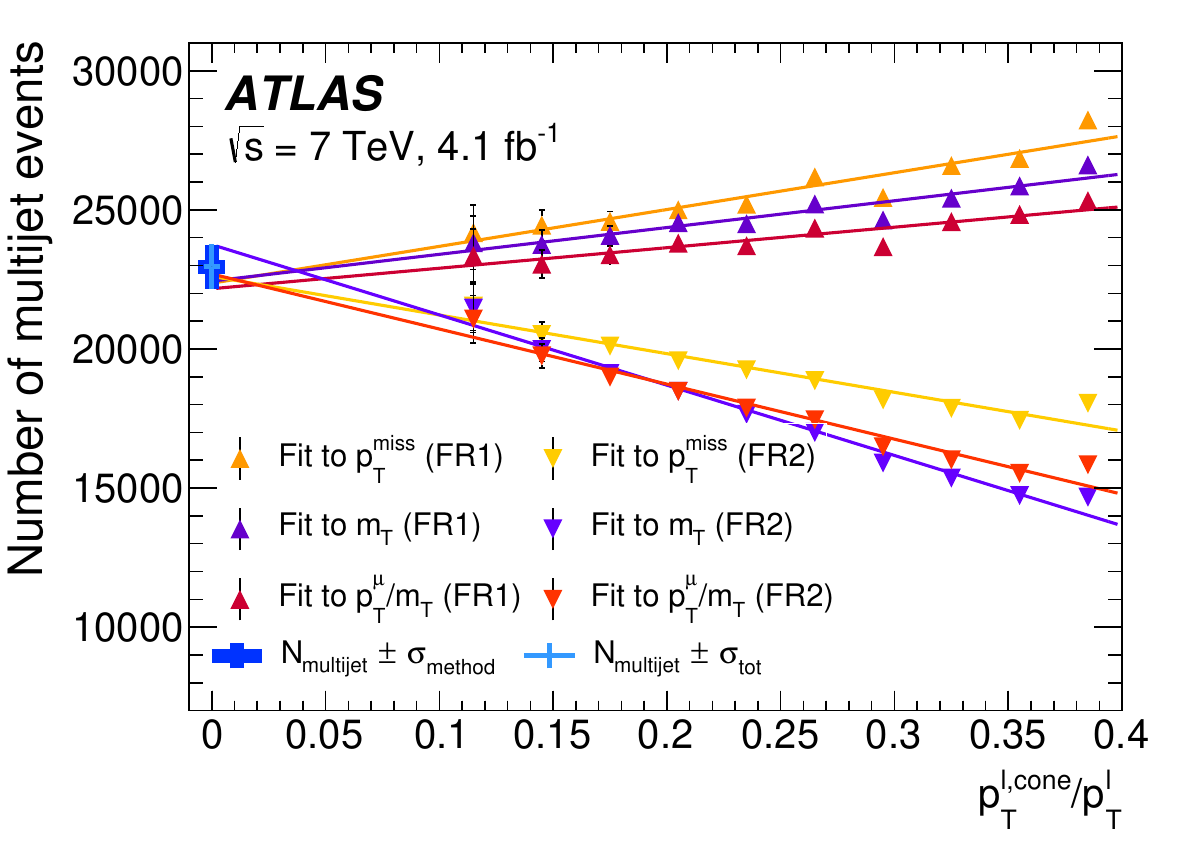}}
\caption{Estimated number of multijet-background events as a function of the lower bound of the isolation-variable range used to define the control regions, for (a) electron and (b) muon decay channel. The estimation is performed for the two regions FR1 and FR2 and three distributions  \mpt, \mt, and $\pt^\ell/\mt$, as described in the text. The linear extrapolations are indicated by the solid lines.  The thick crosses show the results of the linear extrapolation of the background estimate to the signal region, including uncertainties from the extrapolation only. The thin crosses also include the uncertainty induced by the contamination of the control regions by EW and top-quark processes.}
 \label{Fig:QCDExtrapolation}
\end{center}
\end{figure}

The multijet background normalisation is determined by fitting each of the
\mpt{}, $\mt$, and $\pt^\ell/\mt$ distributions in the two kinematic
regions FR1 and FR2, using templates of these distributions based on multijet events and obtained with several
ranges of the anti-isolation variables. The multijet background in the
signal region is determined by correcting the multijet fraction fitted in
the FR1 and FR2 for the different efficiencies
of the selection requirements of the signal region. In the electron channel,
$c_1$ is varied from $4\GeV$ to $9\GeV$ in steps of $1\GeV$, and $c_2$ is set to $c_2 = c_1+1\GeV$.
In the muon channel, $c_1$ is varied from 0.1 to 0.37 in steps of 0.03,
and $c_2$ is set to $c_2=c_1+0.03$. Example results of template fits
in the electron and muon channels are shown in Figure~\ref{fig:QCDfits}. 
The results corresponding to the various observables and to the
different kinematic regions are linearly extrapolated in the isolation
variables to the signal regions, denoted by $c_1 = 0$.
Figure~\ref{Fig:QCDExtrapolation} illustrates the extrapolation procedure.

The systematic uncertainty in the multijet background fraction is
defined as half of the largest difference between the results extrapolated from the different kinematic regions and observables. The multijet background contribution is estimated separately in all measurement categories.
In the electron channel, the multijet background fraction rises from
0.58$\pm$0.08\% at low $|\eta_\ell|$ to 1.73$\pm$0.19\% in the last
measurement bin, averaging the $W^+$ and $W^-$ channels. In the muon
channel, the charge-averaged multijet background fraction decreases
from 0.72$\pm$0.07\% to 0.49$\pm$0.03\%, when going from low to high
$|\eta_\ell|$. The uncertainties in the multijet background fractions are sufficient to account for the observed residual discrepancies between the fitted distributions and the data (see Figure~\ref{fig:QCDfits}).
The estimated multijet background yields are consistent between $W^+$ and
$W^-$, but the multijet background fraction is smaller in
the $W^+$ channels due to the higher signal yield.

Corrections to the shape of the multijet background contributions and
corresponding uncertainties in the distributions used to measure the
$W$-boson mass are estimated with a similar procedure.
The kinematic distributions in the control regions are
obtained for a set of anti-isolation ranges, and parameterised with
linear functions of the lower bound of the anti-isolation requirement.
The distributions are extrapolated to the
signal regions accordingly. Uncertainties in the extrapolated
distributions are dominated by the statistical uncertainty, which is
determined with a toy MC method by fluctuating within their
statistical uncertainty the bin contents of the histograms in
the various anti-isolation ranges.
The resulting multijet background distribution is propagated to the
templates, and the standard deviation of the determined values of
$m_W$ yields the estimated uncertainty due to the shape of the multijet background.
Uncertainties due to the choice of parameterisation are small in
comparison and neglected.  

Uncertainties in the normalisation of multijet, electroweak, and top-quark
background processes are considered correlated across decay channels,
boson charges and rapidity bins, whereas the uncertainty in the shape of multijet
background is considered uncorrelated between decay channels and boson charges.
The impact of the background systematic uncertainties on the
determination of $m_W$ is summarised in
Table~\ref{tab:SecSelWCutflow}.

\begin{table}[tp]
  \begin{center}
    \begin{tabular}{lcccccccc}
      \toprule
      Kinematic distribution & \multicolumn{4}{c}{$\pt^\ell$}                                                        & \multicolumn{4}{c}{$\mt$ } \\
      Decay channel	     & \multicolumn{2}{c}{$W\rightarrow e\nu$}  & \multicolumn{2}{c}{$W\rightarrow \mu\nu$} & \multicolumn{2}{c}{$W\rightarrow e\nu$}   & \multicolumn{2}{c}{$W\rightarrow \mu\nu$}	\\
      $W$-boson charge       & $W^+$	& $W^-$	                    & $W^+$	& $W^-$	                     & $W^+$	& $W^-$	                         & $W^+$	& $W^-$	\\
      \midrule
      $\delta m_W$~[\MeV]\\
      \,\,\,\, $W\rightarrow\tau\nu$ (fraction, shape)    &   0.1 &  0.1   &   0.1  &  0.2 &   0.1  &  0.2 &  0.1   &  0.3  \\
      \,\,\,\, $Z\rightarrow ee$ (fraction, shape)	  &   3.3 &  4.8   &	--  &  --  &   4.3  &  6.4 &   --   &  --   \\
      \,\,\,\, $Z\rightarrow \mu\mu$ (fraction, shape)	  &   --  &  --    &   3.5  &  4.5 &   --   &  --  &   4.3  &  5.2  \\
      \,\,\,\, $Z\rightarrow\tau\tau$	(fraction, shape) &   0.1 &  0.1   &   0.1  &  0.2 &   0.1  &  0.2 &   0.1  &  0.3  \\
      \,\,\,\, $WW$, $WZ$, $ZZ$ (fraction)                &   0.1 &  0.1   &   0.1  &  0.1 &   0.4  &  0.4 &   0.3  &  0.4  \\
      \,\,\,\, Top (fraction)		                  &   0.1 &  0.1   &   0.1  &  0.1 &   0.3  &  0.3 &   0.3  &  0.3  \\
      \,\,\,\, Multijet (fraction)	                  &   3.2 &  3.6   &   1.8  &  2.4 &   8.1  &  8.6 &  3.7   &  4.6  \\
      \,\,\,\, Multijet (shape)	                          &   3.8 &  3.1   &   1.6  &  1.5 &   8.6  &  8.0 &  2.5   &  2.4  \\
      \midrule
      Total 	                                          &   6.0 &  6.8   &   4.3  &  5.3 &   12.6 &  13.4 &   6.2  &  7.4  \\
      \bottomrule
    \end{tabular}
    \caption{Systematic uncertainties in the $m_W$ measurement due to
      electroweak, top-quark, and multijet background estimation, for
      fits to the $p_\textrm{T}^\ell$ and $\mt$ distributions, in the electron and muon
      decay channels, with positively and negatively charged $W$ bosons. \label{tab:SecSelWCutflow}}
      \end{center}
\end{table}

\section{Measurement of the $W$-boson mass \label{sec:wfits}}

This section presents the determination of the mass of the $W$ boson
from template fits to the kinematic distributions of the $W$-boson
decay products. The final measured value is obtained from the
combination of measurements performed using the
lepton transverse momentum and transverse mass distributions
in categories corresponding to the electron and muon decay
channels, positively and negatively charged $W$ bosons, and absolute
pseudorapidity bins of the charged lepton, as illustrated in
Table~\ref{tab:Categories}.
The number of selected events in each category is shown in Table
\ref{tab:SecSelWCutflowMuE}.

\begin{table}[tp]
  \begin{center}
    \begin{tabular}{lrrrrr}
      \toprule
      $|\eta_\ell|$ range & 0--0.8 & 0.8--1.4 & 1.4--2.0 & 2.0--2.4 & Inclusive  \\
      \midrule
      $W^+\rightarrow \mu^+\nu$  & 1\,283\,332 & 1\,063\,131 & 1\,377\,773 & 885\,582 & 4\,609\,818 \\
      $W^-\rightarrow \mu^-\bar\nu$ & 1\,001\,592 & 769\,876 & 916\,163 & 547\,329 & 3\,234\,960 \\
      \midrule
      $|\eta_\ell|$ range &  0--0.6 &  0.6--1.2  & &1.8--2.4 & Inclusive  \\
      \midrule
      $W^+\rightarrow e^+\nu$& 1\,233\,960 & 1\,207\,136 & & 956\,620 & 3\,397\,716  \\
      $W^-\rightarrow e^-\bar\nu$ & 969\,170 & 908\,327 & & 610\,028 & 2\,487\,525 \\
      \bottomrule
    \end{tabular}
    \caption{Numbers of selected $W^+$ and $W^-$ events in the different decay
    channels in data, inclusively and for the various $|\eta_\ell|$
    categories. \label{tab:SecSelWCutflowMuE}}
  \end{center}
\end{table}

\subsection{Control distributions \label{sec:wcontrol}}
\begin{figure}
  \begin{center}
    \subfloat[]{\includegraphics[width=0.49\textwidth]{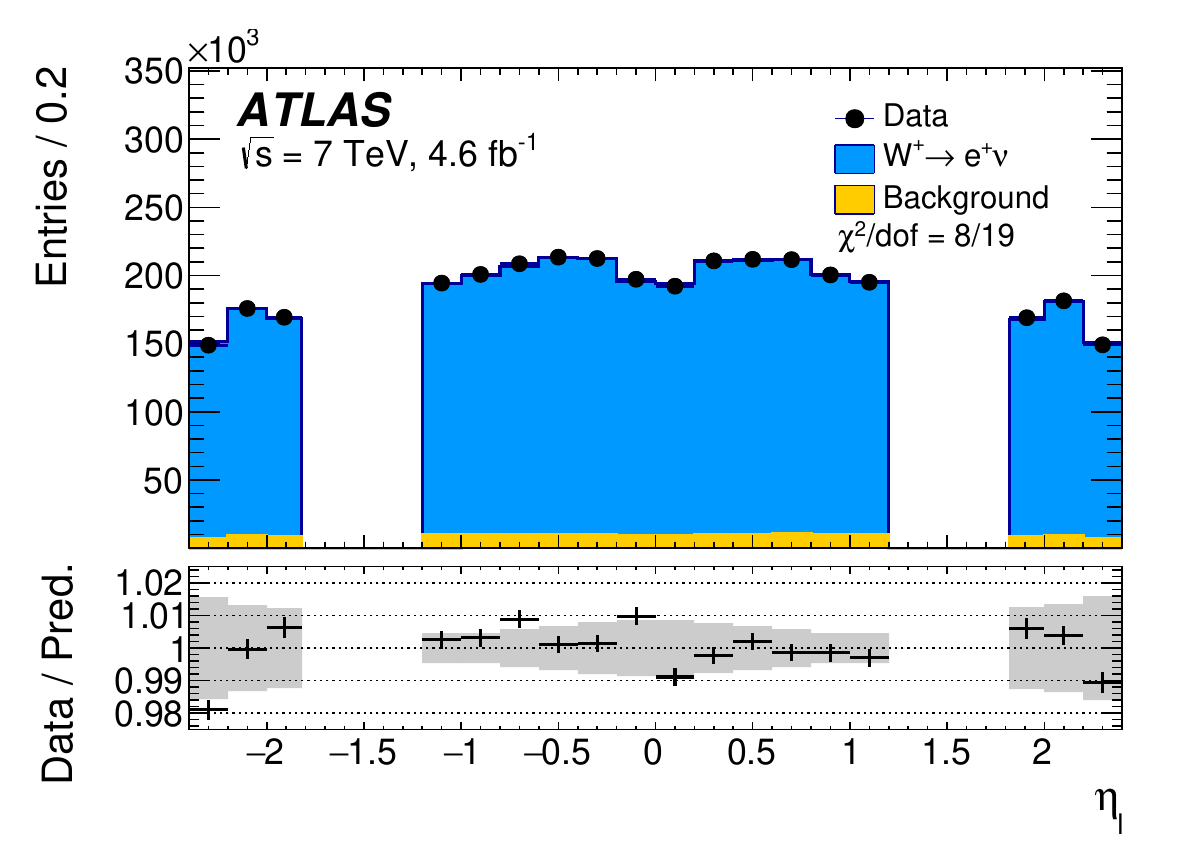}}
    \subfloat[]{\includegraphics[width=0.49\textwidth]{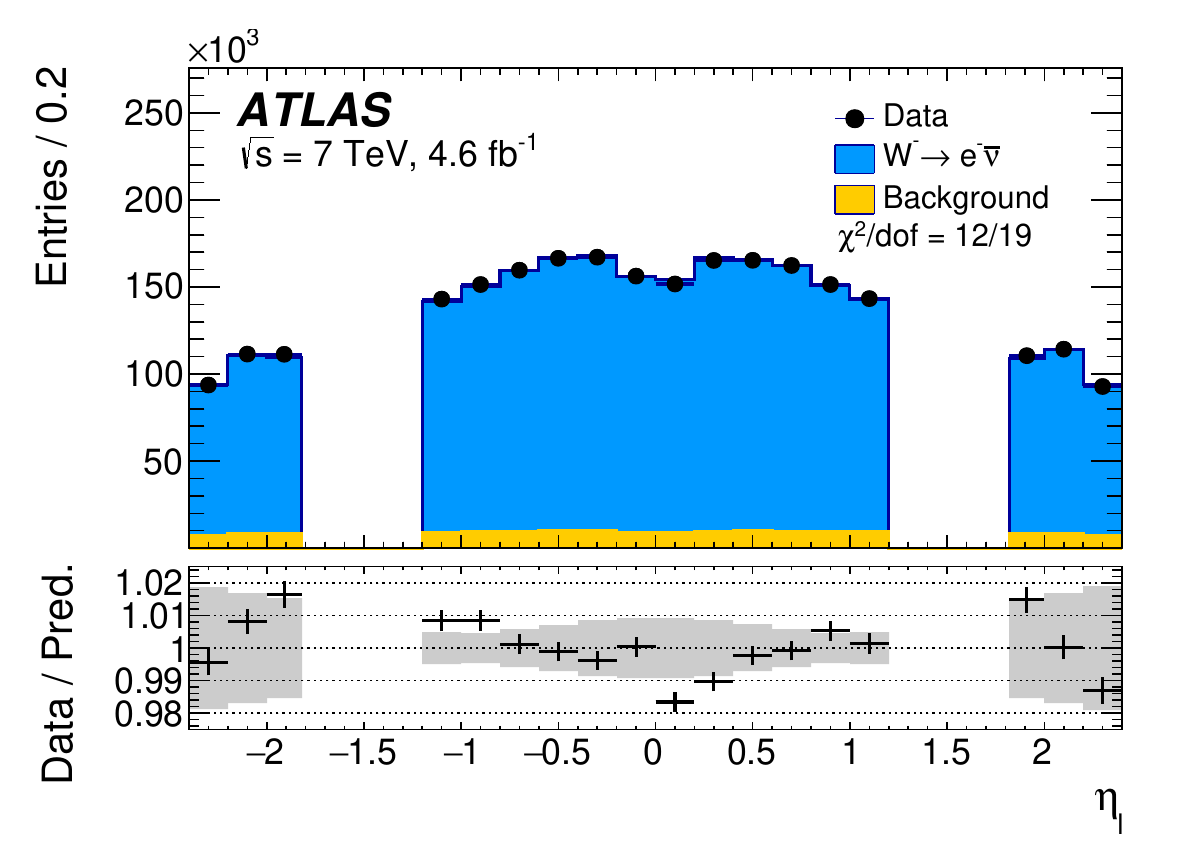}}\\
    \subfloat[]{\includegraphics[width=0.49\textwidth]{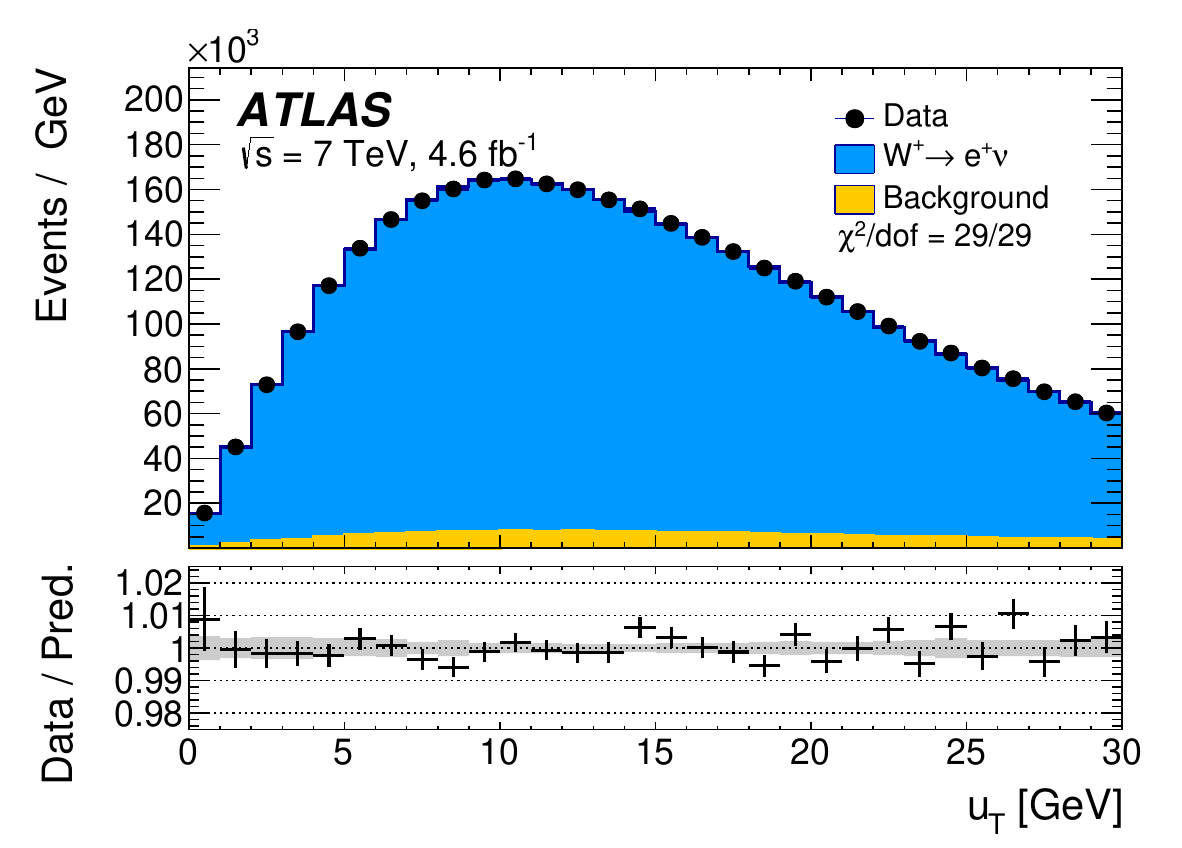}}
    \subfloat[]{\includegraphics[width=0.49\textwidth]{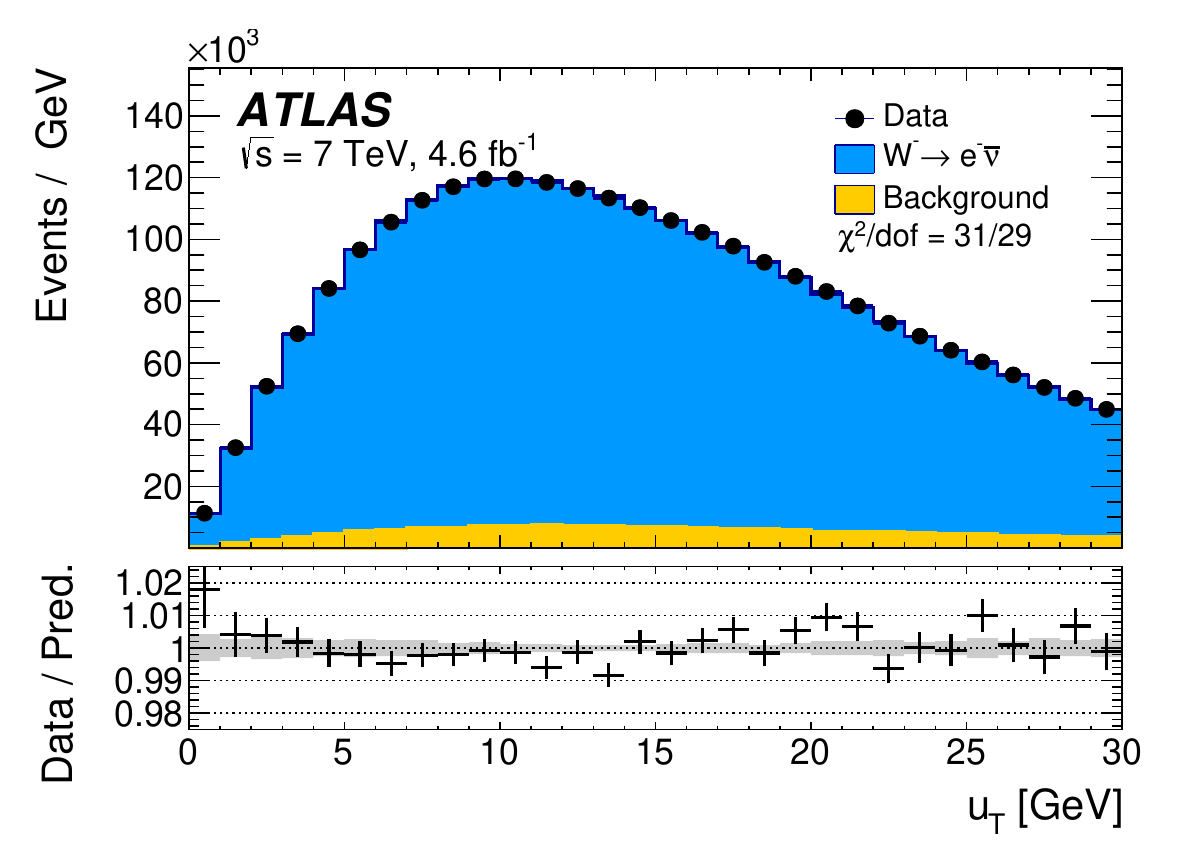}}\\ 
    \subfloat[]{\includegraphics[width=0.49\textwidth]{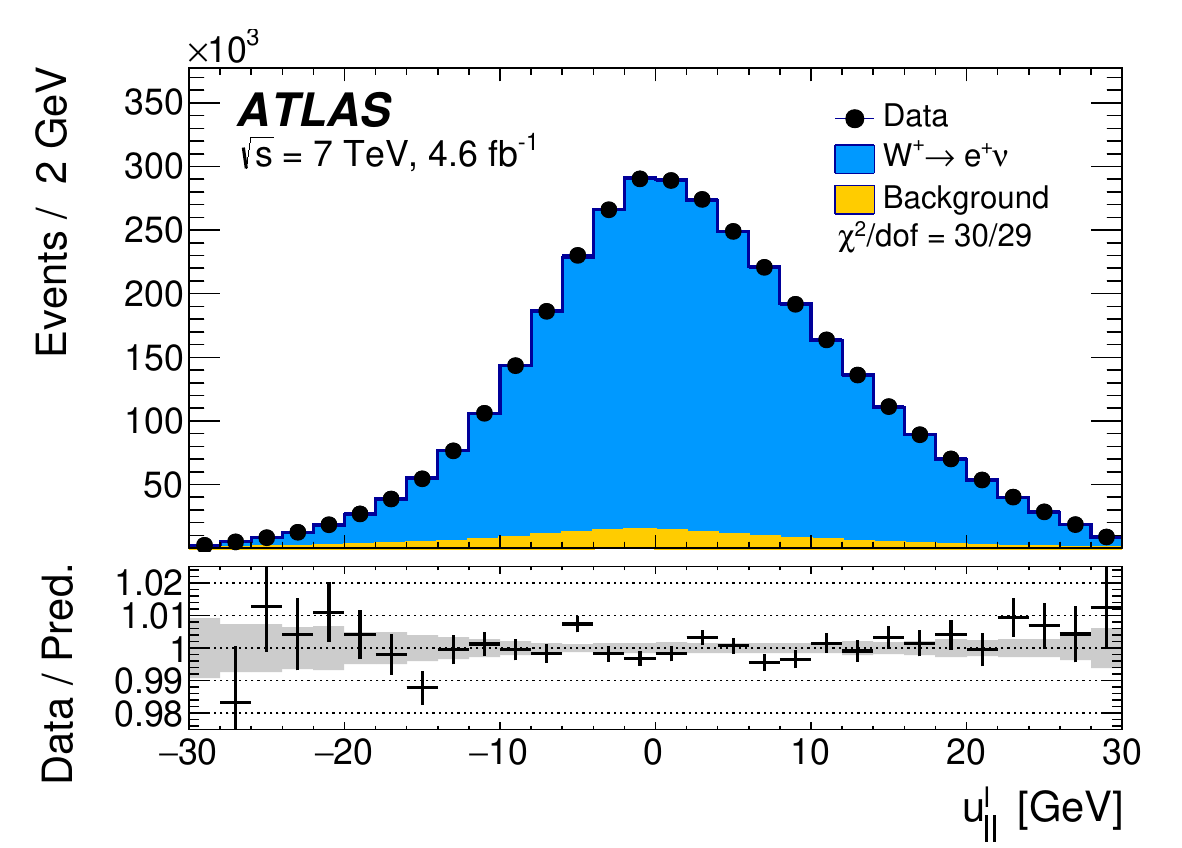}}
    \subfloat[]{\includegraphics[width=0.49\textwidth]{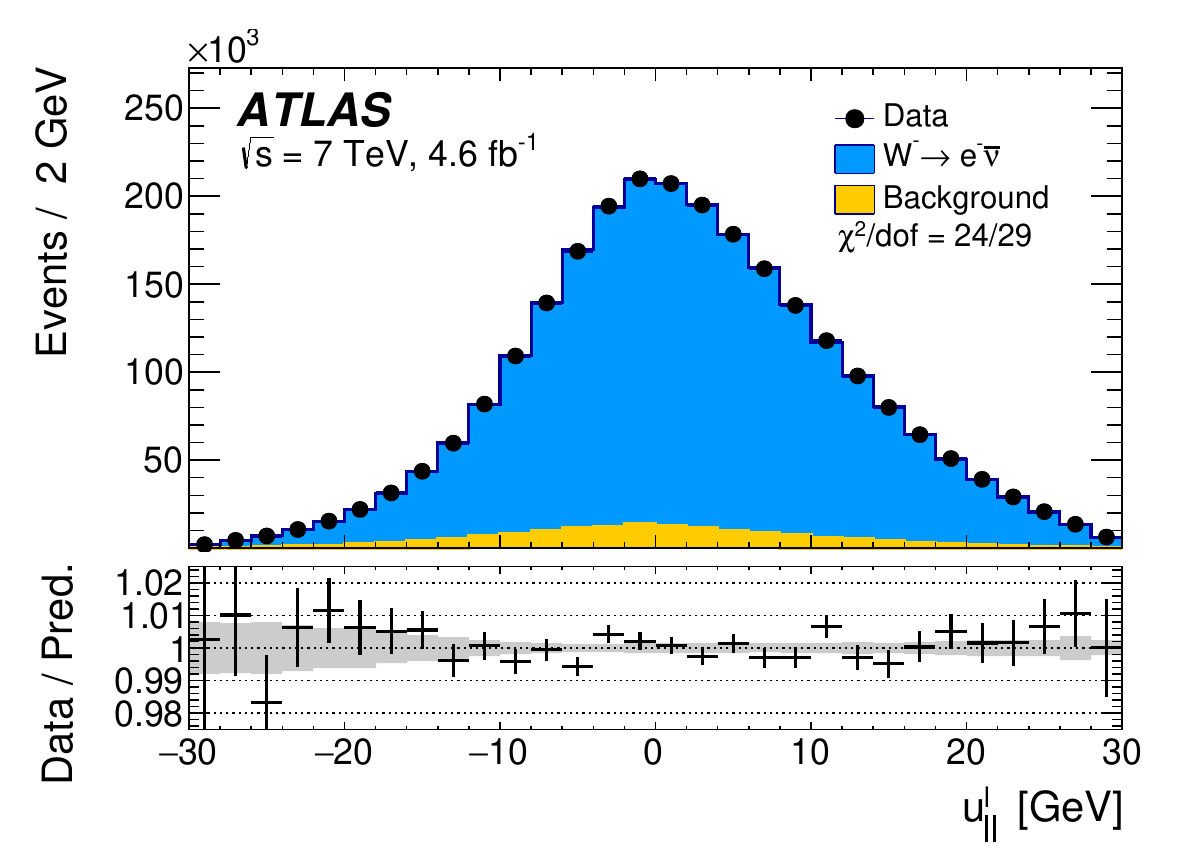}}
  \end{center}
  \caption{The (a,b) $\eta_\ell$, (c,d) \ut, and (e,f) $u_\parallel^\ell$
    distributions for (a,c,e) $W^+$ events and (b,d,f) $W^-$ events in
    the electron decay channel. The data are compared to the simulation including
    signal and background contributions.
    Detector calibration and physics-modelling corrections are applied
    to the simulated events.
    The lower panels show the data-to-prediction ratios, the error bars show the statistical uncertainty,
    and the band shows the systematic uncertainty of the prediction. The $\chi^2$ values displayed in each figure account for all sources of uncertainty and include the effects of bin-to-bin correlations induced by the systematic uncertainties.\label{fig:WSecControlIncE}}
\end{figure}

\begin{figure}
\begin{center}
\subfloat[]{\includegraphics[width=0.49\textwidth]{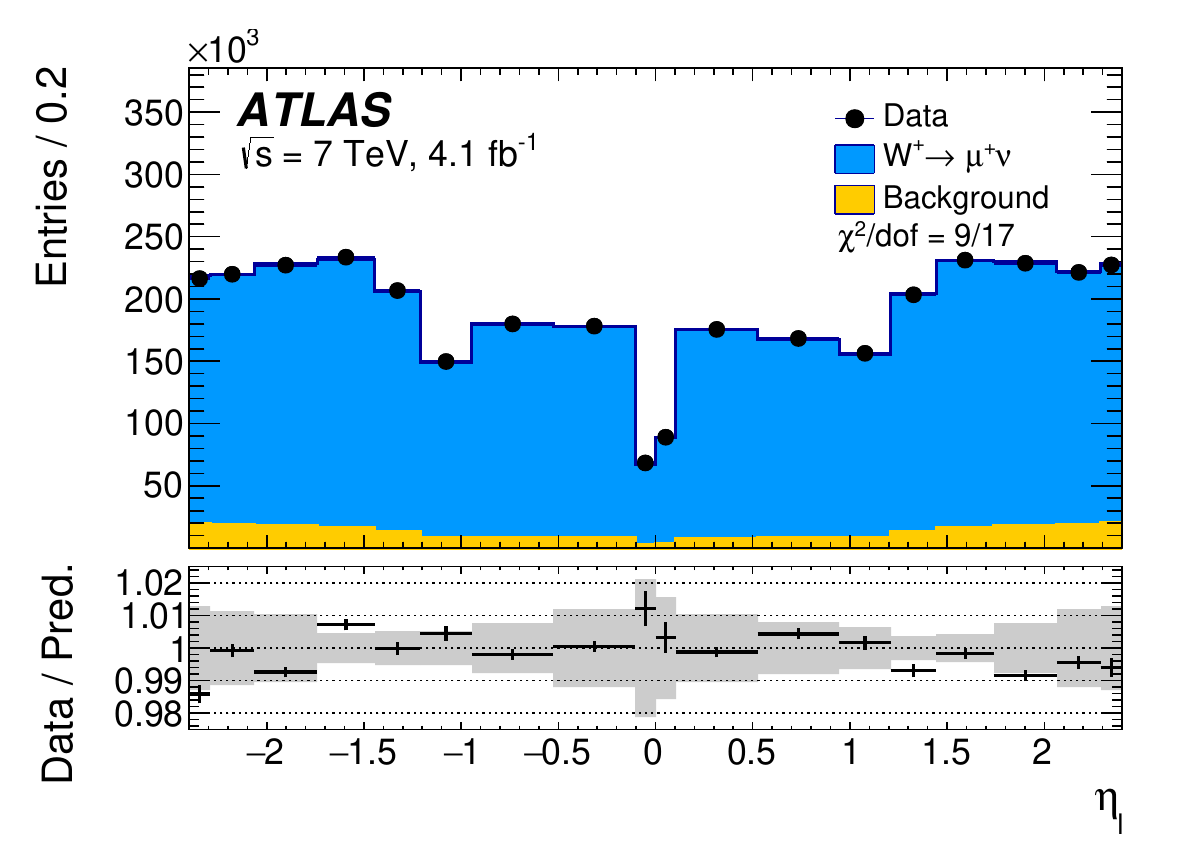}}
\subfloat[]{\includegraphics[width=0.49\textwidth]{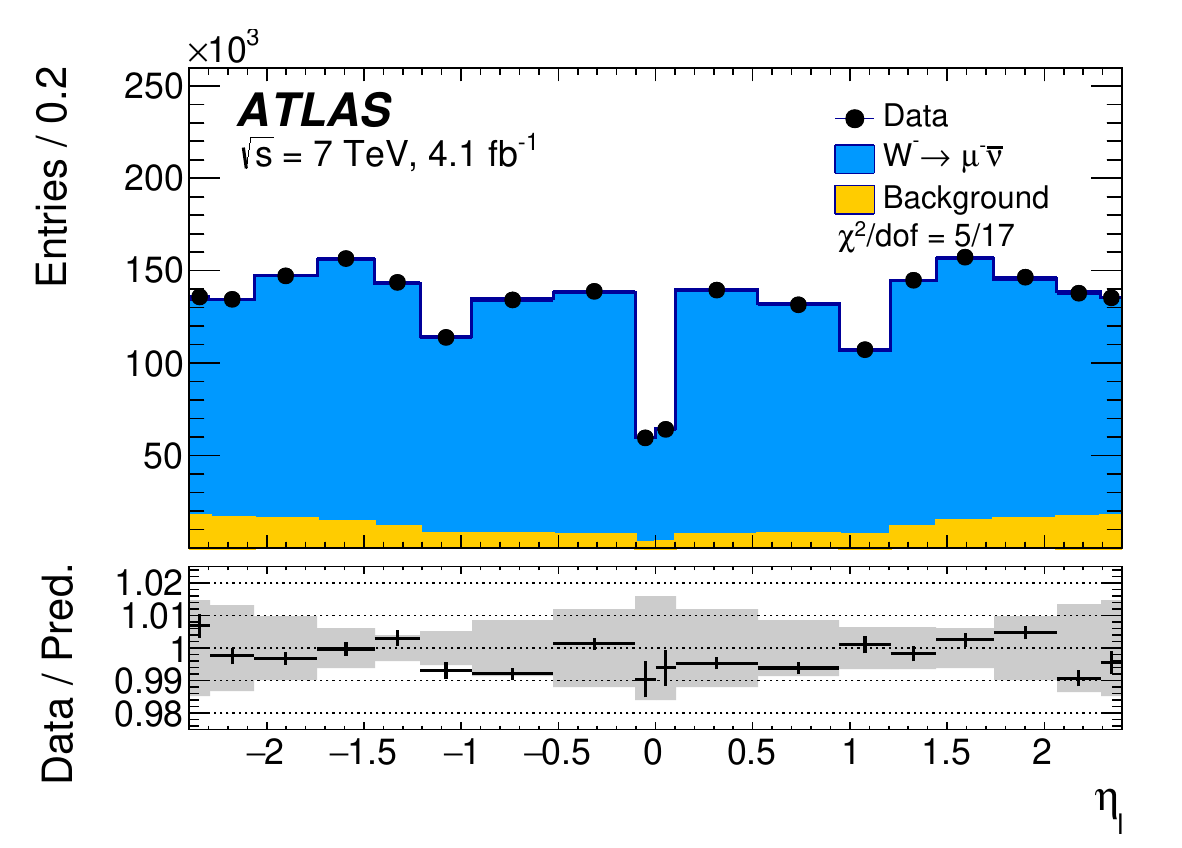}}\\
\subfloat[]{\includegraphics[width=0.49\textwidth]{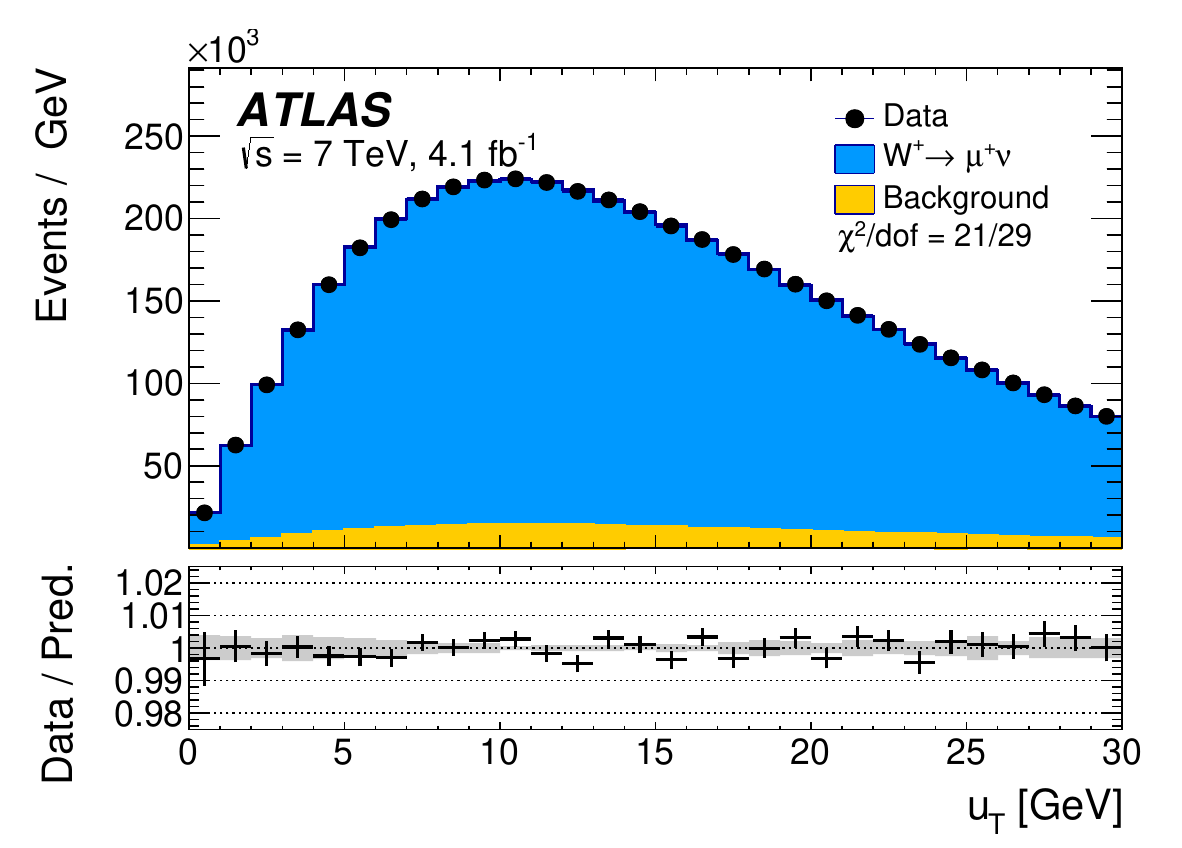}}
\subfloat[]{\includegraphics[width=0.49\textwidth]{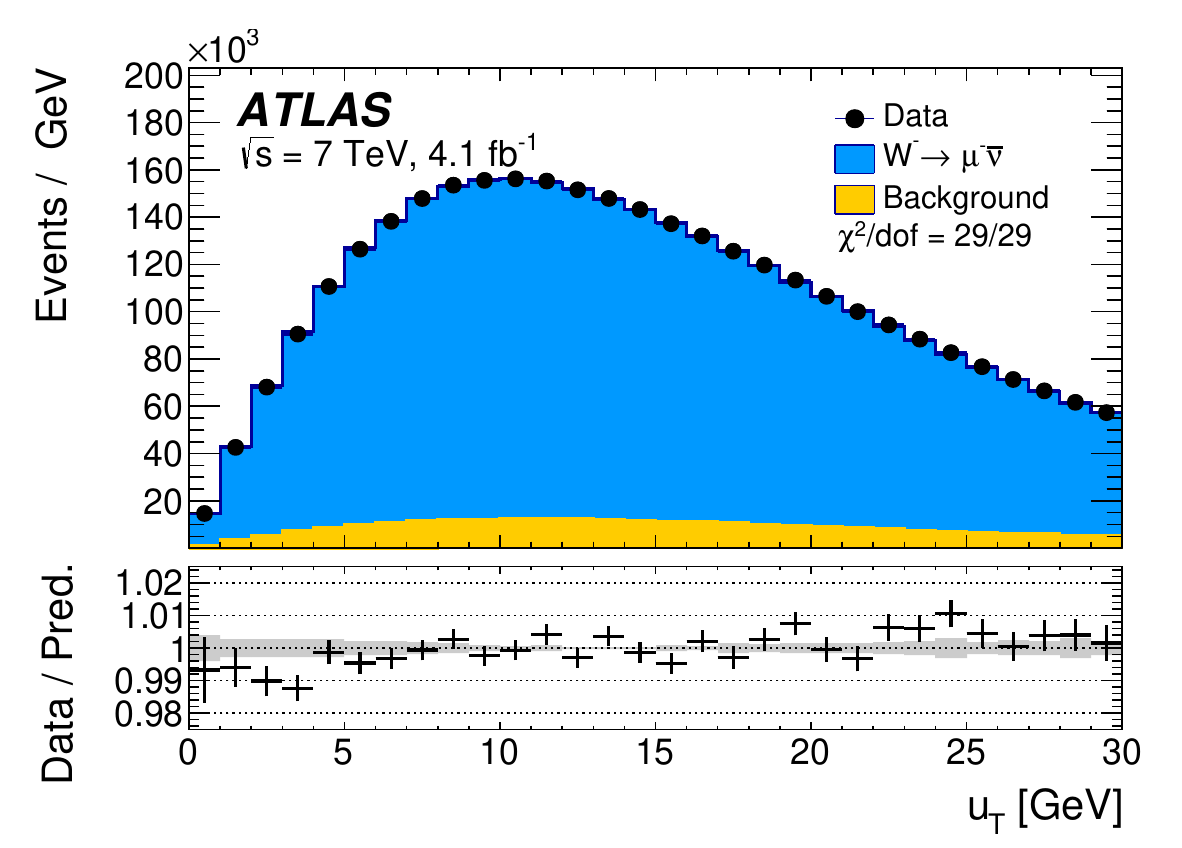}}\\
\subfloat[]{\includegraphics[width=0.49\textwidth]{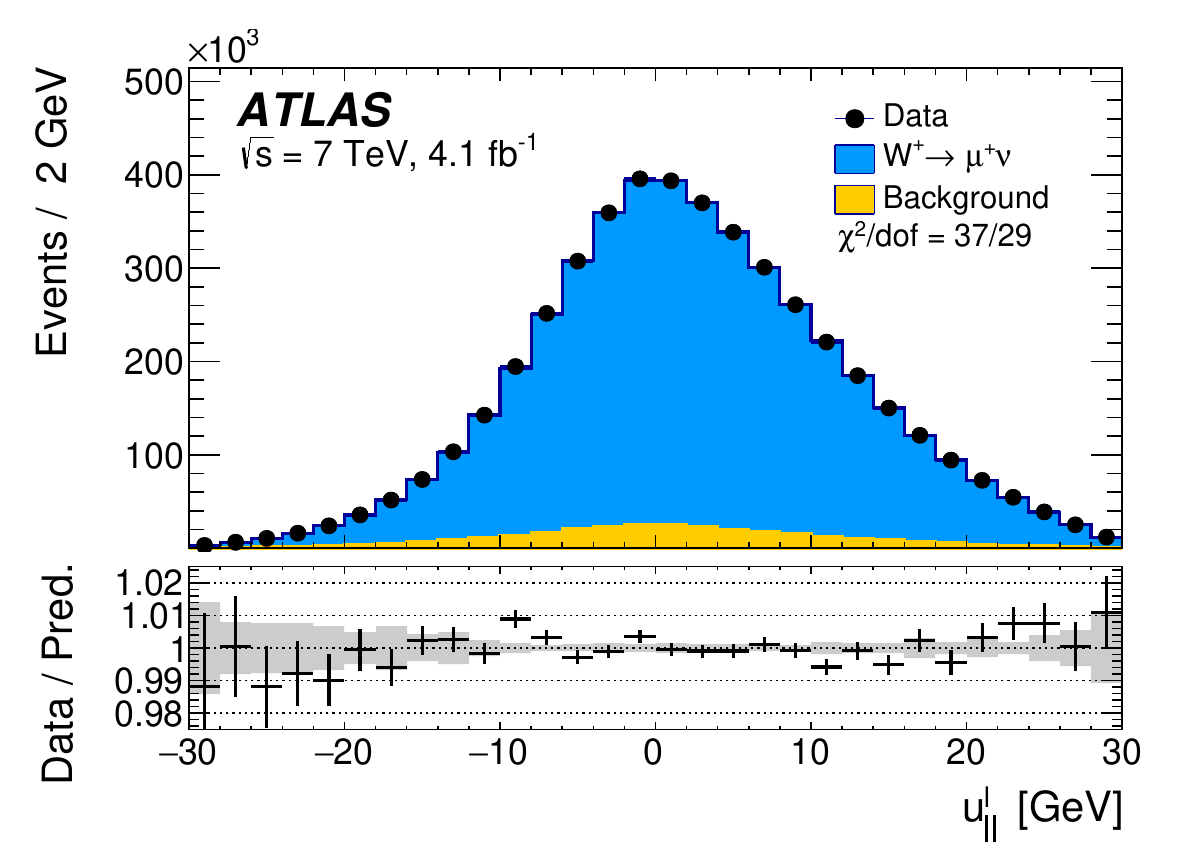}}
\subfloat[]{\includegraphics[width=0.49\textwidth]{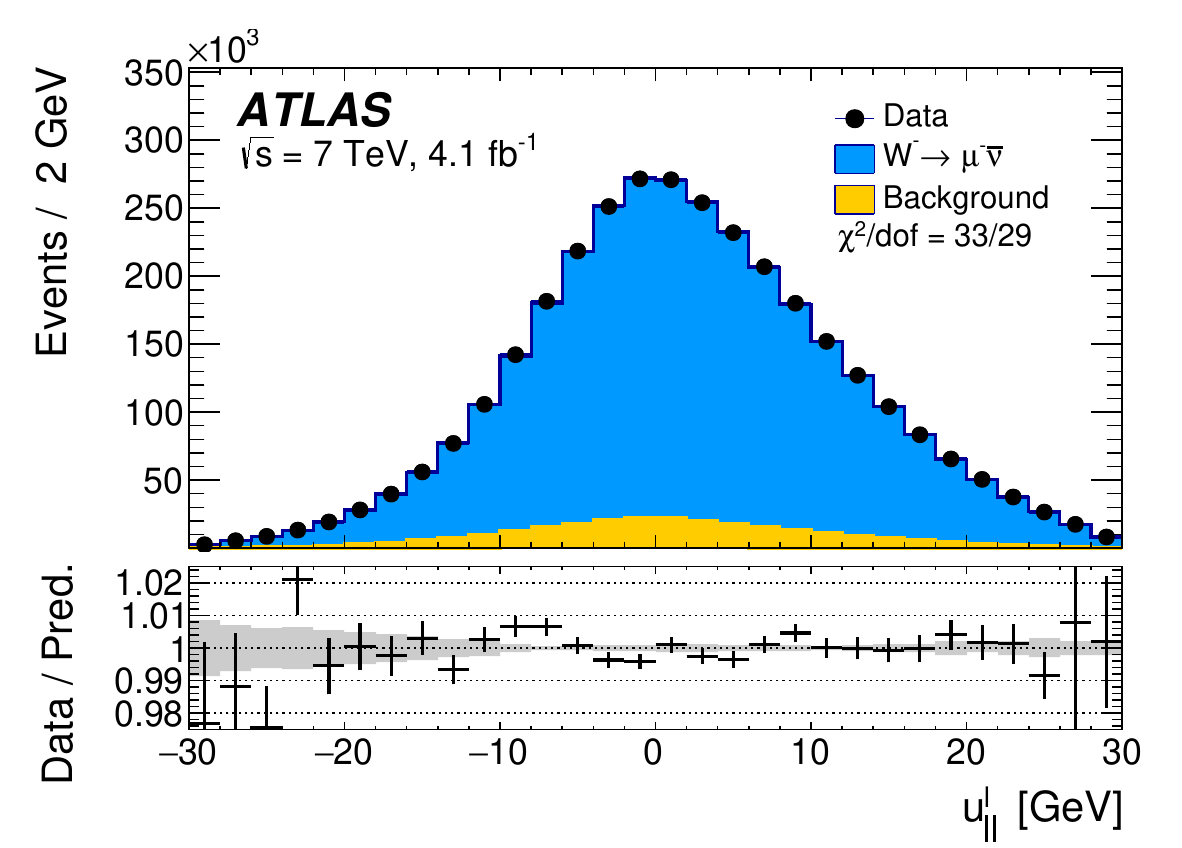}}
\caption{The (a,b) $\eta_\ell$, (c,d) \ut, and (e,f) $u_\parallel^\ell$
  distributions for (a,c,e) $W^+$ events and (b,d,f) $W^-$ events in
  the muon decay channel. The data are compared to the simulation including
  signal and background contributions.
  Detector calibration and physics-modelling corrections are applied
  to the simulated events.
  The lower panels show the data-to-prediction ratios, the error bars show the statistical uncertainty,
  and the band shows the systematic uncertainty of the prediction. The $\chi^2$ values displayed in each figure account for all sources of uncertainty and include the effects of bin-to-bin correlations induced by the systematic uncertainties.\label{fig:WSecControlIncMu}}
\end{center}
\end{figure}

The detector calibration and the physics modelling are validated by
comparing data with simulated $W$-boson signal and backgrounds for several kinematic
distributions that are insensitive to the $W$-boson mass.
The comparison is based on a $\chi^2$ compatibility test, including
statistical and systematic uncertainties, and the bin-to-bin
correlations induced by the latter.
The systematic uncertainty comprises all sources of experimental uncertainty
related to the lepton and recoil calibration, and to the
background subtraction, as well as sources of modelling uncertainty associated
with electroweak corrections, or induced by the helicity fractions of vector-boson
production, the vector-boson transverse-momentum distribution, and the PDFs.
Comparisons of data and simulation for the $\eta_\ell$, \ut, and
$u^\ell_\parallel$ distributions, in positively and
negatively charged $W$-boson events, are shown in
Figures~\ref{fig:WSecControlIncE} and~\ref{fig:WSecControlIncMu} for
the electron and muon decay channels, respectively. 

Data and simulation agree within uncertainties for all distributions, as confirmed by the satisfactory $\chi^2/$dof values. The effect of the residual discrepancies in the \ut\ distributions for $W^-\rightarrow\ell\bar{\nu}$, visible at low values in
Figures~\ref{fig:WSecControlIncE}-(d)~and~\ref{fig:WSecControlIncMu}-(d), is discussed in Section~\ref{sec:mwcrosschecks}. 

\subsection{Data-driven check of the uncertainty in the $\pt^W$ distribution \label{sec:ptwxcheck}}

\begin{figure}
  \begin{center}
  \subfloat[]{\includegraphics[width=0.49\textwidth]{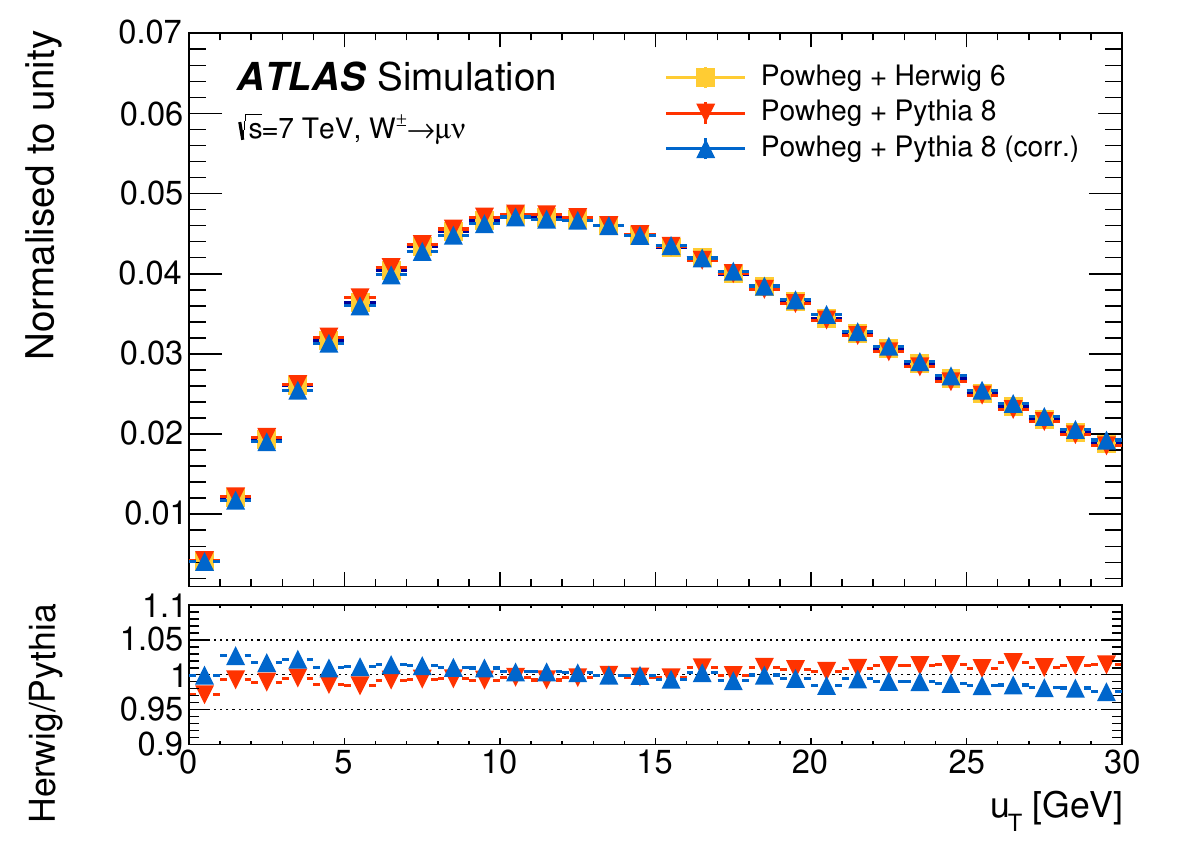}}
  \subfloat[]{\includegraphics[width=0.49\textwidth]{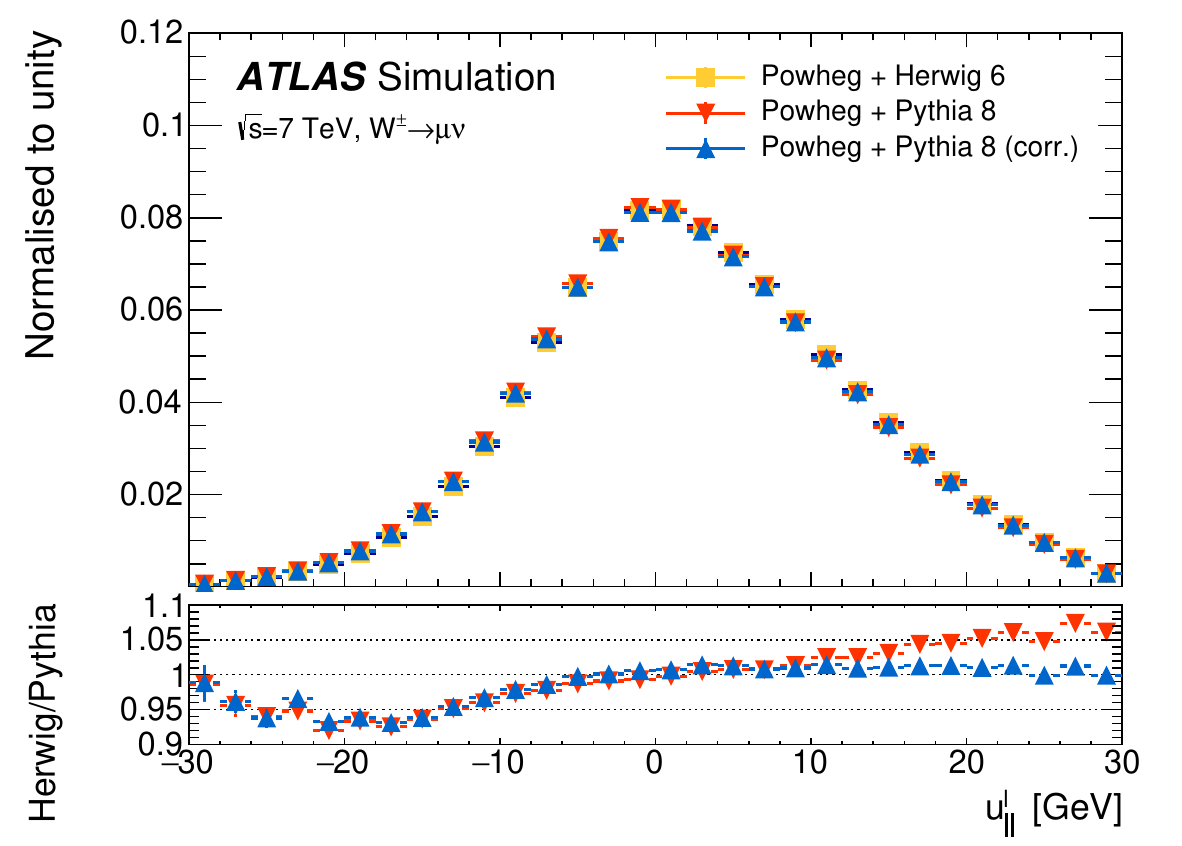}}\\
  \end{center}
  \caption{Distributions of (a) $u_{\textrm{T}}$ and (b) $u_{\parallel}^\ell$ in $W\to\mu\nu$ events simulated using \PowhegPythia and \PowhegHerwig after all analysis selection cuts are applied. The \PowhegPythia distributions are shown before
    and after correction of the recoil response to that of \PowhegHerwig. The lower panels show the ratios of \PowhegHerwig to \PowhegPythia, with and without the recoil response correction in the \PowhegPythia sample. The discrepancy remaining after recoil corrections reflects the different $\pt^W$ distributions.}
  \label{fig:uparsensit}
\end{figure}

The uncertainty in the prediction of the $u^\ell_\parallel$
distribution is dominated by $p_{\textrm{T}}^W$ distribution uncertainties, especially at
negative values of $u^\ell_\parallel$ in the kinematic region
corresponding to $u^\ell_\parallel<-15\GeV$. This is illustrated in Figure~\ref{fig:uparsensit}, which compares the recoil distributions in the \PowhegPythia and \PowhegHerwig samples, before and
after the corrections described in Section~\ref{sec:residcorr} (the $\pt^W$ distribution predicted by \PowhegPythia is not reweighted to that of \PowhegHerwig). As can be seen, the
recoil corrections and the different $\pt^W$ distributions have a comparable effect on the $u_{\textrm{T}}$ distribution. In contrast, the effect of the recoil corrections is small at negative values
of $u^\ell_\parallel$, whereas the difference in the $\pt^W$ distributions has a large impact in this region.

The sensitivity of the $u^\ell_\parallel$ distribution is exploited to validate the
modelling of the $\pt^W$ distribution by \PYTHIA~8~AZ, and its theory-driven uncertainty, described in
Section~\ref{subsubsec:PSunc}, with a data-driven procedure. The
parton-shower factorisation scale $\mu_{\textrm{F}}$ associated with the
$c\bar{q}\to W$ processes constitutes the main source of uncertainty
in the modelling of the $\pt^W$ distribution. Variations of the
$u^\ell_\parallel$ distribution induced by changes in the factorisation scale of
the $c\bar{q}\to W$ processes are parameterised and fitted to the data. 
The $u^\ell_\parallel$ distribution is predicted for the two boundary values of $\mu_{\textrm{F}}$, and
assumed to vary linearly as a function of $\mu_{\textrm{F}}$. 
Variations induced by changes in $\mu_{\textrm{F}}$ are parameterised using a variable $s$ defined in units of the initially allowed range, 
i.e. values of $s=-1,0,+1$ correspond to half the effect\footnote{Half the effect is used because only one of the two quarks in the initial state is heavy, as discussed in Section~\ref{subsubsec:PSunc}.} of changing from $\mu_{\textrm{F}}=m_V$ to $\mu_{\textrm{F}}=m_V/2,m_V,2m_V$ respectively.
The optimal value of $s$ is determined by fitting the
fraction of events 
in the kinematic region $-30<u^\ell_\parallel<-15\GeV$.
The fit accounts for all experimental and modelling uncertainties affecting the $u^\ell_\parallel$ distribution, and gives a value of $s=-0.22 \pm 1.06$. 
The best-fit value of $s$ confirms the good agreement between the the \PYTHIA~8~AZ prediction and the data; its uncertainty is dominated by
PDF and recoil-calibration uncertainties, and matches the variation range of $\mu_{\textrm{F}}$ used for the initial estimation of the $\pt^W$ distribution uncertainty. 

\begin{figure}
  \begin{center}
  \subfloat[]{\includegraphics[width=0.49\textwidth]{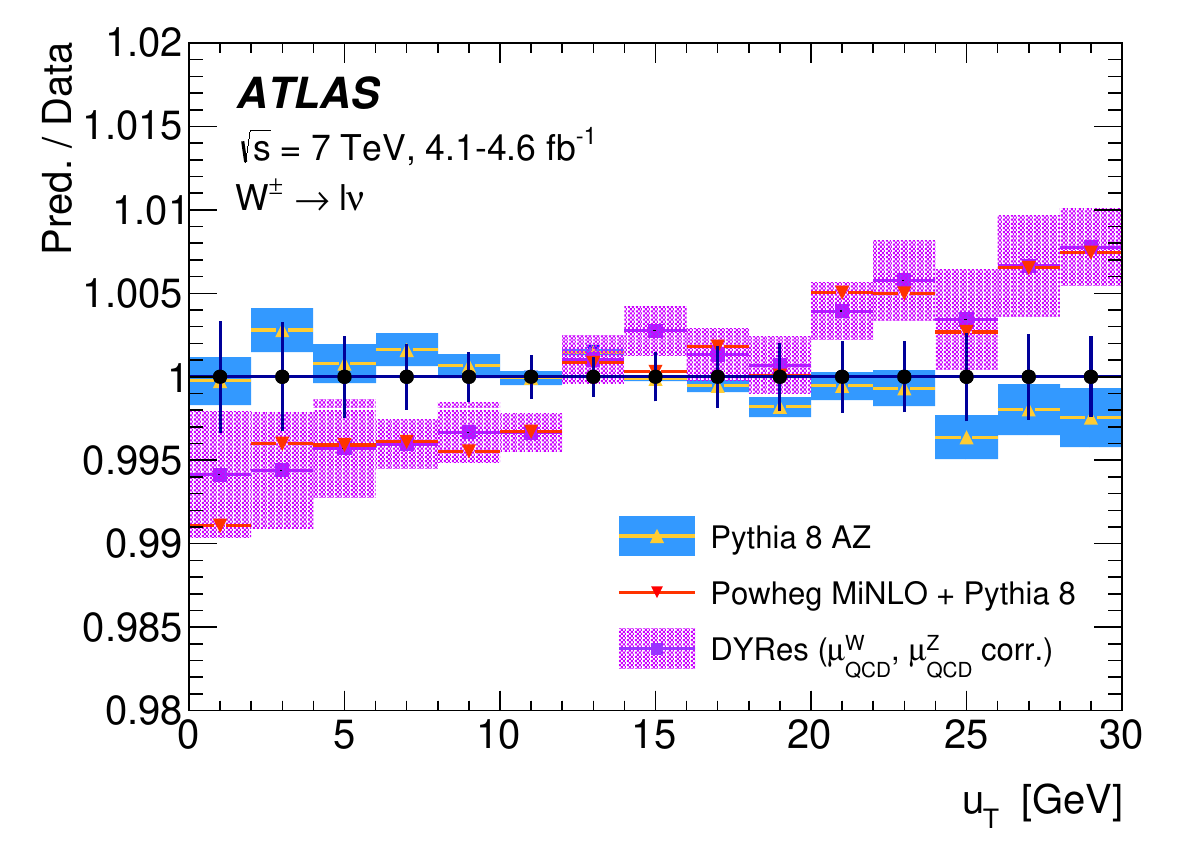}}
  \subfloat[]{\includegraphics[width=0.49\textwidth]{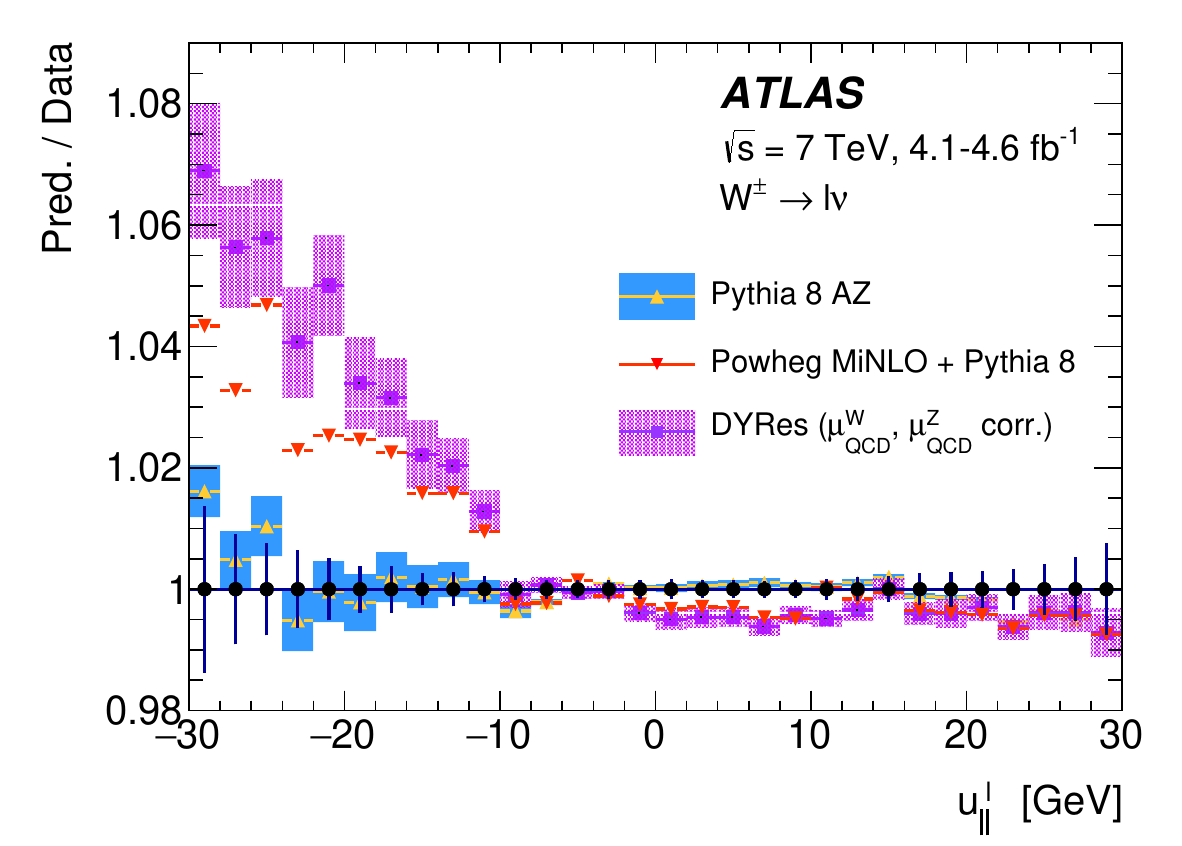}}\\
  \end{center}
  \caption{Ratio between the predictions of \PYTHIA 8~AZ, DYRes and \PowhegMinlo and the data for the (a) $u_{\textrm{T}}$ and (b) $u_{\parallel}^\ell$ distributions in $W\rightarrow \ell\nu$
    events. The $W$-boson rapidity distribution is reweighted according to the NNLO prediction. The error bars on the data points display the total experimental uncertainty, and the band around the \PYTHIA
    8~AZ prediction reflects the uncertainty in the $\pt^W$ distribution. The uncertainty band around the DYRes prediction assumes that uncertainties induced by variations of
    the QCD scales $\mu_\textrm{F}$, $\mu_\textrm{R}$ and $\mu_\textrm{Res}$, collectively referred to as $\mu_\textrm{QCD}$, are fully correlated in $W$ and $Z$ production.}
  \label{fig:uparcomp}
\end{figure}

This validation test supports the \PYTHIA 8~AZ prediction of the $p_{\textrm{T}}^W$ distribution and the theory-driven associated uncertainty estimate. On the other hand, 
as shown in Figure~\ref{fig:uparcomp}, the data disagree with the DYRes and \PowhegMinlo predictions. The latter are obtained reweighting the initial $\pt^W$ distribution in \PowhegPythia according to
the product of the $\pt^Z$ distribution of \PYTHIA~8~AZ, which matches the measurement of Ref.~\cite{STDM-2012-23}, and $R_{W/Z}(\pt)$
as predicted by DYRes and \PowhegMinlo. The uncertainty bands in the DYRes prediction are calculated using variations of the factorization, renormalization and resummation scales $\mu_\textrm{F}$,
$\mu_\textrm{R}$ and $\mu_\textrm{Res}$ following the procedure described in Ref.~\cite{Bozzi:2008bb,Bozzi:2010xn}. The uncertainty obtained applying correlated scale variations in $W$ and $Z$
production does not cover the observed difference with the data. The potential effect of using $R_{W/Z}(\pt)$ as predicted by DYRes instead of \PYTHIA~8~AZ for the determination of $m_W$ is discussed in Section~\ref{sec:mwcrosschecks}.

\subsection{Results for $m_W$ in the measurement categories\label{sec:Unc}}

Measurements of $m_W$ are performed using the $\pt^\ell$ and $\mt$
distributions, separately for positively and negatively charged $W$ bosons, in
three bins of $|\eta_\ell|$ in the electron decay channel, and in four
bins of $|\eta_\ell|$ in the muon decay channel, leading to a total of
28 $m_W$ determinations. In each category, the value of $m_W$ is
determined by a $\chi^2$ minimisation, comparing the $\pt^\ell$ and
$\mt$ distributions in data and simulation for different values of
$m_W$. The templates are generated with values
of $m_W$ in steps of $1$ to $10\MeV$ within a range of $\pm 400\MeV$,
centred around the reference value used in the Monte Carlo signal samples.
The statistical uncertainty is estimated from the half width of the
$\chi^2$ function at the value corresponding to one unit above the
minimum.
Systematic uncertainties due to physics-modelling corrections,
detector-calibration corrections, and background subtraction,
are discussed in Sections~\ref{sec:phymod}--\ref{sec:recoilcalib}
and~\ref{sec:background}, respectively.

The lower and upper bounds of the range of the $\pt^\ell$ distribution
used in the fit are varied from $30$ to $35\GeV$, and from $45$ to $50\GeV$ respectively, in
steps of $1\GeV$. For the \mt\ distribution, the boundaries are varied
from $65$ to $70\GeV$, and from $90$ to $100\GeV$. The total measurement
uncertainty is evaluated for each range, after combining the
measurement categories as described in Section~\ref{sec:WComb} below. The
smallest total uncertainty in $m_W$ is found for the fit ranges $32<\pt^\ell<45\GeV$ and $66<\mt<99\GeV$. The optimisation is
performed before the unblinding of the $m_W$ value and the optimised
range is used for all the results described below.

\begin{figure}
  \begin{center}
    \subfloat[]{\includegraphics[width=0.48\textwidth]{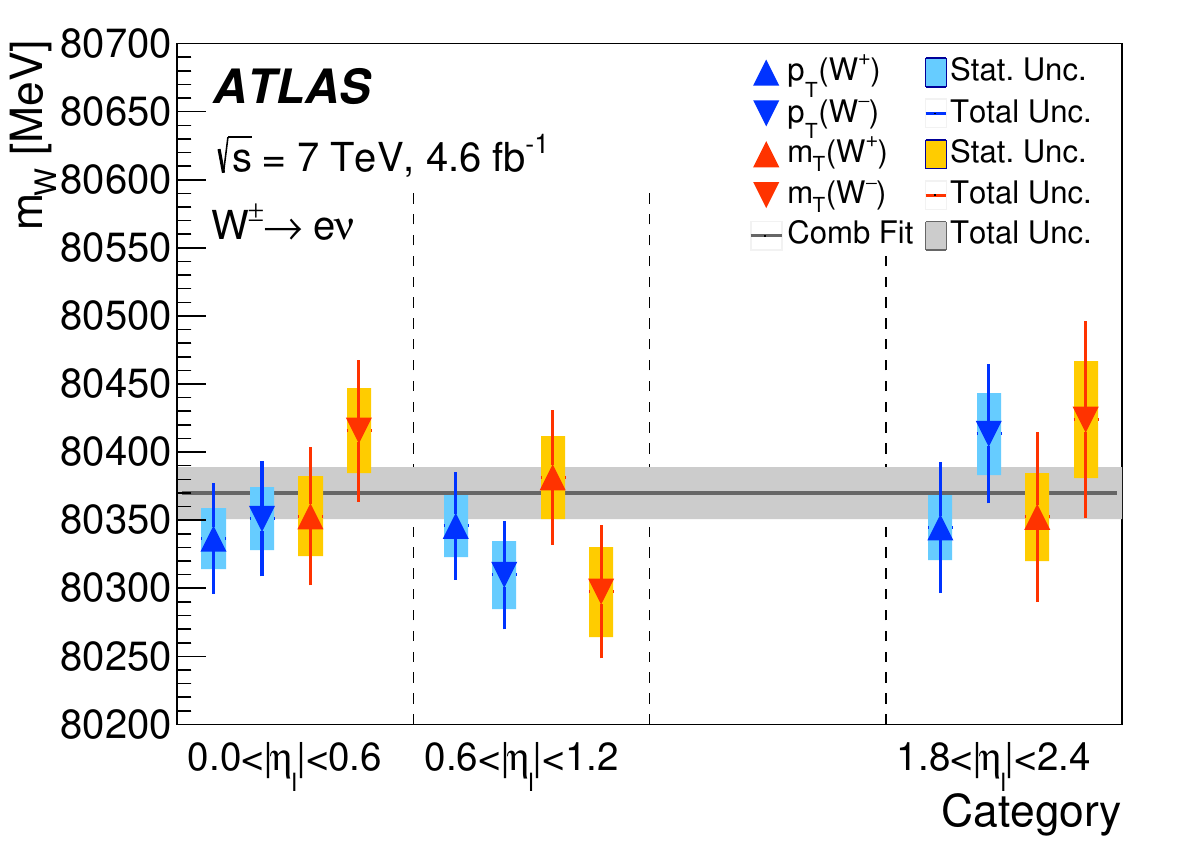}}
    \subfloat[]{\includegraphics[width=0.48\textwidth]{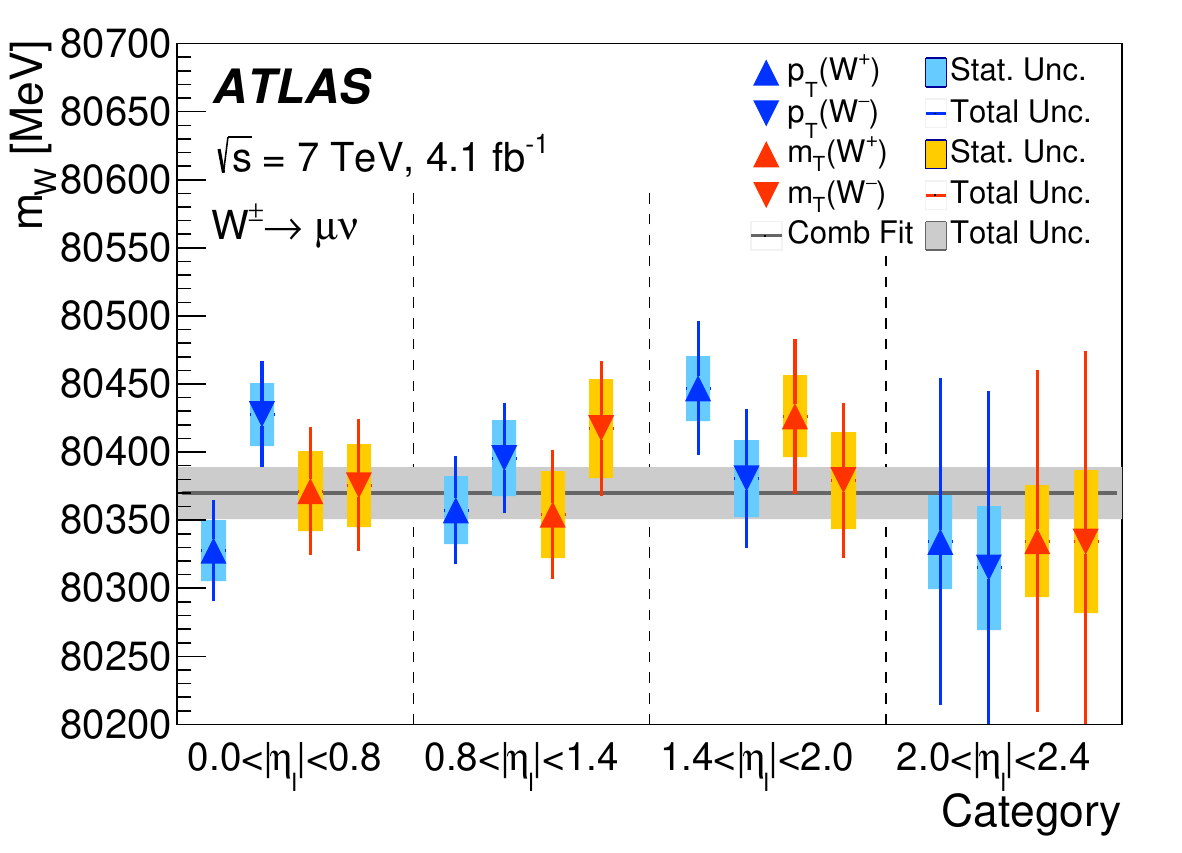}}
  \end{center}
  \caption{Overview of the $m_W$ measurements in the (a) electron and (b) muon decay
    channels. Results are shown for the $\pt^\ell$ and \mt{} distributions, for $W^+$ and $W^-$ events in the different $|\eta_\ell|$ categories. The coloured bands and solid lines
    show the statistical and total uncertainties, respectively. The horizontal line and band show the fully combined result and its uncertainty.}
  \label{fig:WMassFitsOverview}
\end{figure}

The final measurement uncertainty is dominated by modelling uncertainties, with typical values in the range $25$--$35\MeV$ for the various charge and $|\eta_\ell|$
categories.
Lepton-calibration uncertainties are the dominant sources of
experimental systematic uncertainty for the extraction of $m_W$ from
the $\pt^\ell$ distribution. These uncertainties vary from about $15\MeV$ to about $35\MeV$ for most measurement categories, except the highest $|\eta|$ bin in the muon channel where the total uncertainty of
about $120\MeV$ is dominated by the muon momentum linearity uncertainty.
The uncertainty in the calibration of the recoil is the largest source of
experimental systematic uncertainty for the \mt{} distribution, with a typical contribution of about $15\MeV$ for all categories.
The determination of $m_W$ from the $\pt^\ell$ and \mt{} distributions
in the various categories is summarised in
Table~\ref{tab:fitAllResults}, including an overview of statistical
and systematic uncertainties. The results are also shown in Figure~\ref{fig:WMassFitsOverview}.
No significant differences in the values of $m_W$ corresponding to the
different decay channels and to the various charge and $|\eta_\ell|$
categories are observed.

\begin{table}[tp]
  \centering
  \resizebox{\textwidth}{!}{\begin{tabular}{l|c|rrrrrrrr|r}
\toprule
    Channel & $m_W$ & Stat. & Muon & Elec.& Recoil & Bckg. & QCD & EW & PDF& Total  \\
    \mt{}-Fit& [MeV]& Unc. & Unc. & Unc. & Unc. & Unc. & Unc. & Unc. & Unc.  & Unc.  \\
\hline
$W^{+}\rightarrow\mu\nu, |\eta|<0.8$ & 80371.3 & 29.2 & 12.4 & 0.0 & 15.2 & 8.1 & 9.9 & 3.4 & 28.4 & 47.1 \\
$W^{+}\rightarrow\mu\nu, 0.8<|\eta|<1.4$ & 80354.1 & 32.1 & 19.3 & 0.0 & 13.0 & 6.8 & 9.6 & 3.4 & 23.3 & 47.6  \\
$W^{+}\rightarrow\mu\nu, 1.4<|\eta|<2.0$ & 80426.3 & 30.2 & 35.1 & 0.0 & 14.3 & 7.2 & 9.3 & 3.4 & 27.2 & 56.9  \\
$W^{+}\rightarrow\mu\nu, 2.0<|\eta|<2.4$ & 80334.6 & 40.9 & 112.4 & 0.0 & 14.4 & 9.0 & 8.4 & 3.4 & 32.8 & 125.5   \\
\hline
$W^{-}\rightarrow\mu\nu, |\eta|<0.8$ & 80375.5 & 30.6 & 11.6 & 0.0 & 13.1 & 8.5 & 9.5 & 3.4 & 30.6 & 48.5  \\
$W^{-}\rightarrow\mu\nu, 0.8<|\eta|<1.4$ & 80417.5 & 36.4 & 18.5 & 0.0 & 12.2 & 7.7 & 9.7 & 3.4 & 22.2 & 49.7 \\
$W^{-}\rightarrow\mu\nu, 1.4<|\eta|<2.0$ & 80379.4 & 35.6 & 33.9 & 0.0 & 10.5 & 8.1 & 9.7 & 3.4 & 23.1 & 56.9 \\
$W^{-}\rightarrow\mu\nu, 2.0<|\eta|<2.4$ & 80334.2 & 52.4 & 123.7 & 0.0 & 11.6 & 10.2 & 9.9 & 3.4 & 34.1 & 139.9 \\
\hline
$W^{+}\rightarrow e\nu, |\eta|<0.6$ & 80352.9 & 29.4 & 0.0 & 19.5 & 13.1 & 15.3 & 9.9 & 3.4 & 28.5 & 50.8 \\
$W^{+}\rightarrow e\nu, 0.6<|\eta|<1.2$ & 80381.5 & 30.4 & 0.0 & 21.4 & 15.1 & 13.2 & 9.6 & 3.4 & 23.5 & 49.4 \\
$W^{+}\rightarrow e\nu, 1,8<|\eta|<2.4$ & 80352.4 & 32.4 & 0.0 & 26.6 & 16.4 & 32.8 & 8.4 & 3.4 & 27.3 & 62.6 \\
\hline
$W^{-}\rightarrow e\nu, |\eta|<0.6$ & 80415.8 & 31.3 & 0.0 & 16.4 & 11.8 & 15.5 & 9.5 & 3.4 & 31.3 & 52.1 \\
$W^{-}\rightarrow e\nu, 0.6<|\eta|<1.2$ & 80297.5 & 33.0 & 0.0 & 18.7 & 11.2 & 12.8 & 9.7 & 3.4 & 23.9 & 49.0 \\
$W^{-}\rightarrow e\nu, 1.8<|\eta|<2.4$ & 80423.8 & 42.8 & 0.0 & 33.2 & 12.8 & 35.1 & 9.9 & 3.4 & 28.1 & 72.3 \\
\hline
$\pt$-Fit\\
\hline
$W^{+}\rightarrow\mu\nu, |\eta|<0.8$ & 80327.7 & 22.1 & 12.2 & 0.0 & 2.6 & 5.1 & 9.0 & 6.0 & 24.7 & 37.3  \\
$W^{+}\rightarrow\mu\nu, 0.8<|\eta|<1.4$ & 80357.3 & 25.1 & 19.1 & 0.0 & 2.5 & 4.7 & 8.9 & 6.0 & 20.6 & 39.5 \\
$W^{+}\rightarrow\mu\nu, 1.4<|\eta|<2.0$ & 80446.9 & 23.9 & 33.1 & 0.0 & 2.5 & 4.9 & 8.2 & 6.0 & 25.2 & 49.3 \\
$W^{+}\rightarrow\mu\nu, 2.0<|\eta|<2.4$ & 80334.1 & 34.5 & 110.1 & 0.0 & 2.5 & 6.4 & 6.7 & 6.0 & 31.8 & 120.2 \\
\hline
$W^{-}\rightarrow\mu\nu, |\eta|<0.8$ & 80427.8 & 23.3 & 11.6 & 0.0 & 2.6 & 5.8 & 8.1 & 6.0 & 26.4 & 39.0 \\
$W^{-}\rightarrow\mu\nu, 0.8<|\eta|<1.4$ & 80395.6 & 27.9 & 18.3 & 0.0 & 2.5 & 5.6 & 8.0 & 6.0 & 19.8 & 40.5 \\
$W^{-}\rightarrow\mu\nu, 1.4<|\eta|<2.0$ & 80380.6 & 28.1 & 35.2 & 0.0 & 2.6 & 5.6 & 8.0 & 6.0 & 20.6 & 50.9 \\
$W^{-}\rightarrow\mu\nu, 2.0<|\eta|<2.4$ & 80315.2 & 45.5 & 116.1 & 0.0 & 2.6 & 7.6 & 8.3 & 6.0 & 32.7 & 129.6 \\
\hline
$W^{+}\rightarrow e\nu, |\eta|<0.6$ & 80336.5 & 22.2 & 0.0 & 20.1 & 2.5 & 6.4 & 9.0 & 5.3 & 24.5 & 40.7 \\
$W^{+}\rightarrow e\nu, 0.6<|\eta|<1.2$ & 80345.8 & 22.8 & 0.0 & 21.4 & 2.6 & 6.7 & 8.9 & 5.3 & 20.5 & 39.4 \\
$W^{+}\rightarrow e\nu, 1,8<|\eta|<2.4$ & 80344.7 & 24.0 & 0.0 & 30.8 & 2.6 & 11.9 & 6.7 & 5.3 & 24.1 & 48.2  \\
\hline
$W^{-}\rightarrow e\nu, |\eta|<0.6$ &80351.0 & 23.1 & 0.0 & 19.8 & 2.6 & 7.2 & 8.1 & 5.3 & 26.6 & 42.2 \\
$W^{-}\rightarrow e\nu, 0.6<|\eta|<1.2$ & 80309.8 & 24.9 & 0.0 & 19.7 & 2.7 & 7.3 & 8.0 & 5.3 & 20.9 & 39.9 \\
$W^{-}\rightarrow e\nu, 1.8<|\eta|<2.4$ & 80413.4 & 30.1 & 0.0 & 30.7 & 2.7 & 11.5 & 8.3 & 5.3 & 22.7 & 51.0  \\
\bottomrule
  \end{tabular}}
  \caption{Results of the $m_W$ measurements in the electron and muon
    decay channels, for positively and negatively charged $W$ bosons,
    in different lepton-$|\eta|$ ranges, using the $\mt$ and $\pt^\ell$ distributions in the optimised fitting range. 
    The table shows the statistical uncertainties,
    together with all experimental uncertainties, divided into muon-,
    electron-, recoil- and background-related uncertainties, and all
    modelling uncertainties, separately for QCD modelling including scale variations, parton shower and angular
    coefficients, electroweak corrections, and PDFs. All uncertainties are given in \MeV.
    \label{tab:fitAllResults}}

\end{table}

\begin{figure}
  \begin{center}
    \subfloat[]{\includegraphics[width=0.49\textwidth]{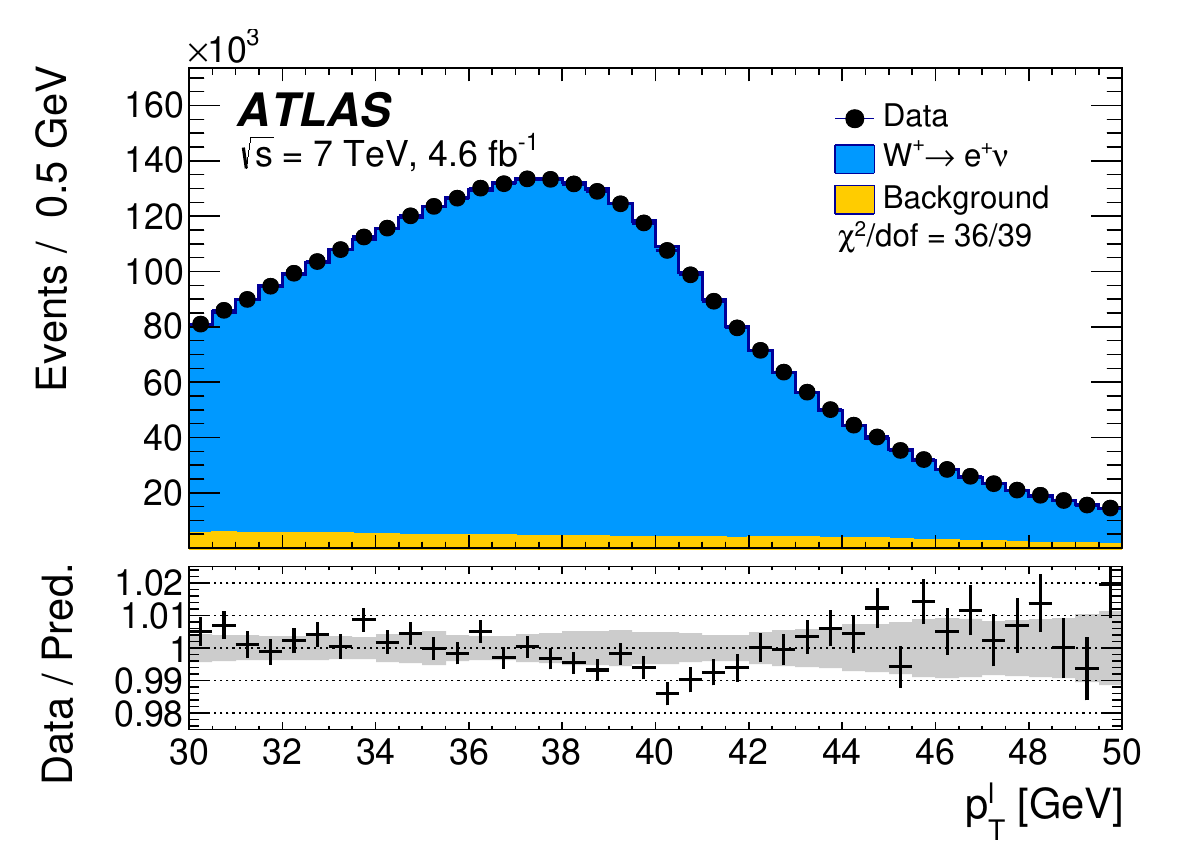}}
    \subfloat[]{\includegraphics[width=0.49\textwidth]{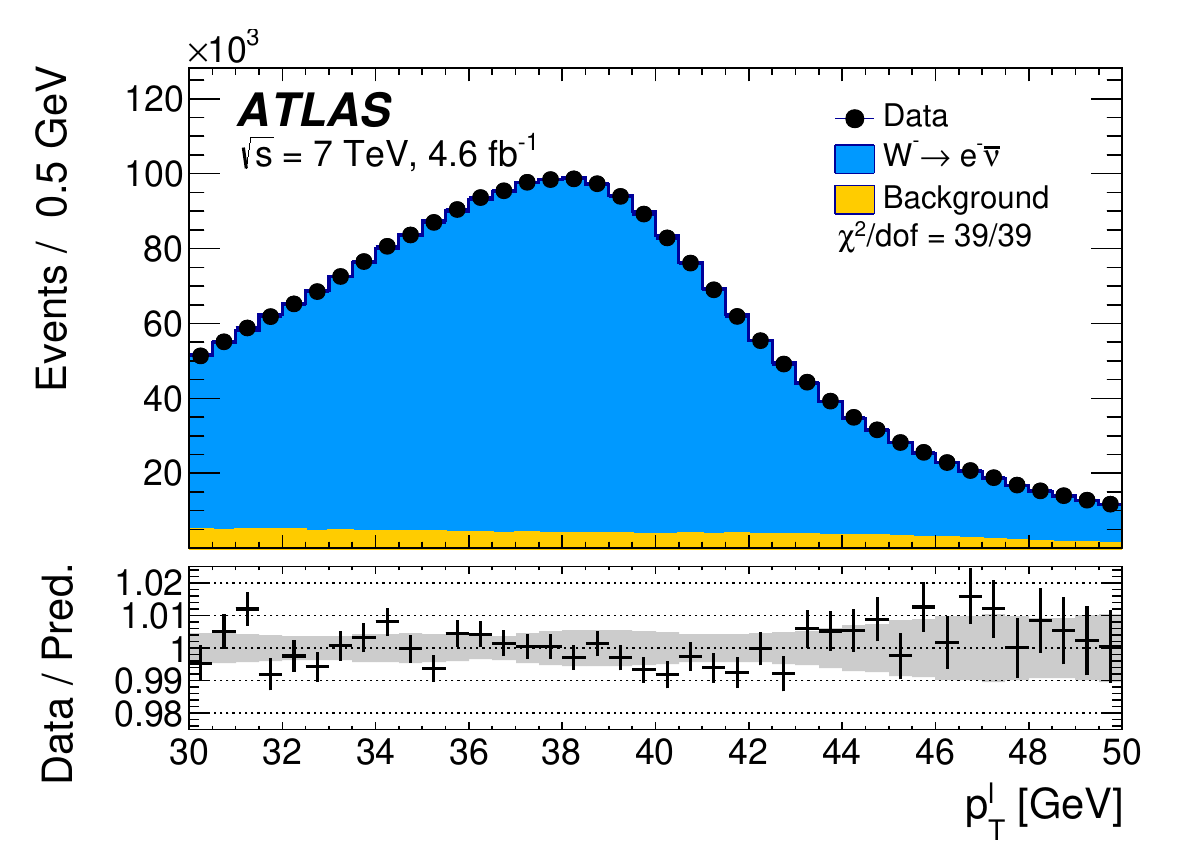}}\\
    \subfloat[]{\includegraphics[width=0.49\textwidth]{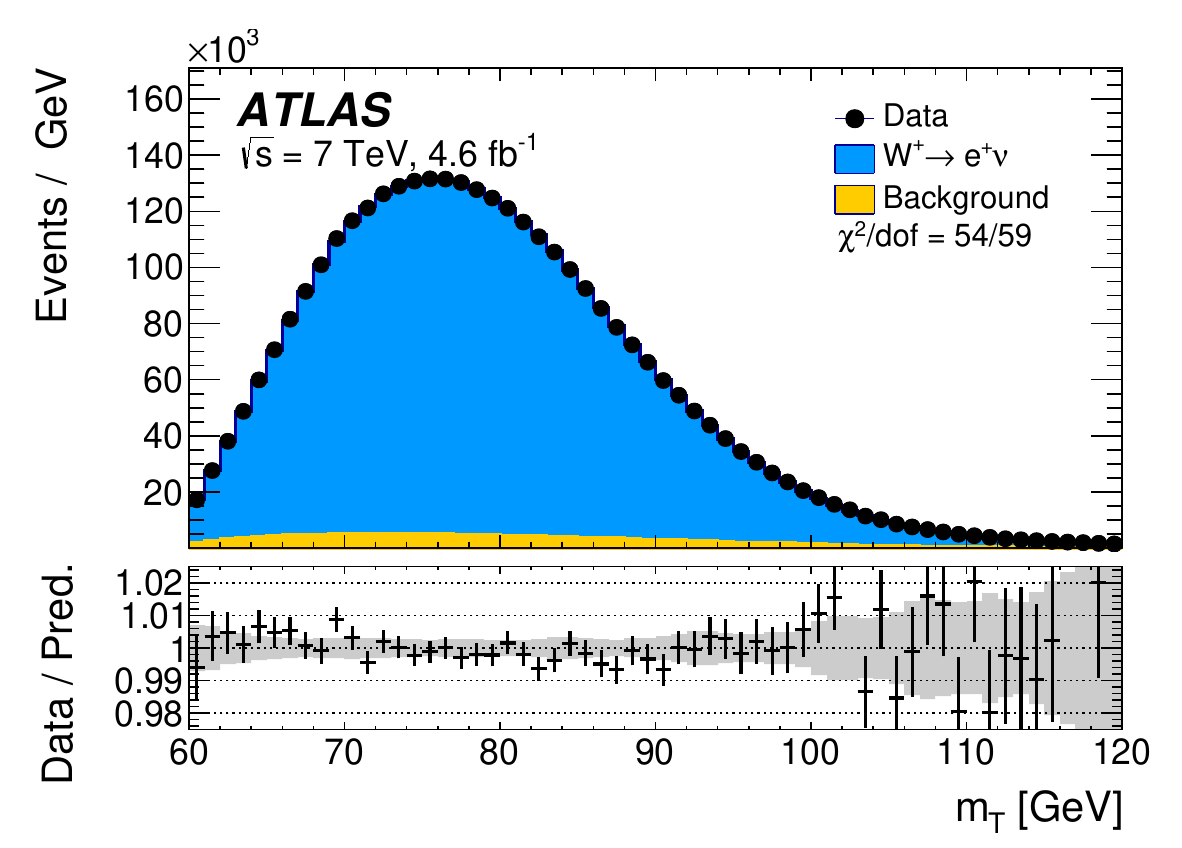}}
    \subfloat[]{\includegraphics[width=0.49\textwidth]{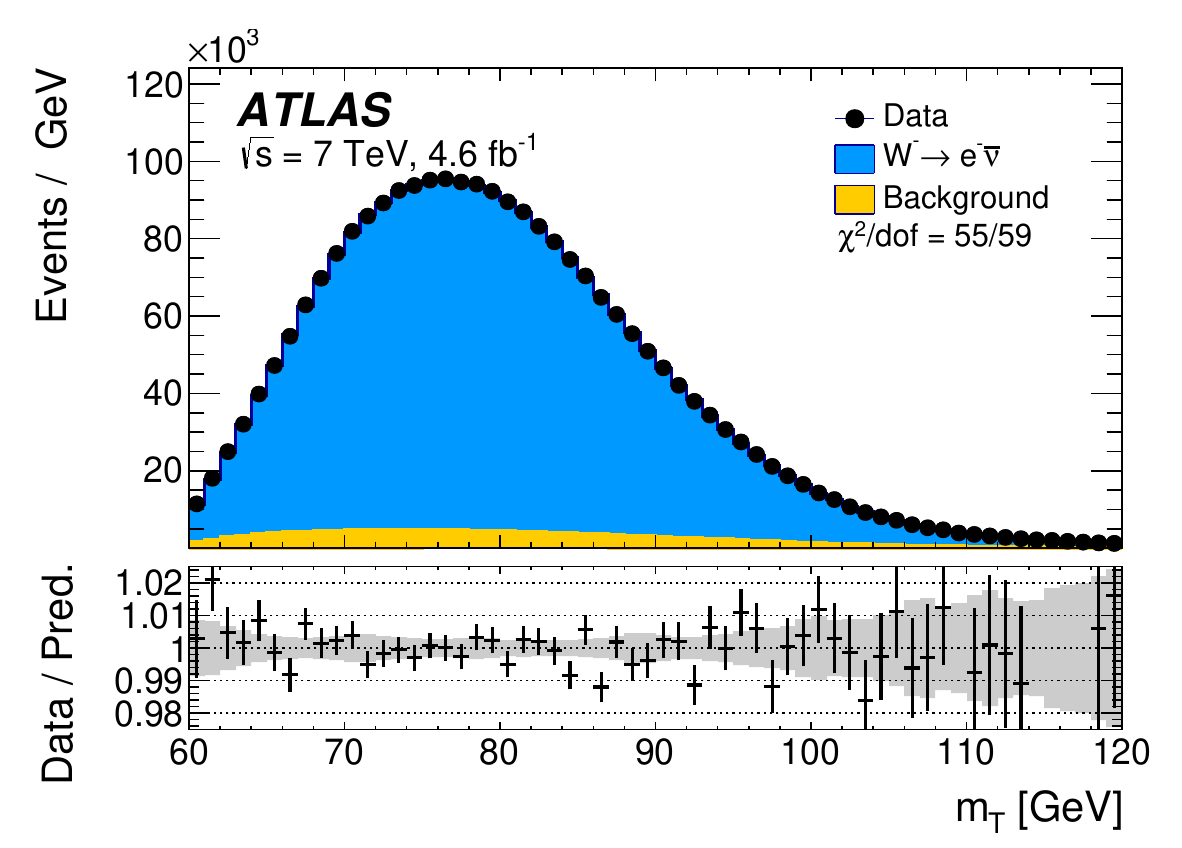}}\\
    \subfloat[]{\includegraphics[width=0.49\textwidth]{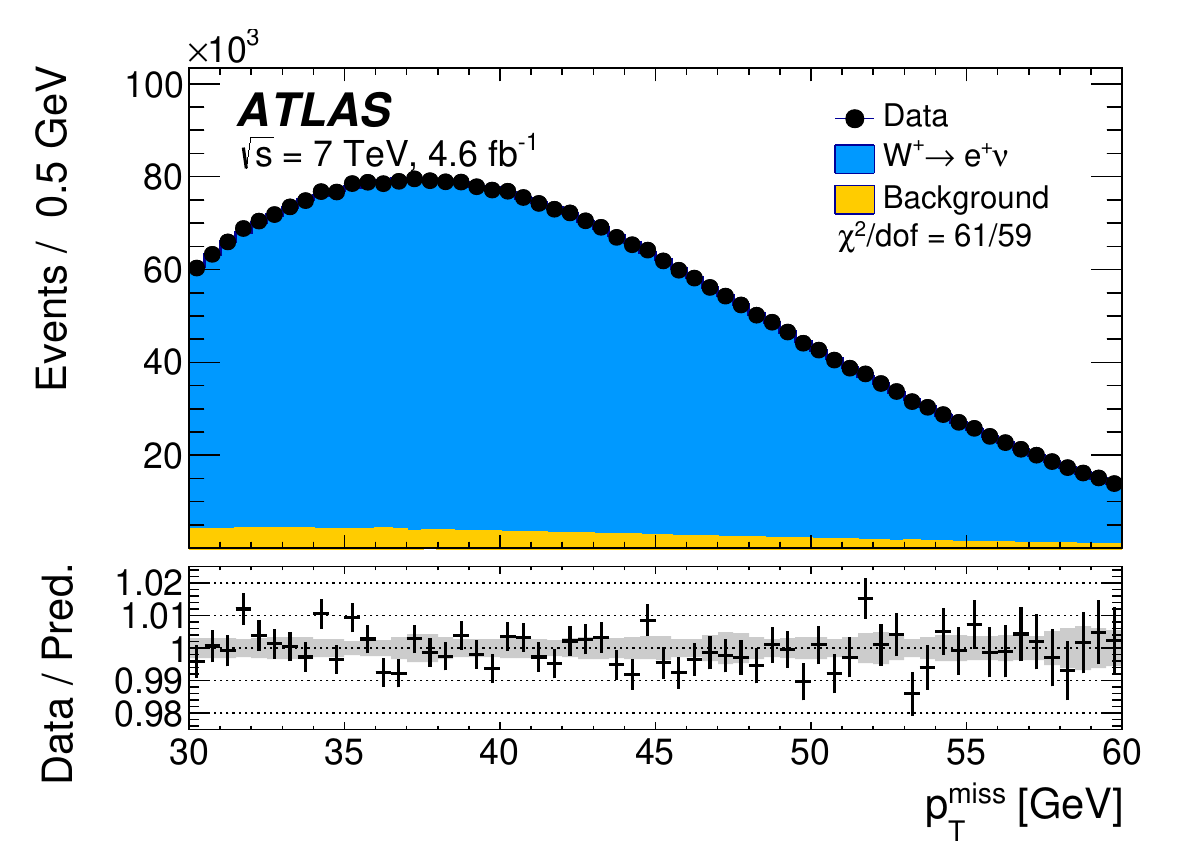}}
    \subfloat[]{\includegraphics[width=0.49\textwidth]{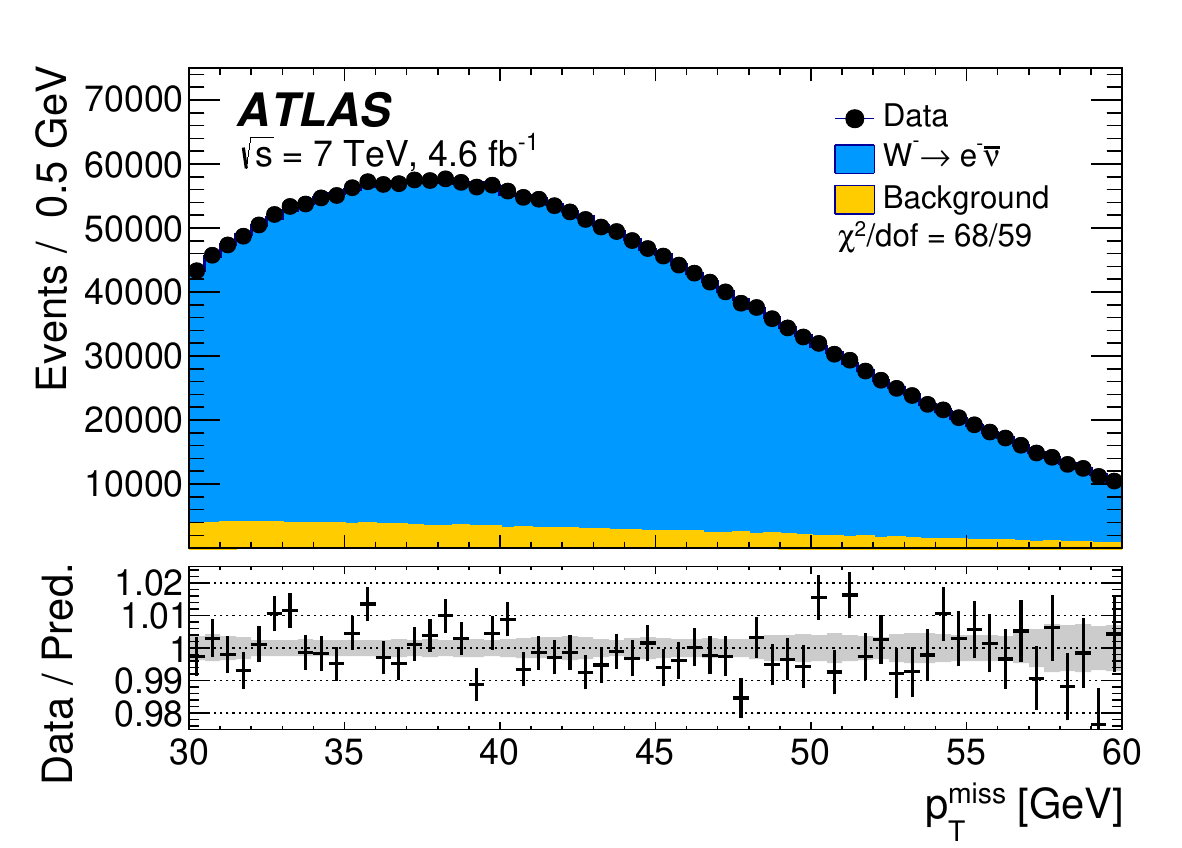}}
    \caption{The (a,b) $\pt^\ell$, (c,d) \mt, and (e,f) \mpt{}
      distributions for (a,c,e) $W^+$ events and (b,d,f) $W^-$ events in
      the electron decay channel. The data are compared to the simulation including
      signal and background contributions.
      Detector calibration and physics-modelling corrections are applied
      to the simulated events. For all simulated distributions, $m_W$ is set according to the overall measurement result.
      The lower panels show the data-to-prediction ratios, the error bars show the statistical uncertainty,
      and the band shows the systematic uncertainty of the prediction. The $\chi^2$ values displayed in each figure account for all sources of uncertainty and include the effects of bin-to-bin correlations induced by the systematic uncertainties. \label{fig:WSecMassIncE}}
  \end{center}
\end{figure}

\begin{figure}
  \begin{center}
    \subfloat[]{\includegraphics[width=0.49\textwidth]{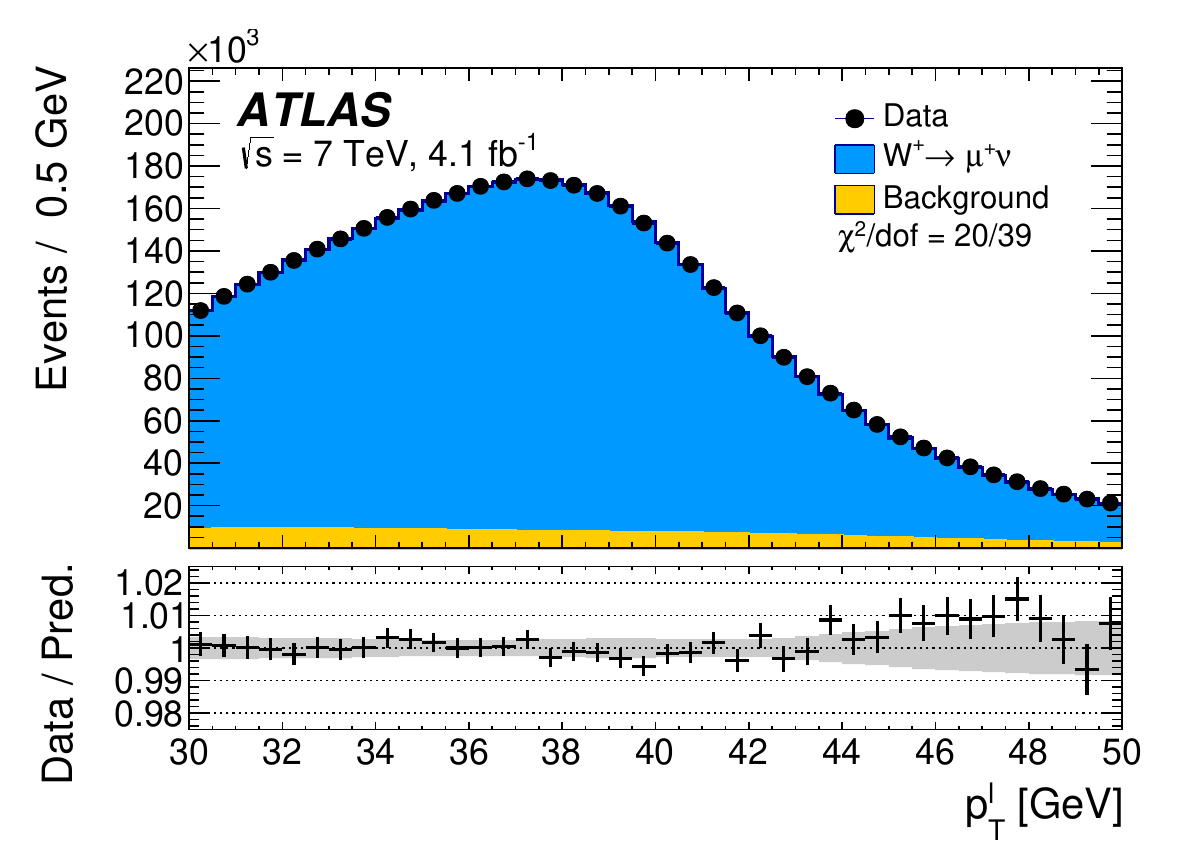}}
    \subfloat[]{\includegraphics[width=0.49\textwidth]{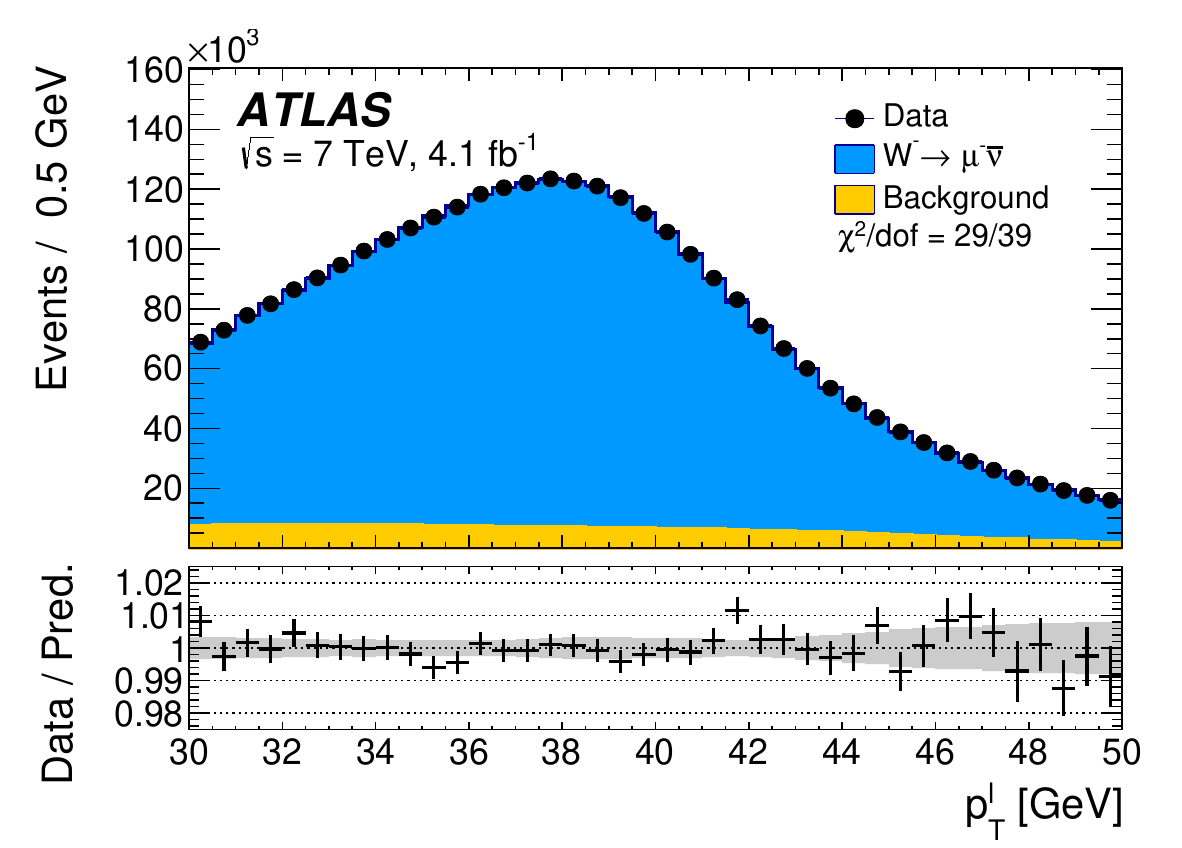}}\\
    \subfloat[]{\includegraphics[width=0.49\textwidth]{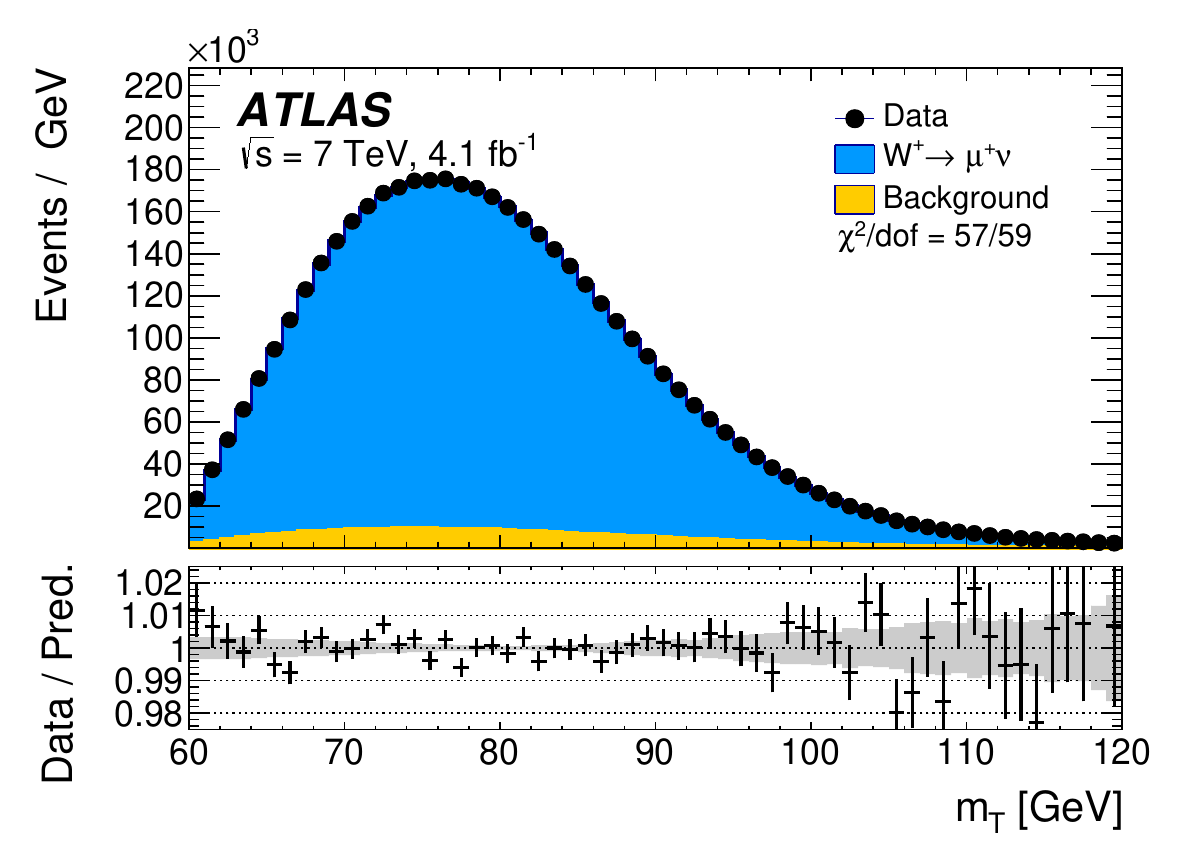}}
    \subfloat[]{\includegraphics[width=0.49\textwidth]{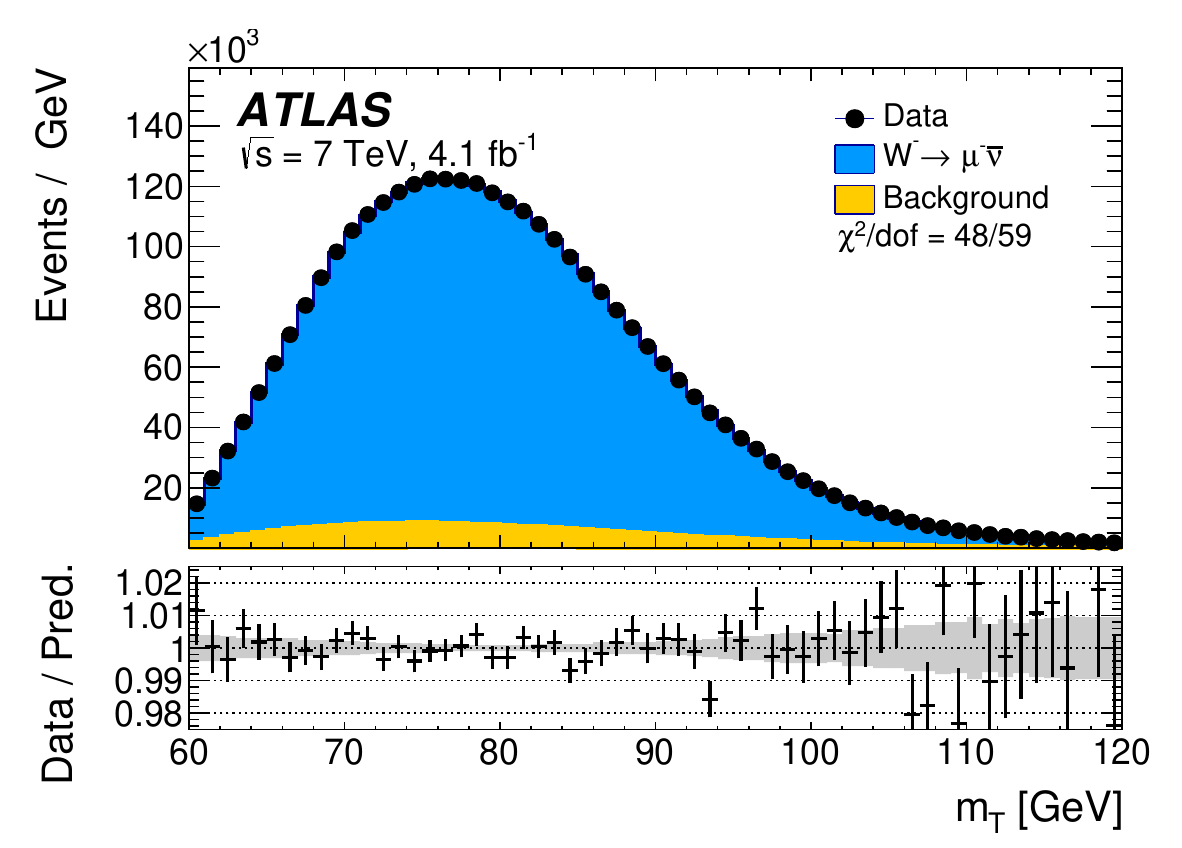}}\\
    \subfloat[]{\includegraphics[width=0.49\textwidth]{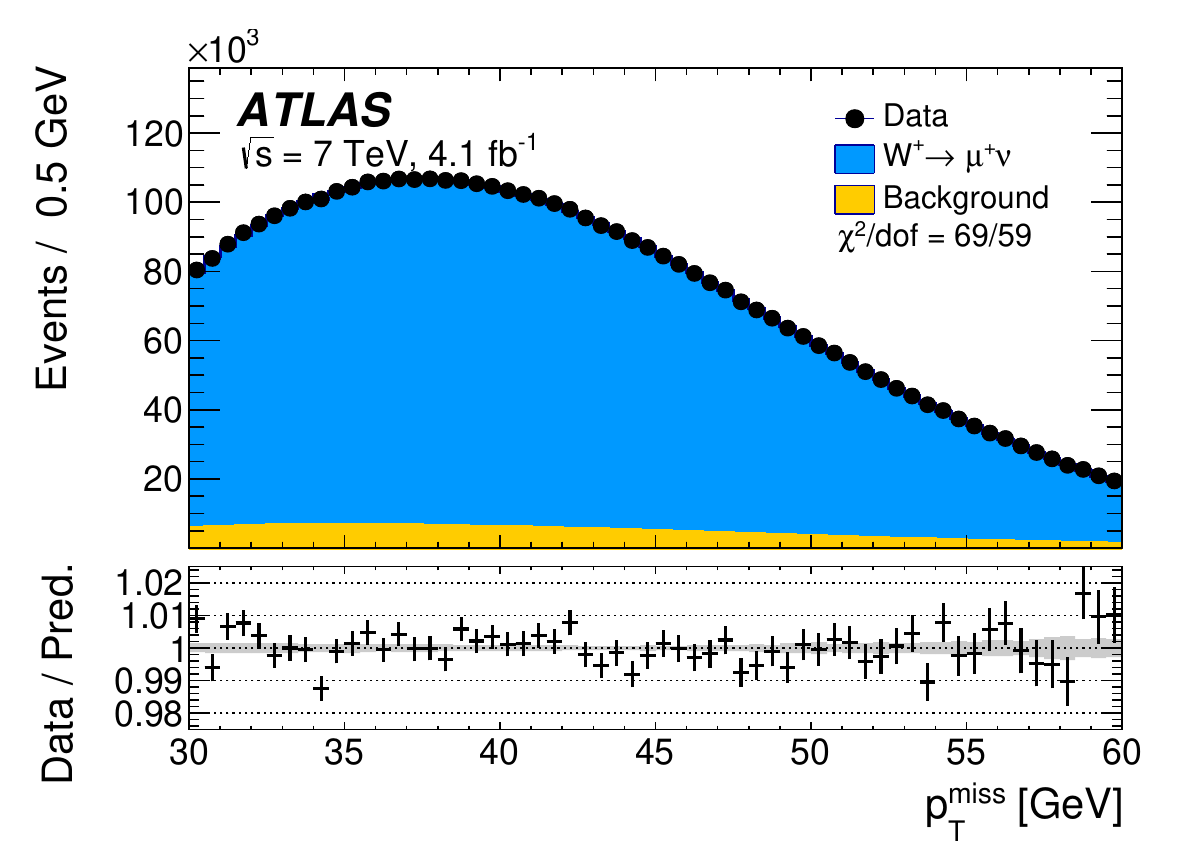}}
    \subfloat[]{\includegraphics[width=0.49\textwidth]{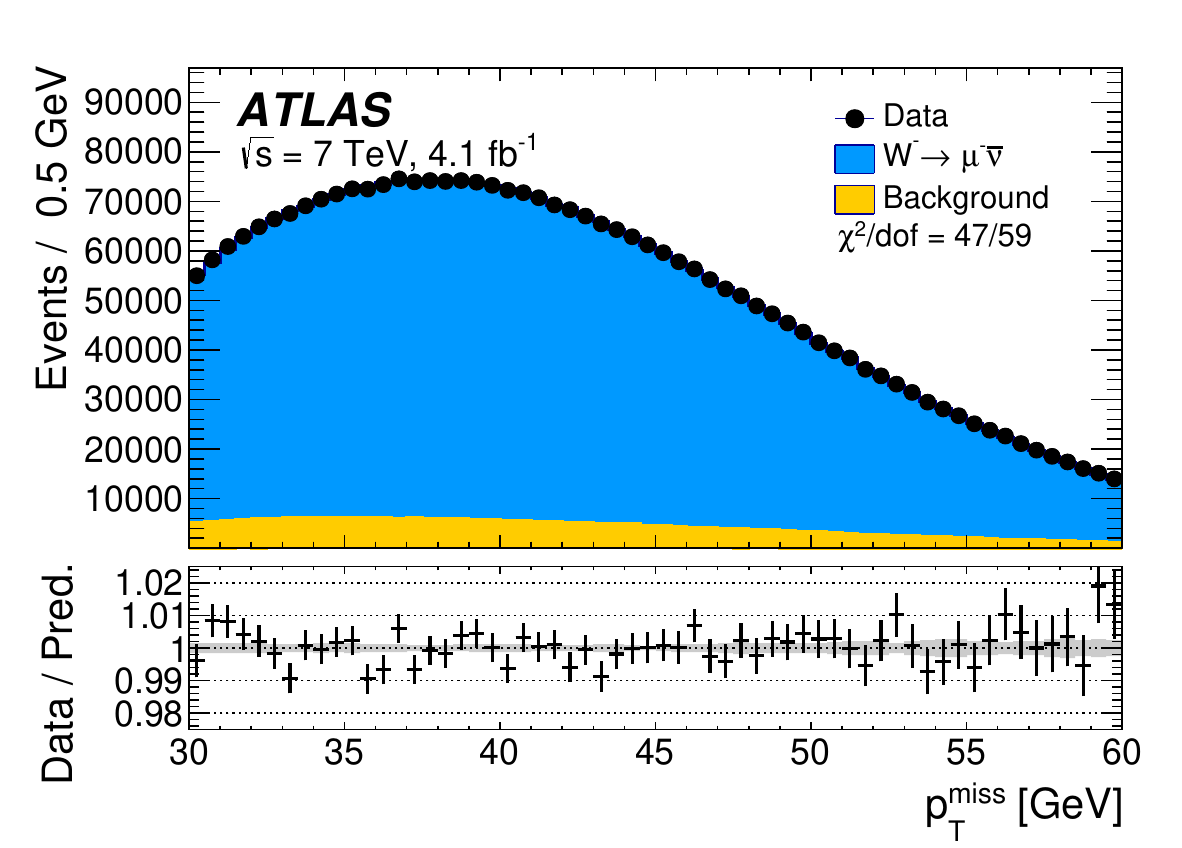}}
    \caption{The (a,b) $\pt^\ell$, (c,d) \mt, and (e,f) \mpt{}
      distributions for (a,c,e) $W^+$ events and (b,d,f) $W^-$ events in
      the muon decay channel. The data are compared to the simulation including
      signal and background contributions. Detector calibration and physics-modelling corrections are applied
      to the simulated events. For all simulated distributions, $m_W$ is set according to the overall measurement result.
      The lower panels show the data-to-prediction ratios, the error bars show the statistical uncertainty,
      and the band shows the systematic uncertainty of the prediction. The $\chi^2$ values displayed in each figure account for all sources of uncertainty and include the effects of bin-to-bin correlations induced by the systematic uncertainties. \label{fig:WSecMassIncMu}}
  \end{center}
\end{figure}

The comparison of data and simulation for kinematic distributions
sensitive to the value of $m_W$
provides further validation of the detector calibration and physics
modelling. The comparison is performed in all measurement categories.
The $\eta$-inclusive $\pt^\ell$, \mt{} and $\mpt$ distributions for positively and
negatively charged $W$ bosons are shown in Figures~\ref{fig:WSecMassIncE} and~\ref{fig:WSecMassIncMu} for the electron and muon decay
channels, respectively. The value of $m_W$ used in the predictions is set to the overall measurement result presented in the next section.
The $\chi^2$ values quantifying the comparison between data and prediction are calculated over the full histogram range and account for all sources of uncertainty. The bin-to-bin correlations induced by
the experimental and physics-modelling systematic uncertainties are also accounted for. Overall, satisfactory agreement is observed. The deficit of data visible for $\pt^\ell\sim 40$--$42\GeV$ in the
$W^+\rightarrow e^+ \nu$ channel does not strongly affect the mass measurement, as the observed effect differs from that expected from $m_W$ variations. Cross-checks of possible sources of this effect were performed, and its impact on the mass determination was shown to be within the corresponding systematic uncertainties.

\subsection{\label{sec:WComb}Combination and final results}

The measurements of $m_W$ in the various categories are combined accounting for
statistical and systematic uncertainties and their correlations.
The statistical correlation of the $m_W$ values determined from the
$\pt^\ell$ and \mt{} distributions is evaluated with the bootstrap
method~\cite{efron1979}, and is approximately 50\% for all measurement categories.

The systematic uncertainties have specific correlation patterns across the $m_W$ measurement categories. Muon-momentum and electron-energy
calibration uncertainties are uncorrelated between the different decay
channels, but largely correlated between the $\pt^\ell$ and \mt{}
distributions. Recoil-calibration uncertainties are
correlated between electron and muon decay channels, and they are small for $\pt^\ell$ distributions. The PDF-induced
uncertainties are largely correlated between electron and muon decay
channels, but significantly anti-correlated between positively and
negatively charged $W$ bosons, as discussed in Section~\ref{sec:phymod}. 
Due to the different balance of systematic uncertainties and to the variety of correlation
patterns, a significant reduction of the uncertainties in the
measurement of $m_W$ is achieved by combining the different decay channels and the charge and $|\eta_\ell|$
categories.

As discussed in Section~\ref{sec:strategy}, the comparison of the results from the $\pt^\ell$ and \mt{} distributions, from the
different decay channels, and in the various charge and $|\eta_\ell|$
categories, provides a test of the experimental and physics modelling corrections.
Discrepancies between the positively and negatively charged lepton categories, or in the various $|\eta_\ell|$ bins would primarily
indicate an insufficient understanding of physics-modelling effects,
such as the PDFs and the $\pt^W$ distribution.
Inconsistencies between the electron and muon channels could indicate problems in the calibration of the
muon-momentum and electron-energy responses.
Significant differences between results from the $\pt^\ell$ and $\mt$ distributions would point to either problems in
the calibration of the recoil, or to an incorrect modelling
of the transverse-momentum distribution of the $W$ boson.
Several measurement combinations are performed, using the best linear
unbiased estimate (BLUE) method
~\cite{Lyons:1988rp,Valassi:2003mu}. The results of the combinations
are verified with the \texttt{HERAverager}
program~\cite{Aaron:2009bp}, which gives very close results.

Table~\ref{tab:fitIndividualResults} shows an overview of partial
$m_W$ measurement combinations.
In the first step, determinations of $m_W$ in the electron and muon
decay channels from the $\mt$ distribution are combined separately for the positive- and
negative-charge categories, and together for both $W$-boson charges. The results are compatible, and the
positively charged, negatively charged, and charge-inclusive
combinations yield values of $\chi^2/$dof corresponding to $2/6$,
$7/6$, and $11/13$, respectively.
Compatibility of the results is also observed for the corresponding
combinations from the $\pt^\ell$ distribution, with values of
$\chi^2/$dof of $5/6$, $10/6$, and $19/13$, for positively charged,
negatively charged, and charge-inclusive combinations, respectively.
The $\chi^2$ compatibility test validates the consistency of the
results in the $W\rightarrow e\nu$ and $W\rightarrow \mu\nu$ decay
channels. The precision of the determination of $m_W$ from the $\mt$
distribution is slightly worse than the result obtained from the $\pt^\ell$ distribution, due to the larger uncertainty induced by the recoil
calibration. In addition, the impact of PDF- and $\pt^W$-related uncertainties on the $\pt^\ell$ fits is limited by the optimisation of the fitting range. 
In the second step, determinations of $m_W$ from the $\pt^\ell$ and
$\mt$ distributions are combined separately for the electron and the
muon decay channels. The results are compatible, with
values of $\chi^2/$dof of 4/5 and 8/5 in the
electron channel for the $\pt^\ell$ and \mt{} distributions,
respectively, and values of 7/7 and 3/7 in the muon channel for the
$\pt^\ell$ and \mt{} distributions, respectively. The $m_W$ determinations in the electron and in
the muon channels agree, further validating the consistency of the
electron and muon calibrations. Agreement between the $m_W$ determinations
from the $\pt^\ell$ and \mt{} distributions supports the
calibration of the recoil, and the modelling of the
transverse momentum of the $W$ boson.

 \begin{table}[tp]
  \centering
  \resizebox{\textwidth}{!}{\begin{tabular}{l|c|rrrrrrrrr|c}
      \toprule
      Combined  & Value & Stat. & Muon & Elec.& Recoil & Bckg. & QCD & EW & PDF & Total  &$\chi^2/$dof\\
    categories& [\MeV]& Unc. & Unc.& Unc.& Unc. & Unc. & Unc. & Unc. & Unc. & Unc.  & of Comb.\\
\midrule
$\mt$, $W^+$, $e$-$\mu$                   & 80370.0       & 12.3  & 8.3   & 6.7   	& 14.5  	& 9.7   	& 9.4   	& 3.4   	& 16.9  	& 30.9  	& 2/6   \\
$\mt$, $W^-$, $e$-$\mu$                   & 80381.1       & 13.9  & 8.8   & 6.6   	& 11.8  	& 10.2   	& 9.7   	& 3.4   	& 16.2  	& 30.5  	& 7/6   \\
$\mt$, $W^\pm$, $e$-$\mu$                 & 80375.7       & 9.6   & 7.8   & 5.5   	& 13.0  	& 8.3   	& 9.6   	& 3.4   	& 10.2   	& 25.1  	& 11/13 \\
\midrule
$\pT^\ell$, $W^+$, $e$-$\mu$              	& 80352.0       & 9.6   & 6.5   	& 8.4   	& 2.5        	& 5.2   	& 8.3   	& 5.7   	& 14.5  	& 23.5  	& 5/6\\
$\pT^\ell$, $W^-$, $e$-$\mu$              	& 80383.4       & 10.8  & 7.0   	& 8.1   	& 2.5        	& 6.1   	& 8.1   	& 5.7   	& 13.5  	& 23.6  	& 10/6   \\
$\pT^\ell$, $W^\pm$, $e$-$\mu$           & 80369.4       	& 7.2   & 6.3        	& 6.7        	& 2.5   	& 4.6   	& 8.3   	& 5.7   	& 9.0   	& 18.7      	& 19/13\\
\midrule
$\pt^\ell$, $W^\pm$, $e$			& 80347.2 	& 9.9   	& 0.0 		& 14.8	& 2.6 	& 5.7 	& 8.2 	& 5.3 	& 8.9 	& 23.1 	& 4/5	\\
$\mt$, $W^\pm$, $e$			& 80364.6 	& 13.5 	& 0.0 		& 14.4 	& 13.2	& 12.8 	& 9.5 	& 3.4 	& 10.2 	& 30.8 	& 8/5	\\
$\mt$-$\pt^\ell$, $W^+$, $e$               & 80345.4       & 11.7       & 0.0   	& 16.0      	& 3.8  	& 7.4   	& 8.3   	& 5.0   	& 13.7  	& 27.4  	& 1/5\\
$\mt$-$\pt^\ell$, $W^-$, $e$              	& 80359.4       & 12.9   	& 0.0   		& 15.1   	& 3.9  	& 8.5   	& 8.4   	& 4.9   	& 13.4  	& 27.6  	& 8/5\\ 
$\mt$-$\pt^\ell$, $W^\pm$, $e$            	& 80349.8      	& 9.0   	& 0.0   		& 14.7  	& 3.3        & 6.1   	& 8.3   	& 5.1   	& 9.0   	& 22.9  	& 12/11\\
\midrule
$\pt^\ell$, $W^\pm$, $\mu$		& 80382.3 	& 10.1 	& 10.7 	& 0.0	 	& 2.5		& 3.9 	& 8.4 	& 6.0 	& 10.7 	& 21.4 	& 7/7	\\
$\mt$, $W^\pm$, $\mu$		        & 80381.5 	& 13.0 	& 11.6 	& 0.0	 	& 13.0	& 6.0 	& 9.6 	& 3.4 	& 11.2 	& 27.2 	& 3/7	\\
$\mt$-$\pt^\ell$, $W^+$, $\mu$         	& 80364.1       & 11.4   	& 12.4   	& 0.0	   	& 4.0  	& 4.7   	& 8.8   	& 5.4   	& 17.6  	& 27.2  	& 5/7\\ 
$\mt$-$\pt^\ell$, $W^-$, $\mu$         	& 80398.6       & 12.0   	& 13.0   	& 0.0  		& 4.1  	& 5.7   	& 8.4   	& 5.3   	& 16.8  	& 27.4  	& 3/7\\ 
$\mt$-$\pt^\ell$, $W^\pm$, $\mu$     	& 80382.0       & 8.6   	& 10.7  	& 0.0   		& 3.7        	& 4.3   	& 8.6   	& 5.4        & 10.9  	& 21.0  & 10/15\\
\midrule
$\mt$-$\pt^\ell$, $W^+$, $e$-$\mu$ 	& 80352.7	& 8.9 	& 6.6 	& 8.2 	& 3.1		& 5.5 	& 8.4 	& 5.4 	& 14.6 	& 23.4 	& 7/13\\
$\mt$-$\pt^\ell$, $W^-$, $e$-$\mu$ 	& 80383.6	& 9.7 	& 7.2 	& 7.8 	& 3.3		& 6.6 	& 8.3 	& 5.3		& 13.6 	& 23.4 	& 15/13\\
\midrule
$\mt$-$\pt^\ell$, $W^\pm$, $e$-$\mu$  & 80369.5 	& 6.8 	& 6.6 	& 6.4 	& 2.9 	& 4.5 	& 8.3 	& 5.5 	& 9.2 	& 18.5 	& 29/27  \\
\bottomrule
  \end{tabular}}
  \caption{Results of the $m_W$ measurements for various combinations
    of categories. The table shows the statistical uncertainties,
    together with all experimental uncertainties, divided into muon-,
    electron-, recoil- and background-related uncertainties, and all
    modelling uncertainties, separately for QCD
    modelling including scale variations, parton shower and angular
    coefficients, electroweak corrections, and PDFs. All uncertainties are given in \MeV.
  \label{tab:fitIndividualResults}}
\end{table}

\begin{figure}
  \begin{center}
    \includegraphics[width=0.85\textwidth]{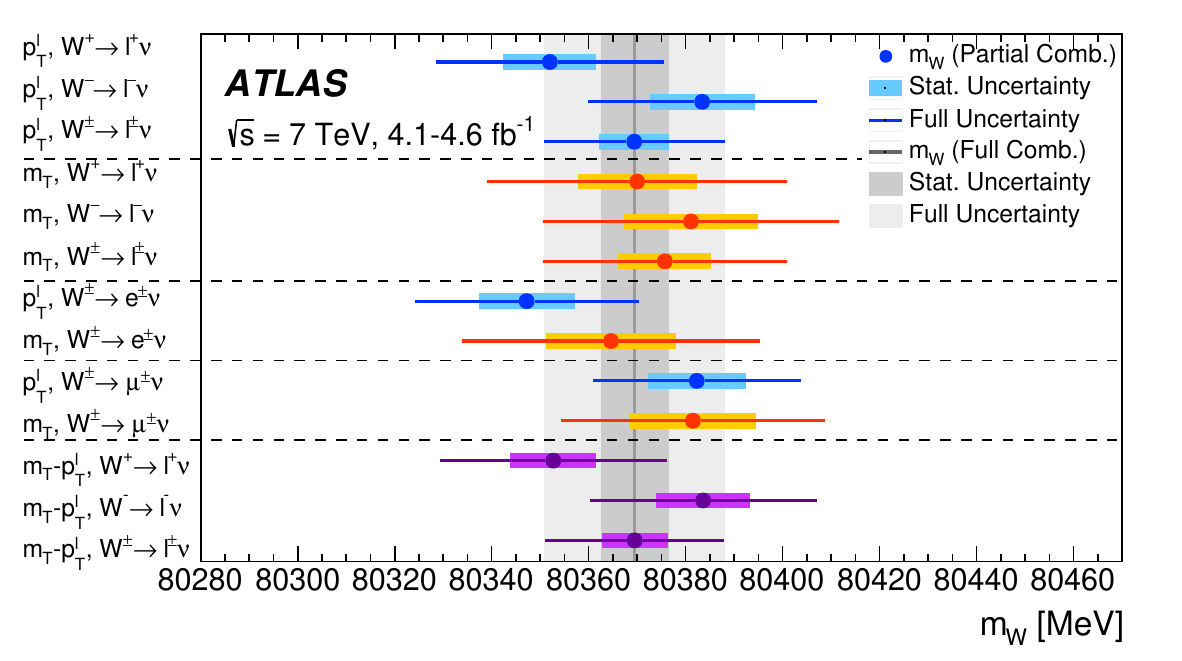}
  \end{center}
  \caption{Overview of the $m_W$ determinations from the $\pt^\ell$ and
    $\mt$ distributions, and for the combination of the $\pt^\ell$ and $\mt$ distributions, in the muon and electron decay channels and
    for $W^+$ and $W^-$ events. The horizontal lines and bands show the statistical and total uncertainties of the individual $m_W$
    determinations. The combined result for $m_W$ and its statistical and total uncertainties
    are also indicated (vertical line and bands).}
\label{fig:WMassCombinationTests}
\end{figure}

The results are summarised in Figure~\ref{fig:WMassCombinationTests}.
The combination of all the determinations of $m_W$ reported in
Table~\ref{tab:fitAllResults} has a value of $\chi^2/$dof of $29/27$, and yields a final result of
\begin{eqnarray}
\nonumber m_W &=& 80369.5 \pm 6.8 \MeV (\textrm{stat.}) \pm 10.6 \MeV (\textrm{exp. syst.}) \pm 13.6 \MeV (\textrm{mod. syst.}) \\
\nonumber &=& 80369.5 \pm  18.5 \MeV,
\end{eqnarray}
\noindent where the first uncertainty is statistical, the second
corresponds to the experimental systematic uncertainty, and the third
to the physics-modelling systematic uncertainty. The latter dominates the total measurement uncertainty, and it itself dominated by strong interaction uncertainties. The experimental systematic
uncertainties are dominated by the lepton calibration; backgrounds and the recoil calibration have a smaller impact. In the final combination, the muon decay channel has a weight of 57\%, and the $\pt^\ell$ fit dominates the measurement with a weight of 86\%. Finally, the charges contribute similarly with a weight of 52\% for $W^{+}$ and of 48\% for $W^{-}$. 

The result is in agreement with the current world average of $m_W = 80385 \pm 15\MeV$~\cite{Agashe:2014kda}, and has a precision comparable to the currently most precise single measurements of the CDF and D0
collaborations~\cite{Aaltonen:2012bp,Abazov:2012bv}.

\subsection{Additional validation tests\label{sec:mwcrosschecks}}

The final combination of $m_W$, presented above, depends only on template fits to the $\pt^\ell$ and \mt{}
distributions. As a validation test, the value of $m_W$ is determined
from the $\mpt$ distribution, performing a fit in the range $30<\mpt<60\GeV$. Consistent results are observed in all
measurement categories, leading to combined results of $80364\pm26$~(stat)~\MeV and $80367 \pm 23$~(stat)~\MeV for the electron and
muon channels, respectively. 

Several additional studies are performed to validate the stability of the $m_W$
measurement. The stability of the result with respect to different pile-up conditions is tested by dividing the event
sample into three bins of $\avg{\mu}$, namely $[2.5,6.5]$, $[6.5,9.5]$, and
$[9.5,16]$. In each bin, $m_W$ measurements are performed
independently using the $\pt^\ell$ and \mt{} distributions. This
categorisation also tests the stability of $m_W$ with respect to
data-taking periods, as the later data-taking periods have on average
more pile-up due to the increasing LHC luminosity. 

The calibration of the recoil and the modelling of the $\pt^W$ distribution are tested by
performing $m_W$ fits in two bins of the recoil corresponding to $[0,15]\GeV$ and
$[15,30]\GeV$, and in two regions corresponding to positive and negative values of $u_\parallel^\ell$. The analysis is also repeated with the $\mpt$ requirement removed from the signal selection, leading to a lower
recoil modelling uncertainty but a higher multijet background
contribution. The stability of the $m_W$ measurements upon removal of
this requirement is studied, and consistent results are obtained. All $m_W$ determinations are consistent
with the nominal result. An overview of the validation tests is shown
in Table~\ref{tab:CrossCheckWmFits}, where only statistical uncertainties are given. Fitting ranges of $30<\pt^\ell<50\GeV$ and $65<\mt<100\GeV$ are used for all these validation tests, to minimise the statistical uncertainty. 

\begin{table}[tp]
  \begin{center}
    \begin{tabular}{lrrrrrr}
      \toprule
      Decay channel  & \multicolumn{2}{c}{\hspace{1cm}$W\rightarrow e\nu$} &  \multicolumn{2}{c}{\hspace{1cm}$W\rightarrow \mu\nu$} &  \multicolumn{2}{c}{\hspace{1cm}Combined} \\ 
      Kinematic distribution & $\pt^\ell$ & $\mt$& $\pt^\ell$& $\mt$& $\pt^\ell$& $\mt$\\
      \midrule
      $\Delta m_W$ [\MeV] \\
      \,\,\,\,$\avg{\mu}$ in $[2.5, 6.5]$ & $8\pm14$& $14\pm18$& $-21\pm12$& $0\pm16$ &	 $-9\pm \;\,9$ & $6\pm12$\\
      \,\,\,\,$\avg{\mu}$ in $[6.5, 9.5]$ & $-6\pm16$& $6\pm23$& $12\pm15$& $-8\pm22$ &		 $4\pm 11$ & $-1\pm16$\\
      \,\,\,\,$\avg{\mu}$ in $[9.5, 16]$ & $-1\pm16$& $3\pm27$& $25\pm16$& $35\pm26$ &		 $12\pm 11$ & $20\pm19$\\
      \midrule
      \,\,\,\,$\ut$ in $[0,15]\GeV$ & $0\pm11$& $-8\pm13$& $5\pm10$& $8\pm12$ &			 $3\pm \;\,7$ & $-1\pm\;\,9$\\
      \,\,\,\,$\ut$ in $[15,30]\GeV$ & $10\pm15$& $0\pm24$& $-4\pm14$& $-18\pm22$ &		 $2\pm 10$ & $-10\pm16$\\
      \midrule
      \,\,\,\,$u_\parallel^\ell<0\,\GeV$& $8\pm15$& $20\pm17$& $3\pm13$& $-1\pm16$ &		 $5\pm 10$ & $9\pm12$\\
      \,\,\,\,$u_\parallel^\ell>0\,\GeV$& $-9\pm10$& $1\pm14$& $-12\pm10$& $10\pm13$ &		 $-11\pm \;\,7$ & $6\pm10$\\
       \midrule
      \,\,\,\,No $\mpt$-cut& $14\pm\;\,9$& $-1\pm13$& $10\pm\;\,8$& $-6\pm12$ &				 $12\pm\;\,6$ & $-4\pm\;\,9$\\
      \bottomrule
    \end{tabular}
  \end{center}
  \caption{Summary of consistency tests for the determination of $m_W$ in several additional
    measurement categories. The $\Delta m_W$ values correspond to the difference between the result for each category and the inclusive result for the corresponding observable ($\pt^\ell$ or \mt). The uncertainties correspond to the statistical uncertainty of the fit to the data of each category alone. Fitting ranges of $30<\pt^\ell<50\GeV$ and $65<\mt<100\GeV$ are used.} 
  \label{tab:CrossCheckWmFits}
\end{table}

\begin{figure}
  \begin{center}
    \subfloat[]{\includegraphics[width=0.49\textwidth]{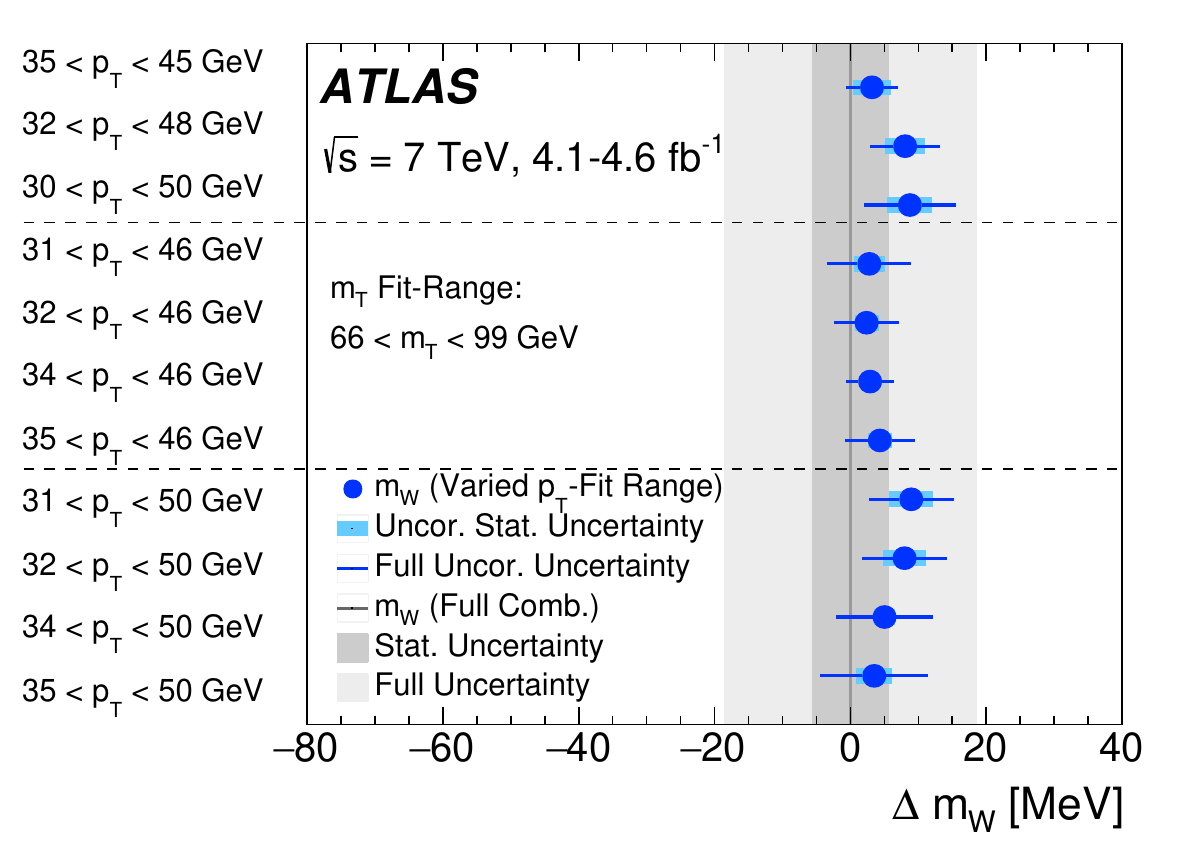}}
    \subfloat[]{\includegraphics[width=0.49\textwidth]{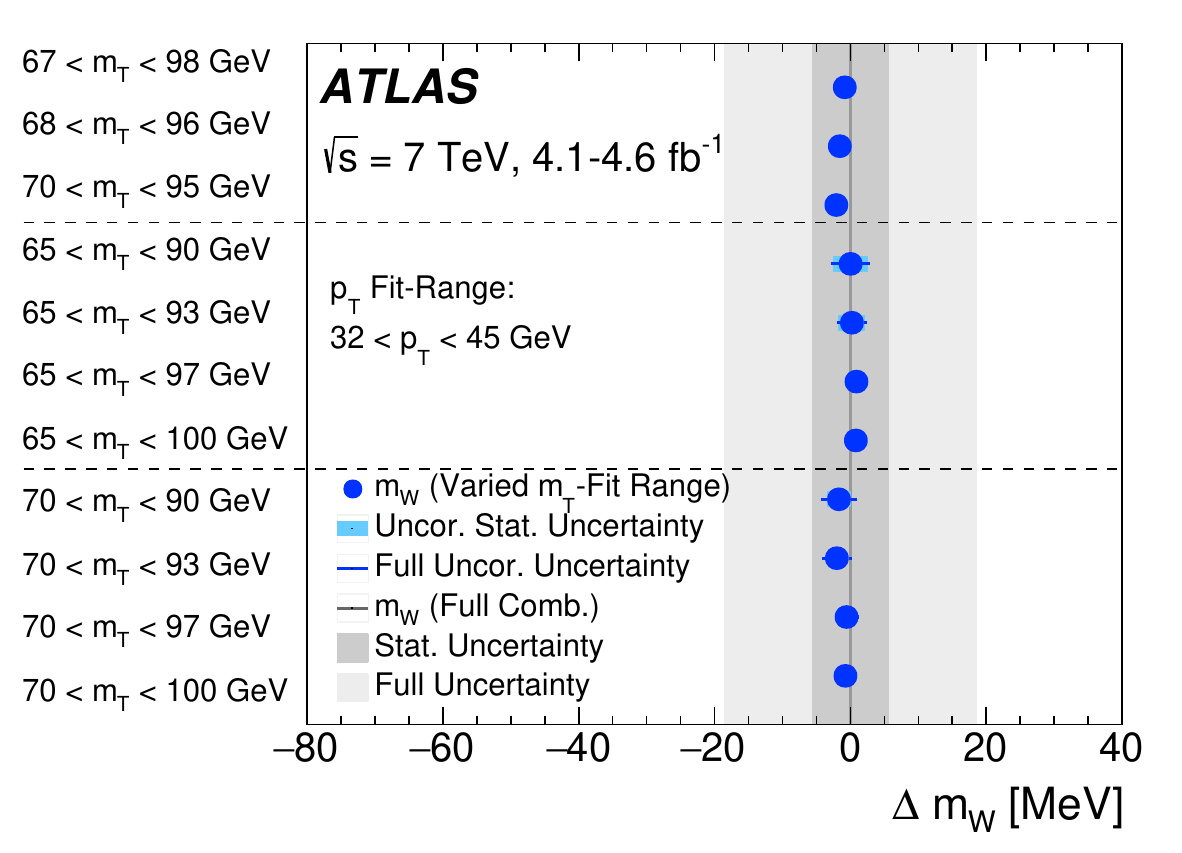}}
  \end{center}
  \caption{Stability of the combined measurement of $m_W$ with respect to variations
    of the kinematic ranges of (a) $\pt^\ell$ and (b) $\mt$ used for the template fits. The optimal \mt\ range is used for the $\pt^\ell$ variations, and the optimal $\pt^\ell$ range is used for the \mt\
    variations. The effect on the result of symmetric variations of the fitting range boundaries, and its dependence on variations of the lower (upper) boundary for two values of the upper (lower) boundary for $\pt^\ell$ ($\mt$) are shown. The bands and solid lines respectively show the statistical and total uncertainty on the difference with the
    optimal result.}
\label{Fig:FitRangeChecks}
\end{figure}

The lower and upper bounds of the range of the $\pt^\ell$ and \mt{}
distributions are varied as in the optimisation procedure described in
Section~\ref{sec:Unc}. The statistical and systematic uncertainties
are evaluated for each range, and are only partially correlated
between different ranges.
Figure~\ref{Fig:FitRangeChecks} shows measured values of $m_W$ for
selected ranges of the $\pt^\ell$ and $\mt$ distributions, where only
the uncorrelated statistical and systematic uncertainties with respect
to the optimal range are shown. The observed variations are all within
two standard deviations of the uncorrelated uncertainties, and small
compared to the overall uncertainty of the measurement, which is
illustrated by the band on Figure~\ref{Fig:FitRangeChecks}.
The largest dependence on the kinematic ranges used for the fits is
observed for variations of the upper bound of the $\pt^\ell$
distribution in the $W^+\rightarrow e^+\nu$ channel, and is related to the
shape of the data-to-prediction ratio for this distribution in the
region $40 < \pt^\ell < 42\GeV$, as discussed in Section~\ref{sec:Unc}.

The effect of the residual discrepancies in the \ut\ distributions for $W^-\rightarrow\ell\bar{\nu}$, visible at low values in Figures~\ref{fig:WSecControlIncE}-(d)~and~\ref{fig:WSecControlIncMu}-(d),
is estimated by adjusting, in turn, the particle-level $\pt^W$ distribution and the recoil calibration corrections to optimize the agreement between data and simulation. The impact of these variations
on the determination of $m_W$ is found to be small compared to the assigned $\pt^W$ modelling and recoil calibration uncertainties, respectively.

When assuming $R_{W/Z}(\pT)$ as predicted by DYRes, instead of \PYTHIA~8~AZ, to model the $\pt^W$ distribution, deviations of about 3\% appear in the distribution
ratios of Figures~\ref{fig:WSecMassIncE} and~\ref{fig:WSecMassIncMu}. This degrades the quality of the mass fits, and shifts the fitted values of $m_W$ by about $-20$ to $-90$~MeV,
depending on the channels, compared to the results of Table~\ref{tab:fitIndividualResults}. Combining all channels, the shift is about $-60$~MeV. Since DYRes does not model the data distributions
sensitive to $\pt^W$, as shown in Figure~\ref{fig:uparcomp}, these shifts are given for information only and are not used to estimate the uncertainty in $m_W$.

\subsection{Measurement of $m_{W^+}-m_{W^-}$}

The results presented in the previous sections can be used to derive a measurement of the mass difference between the positively and negatively charged $W$ bosons, $m_{W^+}-m_{W^-}$. 
Starting from the $m_W$ measurement results in the 28 categories described above, 14 measurements of $m_{W^+}-m_{W^-}$ can be constructed by subtraction of the results obtained from the $W^+$ and $W^-$
samples in the same decay channel and $|\eta|$ category. In practice, the $m_W$ values measured in $W^+$ and $W^-$ events are subtracted linearly, as are the effects of systematic uncertainties on these measurements, while the uncertainty contributions of a statistical nature are added in quadrature. Contrarily to the $m_W$ measurement discussed above, no blinding procedure was applied for the measurement of $m_{W^+}-m_{W^-}$.

In this process, uncertainties that are anti-correlated between $W^+$ and $W^-$ and largely cancel for the $m_W$ measurement become dominant when measuring $m_{W^+}-m_{W^-}$. On the physics-modelling side, the
fixed-order PDF uncertainty and the parton shower PDF uncertainty give the largest contributions, while other sources of uncertainty only weakly depend on charge and tend to cancel. Among the sources
of uncertainty related to lepton calibration, the track sagitta correction dominates in the muon channel, whereas several residual uncertainties contribute in the electron channel. Most lepton and
recoil calibration uncertainties tend to cancel. Background systematic uncertainties contribute as the $Z$ and multijet background fractions differ in the $W^+$ and $W^-$ channels. The dominant
statistical uncertainties arise from the size of the data and Monte Carlo signal samples, and of the control samples used to derive the multijet background. 

The $m_{W^+}-m_{W^-}$ measurement results are shown in Table~\ref{tab:mdiff} for the electron and muon decay channels, and for the combination. The electron channel measurement combines six categories ($\pt^\ell$ and
\mt\ fits in three $|\eta_\ell|$ bins), while the muon channel has four $|\eta_\ell|$ bins and eight categories in total. The fully combined result is 
\begin{eqnarray}
\nonumber m_{W^+}-m_{W^-} &=& -29.2 \pm 12.8 \MeV (\textrm{stat.}) \pm 7.0 \MeV (\textrm{exp. syst.}) \pm 23.9 \MeV (\textrm{mod. syst.}) \\
\nonumber &=& -29.2 \pm 28.0 \MeV,
\end{eqnarray}
\noindent where the first uncertainty is statistical, the second corresponds to the experimental systematic uncertainty, and the third to the physics-modelling systematic uncertainty. 

\begin{table}[tp]
  \centering
  \resizebox{\textwidth}{!}{\begin{tabular}{c|c|crcccccc|c}
\toprule
    Channel & $m_{W^+}-m_{W^-}$ & Stat. & Muon & Elec.& Recoil & Bckg. & QCD & EW & PDF & Total  \\
            & [\MeV]& Unc. & Unc.& Unc.& Unc. & Unc.& Unc.& Unc. & Unc. & Unc.  \\
\midrule
$W\rightarrow e\nu$ & $-29.7$ & $17.5$ & $0.0~$ & $4.9$ & $0.9$ & $5.4$ & $0.5$ & $0.0$ & $24.1$ & $30.7$ \\
$W\rightarrow\mu\nu$ & $-28.6$ & 16.3 & 11.7~~& 0.0 & 1.1 & 5.0 & 0.4 & 0.0 & 26.0 & 33.2 \\
\midrule
Combined  & $-29.2$ & 12.8 & 3.3~~& 4.1 & 1.0 & 4.5 & 0.4 & 0.0 & 23.9 & 28.0 \\
\bottomrule
\end{tabular}}
\caption{Results of the $m_{W^+}-m_{W^-}$ measurements in the electron and muon
    decay channels, and of the combination. The table shows the statistical uncertainties;
    the experimental uncertainties, divided into muon-,
    electron-, recoil- and background-uncertainties; and the
    modelling uncertainties, separately for QCD modelling including scale variations, parton shower and angular
    coefficients, electroweak corrections, and PDFs. All uncertainties are given in \MeV.\label{tab:mdiff}}
\end{table}

\section{Discussion and conclusions \label{sec:conclusions}}

This paper reports a measurement of the $W$-boson mass with the ATLAS detector, obtained through template fits to the kinematic properties
of decay leptons in the electron and muon decay channels.
The measurement is based on proton--proton collision data recorded in
2011 at a centre-of-mass energy of $\sqrt{s} = 7\TeV$ at the LHC,
and corresponding to an integrated luminosity of 4.6~fb$^{-1}$.
The measurement relies on a thorough detector calibration based on the
study of $Z$-boson events, leading to a precise modelling of the detector response to
electrons, muons and the recoil.
Templates for the $W$-boson kinematic distributions are obtained from
the NLO MC generator \POWHEG, interfaced to \textsc{Pythia8} for the parton shower. The
signal samples are supplemented with several additional physics-modelling corrections
allowing for the inclusion of higher-order QCD and
electroweak corrections, and by fits to measured distributions,
so that agreement between the data and the model in the kinematic distributions is improved. 
The $W$-boson mass is obtained from the transverse-momentum distribution of
charged leptons and from the transverse-mass distributions, for
positively and negatively charged $W$ bosons, in the electron and muon
decay channels, and in several kinematic categories. The
individual measurements of $m_W$ are found to be consistent and their combination yields a value of
\begin{eqnarray}
\nonumber m_W &=& 80370 \pm 7~(\textrm{stat.}) \pm 11~(\textrm{exp. syst.}) \pm 14~(\textrm{mod. syst.}) \MeV \\
\nonumber &=& 80370 \pm 19 \MeV,
\end{eqnarray}
\noindent where the first uncertainty is statistical, the second
corresponds to the experimental systematic uncertainty, and the third
to the physics-modelling systematic uncertainty. A measurement of the $W^+$ and $W^-$ mass difference yields $m_{W^+}-m_{W^-} = -29 \pm 28\MeV$.

\begin{figure}[tb]
  \begin{center}
\includegraphics[width=0.70\textwidth]{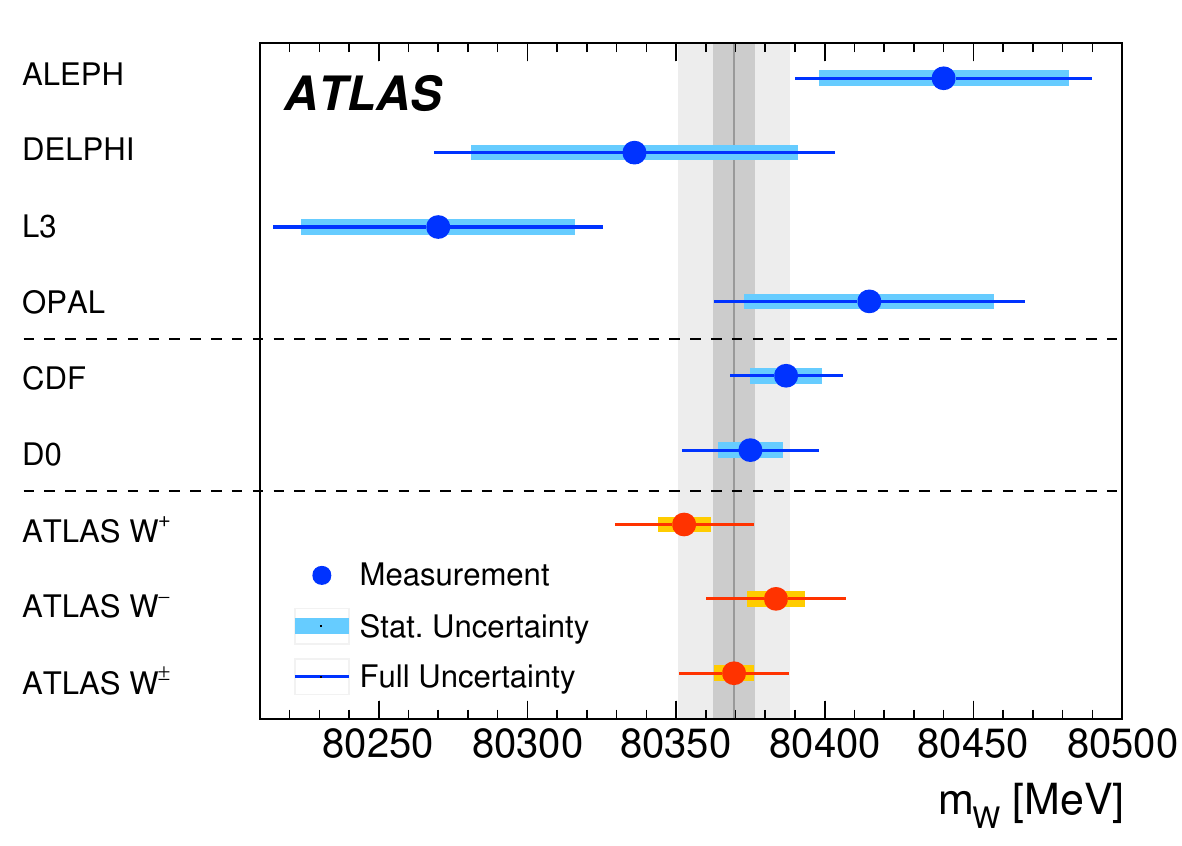}
  \end{center}
\caption{The measured value of $m_W$ is compared to other
  published results, including measurements from the
  LEP experiments ALEPH, DELPHI, L3 and
  OPAL~\cite{Schael:2006mz,Abdallah:2008ad,Achard:2005qy,Abbiendi:2005eq},
  and from the Tevatron collider experiments CDF and
  D0~\cite{Aaltonen:2012bp,Abazov:2012bv}. The vertical bands show the statistical and total uncertainties of the ATLAS
  measurement, and the horizontal bands and lines show the statistical and
  total uncertainties of the other published results. Measured values of
  $m_W$ for positively and negatively charged $W$ bosons are also
  shown.}
\label{Fig:WAFinal}
\end{figure}

\begin{figure}[t!b]
\begin{minipage}[hbt]{0.49\textwidth}
	\centering
	\includegraphics[width=0.99\textwidth]{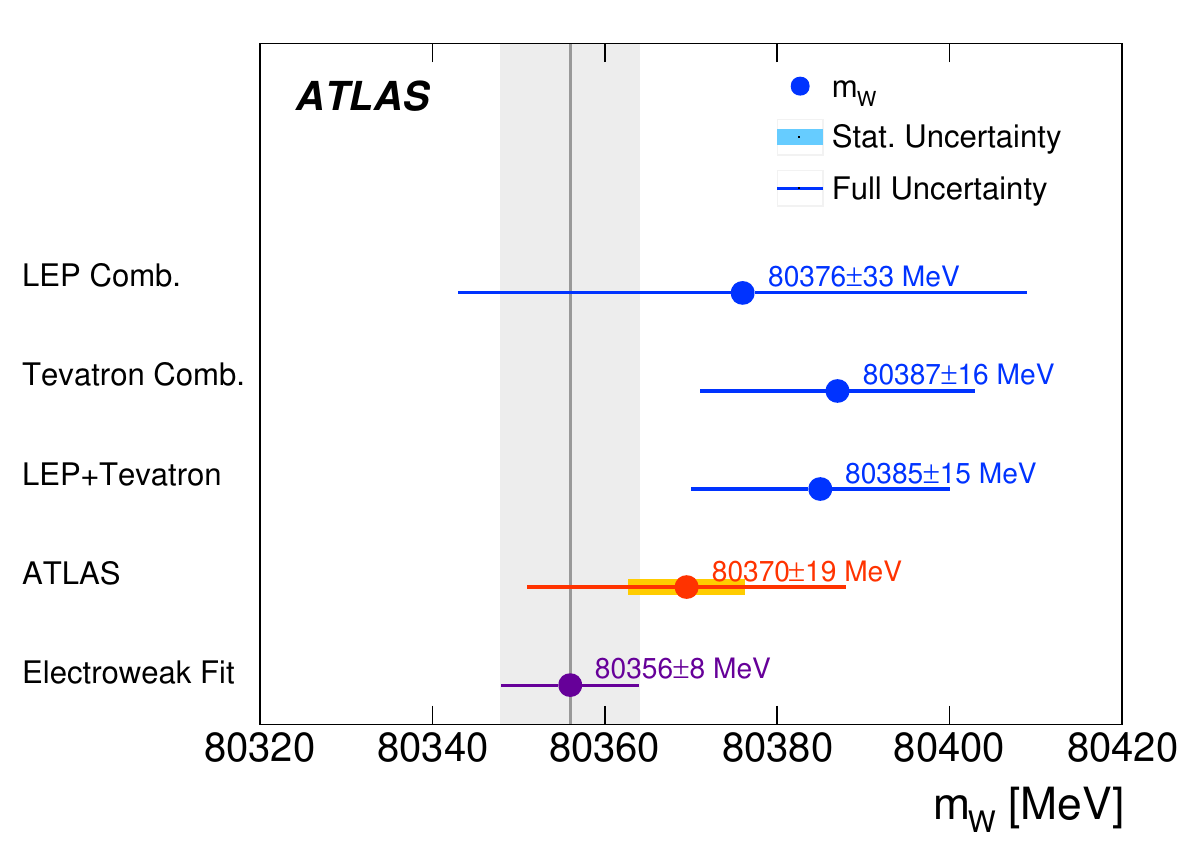}
\caption{The present measurement of $m_W$ is compared to the
  SM prediction from the global electroweak fit~\cite{Baak:2014ora} updated using recent 
  measurements of the top-quark and Higgs-boson masses, $m_t=172.84 \pm 0.70 \GeV$~\cite{Aaboud:2016igd} and $m_{H}=125.09 \pm 0.24 \GeV$~\cite{HIGG-2014-14}, and to the
  combined values of $m_W$ measured at LEP~\cite{Schael:2013ita} and
  at the Tevatron collider~\cite{Aaltonen:2013iut}. 
  \vspace{0.44 cm}\label{Fig:Gfitter1}}

\end{minipage}
\hspace{0.1cm}
\begin{minipage}[hbt]{0.49\textwidth}
	\centering
	\includegraphics[width=0.99\textwidth]{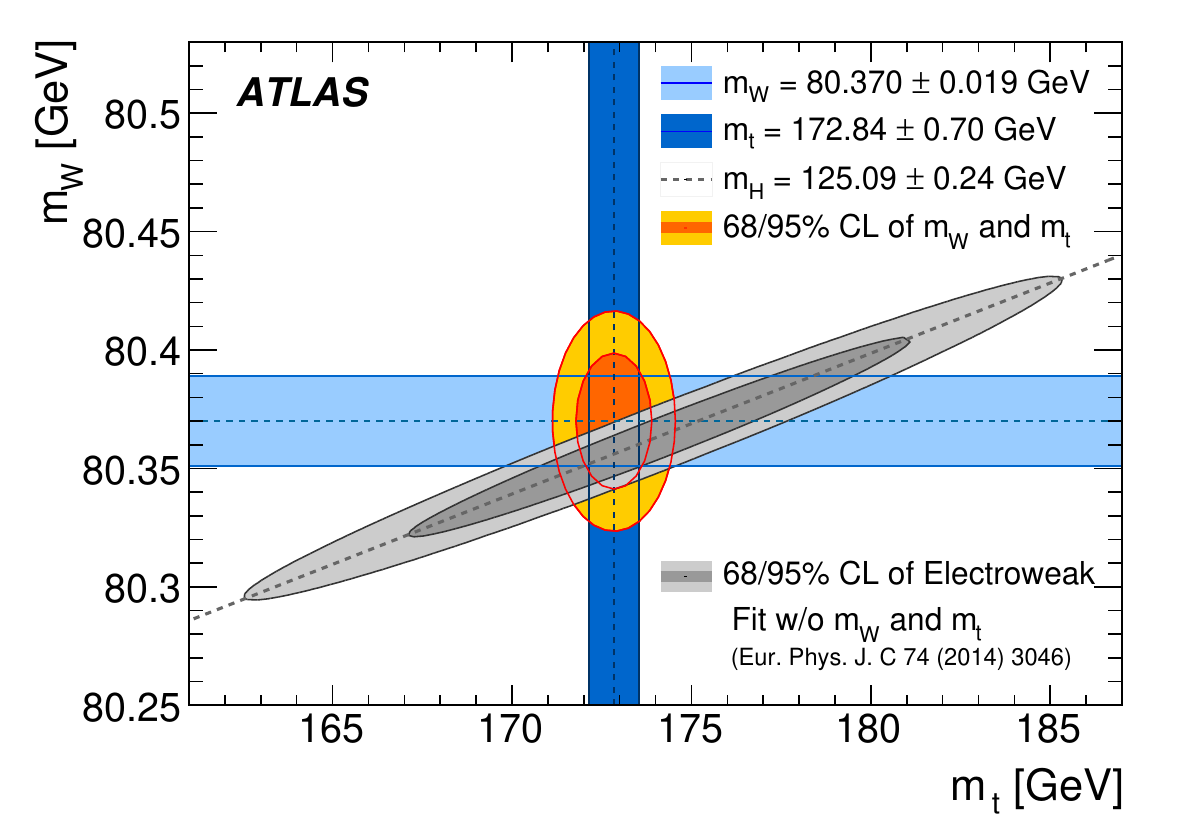}
\caption{The 68\% and 95\% confidence-level contours of the $m_W$ and
  $m_t$ indirect determination from the global electroweak
  fit~\cite{Baak:2014ora} are compared to the
  68\% and 95\% confidence-level contours of the ATLAS measurements of
  the top-quark and $W$-boson masses.
  The determination from the electroweak fit uses as input the LHC
  measurement of the Higgs-boson mass, $m_{H}=125.09 \pm 0.24 \GeV$~\cite{HIGG-2014-14}.\label{Fig:Gfitter2}}
\end{minipage}
\end{figure}

The $W$-boson mass measurement is compatible with the current world average of $m_W = 80385
\pm 15 \MeV$~\cite{Agashe:2014kda}, and similar in precision to the currently leading
measurements performed by the CDF and D0 collaborations~\cite{Aaltonen:2012bp,Abazov:2012bv}. 
An overview of the different $m_W$ measurements is shown in
Figure~\ref{Fig:WAFinal}. 
The compatibility of the measured value of $m_W$ in the context of
the global electroweak fit is illustrated in
Figures~\ref{Fig:Gfitter1} and~\ref{Fig:Gfitter2}.
Figure~\ref{Fig:Gfitter1} compares the present measurement with earlier results, and with the SM prediction updated with regard to Ref.~\cite{Baak:2014ora} using recent measurements of the top-quark and Higgs boson masses, $m_t=172.84 \pm 0.70\GeV$~\cite{Aaboud:2016igd} and $m_{H}=125.09 \pm 0.24\GeV$~\cite{HIGG-2014-14}. This update gives a numerical value for the SM prediction of $m_W=80356\pm 8\MeV$.  
The corresponding two-dimensional 68\% and 95\% confidence limits for $m_W$ and $m_t$ are shown in
Figure~\ref{Fig:Gfitter2}, and compared to the present measurement of $m_W$ and the average of the top-quark mass determinations performed by ATLAS~\cite{Aaboud:2016igd}.

The determination of the $W$-boson mass from the global fit of the
electroweak parameters has an uncertainty of $8\MeV$, which sets a
natural target for the precision of the experimental measurement of the
mass of the $W$ boson.
The modelling uncertainties, which currently dominate the overall uncertainty of the $m_W$ measurement presented in this paper, need to be reduced in order to fully exploit the larger data samples available at centre-of-mass energies of $8$ and $13\TeV$. Better knowledge of the PDFs, as achievable with the inclusion in PDF
fits of recent precise measurements of $W$- and $Z$-boson rapidity
cross sections with the ATLAS detector~\cite{WZ2011}, and improved
QCD and electroweak predictions for Drell--Yan production, are therefore crucial for future
measurements of the $W$-boson mass at the LHC.

\FloatBarrier

\section*{Acknowledgements}


We thank CERN for the very successful operation of the LHC, as well as the
support staff from our institutions without whom ATLAS could not be
operated efficiently.

We acknowledge the support of ANPCyT, Argentina; YerPhI, Armenia; ARC, Australia; BMWFW and FWF, Austria; ANAS, Azerbaijan; SSTC, Belarus; CNPq and FAPESP, Brazil; NSERC, NRC and CFI, Canada; CERN; CONICYT, Chile; CAS, MOST and NSFC, China; COLCIENCIAS, Colombia; MSMT CR, MPO CR and VSC CR, Czech Republic; DNRF and DNSRC, Denmark; IN2P3-CNRS, CEA-DSM/IRFU, France; SRNSF, Georgia; BMBF, HGF, and MPG, Germany; GSRT, Greece; RGC, Hong Kong SAR, China; ISF, I-CORE and Benoziyo Center, Israel; INFN, Italy; MEXT and JSPS, Japan; CNRST, Morocco; NWO, Netherlands; RCN, Norway; MNiSW and NCN, Poland; FCT, Portugal; MNE/IFA, Romania; MES of Russia and NRC KI, Russian Federation; JINR; MESTD, Serbia; MSSR, Slovakia; ARRS and MIZ\v{S}, Slovenia; DST/NRF, South Africa; MINECO, Spain; SRC and Wallenberg Foundation, Sweden; SERI, SNSF and Cantons of Bern and Geneva, Switzerland; MOST, Taiwan; TAEK, Turkey; STFC, United Kingdom; DOE and NSF, United States of America. In addition, individual groups and members have received support from BCKDF, the Canada Council, CANARIE, CRC, Compute Canada, FQRNT, and the Ontario Innovation Trust, Canada; EPLANET, ERC, ERDF, FP7, Horizon 2020 and Marie Sk{\l}odowska-Curie Actions, European Union; Investissements d'Avenir Labex and Idex, ANR, R{\'e}gion Auvergne and Fondation Partager le Savoir, France; DFG and AvH Foundation, Germany; Herakleitos, Thales and Aristeia programmes co-financed by EU-ESF and the Greek NSRF; BSF, GIF and Minerva, Israel; BRF, Norway; CERCA Programme Generalitat de Catalunya, Generalitat Valenciana, Spain; the Royal Society and Leverhulme Trust, United Kingdom.

The crucial computing support from all WLCG partners is acknowledged gratefully, in particular from CERN, the ATLAS Tier-1 facilities at TRIUMF (Canada), NDGF (Denmark, Norway, Sweden), CC-IN2P3 (France), KIT/GridKA (Germany), INFN-CNAF (Italy), NL-T1 (Netherlands), PIC (Spain), ASGC (Taiwan), RAL (UK) and BNL (USA), the Tier-2 facilities worldwide and large non-WLCG resource providers. Major contributors of computing resources are listed in Ref.~\cite{ATL-GEN-PUB-2016-002}.


\printbibliography

\clearpage
 
\begin{flushleft}
{\Large The ATLAS Collaboration}

\bigskip

M.~Aaboud$^\textrm{\scriptsize 34d}$,    
G.~Aad$^\textrm{\scriptsize 99}$,    
B.~Abbott$^\textrm{\scriptsize 125}$,    
J.~Abdallah$^\textrm{\scriptsize 75}$,    
O.~Abdinov$^\textrm{\scriptsize 13,*}$,    
B.~Abeloos$^\textrm{\scriptsize 129}$,    
S.H.~Abidi$^\textrm{\scriptsize 164}$,    
O.S.~AbouZeid$^\textrm{\scriptsize 143}$,    
N.L.~Abraham$^\textrm{\scriptsize 153}$,    
H.~Abramowicz$^\textrm{\scriptsize 158}$
% ...
\bigskip
\\

% ...
$^{13}$Institute of Physics, Azerbaijan Academy of Sciences, Baku; Azerbaijan.\\
% ...
$^{34}$$^{(a)}$Facult\'e des Sciences Ain Chock, R\'eseau Universitaire de Physique des Hautes Energies - Universit\'e Hassan II, Casablanca;$^{(b)}$Centre National de l'Energie des Sciences Techniques Nucleaires (CNESTEN), Rabat;$^{(c)}$Facult\'e des Sciences Semlalia, Universit\'e Cadi Ayyad, LPHEA-Marrakech;$^{(d)}$Facult\'e des Sciences, Universit\'e Mohamed Premier and LPTPM, Oujda;$^{(e)}$Facult\'e des sciences, Universit\'e Mohammed V, Rabat; Morocco.\\
% ...
$^{75}$University of Iowa, Iowa City IA; United States of America.\\
% ...
$^{99}$CPPM, Aix-Marseille Universit\'e, CNRS/IN2P3, Marseille; France.\\
% ...
$^{125}$Homer L. Dodge Department of Physics and Astronomy, University of Oklahoma, Norman OK; United States of America.\\
% ...
$^{129}$LAL, Universit\'e Paris-Sud, CNRS/IN2P3, Universit\'e Paris-Saclay, Orsay; France.\\
% ...
$^{143}$Santa Cruz Institute for Particle Physics, University of California Santa Cruz, Santa Cruz CA; United States of America.\\
% ...
$^{153}$Department of Physics and Astronomy, University of Sussex, Brighton; United Kingdom.\\
% ...
$^{158}$Raymond and Beverly Sackler School of Physics and Astronomy, Tel Aviv University, Tel Aviv; Israel.\\
% ...
$^{164}$Department of Physics, University of Toronto, Toronto ON; Canada.\\
% ...
$^{a}$ Also at  Department of Physics, University of Malaya, Kuala Lumpur; Malaysia.\\
$^{b}$ Also at Academia Sinica Grid Computing, Institute of Physics, Academia Sinica, Taipei; Taiwan.\\
$^{c}$ Also at Borough of Manhattan Community College, City University of New York, NY; United States of America.\\
$^{d}$ Also at Centre for High Performance Computing, CSIR Campus, Rosebank, Cape Town; South Africa.\\
$^{e}$ Also at CERN, Geneva; Switzerland.\\
% ...
$^{*}$ Deceased

\end{flushleft}


\end{document}